\PassOptionsToPackage{unicode}{hyperref}
\PassOptionsToPackage{hyphens}{url}
\PassOptionsToPackage{dvipsnames,svgnames,x11names}{xcolor}
\documentclass[
  12pt]{article}

\usepackage{amsmath,amssymb}
\usepackage{iftex}
\ifPDFTeX
  \usepackage[T1]{fontenc}
  \usepackage[utf8]{inputenc}
  \usepackage{textcomp} 
\else 
  \usepackage{unicode-math}
  \defaultfontfeatures{Scale=MatchLowercase}
  \defaultfontfeatures[\rmfamily]{Ligatures=TeX,Scale=1}
\fi
\usepackage{lmodern}
\ifPDFTeX\else  
\fi
\IfFileExists{upquote.sty}{\usepackage{upquote}}{}
\IfFileExists{microtype.sty}{
  \usepackage[]{microtype}
  \UseMicrotypeSet[protrusion]{basicmath} 
}{}
\makeatletter
\@ifundefined{KOMAClassName}{
  \IfFileExists{parskip.sty}{%
    \usepackage{parskip}
  }{
    \setlength{\parindent}{0pt}
    \setlength{\parskip}{6pt plus 2pt minus 1pt}}
}{
  \KOMAoptions{parskip=half}}
\makeatother
\usepackage{xcolor}
\makeatletter
\ifx\paragraph\undefined\else
  \let\oldparagraph\paragraph
  \renewcommand{\paragraph}{
    \@ifstar
      \xxxParagraphStar
      \xxxParagraphNoStar
  }
  \newcommand{\xxxParagraphStar}[1]{\oldparagraph*{#1}\mbox{}}
  \newcommand{\xxxParagraphNoStar}[1]{\oldparagraph{#1}\mbox{}}
\fi
\ifx\subparagraph\undefined\else
  \let\oldsubparagraph\subparagraph
  \renewcommand{\subparagraph}{
    \@ifstar
      \xxxSubParagraphStar
      \xxxSubParagraphNoStar
  }
  \newcommand{\xxxSubParagraphStar}[1]{\oldsubparagraph*{#1}\mbox{}}
  \newcommand{\xxxSubParagraphNoStar}[1]{\oldsubparagraph{#1}\mbox{}}
\fi
\makeatother

\usepackage{longtable,booktabs,array}
\usepackage{calc} 
\usepackage{etoolbox}
\makeatletter
\patchcmd\longtable{\par}{\if@noskipsec\mbox{}\fi\par}{}{}
\makeatother
\IfFileExists{footnotehyper.sty}{\usepackage{footnotehyper}}{\usepackage{footnote}}
\makesavenoteenv{longtable}
\usepackage{graphicx}
\makeatletter
\def\maxwidth{\ifdim\Gin@nat@width>\linewidth\linewidth\else\Gin@nat@width\fi}
\def\maxheight{\ifdim\Gin@nat@height>\textheight\textheight\else\Gin@nat@height\fi}
\makeatother
\setkeys{Gin}{width=\maxwidth,height=\maxheight,keepaspectratio}
\makeatletter
\def\fps@figure{htbp}
\makeatother

\makeatletter
\@ifpackageloaded{caption}{}{\usepackage{caption}}
\AtBeginDocument{%
\ifdefined\contentsname
  \renewcommand*\contentsname{Table of contents}
\else
  \newcommand\contentsname{Table of contents}
\fi
\ifdefined\listfigurename
  \renewcommand*\listfigurename{List of Figures}
\else
  \newcommand\listfigurename{List of Figures}
\fi
\ifdefined\listtablename
  \renewcommand*\listtablename{List of Tables}
\else
  \newcommand\listtablename{List of Tables}
\fi
\ifdefined\figurename
  \renewcommand*\figurename{Figure}
\else
  \newcommand\figurename{Figure}
\fi
\ifdefined\tablename
  \renewcommand*\tablename{Table}
\else
  \newcommand\tablename{Table}
\fi
}
\@ifpackageloaded{float}{}{\usepackage{float}}
\floatstyle{ruled}
\@ifundefined{c@chapter}{\newfloat{codelisting}{h}{lop}}{\newfloat{codelisting}{h}{lop}[chapter]}
\floatname{codelisting}{Listing}

\makeatother
\makeatletter
\@ifpackageloaded{caption}{}{\usepackage{caption}}
\@ifpackageloaded{subcaption}{}{\usepackage{subcaption}}
\makeatother

\usepackage[]{natbib}
\usepackage{bookmark}

\IfFileExists{xurl.sty}{\usepackage{xurl}}{} 
\hypersetup{
  pdftitle={Bayesian computation for high-dimensional Gaussian Graphical Models with spike-and-slab priors},
  pdfauthor={D\'eborah Sulem; Jack Jewson; David Rossell},
  pdfkeywords={3 to 6 keywords, that do not appear in the title},
  colorlinks=true,
  linkcolor={blue},
  filecolor={Maroon},
  citecolor={Blue},
  urlcolor={Blue},
  pdfcreator={LaTeX via pandoc}}

\newcommand{\anon}{1}

\usepackage{graphicx} 
\usepackage{url}            
\usepackage{booktabs}       
\usepackage{nicefrac}       
\usepackage{microtype}      
\usepackage{lipsum}
\usepackage{bbm}
\usepackage{bm}
\usepackage{float}
\usepackage{epstopdf}
\usepackage{parskip}
\usepackage{caption}
\usepackage{subcaption}
\usepackage{amsthm,amsmath,amsfonts,amssymb,dsfont,amstext,mathtools}
\usepackage{latexsym}
\usepackage{ifdraft}
\usepackage{xcolor}
\usepackage{verbatim}
\usepackage{enumitem}  
\usepackage{tabularx}
\usepackage{tablefootnote}
\usepackage{multirow}
\usepackage{makecell}
\usepackage{placeins} 
\usepackage[english]{babel}
\usepackage{mathrsfs}
\usepackage{soul}
\usepackage[linesnumbered,ruled,vlined]{algorithm2e}

\newcommand{\Y}{\boldsymbol{Y}}
\newcommand{\R}{\mathbb{R}}
\newcommand{\1}{\mathds{1}}

\newcommand{\ds}[1]{{\textcolor{black}{{{}}#1}}{}}
\newcommand{\deb}[1]{{\textcolor{black}{{{}}#1}}{}}
\newcommand\jack[1]{\textcolor{green}{#1}}

\newtheorem{thm}{Theorem}
\newtheorem{lemma}{Lemma}
\newtheorem{cor}{Corollary}
\newtheorem{prop}{Proposition}

\newcommand{\PSD}{\mathcal{M}_+(p)}
\newcommand{\PSDD}{\mathcal{M}_+(p,D)}

\newtheorem{remark}{Remark}
\newtheorem{assumption}{Assumption}

\begin{document}

\def\spacingset#1{\renewcommand{\baselinestretch}%
{#1}\small\normalsize} \spacingset{1}


\if1\anon
{
  \title{\bf Bayesian computation for high-dimensional Gaussian Graphical Models with spike-and-slab priors}
  \author{D\'eborah Sulem 
    \hspace{.2cm}\\
    Faculty of Informatics, Università della Svizzera Italiana\\
    and \\
    Jack Jewson \\
    Department of Econometrics and Business Statistics, Monash University\\
    and\\
    David Rossell \\
    Department of Business and Economics, Universitat Pompeu Fabra \\
    Data Science Center, Barcelona School of Economics}
  \maketitle
} \fi

\if0\anon
{
  \bigskip
  \bigskip
  \bigskip
  \begin{center}
    {\LARGE\bf Bayesian computation for high-dimensional Gaussian Graphical Models with spike-and-slab priors}
\end{center}
  \medskip
} \fi

\bigskip
\begin{abstract}
Gaussian graphical models are widely used to infer dependence structures. 
    Bayesian methods are appealing to quantify uncertainty  associated with structural learning, i.e., the plausibility of conditional independence statements given the data, and parameter estimates.
    However, computational demands have limited their application when the number of variables is large, which prompted the use of pseudo-Bayesian approaches.
    We propose fully Bayesian algorithms that provably scale to high dimensions when the data-generating precision matrix is sparse, at a similar cost to the best pseudo-Bayesian methods.
    First, a Metropolis-Hastings-within-Block-Gibbs algorithm that allows row-wise updates of the precision matrix, using local moves. 
    Second, a global proposal that enables adding or removing multiple edges within a row, which can help explore multi-modal posteriors.    
    We obtain 
    mixing bounds  for both samplers    relative to an ideal Gibbs sampler \color{black} that are dimension-free under suitable settings,    and prove that this ideal Gibbs sampler is geometrically ergodic. \color{black}
    We also provide worst-case polynomial bounds on per-iteration costs,
    though in practice sparse linear algebra lowers these.
    Our examples show that the methods extend the applicability of exact Bayesian inference from roughly 100 to 1,000 variables (equivalently, from 5,000 to 500,000 potential edges).
\end{abstract}
\noindent%
\vfill

\newpage
\spacingset{1.8} 

\section{Introduction}\label{sec-intro}

An undirected graphical model describes conditional dependencies in a $p$-dimensional random vector.
We focus on the Gaussian graphical model (GGM) case with $p \times p$ precision (inverse covariance) matrix $\Omega$,
where $\Omega_{ij}= 0$ if and only if variables $(i,j)$ are conditionally independent given all other variables. 
Learning the graphical model corresponds to inferring which entries of $\Omega$ are non-zero. 
In many applications
$p$ is large,    therefore the number of 
potential graphical models $2^{p(p-1)/2}$ is also large. \color{black}
It is then common to enforce sparsity to improve the accuracy of inference and the interpretability of the estimated graphical model. 


\begin{figure}
\begin{center}
\includegraphics[width =0.325\linewidth]{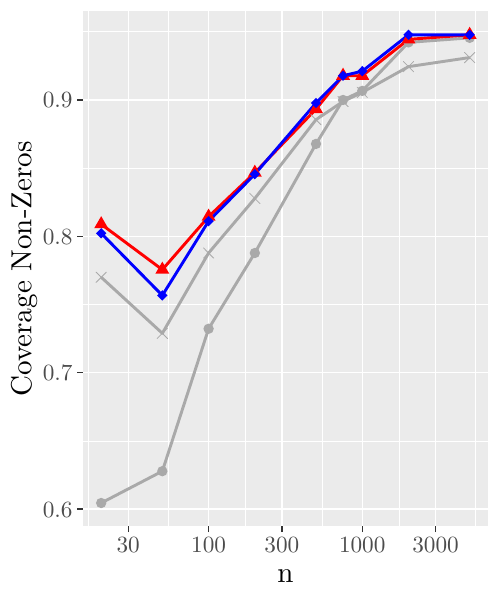}
\includegraphics[width =0.325\linewidth]{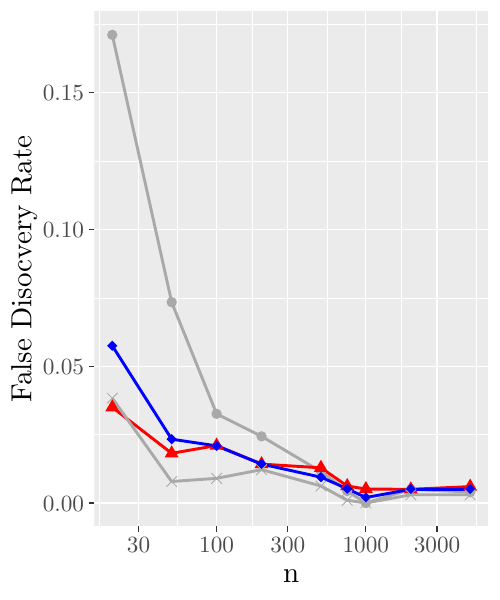}
\includegraphics[width =0.325\linewidth]{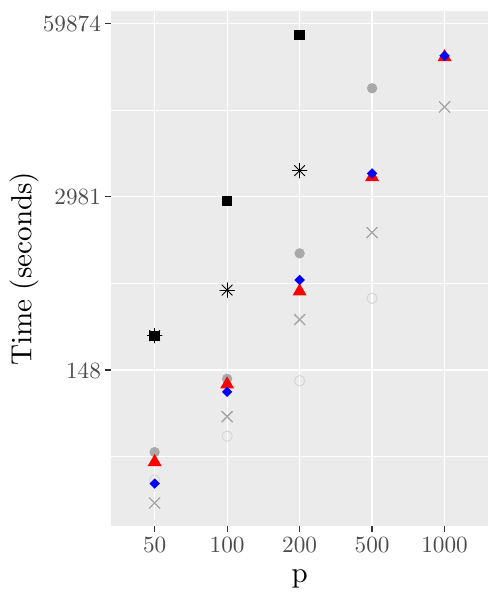}\\
\includegraphics[width =\linewidth, trim=0.cm 0.cm 0cm  9cm,clip]{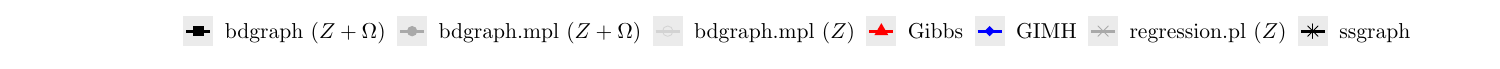}
\caption{Coverage of 95\% credibility intervals for truly non-zero precision matrix entries (left) and false discovery rate (center) 
of our algorithms (Gibbs in blue, GIMH in red) and    pseudo-likelihood methods  bdgraph.mpl 
and regression.pl 
(gray) \color{black} for a tri-diagonal $\Omega$, $p = 10$ and growing $n$.
Right: time to produce 4,000 posterior samples as $p$ grows and $n=2p$, also including    exact Bayesian  methods ssgraph 
and bdgraph}
\label{fig:exact_vs_pseudo}
\end{center}
\end{figure}

Popular likelihood penalties
to estimate $\Omega$ in high dimensions include
LASSO \citep{meinshausen2006high} and
graphical LASSO (GLASSO) \citep{friedman2008sparse,yuan2007model}, 
among others.
These methods use efficient optimisation to obtain point estimates. 
In contrast, we focus on fully Bayesian methods 
that provide a posterior distribution that quantifies uncertainty on the GGM's structure and on $\Omega$ given the observed data.
Capturing this uncertainty is critical for Bayesian inference. 
Alas, the computational cost of posterior sampling has previously proven prohibitive in examples    where $p$ 
is \color{black} more than a few hundreds, posing a major bottleneck. 
An interesting alternative
is to 
use pseudo-likelihoods to simplify calculations \citep{atchade2019quasi, mohammadi2025scalable}. 
In our examples these attained excellent trade-offs between inference and computation, but
their output differs from an exact Bayesian posterior.
Figure \ref{fig:exact_vs_pseudo} shows a simple $p=10$ example, where pseudo-likelihood methods (gray lines) have lower frequentist coverage, particularly for moderate $n$.
The right panel shows that the run times of our algorithms (blue and red) are comparable to those of pseudo-likelihoods, and much faster than the best available exact Bayesian algorithms (black points). 
  
Figure \ref{fig:exact_vs_pseudo2} additionally investigates the power and calibration of the exact and pseudo-Bayesian methods.  
\color{black}
In our real data example with $p = 332$, we further observed better mixing and out-of-sample predictive performance of our exact Bayesian algorithms compared with pseudo-likelihood approaches. 
See Sections \ref{sec:experiments} and \ref{sec:COVID} for a full description. 

We discuss selected literature on fully Bayesian methods.
The first approach relied on the G-Wishart distribution \citep{roverato2002hyper}. 
A former challenge was that    the G-Wishart normalising constants associated to each graphical structure are unavailable, \color{black}
but there are now efficient approximations to their ratios between models that differ by one edge \citep{mohammadi2023accelerating, van2022g}. Model exploration 
is then restricted to moving between such neighbors. 
A popular alternative is to consider
priors with density
\begin{align}\label{eq:factorisable-prior}
    \pi(\Omega) \propto \prod_{i \leq j} \pi_{ij}(\Omega_{ij}) \mathds{1}_{\Omega \in \mathcal{M}_+(p)}
\end{align}
for some 
univariate densities $(\pi_{ij})_{i\leq j}$, 
where $\mathds{1}$ is an indicator and $\mathcal{M}_+(p)$ the space of $p \times p$ symmetric positive-definite matrices.
Examples include continuous shrinkage priors such as the
continuous spike-and-slab   \citep{Wang2015ScalingIU, banerjee2015bayesian, gan2018},
Laplace \citep{wang2012bayesian} 
and horseshoe priors \citep{li2019graphical}.
An argument for these priors over the G-Wishart is that at each Markov Chain Monte Carlo (MCMC) iteration one can update the $p-1$ off-diagonal entries in a row of $\Omega$ \citep{Wang2015ScalingIU}. 
Despite these notable advances, MCMC for continuous shrinkage priors 
faces major limitations.
Each iteration requires sampling all parameters, e.g., $p-1$ if a row of $\Omega$ is updated, 
even if many are effectively zero.
This has a significant computational and storage cost.
Further, there are no guarantees on mixing times, and we show examples where previous algorithms mix poorly.

We instead consider a discrete spike-and-slab prior where entries in $\Omega$ have positive probability of being exactly zero.
Computations then scale better in sparse settings, intuitively one only samples the non-zero coefficients, which are much less than $p-1$.
More formally, 
\cite{yang2016computational, zhou2022dimension} proved fast MCMC mixing in sparse linear regression 
for discrete spike-and-slab priors.
They considered settings where the marginal likelihood of each model 
is available.
   For us, the model is the conditional independence structure whose marginal likelihood is not available, \color{black}
hence their results do not directly apply.
We first show that conditional marginal likelihoods are available for updating row $j$ of $\Omega$, 
given the current value of $\Omega_{-j,-j}$    (the submatrix resulting from removing the $j$-th row and column of $\Omega$). \color{black} Hence, row-wise updates are computationally tractable.
Even then, a challenge is that \cite{yang2016computational, zhou2022dimension} rely on the  posterior concentrating on a certain model (typically, the data-generating model), whereas here we study  conditional posteriors, and these concentrate on a model 
that depends on $\Omega_{-j,-j}$.
A further reason for discrete over continuous shrinkage priors is that the latter place posterior probability one to all entries in $\Omega$ being non-zero, and hence they do not directly quantify uncertainty on the graphical structure.
The continuous spike-and-slab does provide approximate posterior probabilities on models, but as we discuss later 
the posterior probability of the data-generating graph does not converge to 1 as the sample size $n \rightarrow \infty$. 

Our main contributions are developing theory and algorithms for the discrete spike-and-slab prior in the context of GGMs:
\begin{itemize}
\item a column-wise Metropolis-within-Block-Gibbs 
algorithm;
    within each column $\Omega_{\cdot j}$, one can    update edges using \color{black} 
    closed-form expressions and standard local proposals, such as Gibbs, birth-death-swap and locally-informed proposals \citep{zhou2022dimension};

    \item    a global proposal that \color{black} can   
    update multiple edges in $\Omega_{\cdot j}$; 
       the proposal uses a linear regression posterior that \color{black}
    concentrates on the same models as the GGM, provided that the conditioned upon $\Omega_{-j,-j}$ is close to the data-generating $\Omega_{-j,-j}^0$; 
    
    

\item we prove fast mixing (dimension-free) for both proposals, under sparsity assumptions,    relative to an (computationally infeasible) ideal Gibbs sampler that draws $\Omega_{\cdot j}$ exactly from its conditional. We also prove that the ideal Gibbs is geometrically ergodic. \color{black}

\item We show empirically that our algorithms enable exact Bayesian inference at a computational efficiency (accuracy vs. clock time)
that is comparable to pseudo-likelihood methods, 
and an order-of-magnitude better than the fastest exact Bayesian algorithms.
    
\end{itemize}

In Section \ref{sec:spike-and-slab}, we discuss continuous and discrete spike-and-slab priors and the properties of the associated posteriors. In Section \ref{sec:posterior-sampling} we present our local and global MCMC algorithms and analyse their mixing times in Section \ref{sec:algorithmic_guarantees}.
In Sections \ref{sec:experiments} and \ref{sec:COVID}, we show numerical comparisons with state-of-the-art algorithms on simulated and real data.
Section \ref{sec:discussion}    discusses our empirical findings and the relative advantages of our algorithms, and then concludes. \color{black}


\section{Spike-and-slab priors}\label{sec:spike-and-slab}

Let $y_i \sim \text{N}(\mathbf{0}_p, [\Omega^{0}]^{-1})$ independently across $i=1,\ldots,n$,
where $\Omega^0 \in \PSD$ is the data-generating precision matrix. 
We denote by
$\mathbb P$ the data distribution under $\Omega^0$,
and by $Z^0$ the    $p \times p$ binary matrix \color{black}
where $Z^{0}_{ij}= \mathds{1}_{\Omega^0_{ij} \neq 0}$.
   We refer to $Z^0$ as the graphical model associated to the data-generating $\Omega^0$. \color{black}
Let $\Y^T= [y_1,\dots, y_n] \in \R^{p\times n}$ and $S := \Y^T \Y \in \R^{p\times p}$. 

{\bf Notation.} 
Let $|A|$ be the determinant of a matrix $A$, 
$\lambda_1(A)$ and $\lambda_p(A)$ its smallest and largest eigenvalues,
and $\| A \|= \lambda_p(A)$ its spectral norm.
We denote by $A_{i \cdot}$
its row $i$, by $A_{\cdot j}$ its column $j$, and by $A_{-j, -j}$ the sub-matrix without row and column $j$.  
For $p \in \mathbb N \backslash \{0\}$, let $[p] := \{1, \dots, p\}$. 
Given two subsets $A,B \subset [p]$, we denote by $M_{AB}$  the submatrix selecting the rows and columns in $A$ and $B$  in the matrix $M \in \R^{p \times p}$. We also use the notation $M_{z_A z_B}$ for $z_A,z_B\in \{0,1\}^p$ indicator vectors of rows included in $A$ and columns in $B$. We adopt a similar notation for selecting the entries of vector $v$, and we denote
by $|v|_0 := \sum_{i=1}^p \mathds{1}_{v_i \neq 0}$ its $\ell_0$-norm.
We also denote 
by $\Sigma_{-j \mid j}= [\Omega_{-j,-j}]^{-1}= \mbox{Cov}(y_{i,-j} \mid y_{i,j})$    the 
conditional covariance of $y_{i,-j} = (y_{i,1}, \dots,y_{i,j-1}, y_{i,j+1},\dots, y_{i,p})$ given $y_{i,j}$, \color{black} and by $\Sigma_{zz \mid j} = [\Sigma_{-j \mid j}]_{zz}$ that for the sub-vector $y_{i,z}$.
We let $a \vee b = \max\{a,b\}$, $a \wedge b = \min\{ a,b \}$ 
and, for two sequences $u_n, v_n>0$, we write $u_n= O(v_n)$ 
if $u_n/v_n < C$ for some constant $C$ and all $n$ large enough.

\subsection{Prior formulation}
\label{ssec:prior-formulation}

\cite{Wang2015ScalingIU} considered a prior as in \eqref{eq:factorisable-prior} where
$\pi_{ii}(\Omega_{ii}) = \text{Exp}\left(\Omega_{ii}; \lambda/2 \right)$
   is an Exponential distribution with rate $\lambda/2 > 0$, \color{black} and
\begin{align}\label{eq:prior-cont-offdiag}
    &\pi_{ij}(\Omega_{ij}) = \theta \text{N}(\Omega_{ij}; \: 0, g_1^2) + (1 - \theta) \text{N}(\Omega_{ij}; \: 0, g_0^2), & i <j,
\end{align}
is a spike-and-slab prior \citep{george1993variable}, 
$g_0^2>0$ is the spike's variance, $g_1^2 > g_0^2$ the slab's variance, and $\theta \in [0,1]$ the slab probability.
\cite{Wang2015ScalingIU} designed a block Gibbs sampler that represented a major advance: it is simple to implement and it updates a whole row of $\Omega$ in one iteration. There are however two important limitations. 
First, even if the data-generating $\Omega^0$ is very sparse,
one must sample from a $(p-1)$-dimensional Gaussian. 
Second,  the update of $\Omega$ uses a model augmentation with 
spike-and-slab indicators 
which leads to poor mixing if the spike's variance $g_0^2$ is small. 
\cite{Wang2015ScalingIU} reports that, when $g_0 < 0.01$, the sampler mixes poorly (as occurs in regression \cite{george1993variable}).

Alas,    for uncertainty quantification on the graphical structure \color{black} it is desirable to set $g_0=0$, i.e., a discrete spike-and-slab where the spike is a Dirac measure at 0, denoted by $\delta_0(\cdot)$.
We next outline such a prior, 
   and defer further discussion to Section \ref{ssec:prior-features}. \color{black}
For $\Omega \in \PSD$, let $Z$ be the $p \times p$ the associated symmetric binary matrix with $Z_{ij} = \mathds{1}_{\Omega_{ij} \neq 0}$ for $i \leq j$,
and $d(Z)= \max_{i=1, \dots, p} |Z_{.i}|_0$ be the largest edge degree. 
We consider the prior 
  
\begin{align}\label{eq:prior-omega}
\pi(\Omega) \propto 
\prod_{i=1}^p \text{Exp}\left(\Omega_{ii}; \frac{\lambda}{2}\right) \prod_{i<j} \left\{\theta \text{N}(\Omega_{ij}; \: 0, g_1^2)  + (1-\theta) \delta_0(\Omega_{ij}) \right\}\mathds{1}_{\Omega \in \PSD}  \mathds{1}_{d(Z) \leq \bar d},
\end{align}
\color{black} where $\theta \in [0,1]$ and $\bar d \leq p-1$ is an upper-bound on the maximal degree.    Equivalently, $\pi(\Omega)$ is the marginal distribution of $\pi(\Omega, Z) = \pi(\Omega \mid Z) \pi(Z)$, where
\color{black}
\begin{align} \label{eq:prior-z}
    &\pi(\Omega \mid Z) = C_1(Z, g_1, \lambda) \left \{ \prod_{i=1}^p \text{Exp}\left(\Omega_{ii}; \frac{\lambda}{2}\right) \prod_{i<j, Z_{ij = 1}} \text{N}(\Omega_{ij}; \: 0, g_1^2)  \prod_{i<j, Z_{ij = 0}} \delta_0(\Omega_{ij}) \right \} \mathds{1}_{\Omega \in \PSD} \\
    &\pi(Z) \propto \frac{1}{C_1(Z, g_1, \lambda)}   
    \prod_{i<j} \theta^{Z_{ij}} (1 - \theta)^{1 - Z_{ij}}  
    \mathds{1}_{d(Z) \leq \bar d}, \nonumber
\end{align}
and
$C_1(Z,  g_1, \lambda)$ is the normalising constant of $\pi(\Omega \mid Z)$.
As in \cite{Wang2015ScalingIU},
$C_1$    does not feature in $\pi(\Omega)$ and hence neither in the posterior distribution $\pi(\Omega \mid \Y)$.
This is not only computationally convenient, but also facilitates studying the uncertainty quantification and posterior concentration of $\pi(\Omega \mid \Y)$.
However, $C_1$ featuring in the prior of the graphical model structure $\pi(Z)$, makes the prior beliefs on $Z$ hard to interpret. 
In fact, \cite{carter:2026} showed that $C_1$ decreases as one adds edges to $Z$, hence \eqref{eq:prior-z} assigns higher prior probability to sparser models than an alternative prior $\tilde{\pi}(Z) \propto
    \prod_{i<j} \theta^{Z_{ij}} (1 - \theta)^{1 - Z_{ij}}  
    \mathds{1}_{d(Z) \leq \bar d}$.
These authors also showed that to keep $C_1$ close to 1 
it suffices to set the off-diagonal mean $2/\lambda$ to a large enough value, and provided lower bounds on $C_1$. 
In particular, if one takes $\lambda=0$ (a flat improper prior $\Omega_{ii}$) then $C_1=1$.
In our examples we took $2/\lambda= 100$ to ensure a proper prior proper, but our results remained very similar for smaller $\lambda$ (see Section \ref{sec:pdtrunc_prior}). 
\color{black}

Additionally, $\pi(Z)$ constrains the maximal degree $d(Z)$ below some pre-specified $\bar d$. 
Such sparsity constraints are common in linear regression  \citep{yang2016computational,zhou2022dimension}
to ensure computational scalability and
that the posterior on the models $\pi(Z \mid \Y)$ concentrates on the data-generating truth.
We shall see in Sections \ref{sec:posterior-sampling} and \ref{sec:algorithmic_guarantees} that $\bar d$ features in our per-iteration cost bounds, MCMC mixing and posterior concentration results.
We remark that we only use $\bar d$ for our theory; in our examples we set the largest possible $\bar d = p-1$.

The prior slab probability $\theta$ can be any user-specified value,
but in our theory we set $\theta=1/p^\alpha$ for sufficiently large $\alpha > 0$ to guarantee posterior concentration and good mixing (see Section \ref{sec:algorithmic_guarantees}).
However in our examples we set $\theta = 2/(p-1)$, and this simple choice gave better inference than larger $\alpha$ in most cases.
   We refer to $\pi(Z)$ in \eqref{eq:prior-z} as the Binomial prior. Given that users may be uncomfortable with choosing $\theta$, we also consider an alternative where one sets a hyper-prior $\theta \sim \mbox{Beta}(a_\theta, b_\theta)$, which results in the marginal prior $\pi(Z) \propto$
\begin{align} \label{eq:prior-z-bbin}
    &\frac{\int_0^1 \prod_{i<j} \theta^{Z_{ij}} (1 - \theta)^{1 - Z_{ij}} \mbox{Beta}(\theta; a_\theta, b_\theta) d\theta}{C_1(Z, g_1, \lambda)}  
   \propto \frac{\Gamma(a_\theta + |Z|_0) \Gamma(b_\theta + p(p-1)/2 - |Z|_0)}{C_1(Z, g_1, \lambda)} 
\end{align}
if $d(Z) \leq \bar d$, and $\pi(Z)=0$ otherwise.
This sets a Beta-Binomial$(a_\theta,b_\theta)$ prior on the number of edges $|Z|_0$, and assigns equal prior probability to all models with the same $|Z|_0$. In our examples, the simple default $a_\theta=b_\theta=1$ resulted in good inference and computation.

A third option is estimating $\theta$ via empirical Bayes. Briefly, in our software we implemented a stochastic EM algorithm, using Theorem 3 and Proposition 1 in \cite{rognon:2026} to express
the log-posterior's gradient $\nabla_\theta \log \pi(\theta \mid \Y)$ as a sum over $p(p-1)/2$ edges (instead of $2^{p(p-1)/2}$ graphical structures). See Section \ref{sec:ebayes} for details.

Regarding the slab variance $g_1$ in \eqref{eq:prior-z}, by default we recommend $g_1=1$, under the common assumption that the columns in $\Y$ are standardised to have unit sample variances. 
This default is in analogy to Zellner's unit information prior, a classical choice that defines a minimally informative prior and has a well-understood behavior for Bayesian model selection. In fact, our theory shows that for $g_1=1$, Bayes factors associated to $\pi(Z_{\cdot j} \mid \Y, \Omega_{-j,-j})$ behave analogously to those for Zellner's prior in linear regression. 
Further, $g_1=1$ leads to a prior that is as minimally informative as Zellner's. Briefly, when regressing $\Y_{\cdot j}$ on $\Y_{\cdot Z_{-j, j}}$ Zellner's prior has marginal precision $\Y_{\cdot Z_{-j,j}}^T \Y_{\cdot Z_{-j, j}}/n$. Taking $\mbox{tr}(\Y_{\cdot Z_{-j, j}}^T \Y_{\cdot Z_{-j, j}}/n)= |Z_{-j, j}|_0$ as a measure of prior informativeness, which matches the precision of our prior $\Omega_{. Z_{-j,j}} \mid Z_{-j,j} \sim \text{N}(\mathbf{0}, I)$ when $g_1=1$. In our examples, sensitivity analyses found that inference was robust as long as $g_1$ is not too far from the default, say within the range $[0.2, 5]$. For completeness, in our theory we keep $g_1$ generic, and we allow $g_1$ to depend on $(p,n)$. \color{black}


\color{black}

\subsection{Features of discrete spike-and-slab priors}
\label{ssec:prior-features}

Discrete spike-and-slab priors have appealing theoretical and computational properties.
\cite{banerjee2015bayesian} proved optimal minimax posterior contraction rates in 
high-dimensional GGMs for a discrete prior using Laplace slabs.
More critically, a minimal requirement for Bayesian uncertainty quantification is that
$\pi(Z \mid\Y)$ concentrates on 
$Z^0$ as $n \to \infty$. 
However, for any fixed $g_0>0$, the posterior
implied by the continuous spike-and-slab \eqref{eq:prior-cont-offdiag} is provably not consistent, 
even in 
regression with fixed $p$ \citep{narisetty2014bayesian}.
Intuitively, for $g_0>0$, small truly non-zero coefficients $\Omega_{ij}^0 \neq 0$ are asymptotically assigned to the spike (
see Figure \ref{fig:comparison-wang-inf-1} for an illustrative example).
The figure 
also shows that if one decreases $g_0$ from 0.05 to $10^{-6}$, MCMC mixing deteriorates significantly (see Section \ref{app:simulation-figure}). 

In contrast, Proposition \ref{prop:posterior-consistency} shows that the posterior distribution under \eqref{eq:prior-z} consistently recovers the data-generating $Z^0$, when $p$ is fixed and $n \to \infty$. 
   The message is that continuous spike-and-slab priors are not only less  computationally-convenient than their discrete counterparts when $\Omega^0$ is sparse, but that they also fail in basic structural inference properties,  even for fixed $p$. 
Extensions to high-dimensional $p$ should be possible but they are beyond our scope, as they are not central to our computational arguments.
\color{black}

\begin{prop}[Posterior consistency]\label{prop:posterior-consistency}
Suppose that $p$ is fixed, there is $\epsilon_0 > 0$ such that $\Omega^0 \in \mathcal{M}_p(\epsilon_0) 
:= \{ \Omega \in \mathcal{M}_+(p) ;  \:  0 < \epsilon_0 \leq \lambda_1(\Omega )  \leq \lambda_p(\Omega ) \leq \epsilon_0^{-1}  \}$
(i.e., the eigenvalues of  $\Omega^0 $ are bounded) and $d(Z^0) \leq \bar d$. Then, under our prior 
  \eqref{eq:prior-omega}
   with $g_1$ satisfying $1/g_1^2 = o(n)$ and $|\log g_1| = o(n)$ \color{black}
and either 
fixed $\theta > 0$    or with $\pi(Z)$ in \eqref{eq:prior-z-bbin} with fixed $(a_\theta, b_\theta)$, \color{black} then
$\pi(Z^0 \mid \Y ) \xrightarrow[n \to \infty]{\mathbb P_0} 1$.
\end{prop}

Proposition \ref{prop:wang_exactzeroes_notation} gives the conditional posterior under \eqref{eq:prior-z} 
for sampling 
$(Z_{-j,j},\Omega_{\cdot j})$ given $\Omega_{-j,-j}$ and $\Y$.
Recall that for $z \in \{0,1\}^{p-1}$, $\Sigma_{zz \mid j}= [\Omega_{-j,-j}^{-1}]_{zz}= \mbox{Cov}(y_{i,z} \mid y_{i,j})$ and $S = \Y^T \Y$.

\begin{prop} \label{prop:wang_exactzeroes_notation}
Let $z= Z_{-j, j} \in \{0,1\}^{p-1}$, 
$u_1 =  - \Omega_{z j}$ and 
$u_2 = \Omega_{jj} - \Omega_{z j}^T \Sigma_{zz \mid j} \Omega_{z j}$.
Then
\begin{align}
\pi(u_1, u_2, z \mid \Y,\Omega_{-j,-j}) &\propto \text{N}(u_1; m_{z}, U_{z}
^{-1}) \text{Ga}\left(u_2; \frac{n}{2}+1, \frac{S_{jj} + \lambda}{2}\right) \frac{e^{\frac{m_{z}^T U_{z} m_{z}}{2}}}{g_1^{|z|_0}|U_{z}|^{\frac{1}{2}}}
h(z)
\nonumber,
\end{align}
if $|z|_0 \leq \bar d$, and 0 otherwise, where $U_{z} :=(S_{jj}+\lambda) \Sigma_{zz \mid j} + g_1^{-2} I_{z}$, $m_z :=U_{z}^{-1} S_{zj}$,    $\mbox{Ga}(a,b)$ denotes a Gamma distribution with shape $a$ and rate $b$,
$$
h(z)= 
\begin{cases}
\theta^{|z|_0} (1 - \theta)^{p-1-|z|_0} \mbox{, for the Binomial prior in \eqref{eq:prior-z}}
\\
    \Gamma(a_\theta + |Z|_0) \Gamma(b_\theta + p(p-1)/2 - |Z|_0)
    \mbox{, for the Beta-Binomial}(a_\theta,b_\theta) \mbox{ in \eqref{eq:prior-z-bbin}}
    \end{cases}   
$$
  
and $|Z|_0$ is the total number of edges across all columns in $(z,\Omega_{-j,-j})$ and $|z|_0$ that in column $j$. \color{black}
Above, if $z=0$ then $u_1$ has length 0 and $u_2=\Omega_{jj}$.
\end{prop}

Proposition \ref{prop:wang_exactzeroes_notation} implies that edge inclusion indicators can be sampled from
\begin{align}\label{eq:post_submodel}
&\pi(z \mid \Y, \Omega_{-j,-j}) \propto
\frac{e^{\frac{m_{z}^T U_z m_z}{2}}}{g_1^{|z|_0}|U_z|^{\frac{1}{2}}}
h(z) \mathds 1_{|z|_0 \leq \bar d},
\end{align}
the non-zero entries $\Omega_{z j}$ from a $|z|_0$-dimensional Gaussian, and $\Omega_{jj}$ is a simple function of a Gamma draw and $\Omega_{zj}$. 
Proposition \ref{prop:wang_exactzeroes_notation} 
is analogous to Proposition 1 in \cite{Wang2015ScalingIU} (reproduced as Proposition \ref{prop:wang}) for the continuous spike-and-slab.
The main difference is that here the dimension of $\Omega_{zj}$, and hence of $U_{z}$, is much smaller than $\mbox{dim}(\Omega_{\cdot j})=p-1$ in sparse regimes, providing significant savings    (see Section \ref{ssec:iteration_cost}). \color{black}
As in \cite{Wang2015ScalingIU}, 
$\Omega$ constructed from $(\Omega_{-j,-j}, u_1,u_2, z)$ 
sampled as in Proposition \ref{prop:wang_exactzeroes_notation} is guaranteed to be positive-definite.

\section{Posterior sampling}\label{sec:posterior-sampling}

\begin{algorithm}
\caption{Ideal random scan Gibbs sampler}\label{alg:gibbs_GGM_zeros}
\KwIn{ $\Y$, prior parameters ($\theta,g_1,\lambda)$, number of iterations $T$, initial value $\Omega^{(0)}$.}
\KwOut{$\{(\Omega^{(t)}, Z^{(t)})\}_{t \in [T]}$.} 
\For{$t=1, \ldots, T$}{
Let $\Omega= \Omega^{(t-1)}$ and choose $j \in [p]$ uniformly at random.

Denote $z= Z_{\cdot j}$. 
Sample $z \sim \pi(z \mid \Y, \Omega_{-j,-j})$. \\
Sample $u_1 \sim \pi(u_1\mid z, \Y, \Omega_{-j,-j}) = \text{N}(u_1; \: m_{z}, U_{z}^{-1})$, for $(m_z,U_z)$ in Proposition \ref{prop:wang_exactzeroes_notation}. \\
Sample $u_2 \sim \pi(u_2 \mid \Y) = \text{Ga}(u_2; \frac{n}{2}+1, \frac{S_{jj} + \lambda}{2})$. \\
Update $\Omega_{\cdot j}$ setting 
$\Omega_{z,j} = \Omega_{j,z} = -u_1$, 
other entries in $\Omega_{-j,j}$ to 0, 
and $\Omega_{jj} = u_2 + u_1^T \Sigma_{zz \mid j} u_1$.
Set $\Omega^{(t)} = \Omega$.
}
\end{algorithm}


   We propose algorithms to sample from $\pi(\Omega \mid \Y)$, specifically \color{black}
block-Gibbs algorithms that sample columns $\Omega_{.j}$ given $\Omega_{-j,-j}$ and $\Y$.
   Each iteration $t$ returns a $p \times p$ matrix $\Omega^{(t)}$, and the corresponding edge inclusion indicators $Z^{(t)}$. \color{black}
We first present Algorithm \ref{alg:gibbs_GGM_zeros}, an ideal random scan Gibbs sampler that    at each iteration chooses a column $j$ randomly and then \color{black}
updates the edges $z=Z_{.j}$ from their conditional posterior $\pi(z \mid \Y,\Omega_{-j,-j})$ in \eqref{eq:post_submodel},
and then the non-zero entries from their exact posterior $\pi(\Omega_{.j} \mid z, \Y, \Omega_{-j,-j})$ in Proposition \ref{prop:wang_exactzeroes_notation}.
We call it ideal because for large $p$ direct sampling from \eqref{eq:post_submodel} is unfeasible, as one cannot enumerate all $2^{p-1}$ models.
   In practice, our algorithms use Metropolis-within-Block-Gibbs (MH-within-Gibbs): for each column $j$, we sample from \eqref{eq:post_submodel} by performing $M \geq 1$ Metropolis-Hastings updates. \color{black} 
The ideal sampler serves as a benchmark for our theory in Section \ref{sec:algorithmic_guarantees}.
   Therein we analyse random scan algorithms where at each iteration a column $j$ is randomly chosen as in Algorithm \ref{alg:gibbs_GGM_zeros}, as these are simpler to analyse, 
but our software actually uses random sequence Gibbs where at each iteration columns are updated in a randomly permuted order.  \color{black}

Section \ref{sec:serial-gibbs} discusses using Metropolis-Hastings with local moves, and Section \ref{sec:par-gibbs} considers global moves.
   Finally, Section \ref{ssec:iteration_cost} discusses the per-iteration cost of our algorithms. 
We also discuss defaults for $M$, and that
taking a large $M>1$ to improve mixing (as $M \to \infty$, MH-within-Gibbs becomes equivalent to ideal Gibbs) requires a negligible number of operations. \color{black}

 
\subsection{Local moves}\label{sec:serial-gibbs}


To sample from $\pi(z \mid \Y, \Omega_{-j,-j})$ in \eqref{eq:post_submodel},
we obtain $M \geq 1$ samples $z^{(1)},\ldots,z^{(M)}$ using an MCMC kernel with invariant distribution $\pi(z \mid \Y, \Omega_{-j,-j})$, and set $z=z^{(M)}$ in Algorithm \ref{alg:gibbs_GGM_zeros}.
Expression \eqref{eq:post_submodel} has a structure similar to the posterior distribution on variable inclusion indicators in sparse linear regression, so there are many MCMC algorithms that one can adapt.
A simple kernel that was surprisingly effective in our examples is Gibbs sampling \citep{george1997approaches}, and we refer to its use within Algorithm \ref{alg:gibbs_GGM_zeros} simply as Gibbs.
There, in our examples, we set $M=p-1$, so that all entries in $z$ are updated (in random order).
We also consider a birth-death-swap algorithm (BDMH, \cite{yang2016computational})
and the locally-informed and thresholded algorithm (LIT, \cite{zhou2022dimension}).
Briefly, LIT proposes new models by evaluating \eqref{eq:post_submodel} for all models that add/remove/swap one edge from the current $z$.
LIT 
enjoys a dimension-free spectral gap in linear regression \citep{zhou2022dimension}.
In Section \ref{sec:spectral-gap-bdmh} we obtain a similar dimension-free gap for BDMH with $M=p$ in GGMs. Further, its computational cost is the same as for LIT with $M=1$, see Section \ref{ssec:iteration_cost}.
In most of our examples
BDMH and Gibbs performed better than LIT, 
see Section \ref{sec:mcmc_algorithms} for a full description. 
Based on our experiments, in BDMH, by default we recommend $M=\sqrt{p}$ updates, whereas in LIT, $M=1$ suffices.


\subsection{Global moves} \label{sec:par-gibbs}

   An alternative to sample from $\pi(z \mid Y, \Omega_{-j,-j} )$ is to \color{black} consider a globally-informed independent Metropolis-Hastings (GIMH) algorithm.
For this, we use the well-known relation between $\Omega_{\cdot j}$ and the coefficients from regressing $\Y_{\cdot j}$ on $\Y_{\cdot, -j}$,
also used by \cite{meinshausen2006high,zhou2011high,atchade2019quasi, mohammadi2025scalable}
to obtain pseudo-likelihoods.
Ours is its first use for    sampling from the exact posterior distribution. \color{black}    Besides,  based on our experiments and theory, 
$M=\sqrt{p}$ 
is a sensible default for GIMH. \color{black} 

The conditional distribution $\Y_{\cdot j} \mid \Y_{\cdot, -j}, \Omega$
is    Gaussian, with a mean that is linear in $\Y_{\cdot -j}$ and variance that does not depend on $\Y_{\cdot j}$, that is \color{black}
a linear regression where 
$\Y_{\cdot, -j}$ are the covariates,
the regression coefficients are a re-scaled version of $-\Omega_{-j, j}$,
and $z= Z_{\cdot j}$ 
are variable inclusion indicators.
Then under a conjugate prior (Lemma \ref{lem:posterior_linreg_notation})
the linear regression posterior on $z$ is
\begin{align}\label{eq:post-kappa-lr}
    \pi^{LR}_{j}(z \mid \Y) \propto  \frac{   h(z) \color{black}}{g_1^{|z|_0} |W_z|^{\frac{1}{2}}} \left(\frac{1}{ \lambda  + S_{jj} - S_{zj}^T W_z^{-1} S_{zj} }\right)^{\frac{n}{2}+1} \mathds 1_{|z|_0 \leq \bar d},
\end{align}
with    $h(z)$ as in Proposition \ref{prop:wang_exactzeroes_notation} and \color{black} $W_z= [S_{-j,-j}]_{zz}+ g_1^{-2}  I_z$. 
See Section \ref{Sec:SSPriorLR} for details.

The motivation for considering \eqref{eq:post-kappa-lr} to build a proposal for $\pi(z \mid \Y, \Omega_{-j,-j})$ is as follows. 
Firstly, there are efficient algorithms to sample from \eqref{eq:post-kappa-lr}, using Gibbs, BDMH or LIT. 
   In fact, sampling from \eqref{eq:post-kappa-lr} does not require the $U_z$ matrix featuring in $\pi(z \mid \Y, \Omega_{-j,-j})$, which in turn requires updating inverses of $\Omega_{-j,-j}$ at each iteration. Hence, the cost of drawing proposal samples is relatively small. \color{black}
Secondly, the proposal is independent of the current state $\Omega^{(t)}$. Hence, proposal samples for all columns $j \in [p]$ can be pre-computed in parallel, before running Algorithm \ref{alg:gibbs_GGM_zeros}. 
And critically, \eqref{eq:post-kappa-lr} is expected to roughly match the target \eqref{eq:post_submodel}.
More precisely, 
we use a tempered proposal where \eqref{eq:post-kappa-lr} is raised to a power $\upsilon \in [0,1]$. 
When $\Omega_{-j,-j}$ is close to the data-generating $\Omega_{-j,-j}^0$,
the tempered proposal concentrates on the same models as \eqref{eq:post_submodel} at a rate that guarantees good mixing.
Section \ref{sec:algorithmic_guarantees} outlines formal conditions on $\upsilon$ to obtain such guarantees but,
in practice, the simple default $\upsilon= 0.75$ worked best in our experiments.
Specifically, our GIMH algorithm 
proposes $z^*$ from a $\upsilon$-tempered linear regression posterior \eqref{eq:post-kappa-lr} and accepts it with probability $b_{acc} =$ 
    \begin{align}\label{eq:acc-ratio-lr} 
    &  1 \wedge \frac{\pi( z^* \mid \Y, \Omega_{-j,-j}) \left[ \pi_{j}^{LR}(z \mid \Y) \right]^\upsilon}{\pi(z \mid \Y, \Omega_{-j,-j}) \left[ \pi^{LR}_{j}(z^* \mid \Y) \right]^\upsilon}=
    1 \wedge \left[ \frac{|W_{z^*}|^\upsilon |U_{z}|}{|W_{z}|^\upsilon |U_{z^*}|} \frac{e^{S_{z^*j}^T U_{z^*}^{-1}S_{z^*j}}}{e^{S_{zj}^TU_{z}^{-1}S_{zj} }} \right]^{\frac{1}{2}}
        \left[ \frac{g_1^{|z|_0 - |z^*|_0}    h(z^*) \color{black}}{   h(z) \color{black}} \right]^{(1-\upsilon) }
        \\
    & \times \left( \lambda + S_{jj} - S_{z^*j}^T W_{z^*}^{-1}S_{z^*j}\right)^{(\frac{n}{2} + 1)\upsilon} /
   \left( \lambda + S_{jj} - S_{zj}^T W_{z}^{-1}S_{zj}\right)^{(\frac{n}{2} + 1)\upsilon}. \nonumber
    \end{align}
where $z$ is the current model, see Algorithm \ref{alg:metropolis_gibbs}.
   We remark that this accept-reject step assumes that $z^*$ is an independent proposal from $Q_{\upsilon}^j(z) \propto [\pi_j^{LR}(z \mid \Y)]^\nu$, which holds if one obtained sufficiently many samples from $Q_{\upsilon}^j(z)$
to ensure convergence as a pre-computation step, and used thinning so that samples are effectively independent. As an alternative, in Section \ref{ssec:global_proposal_dependent} we present local-global and swap proposals that allow using dependent samples, by adjusting \eqref{eq:acc-ratio-lr} accordingly. 
\color{black}

\subsection{Per-iteration cost analysis}
\label{ssec:iteration_cost}

We discuss the number of operations required by our algorithms, see Section \ref{sec:mcmc_algorithms} for further details. 
A single iteration of the BDMH and random scan Gibbs algorithms performing $M \geq 1$ updates per column
requires $O(d_{max}^3 + M d_{max}^2)$= $O(\bar d^3 + M \bar d^2)$ operations,
where $d_{max}$ is the maximum 
degree across the models considered in that iteration.
This is because evaluating $\pi(z \mid \Y, \Omega_{-j,-j})$ in \eqref{eq:post_submodel} requires $O(|z|_0^3)$ operations but, when \eqref{eq:post_submodel} was already computed for a neighbor model (adding and/or removing 1 edge), rank 1 Cholesky updates return \eqref{eq:post_submodel} in $O(|z|_0^2)$ operations.
Importantly, \eqref{eq:post_submodel} is evaluated only once for any given $z$,
when $z$ is first visited. 
This brings significant savings when doing $M>1$ updates per column, because as $\pi(z \mid \Y, \Omega_{-j,-j})$ concentrates on a few $z$'s, many $z$'s are revisited.

Accounting for $p^3$ operations in the first iteration, the total cost for $T$ iterations of the BDMH and random scan Gibbs algorithms is $O(p^3 + T[p^2 + \bar d^3 + M \bar d^2])$.
In contrast, the continuous spike-and-slab requires $O(T p^3)$ operations.
An iteration of LIT with $M=1$ requires $O(p^2 + \bar{z}_0^3 + p \bar{z}_0^2)$= $O(p^2 + \bar d^3 + p \bar d^2)$ operations, the same as BDMH and Gibbs for $M=p$ updates.
Regarding GIMH, computing \eqref{eq:acc-ratio-lr}
requires $O(p^2 + \max\{|z|_0^3,|z^*|_0^3\})$ operations. After the proposal samples from \eqref{eq:post-kappa-lr} are obtained, the total complexity for $T$ iterations is the same as for our local samplers.
The extra cost to obtain $T$ samples from \eqref{eq:post-kappa-lr} for all $p$ columns using Gibbs or BDMH requires $O(p T (\bar{d}^3 + p \bar{d}^2))$ operations
but here the $p$ chains can be run in parallel (i.e., with $p$ cores, each performs $O(T (\bar{d}^3 + p \bar{d}^2))$ operations). 
\color{black}

\section{Efficiency of the MCMC algorithms}
\label{sec:algorithmic_guarantees}

   The efficiency of our MCMC algorithms depend on the per-iteration cost (which is low in sparse settings, Section \ref{ssec:iteration_cost}) and its mixing, which measures the number of iterations required for MCMC convergence. We now study mixing by bounding \color{black}
conditional spectral gaps and $s$-conductances (see Section \ref{sec:background-mcmc} for an overview).
Briefly, the former bounds the total variation distance to the stationary distribution $\pi(\Omega \mid \Y)$ when the chain is started at an arbitrary  $\Omega^{(0)}$,
and the latter when it is given a warm start \citep{ascolani2024scalability}.
   More specifically, the $s$-conductance of a Markov transition kernel $P$, denoted by $\phi_s(P)$ for $s \in (0, 1/2)$, measures the probability of leaving any set $S$ 
such that
$\pi(\Omega \in S\mid \Y) \in (s,1/2)$.
Sections \ref{sec:spectral-gap-bdmh}-\ref{sec:gimh} show that our local and global 
MH-within-Gibbs samplers targeting $\pi(\Omega_{\cdot j} \mid \Y, \Omega_{-j,-j}^{(t)})$ mix similarly to ideal Gibbs sampling from $\pi(\Omega_{\cdot j} \mid \Y, \Omega_{-j,-j}^{(t)})$ exactly, upon sparsity conditions, assumptions on $\Omega^{(t)}$ and as $n \to \infty$. Section \ref{ssec:ideal_gibbs} shows that ideal Gibbs is geometrically ergodic and, while it does not provide sharp rates, it does not require assumptions on $\Omega^{(t)}$ and holds for any $n$.
\color{black}
We focus on the BDMH and GIMH algorithms, and leave the LIT algorithm for future work, for brevity and because the BDMH spectral bound is already dimension-free for a computational cost roughly equal to LIT (Section \ref{sec:spectral-gap-bdmh}).

Our theory applies both to the Binomial prior in \eqref{eq:prior-z} with edge inclusion probability $\theta= 1/p^{\alpha}$, and to the Beta-Binomial in \eqref{eq:prior-z-bbin} taking $a_\theta=1$ and $b_\theta= p^\alpha$, under conditions that require $\alpha \geq 0$ to be large enough (i.e., the prior to be sufficiently sparse). 
Throughout, we define $\tilde{\alpha}:=\alpha$ 
for the Binomial prior 
and $\tilde{\alpha} := \max\{2, \alpha\}$ for the Beta-Binomial, and we make generic statements involving $\tilde{\alpha}$ that apply to both priors. 

To alleviate notation let $\Omega_{-j,-j}= \Omega_{-j,-j}^{(t)}$ be the currently conditioned-upon value.
Here we study random scan Gibbs algorithms that choose $j$ uniformly at random. 
In our software we use random sequence Gibbs, which updates all columns in random order and worked slightly better in practice.
We denote by 
$P_{BD}^{j,\Omega_{-j,-j}, M}$ and $P_{GI}^{j,\Omega_{-j,-j},M}$ the lazy conditional kernels performing $M$ birth-death-swap and globally-informed (respectively) updates  for edge indicators $z = Z_{-j,j}$ conditionally on $\Omega_{-j, -j}$, for a given column $j$. 
sampling from $\pi(\Omega_{.j} \mid \Y, \Omega_{-j,-j})$ exactly. By lazy, we mean that $P_{\cdot}^{j,\Omega_{-j,-j}, M}(z,z') = \frac{1}{2}\delta_z(z') + \frac{1}{2} \Tilde P_{\cdot}^{j,\Omega_{-j,-j}, M}$ with $\Tilde P_{\cdot}^{j,\Omega_{-j,-j}, M}$ the kernel performing $M$ (local or global) updates for $z$.
\color{black}
Finally, denote the  lazy BDMH kernel that jointly updates $Z$ and $\Omega$ by
\begin{align*}
  P^{M}_{BD}(\Omega, \Omega') = \frac{1}{2} \delta_{\Omega} (\Omega') + \frac{1}{2p} \sum_{j=1}^p P^{j,\Omega_{-j,-j}, M}_{BD} (z, z')  \pi(\Omega_{\cdot j}' \mid z', \Y, \Omega_{-j,-j}) \delta_{\Omega_{-j,-j}} (\Omega_{-j,-j}'),
\end{align*}
which sets $\Omega'=\Omega$ with probability $1/2$ and otherwise chooses a random column $j$, applies $M \geq 1$ birth-death-swap updates to the edge indicators $Z_{-j,j}$, and then samples $\Omega_{\cdot j} \sim \pi(\Omega_{\cdot j} \mid Z_{-j,j}, \Y, \Omega_{-j,-j})$ exactly. 
  
Define the lazy kernels $P_{GI}^{M}$ and $P_{Gibbs}$, respectively with global proposals and the ideal Gibbs, analogously.
These lazy kernels 
are positive-definite and we use them because they are easier to study and their spectral gaps differ from those of the original kernels by a factor $\leq 2$.
\color{black}




\subsection{Birth-death-swap proposal}\label{sec:spectral-gap-bdmh}

  
Section \ref{ssec:spectral-gap-z-bdmh} bounds the spectral gap of $P_{BD}^{j,\Omega_{-j,-j}, M}$, the kernel only updating $ Z_{-j,j}$. 
Section \ref{ssec:conductance-bdmh} then 
bounds the $s$-conductance $\phi_s(P^{M}_{BD})$ in relation to that of the ideal Gibbs sampler.
\color{black}

\subsubsection{Spectral gap}
\label{ssec:spectral-gap-z-bdmh}

For simplicity we give our result for BDMH with birth, death and swap proposal probabilities such that all models in a neighborhood of $Z_{-j,j}$ have equal proposal probability,
but we obtained a similar result for other 
probabilities, as long as they are 
in (0,1).

We outline three conditions    on $\Omega^{(t)}$ \color{black} required by our main result, Theorem \ref{thm:bdsgap_chang_omegageneral}.
   Intuitively, these describe values of $\Omega^{(t)}$ such that the BDMH sampler is guaranteed to mix well, relative to ideal Gibbs sampling from $\pi(\Omega_{\cdot j} \mid \Y, \Omega_{-j,-j}^{(t)})$. 
Recall that $S = \Y^T \Y$ and 
$\Sigma_{AB |C} = \Sigma_{AB} - \Sigma_{AC} \Sigma_{CC}^{-1} \Sigma_{CA}$ is the conditional covariance for any sets $A,B,C \subseteq [p]$, with $\Sigma= [\Omega^{(t)}]^{-1}$. We define $\Sigma_{AB |C}^0$ similarly with $\Sigma^0 := [\Omega^0]^{-1}$. \color{black}
Assumption \ref{ass-eigen0} sets mild eigenvalue conditions on $\Omega_{-j,-j}    = \Omega_{-j,-j}^{(t)}$. 
Assumptions \ref{ass-maxexpec-omegageneral-overfitted}-\ref{ass-betamin_omegageneral} are sparsity and betamin conditions (respectively) guaranteeing that $\pi(Z_{-j,j} \mid \Y, \Omega_{-j,-j})$ concentrates around an optimal $z^* =\arg\max_{Z_{-j,j}} \mathbb E\left[ \log \pi(Z_{-j,j} \mid \Y, \Omega_{-j,-j})\right]$ 
(see \eqref{eq:z-star} in Section \ref{app:proof-theorem-1} for its exact definition). 
\color{black}
If $\Omega_{-j,-j}$ is close to $\Omega_{-j,-j}^0$ then $z^*= Z_{-j, j}^0$ is the data-generating truth    and our results are analogous to those in sparse linear regression \citep{yang2016computational}. \color{black}
However, more generally we may have that $z^* \neq Z_{-j,j}^0$, e.g., $z^*$ may be less sparse than $Z_{-j, j}^0$,    deviating from the existing literature. Let us define $\mathcal{Z}(\bar d):= \{z \in \{0,1\}^{p-1} : |z|_0 \leq \bar d\}$.
\color{black}

\begin{assumption}\label{ass-eigen0}
There exists $\epsilon_0 \in (0,1)$ such that for any $j \in [p]$, 
$   \epsilon_0 \leq \lambda_{min}(\Omega_{-j,-j}) \leq \lambda_{max}(\Omega_{-j,-j}) \leq \epsilon_0^{-1}$.
   Further, the prior slab variance $g_1^2$ satisfies $\bar d / (n g_1^2)= o(1)$. \color{black}
\end{assumption}

\begin{assumption}\label{ass-maxexpec-omegageneral-overfitted}
For any $j \in [p]$
and $z\in \mathcal{Z}(\bar d)$ such that $z \supset z^*$, there is a model $z'= z \setminus \{k\}$ that drops the edge $k \in z \setminus z^*$ from $z$, such that
\begin{align}
&\lim_{n \to \infty} \sqrt{\frac{\Sigma_{kk \mid z' \cup j}}{\Tilde \Sigma_{kk \mid z' \cup j}}}
\sqrt{2[(   \tilde{\alpha} \color{black} - 1) \log p +     \log(g_1 \sqrt{n}) \color{black} - \log (\bar d +1)]} 
\nonumber \\
& 
- (1 + \epsilon) \frac{\sqrt{n \Sigma_{jj}^0} |\mu_{k, z'}|}{\sqrt{\Tilde \Sigma_{kk \mid z' \cup j}}}
- \mathbb E \left[ \max_{z \supset z^*} \frac{|d_{k, z'}|  - \mathbb E |d_{k, z'}|}{\sqrt{\Tilde \Sigma_{kk \mid z' \cup j} S_{jj}}} \right]
 - \sqrt{2 \log p} = \infty.
\nonumber
\end{align}
with  $\epsilon \in (0, 1/2)$ a constant independent of $z$ and $k$,
$d_{k, z'} := s_k - s_{z'}^T \Sigma_{z'z' \mid j}^{-1} \Sigma_{k z' \mid j}$, $s := S_{-j,j}$,
$\mu_{k, z'} := (\Sigma_{kj}^0 - \Sigma_{k z' \mid j} \Sigma_{z' z' \mid j}^{-1} \Sigma_{z' k}^0) / \Sigma_{jj}^0$ and
$
\Tilde \Sigma_{kk \mid z' \cup j} :=  \Sigma_{kk \mid j}^{0} + \Sigma_{k z' \mid j} \Sigma_{z'z' \mid j}^{-1}  \Sigma_{z'z' \mid j}^{0} \Sigma_{z'z' \mid j}^{-1}   \Sigma_{ z' k \mid j} - \Sigma_{k z' \mid j}^{0} \Sigma_{z'z' \mid j}^{-1} \Sigma_{z' k \mid j} - \Sigma_{k z' \mid j} \Sigma_{z'z' \mid j}^{-1} \Sigma_{z' k \mid j}^0.
$
\end{assumption}

\begin{assumption}\label{ass-betamin_omegageneral}
For any $j \in [p]$ and $z \in \mathcal{Z}(\bar d)$ such that $z \not\supset z^*$, there is an edge $\ell \in z^* \setminus z$ such that
\begin{align}
    \lim_{n \to \infty} n \left( \frac{1}{2} - \epsilon \right)  \min_{z \in \mathcal{Z}(\bar d), z \not\supset z^*} \frac{ \Sigma_{jj}^0 \mu_{\ell,z}^2}{\Tilde \Sigma_{\ell \ell \mid z \cup j}}  - (\bar d + 2) \log p = \infty.
    \nonumber
\end{align}
for some $\epsilon \in (0,1/2)$
and 
$\tilde \Sigma_{\ell \ell \mid z \cup j}$ and $\mu_{\ell,z}$ as in Assumption \ref{ass-maxexpec-omegageneral-overfitted}. Further,
\begin{align}
\left[\log(\bar d +1) + (   \tilde{\alpha} \color{black} + 1) \log p +    \log (g_1 \sqrt{n}) \color{black} \right]
= o\left( n\min_{z \in \mathcal{Z}(\bar d), z \not\supset z^*} \frac{\Sigma_{jj }^0 \mu_{\ell,z}^2}{\Tilde \Sigma_{\ell \ell \mid z \cup j}} \right).
\nonumber
\end{align}
\end{assumption}

   Assumptions \ref{ass-maxexpec-omegageneral-overfitted}-\ref{ass-betamin_omegageneral} can be stringent, depending on the conditioned-upon $\Omega_{-j,-j}$. \color{black}
Assumption \ref{ass-maxexpec-omegageneral-overfitted} ensures that overfitted models $z \supset z^*$ receive small posterior mass, akin to the sparse projection condition in \cite{yang2016computational} (Assumption B, 7c).
For intuition, $z$ adds an unnecessary edge $k$ to $z'$ (an edge not included in the optimal $z^*$),
and larger $|d_{k,z'}|$ assigns more posterior mass to $z$    (adding the unnecessary edge). \color{black}
If $\Omega_{-j,-j}=\Omega_{-j,-j}^0$, 
then $\tilde{\Sigma}_{kk \mid z' \cup j}=\Sigma_{kk \mid z' \cup j}$, $\mu_{k,z'}=0$
and $|d_{k, z'}|/\sqrt{\Sigma_{kk \mid z' \cup j}S_{jj}}$ is the absolute value of a standard Gaussian (see Lemma \ref{lem:exp-max}). 
Hence, Assumption \ref{ass-maxexpec-omegageneral-overfitted} requires that the expected maximum of absolute standard Gaussians
across the $\leq p^{\bar d}$ overfitted models is small relative to the prior sparsity induced by    $\tilde{\alpha}$. \color{black}
The worst-case scenario would be if the Gaussians were independent, then said expectation is $\leq \sqrt{2 \bar d \log p}$ 
and one must take    $\tilde{\alpha} > \bar d + 2$, a fairly sparse prior. \color{black}
In our case however they 
are dependent, 
as all the $d_{k,z'}$ variables are a function of 
$s=S_{-j,j} \in \mathbb{R}^{p-1}$, and the expectation can be much smaller.
For example, if $\Omega_{-j,-j}=\Omega_{-j,-j}^0$ and under Assumption \ref{ass-eigen0}, the expectation of this maximum is $\leq \epsilon_0^{-2} \sqrt{2 \log p}$ (see Lemma \ref{lem:exp-max}) and $    \tilde{\alpha} \color{black} > 3 + \epsilon_0^{-4}$ suffices. 
   That is, if $\Omega^{(t)}$ is well-conditioned then Assumption \ref{ass-maxexpec-omegageneral-overfitted} is milder. \color{black}
Assumption \ref{ass-betamin_omegageneral} is a betamin condition.
Briefly, for fixed $\Omega_{-j,-j}$ and $\Omega_{-j,-j}^0$ we have fixed $\Sigma_{jj}^0 \mu_{k,z}^2$, $\Sigma_{kk \mid z \cup j}$ and $\Tilde \Sigma_{kk \mid z \cup j}$, for any $z \not \supset z^*$ and $k \in z^* \backslash z$. Hence, Assumption \ref{ass-betamin_omegageneral} requires that $n c - (\bar d + 2) \log p$ diverges to $\infty$ and that $(   \tilde{\alpha} \color{black} + 1) \log p= o(n)$, where $c>0$ is a constant depending on $\Sigma_{jj}^0 \mu_{k,z}^2$ and $\Tilde \Sigma_{kk \mid z \cup j}$. 
   Hence, Assumption \ref{ass-betamin_omegageneral} becomes milder as $n$ grows. \color{black}

Theorem \ref{thm:bdsgap_chang_omegageneral} proves that for $M=1$, the inverse spectral gap of our conditional birth-death-swap kernel is bounded by a linear function of $p$ and the maximal degree $\bar d$. Moreover, if $\bar d$ is fixed and one takes $M=p$ steps, then the gap is dimension-free (that is, independent of $p$). As $M \to \infty$, this gap tends to 1 and BDMH becomes equivalent to the ideal Gibbs sampler.

\begin{thm} \label{thm:bdsgap_chang_omegageneral}
Suppose that $\Omega^{(t)}$ satisfies 
Assumptions \ref{ass-eigen0}, \ref{ass-maxexpec-omegageneral-overfitted} and \ref{ass-betamin_omegageneral}. Then,
for $M=1$,
$$\lim_{n \to \infty} \mathbb P \left[ \bigcap_{j=1}^p \left \{ \mbox{Gap}(  P_{BD}^{j,\Omega^{(t)}_{-j,-j}, 1})^{-1} \leq (4 + \epsilon) p (\bar d + 1) \right\} \right] = 1$$ for any $\epsilon > 0$.
For general $M \geq 1$ and any $\epsilon > 0$, we have that
\begin{align*}
   \lim_{n \to \infty} \mathbb P \left[ \bigcap_{j=1}^p \left \{ \text{Gap}( P^{j,\Omega^{(t)}_{-j,-j} M}_{BD}) \geq 1 - e^{-\frac{M}{( 4 + \epsilon) p (\bar d + 1)}} \right \} \right] = 1. 
\end{align*}
\end{thm}

\subsubsection{$s$-conductance}
\label{ssec:conductance-bdmh}

   Our next result 
relates the $s$-conductance of our BDMH kernel to its spectral gap 
and to the $s$-conductance of the ideal Gibbs sampler. \color{black} 

\begin{cor}\label{cor:s_conductance}
   Let $B= \{ \Omega^{(t)} \in \PSD: 
\mbox{Assumptions \ref{ass-eigen0}-\ref{ass-betamin_omegageneral} hold} \}$. \color{black}
     Then, for any $s \in (0,1/2)$ and $ \epsilon > 0$,
    \begin{align}\nonumber
    \lim_{n \to \infty} \mathbb P \left(
    \phi_s( P^{M}_{BD}) \geq \frac{1}{2}  \phi_s(P_{Gibbs})
    \left[ 1 - e^{-\frac{M}{ (4 + \epsilon) p \bar d}}  - \frac{2\pi(B^c | \Y)}{s \phi_s(P_{Gibbs})} \right] \right) = 1.
\end{align}
\end{cor}

Whereas Theorem \ref{thm:bdsgap_chang_omegageneral} only holds for 
   $\Omega^{(t)} \in B$, \color{black}
Corollary \ref{cor:s_conductance} makes the role of $B$ explicit. In the best scenario, 
$\pi(\Omega \mid \Y)$ concentrates on $B$ and $\pi(B^c \mid \Y) = o_{P}(1)$. 
Then, if one sets $M$ of order $p \bar d$,
the s-conductances 
of $P^{M}_{BD}$ and $P_{Gibbs}$ are of the same order
as long as $s \phi_s(P_{Gibbs}) \geq C   \pi(B^c \mid \Y)$ for some  $C > 2$. 
In a less favorable scenario where $\pi(B \mid \Y) < 1$ even asymptotically,
Corollary \ref{cor:s_conductance} bounds the loss in $s$-conductance relatively to ideal Gibbs.

\subsection{Globally informed proposal}\label{sec:gimh}

We now bound the spectral gap of $ P^{j, \Omega_{-j,-j},M}_{GI}$, our GIMH conditional kernel  with $M \geq 1$ updates, and subsequently the $s$-conductance of the corresponding lazy global kernel $P^{M}_{GI}$. 
For a column $j$, recall that $Q_{\upsilon}^j(z) \propto [\pi_j^{LR}(z \mid \Y)]^\upsilon$ is the global proposal for the edge indicators $z = Z_{-j,j}$, 
where $\pi_j^{LR}$ is the linear regression posterior 
defined in \eqref{eq:post-kappa-lr}, and $\upsilon > 0$ is the tempering parameter.
For simplicity, say $\pi_j^{LR}$ posterior is obtained
under Zellner's prior
$
\pi(\beta_z, \sigma^2 \mid z)= \text{N}(\beta_z; \mathbf{0}, \tau^{-1} \sigma^2 (\Y_{\cdot z}^T \Y_{\cdot z}/n)^{-1}) \times \mbox{IG}(\sigma^2; a/2, \lambda/2),
$
where 
$\beta_z = - \Omega_{zj} / \Omega_{jj}$, $\sigma^2 = (\Omega_{jj})^{-1}$, $ a > 0$,
 and $\tau > 0$ is a precision parameter.
By default $\tau=1$ and $a=2$, so that $\sigma^{-2} \sim \mbox{Exp}(\frac{\lambda}{2})$ matches the prior in \eqref{eq:prior-z}.

Our results require more stringent conditions than the BDMH kernel.
On the other hand the proof is conceptually simpler; it relies on bounding importance weights
\begin{align}\label{eq:weight}
    w^j(z) := \frac{\pi(z | \Omega_{-j,-j}, \Y)}{Q^j_{\upsilon}(z)}=
    \frac{\pi(z | \Omega_{-j,-j}, \Y)}{\pi(z^0 | \Omega_{-j,-j}, \Y)} 
    \frac{\pi(z^0 | \Omega_{-j,-j}, \Y)}{Q^j_{\upsilon}(z^0)} 
    \frac{Q^j_{\upsilon}(z^0)}{Q^j_{\upsilon}(z)},
\end{align}
where $z = Z_{-j,j}$ and $z^0 = Z_{-j,j}^0$. If $\max_{z \in \mathcal{Z}(\bar d)} w^j(z) \leq W$
then the spectral gap of the conditional kernel is 
$\geq 1/W$ \citep{wang2022exact}. 
We obtain a $W$ that does not depend on $p$, 
giving a dimension-free gap.
First, we bound $\pi(z | \Omega_{-j,-j}, \Y) / \pi(z^0 | \Omega_{-j,-j}, \Y)$ and $Q_{\upsilon}^j(z^0) / Q_{\upsilon}^j(z)$. 
Second, since the middle term in \eqref{eq:weight} is $\leq 1/Q_{\upsilon}^j(z^0)$, we show that $Q_{\upsilon}^j(z^0) \stackrel{\mathbb P}{\longrightarrow} 1$. 


We require two sets of conditions listed below. 
First, C0-C4 guarantee that $\pi(Z_{-j,j} \mid \Y, \Omega_{-j,-j})$ concentrates on the data-generating $z^0$, and require $\Omega_{-j,-j}= \Omega_{-j,-j}^{(t)}$ to be close to $\Omega_{-j,-j}^0$.
They are hence stronger counterparts to Assumptions \ref{ass-maxexpec-omegageneral-overfitted}-\ref{ass-betamin_omegageneral}, which did not impose such a requirement. In fact, Lemma \ref{lem:conditions-ass} below shows that C0-C4 suffice for Assumptions \ref{ass-maxexpec-omegageneral-overfitted}-\ref{ass-betamin_omegageneral} to hold.
Second, conditions D1-D4 in Section \ref{sec:regularity_linear_regression} are mild and guarantee that the linear regression proposal also concentrates on $z^0$.
\color{black}
Throughout we use the simplified notation 
$\Omega_{zz} := [\Omega_{-j,-j}]_{zz}$ and $\Omega_{zz}^0 := [\Omega^0_{-j,-j}]_{zz}$. 



\begin{itemize}
    \item[(C0)] The sample size $n$, dimension $p$ and    $\Tilde \alpha $ verify $\Tilde \alpha \log p= o(n)$ \color{black}.
    \item[(C1)] $\Omega^0 \in \mathcal{M}(\epsilon_0)$, i.e. 
    $\epsilon_0 \leq \lambda_{min}(\Omega^0) \leq \lambda_{max}(\Omega^0) \leq \epsilon_0^{-1}$ for some constant $\epsilon_0 \in (0,1)$.
    \item[(C2)] There is 
a  constant $ \omega_{min} > 0$  
such that 
$\min_{j \in [p]} \min_{k \in [p]: Z_{kj}^0=1} (\Omega_{kj}^{0})^2 / (\Omega_{jj}^0)^2 \geq \omega_{min}$.
\item[(C3)] Define
    $\mathcal{B}_T(\Omega^0, \delta, \epsilon_0) := \{ \Omega \in \mathcal{M}(\epsilon_0) :  \|\Omega_{zz} - \Omega^0_{zz}\| \leq \delta, \: j \in [p], z \in \mathcal{Z}(\bar d)\}.$ 
Assume that $\Omega^{(t)} \in \mathcal{B}_T(\Omega^0, \delta, \epsilon_0)$  with
$
    \delta < 
     \frac{\epsilon_0^{6}}{48} \sqrt{\frac{\log n}{2n} + \frac{\log p}{n}}.
$
\item[(C4)]  
It holds that $\tilde{\alpha} > 1 + (\sqrt{2\bar d} +  \sqrt{2} + \frac{1}{4})^2$. 

\end{itemize}

Conditions C0, C1 and C2 are mild sparsity, eigenvalue and betamin conditions (respectively)    on $\Omega^0$. \color{black} 
   C4 is a strong condition that the prior sparsity parameter $\tilde \alpha$ is large enough. 
In our experiments we obtained excellent results with a milder $\tilde \alpha=1$, so we speculate that C4 is stronger than strictly needed. \color{black}
Finally, C3 states that the current  $\Omega^{(t)}$ is close to $\Omega^0$,
specifically that sub-matrices of size at most $\bar d \times \bar d$ lie in a $\delta$-neighborhood of $\Omega^0$ with $\delta$ small enough. 
C3 basically requires that $\pi(\Omega \mid \Y)$ is concentrated around $\Omega^0$. 
For fixed $p$, said concentration holds with probability 1 as $n$ grows (Proposition \ref{prop:posterior-consistency}),
but to our knowledge high-dimensional concentration rates for sub-matrices of $\Omega$ are not known.
In our experiments in Section \ref{sec:experiments}, we assessed our algorithms when C1-C4 do not hold, obtaining good results.



\begin{lemma}\label{lem:conditions-ass}
Assume that C0-C4 hold    and $\frac{g_1}{\bar d+1} > \frac{1}{\sqrt{n}}$. \color{black}
Then Assumptions  \ref{ass-maxexpec-omegageneral-overfitted}-\ref{ass-betamin_omegageneral} hold with  $z^* = Z_{-j,j}^0$, for any  $j \in [p]$ and $n \geq n_0$ with $n_0$ a constant depending only on $(\epsilon_0, \omega_{min})$.
\end{lemma}

\begin{thm}\label{thm:spectral-gap-gimh}
    Suppose that    $\Omega^0$ and $\Omega^{(t)}$ satisfy \color{black} C0-C4 and D1-D4, and that    $\tilde{\alpha}^{LR}$ \color{black} and $\upsilon$ verify
    \begin{align}
        &   \tilde{\alpha}^{LR} > 4 \max(\frac{\epsilon_1}{\epsilon_0^{2}}, 1)  \color{black}
        ,\qquad 
        &\frac{1}{   \tilde{\alpha}^{LR} \color{black}} < \upsilon < \min \left( \frac{c_0}{   \tilde{\alpha}^{LR} \color{black}}, \frac{1}{4} \frac{\epsilon_0^{2}}{\epsilon_1}, \frac{1}{2} \right), \label{eq:ass-beta}
    \end{align}
    with $c_0 = \frac{1 +    \tilde \alpha - ( \sqrt{2\bar d} + \sqrt{2} + \frac{1}{4} \color{black})^2}{2}$.
Then, for any $\epsilon \in (0,1)$, 
and $M \geq 1$,
\begin{align*}
     \lim_{n \to \infty} \mathbb{P} \left(  \bigcap_{j=1}^p \left \{ \text{Gap}(P^{j,\Omega_{-j,-j}^{(t)},M}_{GI}) \geq 1 - e^{-M ( 1 - \epsilon)} \right \} \right) = 1. 
\end{align*}
\end{thm}

Theorem \ref{thm:spectral-gap-gimh} states that
the spectral gap of the conditional MH kernel with global proposal is arbitrarily close to 1 as $n \to \infty$, even with $M=1$ updates. Intuitively this occurs because, as $n$ grows, the global proposal and $\pi(z \mid \Y, \Omega_{-j,-j})$
both concentrate on $z^0$.
The result holds when the tempering parameter $\upsilon$ lies in a suitable range.
Under C4, $   \tilde{\alpha} > 1 + (\sqrt{2\bar d} + \sqrt{2} + \frac{1}{4})^2 \color{black}$, hence 
$1/\tilde \alpha^{LR} < c_0 /    \tilde{\alpha}^{LR} \color{black}$ in \eqref{eq:ass-beta}.
   In practice, $\upsilon= 0.75$ worked best in our experiments. \color{black}
Corollary \ref{cor:s_conductance-gimh} below is analogous to
Corollary \ref{cor:s_conductance} for the GIMH kernel $P^{M}_{GI}$.


\begin{cor}\label{cor:s_conductance-gimh}
     Under the Assumptions of Theorem \ref{thm:spectral-gap-gimh}, let $B := B_T(\Omega^0, \delta)$ defined in C3. For any $s \in (0, 1/2)$ and $\epsilon > 0$, 
     with probability going to 1 as $n \to \infty$ it holds that
    \begin{align}\nonumber
    \phi_s( P_{GI}^{M}) \geq  \frac{1}{2} \phi_s(P_{Gibbs})
    \left[ 1 - e^{-M(1 - \epsilon)}  - \frac{2\pi(B^c | \Y)}{s \phi_s(P^L_{Gibbs})} \right].
\end{align}
\end{cor}


\subsection{Ideal Gibbs sampler}
\label{ssec:ideal_gibbs}

Our final result 
establishes the geometric ergodicity of the ideal Gibbs. 
For simplicity, we prove the result for sampling the off-diagonal entries in $\Omega$ given a fixed $\mbox{diag}(\Omega)$. 

\begin{thm}
The kernel of a full sweep ($p$ updates) of the random scan Gibbs sampler, targeting $\pi(\Omega \mid \Y, \mbox{diag}(\Omega))$, denoted by $P_{Gibbs}^p$, is geometrically ergodic.
That is, there exists a function $M: \PSD \to [1,\infty)$, a constant $K < \infty$, and a rate $r \in (0,1)$ such that,
for every $\Omega^{(0)} \in \PSD$,
$
\| P_{Gibbs}^{pt}(\Omega^{(0)}, \cdot) - \pi(\cdot \mid \Y, \mbox{diag}(\Omega)) \|_{TV} \leq K M(\Omega^{(0)}) r^t.
$
\label{thm:rsgibbs_geomergodic}
\end{thm}

The proof does not provide sharp bounds for the dependence of the geometric rate $r$ on the dimension. This is a challenging problem that deserves a dedicated study elsewhere.
On the other hand, Theorem \ref{thm:rsgibbs_geomergodic} makes no assumptions about $n$, $\Omega^{(t)}$ or $\Omega^0$.

\color{black}

\section{Simulated experiments}\label{sec:experiments}

We assessed the precision in approximating the posterior $\pi(\Omega \mid \Y)$ relative to clock time for our Gibbs, BDMH and GIMH samplers,    under the Binomial and Beta-Binomial model priors, \color{black} and the quality of the inference.
We also assessed the following state-of-the-art methods.
    First, \textit{ssgraph}, a block Gibbs sampler for continuous spike-and-slab priors \citep{Wang2015ScalingIU,ssgraph}.
    Second, \textit{bdgraph}, a continuous-time birth-death sampler for $G$-Wishart priors \citep{mohammadi2019bdgraph,mohammadi2023accelerating}. 
    Third, \textit{bdgraph.mpl}, a marginal pseudo-likelihood sampler for $G$-Wishart priors by \cite{mohammadi2019bdgraph,mohammadi2025scalable}. bdgraph.mpl samples $Z$ but not $\Omega$, hence we sampled $\Omega$ from its conjugate $G$-Wishart posterior.
    Fourth, \textit{regression.pl}, a linear regression pseudo-posterior sampler of \cite{atchade2019quasi}, implemented by us 
    under a discrete Gaussian spike-and-slab prior.
    Finally, GLASSO \citep{friedman2008sparse} setting the regularisation parameter with BIC and EBIC ($\gamma = 0.5$), using R package \texttt{huge} \citep{zhao2012huge}.
All methods are coded in C, hence the run times are fairly comparable.
For our BDMH and GIMH algorithms we used $M=\sqrt{p}$ updates per iteration, and for GIMH we set the tempering parameter to $\upsilon= 0.75$.



For all methods adopting the Binomial model prior we used $\theta=2/(p-1)$ 
to represent the prior sparsity belief that the expected node degree is 2, 
regardless of $p$. Our methods also allow for the Beta-Binomial model prior, and we set $a_{\theta} = b_{\theta} = 1$ (a uniform prior on $\theta$).
   Our empirical Bayes algorithm obtains $\hat{\theta}$ from data as described earlier. \color{black}
We set our prior parameters to the default $g_1=1$ and $\lambda = 0.02$ 
(Section \ref{ssec:prior-formulation}).
In \texttt{ssgraph} we also set 
$g_1^2=1$ and the default spike variance $g_0^2 = 0.0004$. 
In \texttt{regression.pl} we set $g_1^2=1$ and the same prior on the error variances implied by $\lambda$.
Finally, in \texttt{bdgraph} and \texttt{bdgraph.mpl} we set the default $G$-Wishart priors with 3 degrees of freedom and identity scale matrix. 
  
Section \ref{sec:sensitivity} presents a sensitivity analysis for $g_1$ and $\lambda$ showing that posterior sampling efficiency is stable across a reasonable range of these values, and that the quality of inference is stable to $\lambda$ and can deteriorate for $g_1$ much bigger or smaller than our default.
\color{black}

We considered the following scenarios for the data-generating graph $Z^0$.
\begin{itemize}
    \item \textit{Random-$1/p$}: the presence of an edge between any two vertices is drawn uniformly at random with probability $1/p$. That is, on average there are $(p-1)/p$ edges per node.
    
    \item \textit{Tri-diagonal}: each node has an edge to its neighbouring two nodes, i.e., $Z_{jk}^0=1$ for $|j-k|=1$ and $Z_{jk}^0=0$ otherwise. 
    
    \item \textit{Block-diagonal}: $\Omega$ is block-diagonal, with fully-connected blocks of size $4$. 
\end{itemize}
Given a graph $Z^0$, we generated $\Omega_{jj}^0 \sim Ga(3, 1)$ and
$\Omega_{jk}^0 := \rho_{jk}\sqrt{\Omega_{jj}}\sqrt{\Omega_{kk}}$,
where $\rho_{jk}=0$ if there is no edge between $(j,k)$ (i.e., $Z_{jk}^0=0$) and otherwise 
$\rho_{jk} \sim Unif(\{-0.5, -0.4, \ldots, -0.1, 0.1, \ldots, 0.5\})$.
We repeated this process until sampling a positive-definite $\Omega^0$.
We also considered a further simulation setting where the smallest eigenvalue of $\Omega^0$ is very close to 0, and was designed to result in multi-modal  posteriors: 
\begin{itemize}
    \item \textit{Ill-conditioned banded-diagonal}: we set $\Omega_{jj}^0 = 1.5$, $\Omega_{jk}^0 = 0.9$ if $|j-k| = 1$, $\Omega_{jk}^0 = 0.5$ if $|j-k| = 2$, $\Omega_{jk}^0 = 0.35$ if $|j-k| = 3$, and $\Omega_{jk}^0 = 0$ otherwise.
\end{itemize}
We considered $p \in \{50, 100, 200\}$, $n = \{p, 2p, 5p\}$,
and performed $N=20$ independent simulations. 
We also set $p=500$ and $p=1000$ for the time benchmarking shown in Figure 1.
We could only run \texttt{bdgraph} and \texttt{ssgraph} for $p \leq 100$, as for larger $p$ the run times were excessive. 
We hence present figures for $p=100$ here so that all methods are featured.    Observations $\Y$ were standardized to have column variance of 1 before running the MCMC.
\color{black}

\subsection{Posterior Sampling Efficiency}
\label{ssec:sampling_efficiency}

\begin{figure}
\begin{center}
\begin{tabular}{cc}
Mean $|\hat{\mathbb E}^{(1)}(\Omega_{jk} \mid \Y) - \hat{\mathbb E}^{(2)}(\Omega_{jk} \mid \Y)|$  & Mean $|\hat{\pi}^{(1)}(\Omega_{jk} \neq 0 \mid \Y) - \hat{\pi}^{(2)}(\Omega_{jk} \neq 0 \mid \Y)|$ \\
\includegraphics[trim= {0.0cm 2.2cm 0.0cm 0.0cm}, clip,width =0.475\linewidth]{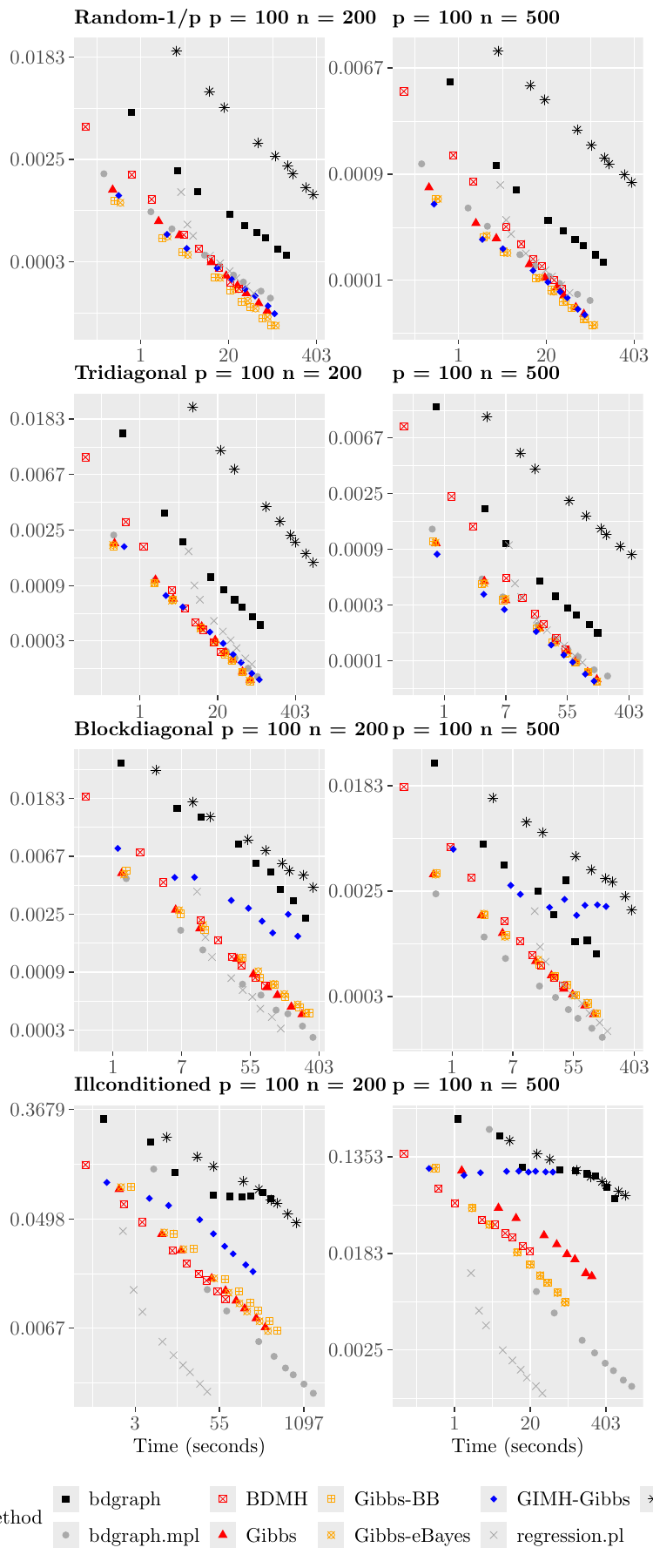} & \includegraphics[trim= {0.0cm 2.2cm 0.0cm 0.0cm}, clip,width =0.475\linewidth]{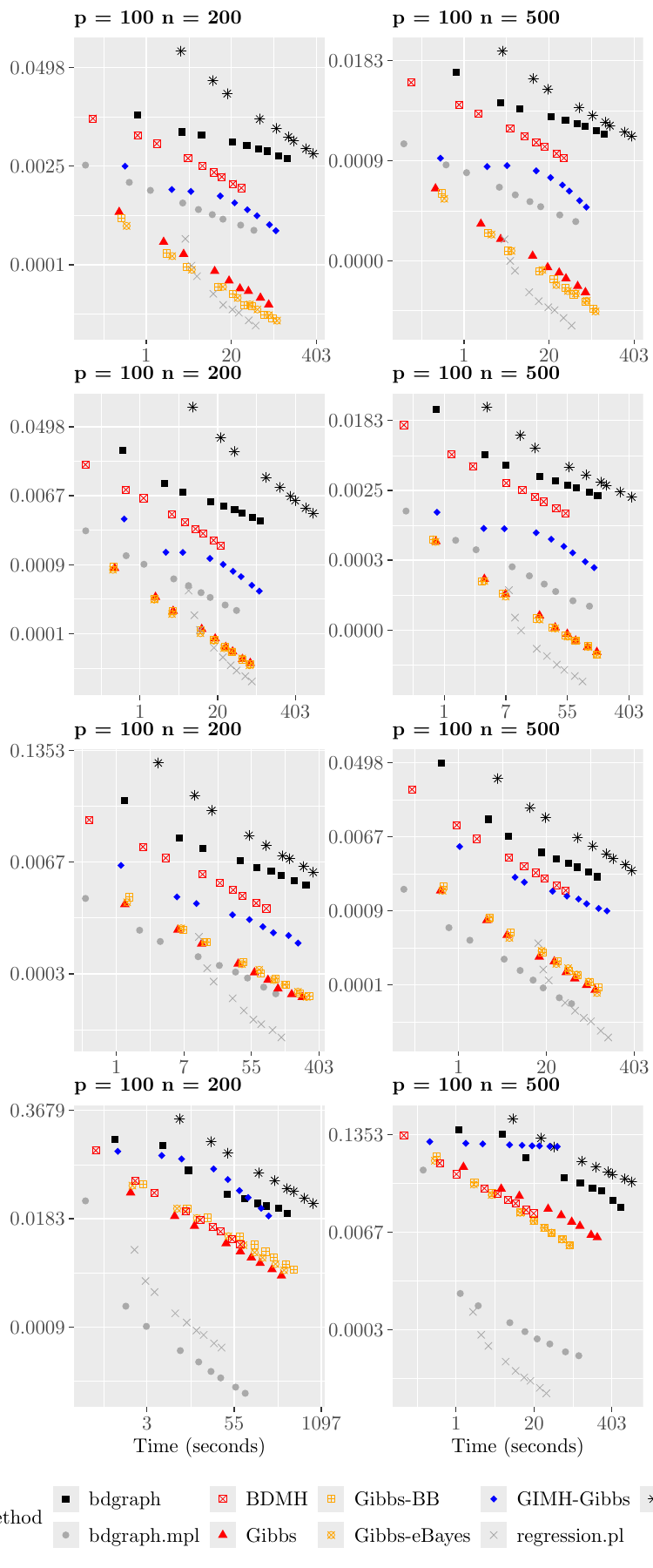}\\
\multicolumn{2}{c}{\includegraphics[trim= {0.0cm 0.4cm 0.0cm 29.2cm}, clip, width =0.95\linewidth]{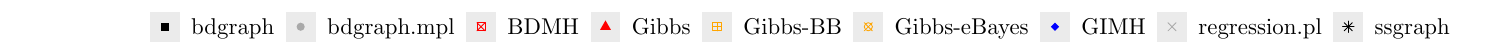}}\\
\end{tabular}
\caption{Difference in posterior mean and edge inclusion probability estimates between two chains initialised at GLASSO-EBIC and diag(1) vs. clock time for $p = 100$, $n \in \{2p, 5p\}$. Black: exact Bayesian methods (ssgraph, bdgraph). Gray: pseudo-likelihood methods (bdgraph.mpl, regression.pl). Red: Gibbs with fixed $\theta$ and BDMH. Orange: Gibbs with $\theta\sim \text{Beta}(1, 1)$ and empirical Bayes. Blue: GIMH ($\upsilon = 0.75$)}
\label{fig:mixing_p100}
\end{center}
\end{figure}

To assess the MCMC accuracy we run each algorithm twice, first initialising $\Omega^{(0)}$ at the GLASSO-EBIC estimate and second at $\Omega^{(0)}=I$.
We then obtained the mean absolute difference in the MCMC estimates $\hat{\mathbb E}(\Omega_{jk} \mid \Y)$ and $\hat{\pi}(\Omega_{jk} \neq 0 \mid \Y)$ between the two chains, as accuracy metrics. 
Posterior means for \texttt{regression.pl} were calculated by averaging the lower and upper triangles as recommended by \cite{atchade2019quasi}. 
Figure \ref{fig:mixing_p100} shows these metrics for all methods 
for $p = 100$ variables and $n\in \{2p, 5p\}$ observations.
Fairly similar results were observed for $p=50$ and $p=200$, see Section \ref{Sec:mixingp50} and \ref{Sec:mixingp200}. 
Figure \ref{fig:mixing_p100_Serial} additionally demonstrates that the Gibbs and BDMH algorithms outperformed the LIT sampler. 

There are two key findings.
First, our algorithms (red, orange and blue dots) significantly improved upon other methods that target an exact Bayesian posterior (\texttt{bdgraph} and \texttt{ssgraph}, black dots).
For the same computation time, 
in most scenarios the difference in $\hat{\mathbb E}(\Omega_{jk} \mid \Y)$ and $\hat{\pi}(\Omega_{jk} \neq 0 \mid \Y)$ for GIMH (blue dots) was between 5 and 50 times smaller. 
The precision in $\hat{\mathbb E}(\Omega_{jk} \mid \Y)$ of
our Gibbs and BDMH algorithms with fixed $\theta$ (red dots)
was equal or slightly higher than for GIMH in most settings, 
and the precision in $\hat{\pi}(\Omega_{jk} \neq 0 \mid \Y)$ was particularly good for Gibbs. 
  
Figure \ref{fig:GIMH_GibbsBDMH_p100} demonstrates that GIMH is improved in the illconditioned example when $\nu = 0.5$.  
The Beta-Binomial and empirical Bayes model priors (orange dots) did not affect the performance of Gibbs to a meaningful extent.
\color{black}
Upon further inspection, the good performance of Gibbs was due to using a Rao-Blackwellised estimator of $\pi(\Omega_{jk} \neq 0 \mid \Y)$. 
That is, rather than using the average of the MCMC edge inclusion indicators, in Figure \ref{fig:mixing_p100} we used the average of the edge inclusion probabilities across MCMC iterations.
Figure \ref{fig:mixing_p100_RB} shows results when using the MCMC average, and there Gibbs is no longer more precise.
That is, it is not that Gibbs is more efficient generally, but that it leads to a particularly efficient Rao-Blackwellisation for $\pi(\Omega_{jk} \neq 0 \mid \Y)$.
The second key finding is that (as expected) the most computationally efficient algorithms were those based on pseudo-likelihoods (regression.pl, bdgraph.mpl).
This said, our exact Bayesian algorithms showed similar run times for a given precision in most cases, the largest differences occurring in the ill-conditioned scenario.

\subsection{Posterior Inference}{\label{Sec:Data}}

We next compare the quality of the inferences across methods, including the GLASSO-BIC and GLASSO-EBIC.
For all Bayesian methods we initialised the chains at the GLASSO-EBIC estimate and performed 15,000 iterations, discarding the first 5,000 for warmup. 



\begin{figure}[hbt!]
\begin{center}
\includegraphics[width =0.95\linewidth]{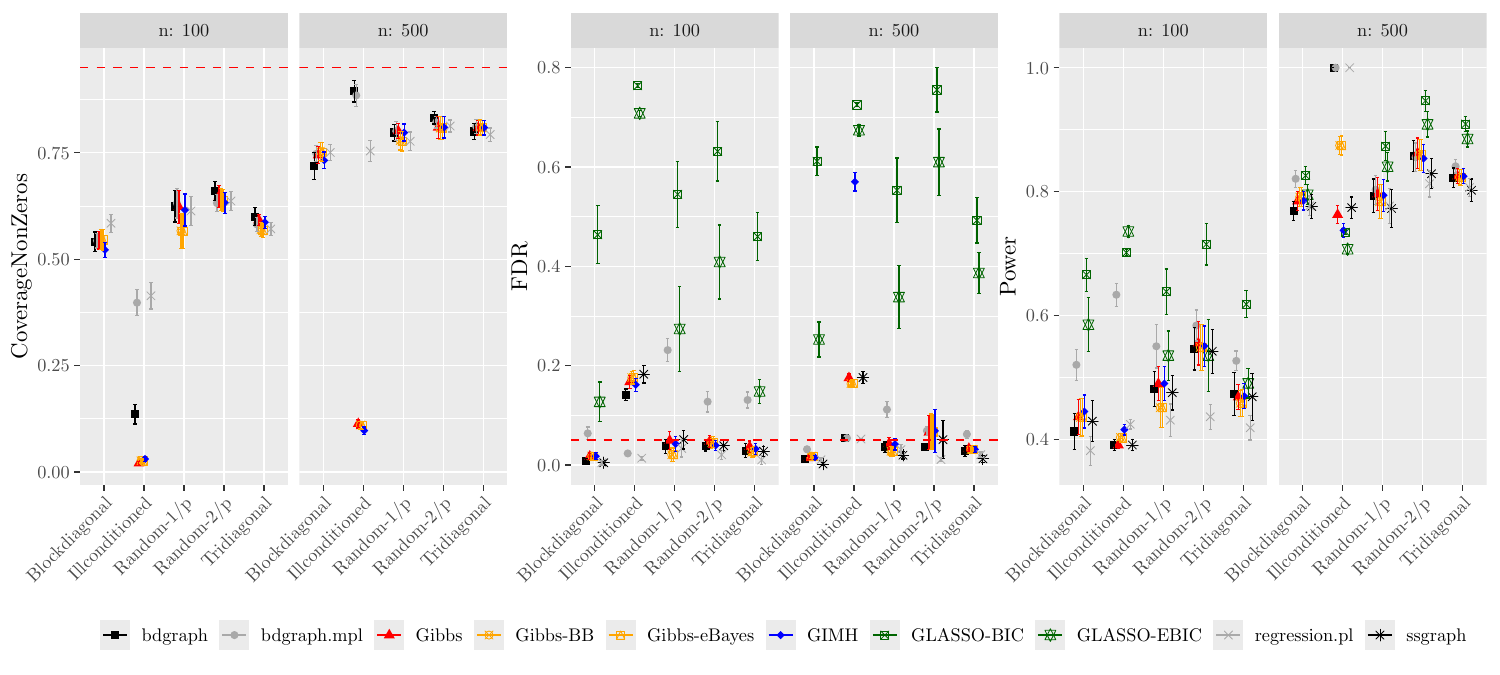}
\caption{
Simulations. Quality of inference for GLASSO-BIC, GLASSO-EBIC, ssgraph, bdgraph, bdgraph.mpl, regression.pl, Gibbs (with fixed $\theta$, $\theta\sim \text{Beta}(1, 1)$ and Empirical Bayes) and GIMH ($w = 0.75$)  for  $p = 100$ and $n \in \{p, 2p, 5p\}$. Higher power, CoverageNonZeros (average coverage of non-zero entries) and smaller FDR are better. Intervals are 2 standard errors. Red dotted lines indicate the target coverage=0.95 and FDR=0.05.
}
\label{fig:Inference_p100}
\end{center}
\end{figure}

For $p = 100$, Figure \ref{fig:Inference_p100} compares the coverage of 95\% posterior credible intervals for the truly non-zero $\Omega_{jk}^0$ (left panel), the frequentist false discovery rate (FDR, middle panel) and power (right panel) when thresholding $\hat{\pi}(\Omega_{jk} \neq 0 \mid \Y)$ such that the average inclusion probability of declared edges was greater than 0.95, controlling the Bayesian FDR below 0.05 \citep{mueller:2004}. 
Figure \ref{fig:Inference_p100_additional}  additionally compares the ROC-AUC and mean absolute error (MAE) of $\hat{\Omega}_{jk}$ of each method. 
Reassuringly, Gibbs and GIMH produce almost identical inferences to the other exact Bayesian bdgraph and ssgraph, while as discussed, being more computationally efficient.
The performance of the pseudo-posterior methods was not drastically different to the exact posterior methods. However, regression.pl showed slightly reduced power and coverage, as was also observed in Figure \ref{fig:exact_vs_pseudo}. 
The FDR is also marginally larger for bdgraph.mpl than the other methods.  
In general, all Bayesian methods provided a good FDR control, as expected given the sparsity induced by the prior specification.
As a consequence power was low, and hence the coverage was also below the nominal 0.95 level. Note that both power and coverage improved noticeably as $n$ grows. 
  
The Beta-Binomial and empirical Bayes priors achieved very similar inference to the Binomial fixed $\theta$ prior, providing marginally lower coverage and power, but marginally improved FDR.
\color{black}


\section{COVID-19 Data}{\label{sec:COVID}}

We consider a dataset from \cite{jewson2024graphical} with COVID-19 infection rates from 01/2020 to 11/2021 
($n = 97$ weeks)
across $p=332$ meta-counties in the US. 
The meta-counties result from grouping neighboring counties that have small populations until all counties had $\geq 500,000$ inhabitants. 
The authors regressed county log-infection rates (log infections divided by the county's population) on temperature, vaccination rates, an index measuring the stringency of pandemic measures, 
a weekly fixed-effect term estimating the mean infections across all counties in that particular week, and a first-order auto-regressive term measuring the infection rate in the previous week. 
The goal was to learn in what counties COVID-19 co-evolved beyond what's explainable by these covariates, i.e., to identify non-zero partial correlations in the regression residuals across counties.
The authors showed the error normality assumption was reasonable, 
see \cite{jewson2024graphical} for details. 


Table \ref{Tab:Covid} presents the results of our Gibbs algorithm    with Binomial, Beta-Binomial and empirical Bayes model priors, \color{black} as well as bdgraph.mpl, regression.pl, GLASSO-BIC and GLASSO-EBIC. The number of MCMC iterations used and convergence diagnostics are presented in Section \ref{Sec:COVID_diagnostics}.
As before, the estimates $\hat{\pi}(\Omega_{jk} \neq 0 \mid \Y)$ are thresholded to control the Bayesian FDR below 0.05 \citep{mueller:2004}.
The Bayesian methods selected many fewer edges than GLASSO-BIC, whereas GLASSO-EBIC selected no edges, i.e., the GLASSO was very sensitive to the regularisation parameter.    The Beta-Binomial and empirical Bayes model priors resulted in marginally more edges than when $\theta$ is fixed. \color{black}
regression.pl estimated fewer edges than our Gibbs and bdgraph.mpl estimated more. 

To evaluate the quality of these inferences we conducted a 10-fold cross validation, using the posterior mean estimated on 90\% of the data to predict a held-out 10\%. We evaluated the average log-likelihood 
of the held-out data
evaluated at $\hat{\Omega}$ obtained in the training data,
and also the squared correlation between observations and out-of-sample predictions, i.e., $R^2=\mbox{cor}(\Y, \hat{\Y})^2$  
where $\hat{y}_{ij} = \sum_{k=1, k\neq j}^p-\frac{\hat{\Omega}_{jk}}{\hat{\Omega}_{jj}}y_{ik}$. 
Our Gibbs algorithm    with fixed $\theta$ \color{black} outperformed the two pseudo-Bayesian methods in both metrics (Table \ref{Tab:Covid}). 
These results support the simulation results that, in some situations, bdgraph.mpl may show inflated FDR and regression.pl reduced power relative to exact Bayesian methods. 
  
Under the Beta-Binomial and empirical Bayes priors Gibbs sampling achieved lower held-out  log-likelihood, but higher $R^2$ than the fixed $\theta$ case.  
\color{black}
GLASSO showed the best out-of-sample predictive accuracy when its regularisation was selected with the BIC, albeit requiring an order of magnitude more edges to do so, and worst when selected with the EBIC. 
These findings agree with our simulations, where GLASSO exhibited good predictive accuracy but its FDR was too high for hypothesis testing standards.

The GLASSO run times were excellent however, which in line with our simulations suggests that if one only seeks sparse point estimates then GLASSO is highly competitive.
The run time of our Gibbs algorithm was comparable to that of the pseudo-Bayesian methods, supporting that it is possible to conduct fully Bayesian inference in settings where only pseudo-Bayesian inference was previously feasible.

\begin{table}[ht]
\centering
\caption{COVID-19 data. Number of selected edges, computation time and 10-fold cross-validated log-likelihood and $R^2$ for our Gibbs sampler, bdgraph.mpl, regression.pl and GLASSO with regularisation parameter chosen with BIC and EBIC}
\begin{tabular}{lcccc}
 \hline
 Method & \# Edges & Time & log-likelihood & $R^2$\\ 
  \hline
 Gibbs  ($\theta = \frac{2}{(p-1)})$     & 270 & 3.62 hours & -416.93 & 0.426\\
 Gibbs  ($\theta \sim BB(1, 1)$)     & 288 & 2.81 hours & -426.04 & 0.427\\
 Gibbs  (e-Bayes)     & 289 & 2.69 hours & -425.36& 0.427\\
 bdgraph.mpl   & 361 & 4.41 hours & -420.34 & 0.403\\ 
 regression.pl & 149 & 1.80 hours & -463.52 & 0.380\\ 
 GLASSO-BIC    & 3728 & 1.4 seconds & -374.26 & 0.651 \\
 GLASSO-EBIC   & 0   & 4.8 seconds & -500.52 & 0\\
  \hline
\end{tabular}
\label{Tab:Covid}
\end{table}

\begin{figure}[!ht]
\centering
\includegraphics[trim= {0.0cm 0.5cm 0.0cm 0.5cm}, clip, width = 0.95\linewidth]{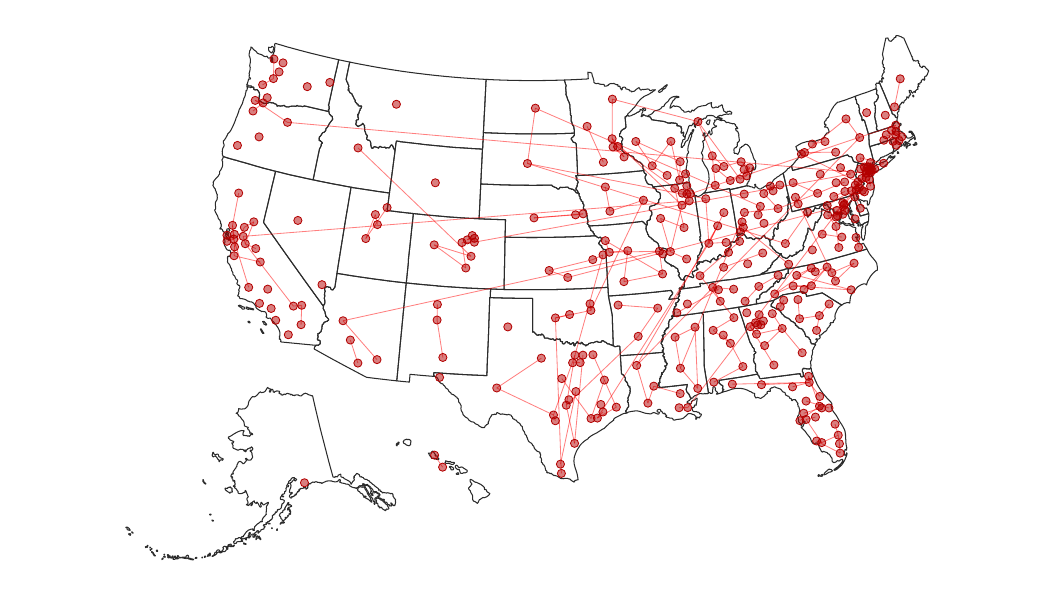}
\caption{Edges identified by  Gibbs (fixed $\theta$) after thresholding the posterior probabilities.}
\label{fig:edges_SerialGibbs}
\end{figure}

Figure \ref{fig:edges_SerialGibbs} plots the edges selected by our Gibbs algorithm on a USA map. Most edges are between geographically close counties, as expected given that COVID-19 spreads locally, but there are also some edges between distant West and East coast counties.
There is literature explaining such distant relations in COVID-19 infections.
For example, \cite{kuchler:2021} argued that they may be related to social media connections, which reflect people having similar backgrounds and being exposed to similar information, and hence possibly having similar attitudes towards health prevention.
Figures \ref{fig:edges_bdgraph.mpl} and \ref{fig:edges_regression.pl} compare these edges to those selected bdgraph.mpl and regression.pl respectively. 
The edges selected by regression.pl are mostly a subset of those selected by Gibbs, which are mostly a subset of those estimated by bdgraph.mpl.
That is, the three solutions differ in their overall sparsity, but they are otherwise structurally similar.

\section{Discussion}
\label{sec:discussion}

We showed that it is possible to conduct fully Bayesian inference in dimensions where previously only pseudo-Bayesian methods were feasible, at similar run times.
While the latter showed excellent performance in most examples, 
they do not provide fully Bayesian uncertainty quantification.
To be clear, we do not argue that one should necessarily prefer fully Bayesian over pseudo-Bayesian inference, simply that there is no longer a computational bottleneck for doing so.
Our algorithms and theory rely on having closed-form expressions for conditional marginal likelihoods of row-wise edge configurations, given the data and the rest of the precision matrix, and on the posterior distribution on edge inclusions being sparse.
This enables local moves changing a single edge and global moves updating multiple edges in a row that have low complexity (number of operations). 
   As future research, one could consider adaptive or delayed-acceptance strategies combining local and global moves. 

Our software implements all these local (BDMH, Gibbs and LIT) and global proposals. Our empirical results suggest that Gibbs is a good default choice.
Global moves may be preferred when the posterior is multi-modal but without such indication a-priori, we recommend to first try Gibbs and if one observes slow mixing, then to consider global moves. 
In our examples, a simple mixing diagnostic was running a single chain and plotting the running estimate of posterior edge probabilities $\hat{\pi}(Z_{ij}=1 \mid \Y)$. 
Effective sample sizes were a less useful diagnostic, for example if $\pi(\Omega_{ij} = 0 \mid \Y)=1$ (the posterior is concentrated) then ideally $\Omega_{ij}^{(t)}=0$ for all MCMC iterations, but the ESS would be 1. 
Another useful diagnostic was running two independent chains initialised at the null model (no edges) and via GLASSO, and comparing the obtained $\hat{\pi}(Z_{ij}=1 \mid \Y)$.
\color{black}


   A main message is that, although discrete problems such as variable selection or edge selection are worst-case NP hard, with high probability they can be solved in polynomial time, when the posterior  concentrates in suitable ways as $n \to \infty$.
To guarantee such concentration, somewhat stringent sparsity and betamin conditions are needed. 
Ours are on par with the best known for regression  \citep{yang2016computational, zhou2022dimension},
and they cannot be improved much in general: there are worst-case settings where they are basically necessary.
In practice one is often not in such a worst-case setting. 
In our examples we observed good mixing, and better inference, for less sparse prior choices, also when the posterior was not concentrated. As is common in the literature, there is a gap between sufficient assumptions and those needed in practice. 
Future research would be welcome.
\color{black}

It would also be interesting to consider graphical models beyond Gaussianity. 
   For example, \cite{florez:2025} extended the GGM continuous spike-and-slabs 
to graphical models for mixed data types. 
Our discrete spike-and-slab can be similarly extended. We hence hope that our framework provides a basic workhorse for graphical models. 
\color{black}


\section*{Acknowledgements}

We thank Reza Mohamadi and Willem van dem Boom for advice on using their R packages. We also thank Samuel Power  and Filippo Ascolani for useful discussions.
JJ was partially funded by Juan de la Cierva Formación de la Agencia Estatal de Investigación  FJC2020-046348-I,
and DR by grant 2025 ICREA 00045 (Dept. of Research and Universities of the Government of Catalonia, Academy of Excellence Programme),
grant PID2022-138268NB-I00 by MCIN/AEI/10.13039/501100011033 / FEDER (EU),
and grant Consolidaci\'on investigadora CNS2022-135963 by the AEI.



{\bf Disclosure statement.} There are no conflicts of interest.

{\bf Data availability statement.} 
The COVID-19 data is available as supplementary material. The raw data is at \url{https://github.com/CSSEGISandData/COVID-19}.

{\bf Supplementary material.} The supplement contains all proofs, implementation details, additional empirical results, and R code to reproduce our examples.









\appendix

\renewcommand{\thefigure}{S\arabic{figure}}
\renewcommand{\thetable}{S\arabic{table}}
\renewcommand{\thesection}{S\arabic{section}}
\setcounter{table}{0}
\setcounter{figure}{0}
\setcounter{section}{0}

\renewcommand{\thelemma}{S\arabic{lemma}}
\renewcommand{\theprop}{S\arabic{prop}}
\renewcommand{\thethm}{S\arabic{thm}}

\section{Effect of positive-definiteness truncation on prior interpretation}
\label{sec:pdtrunc_prior}

Following \cite{Wang2015ScalingIU}, our Binomial and Beta-Binomial priors $\pi(Z)$ in \eqref{eq:prior-z}
\eqref{eq:prior-z-bbin} on the graphical model structure both 
feature a term $1/C_1(Z, g_1, \lambda)=
\tilde \pi(\Omega \in \mathcal{M}_{+}(p) \mid Z, g_1, \lambda)$, the  probability that $\Omega$ is positive-definite under a prior $\tilde \pi$ such that the non-zero entries in $\Omega$ given by $Z$ are drawn independently from the Gaussian slab with variance $g_1^2$, and its diagonal entries from an exponential with rate $\lambda/2$.
For example, recall that our Binomial prior is
$$\pi(Z) \propto \frac{1}{C_1(Z, g_1, \lambda)} \prod_{i < j} \theta^{z_{ij}} (1 - \theta)^{1- z_{ij}} \mathds{1}_{d(Z) \leq \bar d}.$$

Adding $C_1$ to $\pi(Z)$ is computationally convenient because $\pi(\Omega \mid Z) = C_1(Z, g_1, \lambda) \tilde{\pi}(\Omega \mid Z) \mathds{1}_{\Omega \in \PSD}$, and therefore $\pi(\Omega, Z)$ and $\pi(\Omega, Z \mid \Y)$ do not depend on $C_1(Z, g_1, \lambda)$.
This is also appealing in that it facilitates understanding the behavior of $\pi(Z \mid \Y)$.
For example, under an alternative prior $\pi'(Z) \propto \prod_{i < j} \theta^{z_{ij}} (1 - \theta)^{1- z_{ij}} \mathds{1}_{d(Z) \leq \bar d}$ not featuring $C_1$, the posterior probabilities over graphs are 
$\pi'(Z \mid \mathbf Y) \propto C_1(Z, g_1, \lambda) p(\mathbf Y\mid Z) \pi'(Z)$, where $p(\mathbf Y \mid Z)$ is the marginal likelihood under $Z$. Understanding the behavior of $\pi'(Z \mid \Y)$, including posterior concentration and uncertainty quantification, is significantly complicated by $C_1$. 
This said, one may worry that if $C_1$ is either too large or too small, then $\pi(Z)$ and $\pi(\Omega)$ could be dramatically different from $\pi'(Z)$ and $\pi'(\Omega)$. This may be undesirable in that $\pi(Z)$ becomes harder to interpret than $\pi'(Z)$.
To summarise, these issues point towards an interesting trade-off: adding $C_1$ to $\pi(Z)$ makes the prior less interpretable but the posterior easier to characterize, whereas the contrary happens when dropping $C_1$ from $\pi(Z)$.

Fortunately,\cite{carter:2026} provided results that that describe the behavior of $C_1$ and make it simple to specify the prior parameters such that $C_1$ is provably close to 1, and hence that $\pi(Z)$ remains interpretable. 
We discuss first the behavior of $C_1$ and subsequently give a specific recipe for bounding $C_1$.
Recall that $C_1(Z, g_1, \lambda)= 1/\tilde \pi(\Omega \in \mathcal{M}_{+}(p) \mid Z, g_1, \lambda) \geq 1$, hence it suffices to find an upper-bound for $C_1$ that is close to 1.
\cite{carter:2026} showed that:

\begin{enumerate}[leftmargin=*]
    \item $C_1(Z, g_1, \lambda)$ (essentially) increases when adding edges to $Z$. More precisely, if $(Z,Z')$ are such that $z_{ij} \leq z_{ij}'$ for all $(i,j)$, then the expectation of the smallest eigenvalue of $\Omega$ under $Z$ is less or equal than that under $Z'$, and further the distribution of said eigenvalues concentrate on their expectations.
    The implication is that introducing $C_1(Z, g_1, \lambda)$ into $\pi(Z)$ results in assigning higher prior probability to sparser models.

    \item $C_1(Z, g_1, \lambda)$ is a strictly decreasing function of $g_1$ and of $\lambda$.
    Therefore, to keep $C_1$ close to 1 either the variance $g_1^2$ of off-diagonal $\Omega_{ij}$ should be small, or the mean $2/\lambda$ of the diagonal entries $\Omega_{ii}$ should be large.

    \item The total variation distance and Kullback-Leibler divergence between an untruncated prior $\tilde{\pi}(\Omega)$ and its positive-definiteness truncated counterpart $\pi(\Omega)$ is a simple function of $\tilde{\pi}(\Omega \not\in \mathcal{M}_+(p))$. 
    Specifically, let $\tilde{\pi}(\Omega, Z)= \tilde{\pi}(\Omega \mid Z) \tilde{\pi}(Z)$ such that $\tilde{\pi}(\Omega \mid Z)$ does not constrain $\Omega$ to be positive-definite, and $\tilde{\pi}(Z)$ is an arbitrary prior, and let $\tilde{\pi}(\Omega)= \sum_Z \tilde{\pi}(\Omega \mid Z) \tilde{\pi}(Z)$ be the implied marginal prior on $\Omega$. 
Let $\pi(\Omega,Z) \propto \tilde{\pi}(\Omega,Z) \mathds{1}_{\Omega \in \mathcal{M}_+(p)}$ be a positive-definiteness truncated prior such as ours, and $\pi(\Omega)$ the corresponding marginal prior.
Then, 
\begin{align}
&\mbox{TV}(\tilde{\pi}, \pi)=
\tilde{\pi}(\Omega \not\in \mathcal{M}_+(p))
\nonumber \\
&\mbox{KL}(\tilde{\pi}, \pi)= 
-\log \left(\tilde{\pi}(\Omega \in \mathcal{M}_+(p)) \right).
\nonumber
\end{align}
where $\tilde{\pi}(\Omega \in \mathcal{M}_+(p))$ is strictly decreasing in $g_1^2$ and in $\lambda$.
This implies that, by setting small enough $g_1$, the prior beliefs implied by $\tilde{\pi}(\Omega)$ and $\pi(\Omega)$ are similar.

\end{enumerate}

In terms of obtaining explicit bounds for $C_1$ for any problem dimension $p$, Theorem \ref{thm:lowerbound_c1} reproduces Theorem 3 (iii) in \cite{carter:2026}.
We denote by $d(Z)$ the maximum node degree of $Z$ (number of variables connected to a single node).

\begin{thm}[Theorem 3 (iii) in \cite{carter:2026}]
\label{thm:lowerbound_c1}
Let $\Omega_{ii} \sim \mbox{Exp}(1/\mu)$, $\Omega_{ij} \mid Z_{ij} = 1 \sim N(0,g_1^2)$ and $\Omega_{ij} \mid Z_{ij}=0= \delta_0(\Omega_{ij})$ independently across $i \leq j$. Then
$$
\frac{1}{C_1(Z,g_1,\lambda)} \geq 1 - e^{- p t / \mu} + 2 p e^{-\frac{t^2}{2 g_1^2 d(Z)}}
$$
for any $t>0$.
In particular, assume that $ g_1 \sqrt{d(Z)} / \mu \leq 2\sqrt{2}$, then
\begin{align}
\frac{1}{C_1(Z,g_1,\lambda)} 
\geq 1 -   \frac{p g_1 \sqrt{2 d(Z)}}{\mu} \left[ \sqrt{\ln\left(\frac{2\sqrt{2} \mu}{g_1 \sqrt{d(Z)}}\right)} + 1 \right].
\nonumber
\end{align}

Further, if 
$\mu \geq \frac{2 g_1 p \sqrt{d(Z)}}{\alpha} \sqrt{- W_{-1}\left( \frac{- \alpha^2}{32 e^2 p^2} \right)}$
then $\frac{1}{C_1(Z,g_1,\lambda)} \geq 1 - \alpha$, where $W_{-1}$ is the branch of Lambert's $W_q$ function associated to $q=-1$.
\end{thm}

Theorem \ref{thm:lowerbound_c1} implies that as the diagonal mean $\mu$ goes to $\infty$ (equivalently, the rate parameter $\lambda \to 0$), then $C_1$ converges to 1 (keeping all other parameters fixed). That is, for any $Z$ and $g_1$, we have that
\begin{align}
C_1(Z, g_1, \lambda=0) = 1.
    \nonumber
\end{align}
Note that $\lambda=0$ corresponds to setting an improper flat prior on $\Omega_{ii}$, that is $\pi(\Omega_{ii}) \propto 1$.
If one wishes to avoid improper priors, Theorem \ref{thm:lowerbound_c1} shows how to choose sufficiently large $\mu$ such that $1/C_1(Z,g_1,\lambda) > 1 - \alpha$ for any user-specified $\alpha$ and all $Z$ such that $d(Z) \leq \bar d$.



\color{black}

\section{MCMC algorithms and computational complexity}
\label{sec:mcmc_algorithms}

\subsection{Ideal Gibbs sampler and local proposal algorithms}

We first discuss the computational cost of the ideal \deb{Gibbs} sampler in Algorithm \ref{alg:gibbs_GGM_zeros}.
We subsequently outline the Gibbs sampler for $\pi(z \mid \Y, \Omega_{-j-j})$
in Algorithm \ref{alg:nested_gibbs}, 
the BDMH sampler in Algorithm \ref{alg:birth-death} and also outline the LIT sampler. Finally, we discuss their computational cost.

\deb{The ideal random scan Gibbs} Algorithm \ref{alg:gibbs_GGM_zeros} requires a total of
\deb{$O(p^3 + T ( p^2 + \bar{z}_0^3))=
O(p^3 + T  ( p^2 + \bar d^3))$} operations, 
where $\bar{z}_0^3$ is the average of $|z|_0^3$ across the visited models $z$ in all iterations.
In contrast, the continuous spike-and-slab requires $O(T p^4)$ operations.
The $p^3$ term is from inverting $\Omega^{(0)}$, done only in the first iteration.
The remaining cost is from sampling $u_1$ for the $p$ columns.
One obtains the inverse of the Cholesky decomposition of $U_z=(S_{jj}+\lambda)
\Sigma_{zz \mid j} + g_1^{-2} I_z$, where recall that
$\Sigma_{zz \mid j}=([\Omega_{-j,-j}]^{-1})_{zz}$. 
This requires $[\Omega_{-j,-j}]^{-1}$  which, using fast one-rank updates 
(see Lemma \ref{lem:update_inv} in Section \ref{app:lemmas}) requires $O(p^2)$ operations. 
Then, the Cholesky decomposition requires $O(|z|_0^3)=O(\bar d^3)$ operations. 
For the continuous spike-and-slab prior, each column update in \cite{Wang2015ScalingIU} requires sampling from a $(p-1)$-dimensional Gaussian, which has cost $O(p^3)$.

\begin{algorithm}
\caption{Model Gibbs sampling}\label{alg:nested_gibbs}
\KwIn{ Initial state $z$.}
\KwOut{$z^*$, an approximate sample from $\pi( z \mid \Y, \Omega_{-j,-j})$.} 
Initialisation: $z^* = z$. \\
\For{$k=1, \ldots, p-1$}{

Sample $z_k^* \sim \pi(z_k \mid z^*_{-k}, \Y, \Omega_{-j,-j})$ \\ 

}
Return $z^*$
\end{algorithm}

\begin{algorithm}
\caption{Birth-Death Metropolis-Hastings sampling}\label{alg:birth-death}
\KwIn{ Initial state $z$, number of moves $M$.}
\KwOut{$z^*$, an approximate sample from $\pi( z \mid \Y, \Omega_{-j,-j})$.} 
Initialisation: $z^{(0)} = z$. \\
\For{$t =1, \ldots, M$}{
Sample $\bar z \sim Q_{BD}(z \mid z^{(t-1)})$ \\
With probability $b$ given by \eqref{eq:acc-ratio-bd}, set $z^{(t)} = \bar z$, and with probability $1-b$ set $z^{(t)} = z^{(t-1)}$.
}
Set $z^* = z^{(M)}$.
\end{algorithm}

The Gibbs sampler is described in Algorithm \ref{alg:nested_gibbs}.
For each coordinate $k \in [p-1]$, the distribution  $\pi(z_k \mid z_{-k}, \Y, \Omega_{-j,-j})$ is a Bernoulli distribution with mean $\frac{1}{1 + r_k}$ and
\begin{align*}
    r_k &=  \frac{\pi(z^- \mid  \Y, \Omega_{-j,-j})}{\pi(z^+  \mid  \Y, \Omega_{-j,-j})}  =  \frac{g_1(1 - \theta)|U_{z^+}|^{1/2}}{\theta |U_{z^-}|^{1/2}}e^{\frac{m_{z^-}^T U_{z^-} m_{z^-}}{2} - \frac{m_{z^+}^T U_{z^+} m_{z^+}}{2}},
\end{align*}
where $z^- = (z_{1}, \ldots, 0, \ldots, z_{p-1})$, $z^+ = (z_{1}, \ldots, 1, \ldots, z_{p-1})$. 

The BDMH algorithm is given in Algorithm \ref{alg:birth-death}.
A new state $z^*$ is proposed such that it differs from  the previous state $z^{(t-1,r-1)}$ by adding or deleting an edge, or swapping a previously active for a previously inactive edge.
Then $z^*$ is accepted as the new state, i.e., $z^{(t-1,r)} = z^*$, with probability given by
    \begin{align}\label{eq:acc-ratio-bd}
        b = 1 \wedge \frac{\pi( z^* \mid \Y, \Omega_{-j,-j}) Q_{BD}(z^{(t-1,r-1)} \mid z^*)}{\pi(z^{(t-1,r-1)} \mid \Y, \Omega_{-j,-j}) Q_{BD}(z^* \mid z^{(t-1,r-1)})},
    \end{align}
where $Q_{BD}(\cdot \mid\cdot)$ is the Birth-Death-Swap proposal distribution from which $z^*$ is sampled from, and with probability $1-b$,  $z^{(t-1,r)} = z^{(t-1,r-1)}$. 

For clarity, we recall the definition of $Q_{BD}(\cdot \mid\cdot)$. Let $d_H(z,z^{\ast}) := \sum_{k=1}^{p-1}\1(z_k \neq z^{\ast}_k)$ denote the Hamming distance between $z$ and $z^{\ast}$ and
\begin{align}
    Q_{BD}(z^{\ast} \mid z) := \begin{cases}
        \frac{p_{birth}}{p-1-|z|_0} &\textrm{ if } d_H(z,z^{\ast}) = 1 \textrm{, } |z^{\ast}|_0 = 1 + |z|_0 \textrm{ and } |z|_0 < p-1 \\
        \frac{p_{death}}{|z|_0} &\textrm{ if } d_H(z,z^{\ast}) = 1 \textrm{, } |z^{\ast}|_0 = |z|_0 - 1 \textrm{ and } |z|_0 > 0\\
        \frac{1 - p_{birth} - p_{death}}{|z|_0(p-1 - |z|_0)} &\textrm{ if } d_H(z,z^{\ast}) = 2 \textrm{, } |z^{\ast}|_0 = |z|_0 \textrm{ and } 1 < |z|_0 < p-1\\
        0 &\textrm{ otherwise}
    \end{cases} \label{eq:bdmh-proposal}
\end{align}
with $0 < p_{birth}, p_{death} < 1$ the probabilities of birth and death moves such that $p_{birth} + p_{death} \leq 1$.

Finally, in the LIT algorithm the proposal distribution considers the same models as the Birth-Death-Swap kernel \eqref{eq:bdmh-proposal}, but the probability of proposing a new model $z^*$ depends on its posterior probability $\pi(z^* \mid \Y, \Omega_{-j,-j})$. For $z^*$ corresponding to a birth move from $z$, its proposal probability is
\begin{align*}
    \frac{p_{birth} w_{birth}(z^*|z)}{Z_{birth}(z)},
\end{align*}
with $Z_{birth}(z) = \sum_{z' \in \mathcal{N}_{birth}(z)} w_{birth}(z'|z)$ with $\mathcal{N}_{birth}(z)$ the neighboring models of $z$ corresponding to birth moves and 
\begin{align*}
    w_{birth}(z'|z) = p_{birth, min} \vee \frac{\pi(z|\Y, \Omega_{-j,-j}) }{\pi(z|\Y, \Omega_{-j,-j}) } \wedge p_{birth, max},
\end{align*}
where $p_{birth, min} , p_{birth, max} \in [0,1]$, $p_{birth, min}\leq p_{birth, max}$ are minimum and maximum weights given to a neighboring model $z'$ of $z$.
Similar proposal probabilities are derived for death and swap moves. 
\cite{zhou2022dimension} recommend setting $p_{birth, min}$  and  $p_{birth, max}$ to $p^{-k}$ and $p^{k}$ with $k=1$ or 2 (and similarly for death and swap moves), and we used $k=1$ in our examples. 

We now argue why a single iteration of the Gibbs and BDMH algorithms requires
\deb{$O(p^2 + \bar{z}_0^3 + M \bar{z}_0^2)$= $O(p^2 + \bar d^3 + M \bar d^2)$} operations, 
where $\bar{z}_0$ is the maximum degree across the visited models,
and that a single LIT iteration with $M=1$ update requires 
\deb{$O(p^2 + \bar{z}_0^3 + p \bar{z}_0^2)$= $O(p^2 + \bar d^3 + p \bar d^2)$} operations.
The first application of the Gibbs or BDMH kernel requires the Cholesky decomposition of a sub-matrix of $\Omega_{-j,-j}$ of size $|z|_0 \times |z|_0$, which requires $O(|z|_0^3)$ operations. 
Subsequent kernel updates require only order $O(|z|_0^2)$ operations. 
Since each update adds/removes/swaps at most 1 entry of $z$, 
one needs to invert a sub-matrix $\Omega_{-j,-j}$ that differs from that in the previous step by one row/column, hence one may update the Cholesky decomposition using rank 1 updates that require $O(|z|_0^2)$ operations.

For the LIT the reasoning is similar. One must evaluate \eqref{eq:post_submodel} for all neighbor models obtained by adding/deleting one edge from the current $z$, and there are $p-1$ such models. For the first model considered one must obtain a Cholesky decomposition that requires $O(|z|_0^3)$ operations, and for subsequent models one can use rank 1 Cholesky updates, each of which requires $O(|z|_0^2)$ operations.

\subsection{Independent global proposal}
\label{ssec:global_proposal_suppl}

\begin{algorithm}
\caption{Independent global proposal Metropolis-Hastings algorithm}\label{alg:metropolis_gibbs}
\KwIn{ $\Y$, prior parameters $(\theta,g_1,\lambda)$, tempering parameter $\upsilon$, number of iterations $T$, number of proposal iterations $T'$, initial value $\Omega^{(0, 0)}$.}
\KwOut{$\{(\Omega^{(t)}, Z^{(t)})\}_{t \in [T]}$.} 
\SetKwBlock{DoParallel}{do in parallel for each $j = 1, \dots, p$}{end}
\DoParallel{
Run a Markov Chain for $T'$ iterations with invariant distribution $[\pi_j^{LR}(\cdot \mid \Y)]^\upsilon$.

   Denote by $\mathcal{Z}_j \subset \{z_j^{(t)}\}_{t \in [T']}$ a subset of the produced samples such that $\mathcal{Z}_j$ can be viewed as independent draws from $[\pi_j^{LR}(z \mid \Y)]^\upsilon$ (e.g., discarding a burn-in and thinning the remaining draws). \color{black}
}
\For{$t=1,\ldots,T$}{
Sample $j \in [p]$ uniformly at random, and let $\Omega_{-j,-j}=\Omega_{-j,-j}^{(t-1)}$, 
$\Sigma_{-j,-j|j}= [\Omega_{-j,-j}]^{-1}$.

Sample $u_2^* \sim \pi(u_2 \mid \Y) = \text{Ga}(u_2; \frac{n}{2}+1, \frac{S_{jj} + \lambda}{2})$.

Propose $z^*$ by drawing an entry from $\mathcal{Z}_j$ uniformly at random.
With probability $b_{acc}$ given by \eqref{eq:acc-ratio-lr}, set $Z_{-j, j}^{(t)}=z^*$ and otherwise set $Z_{-j, j}^{(t)}= Z_{-j, j}^{(t-1)}$

Sample $u_1^* \sim \pi(u_1\mid Z_{-j, j}^{(t)}, \Y, \Omega_{-j,-j}) = \text{N}(u_1; \: m_{Z_{-j, j}^{(t)}}, U_{Z_{-j,j}^{(t)}}^{-1})$, where $(m_{Z_{-j, j}^{(t)}}, U_{Z_{-j, j}^{(t)}})$ are given in Proposition \ref{prop:wang_exactzeroes_notation}.

Set $\Omega_{\cdot j}^* = (\Omega_{-j,j}^{*}, \Omega_{jj}^{*})$ using 
\begin{align*}
    &\Omega_{Z^{(t)}, j}^{*} = -u^{\ast}_{1} \\
    &\Omega_{kj}^{*} = 0  \quad \text{ if }  Z_{k,j}^{(t)} = 0, \quad k = 1, \dots, p-1. \\
    &\Omega_{jj}^{*} = u^{\ast}_2 + [u^{\ast}_1]^T\Sigma_{Z_{-j, j}^{(t)},Z_{-j, j}^{(t)} \mid j} u^{\ast}_1
\end{align*}
}
\end{algorithm}

Algorithm \ref{alg:metropolis_gibbs} describes the independent global proposal Metropolis-Hasting algorithm (GIMH).
Note that $\Omega_{.j}^*$ is always updated given the edge inclusion indicator $Z_{\cdot j}^{(t)}$, regardless of whether the latter was accepted in the Metropolis-Hastings step.
  
More critically, the algorithm assumes that one has access to a set of independent samples $\mathcal{Z}_j$ from $[\pi_j^{LR}(z \mid \Y)]^\upsilon$. In practice, unless $p$ is small, one must use MCMC to obtain approximate dependent samples from $[\pi_j^{LR}(z \mid \Y)]^\upsilon$. Algorithm \ref{alg:metropolis_gibbs} assumes that this is done as a pre-computation step, where one may use convergence diagnostics and thinning to make the independent samples assumption tenable.
The next section describes an alternative dependent global proposal algorithm that bypasses this requirement.

\subsection{Beyond independent global proposals}
\label{ssec:global_proposal_dependent}

The algorithm in Section \ref{ssec:global_proposal_suppl} requires independent samples the tempered linear regression posterior $[\pi_j^{LR}(\cdot \mid \Y)]^\upsilon$ of each column $j=1,\ldots,p$, where $\upsilon > 0$ is the tempering parameter.
This assumes that, for each column, one has run an MCMC algorithm targeting the linear regression posterior for a sufficiently large time to ensure convergence, and that thinning was used to ensure that the proposal samples are practically independent. 
We now describe two alternative algorithms that allow for global moves and that do not require the linear regression MCMC to have converged or to produce independent samples.
First, Algorithm \ref{alg:local_global} replaces the global proposal by a local-global proposal. It also obtains samples $\mathcal{Z}_j$ from the tempered linear regression proposal for each column $j$ as a pre-computation step, and it computes its re-normalized empirical distribution. At iteration $t$ of the main algorithm, $Z_{-j,j}^*$ is proposed from this empirical distribution with probability $p_{glob}$, and otherwise $Z_{-j,j}^*$ is sampled from a local proposal $h_j$ that depends on $Z_{-j,j}^{(t-1)}$ (e.g., Gibbs, birth-death-swap, or LIT). Relative to Algorithm \ref{alg:metropolis_gibbs}, the MH acceptance probability is adjusted to account for the density of the local-glocal proposal, as usual.
Algorithm \ref{alg:local_global} allows global moves between particles in $\mathcal{Z}_j$. Note that, if the current $Z_{-j,j}^{(t-1)} \not\in \mathcal{Z}_j$ and also not in the set of neighbors of that can be reached by taking a local move $h_j$ from $\mathcal{Z}_j$, then the acceptance probability $b_{acc}=0$.

\begin{algorithm}
\caption{Local-global proposal Metropolis-Hastings algorithm}\label{alg:local_global}
  
\KwIn{ $\Y$, prior parameters $(\theta,g_1,\lambda)$, global proposal probability $p_{glob}$, tempering parameter $\upsilon$, number of iterations $T$, number of proposal iterations $T'$, initial value $\Omega^{(0, 0)}$.}
\KwOut{$\{(\Omega^{(t)}, Z^{(t)})\}_{t \in [T]}$.} 
\SetKwBlock{DoParallel}{do in parallel for each $j = 1, \dots, p$}{end}
\DoParallel{
Run a Markov Chain for $T'$ iterations with invariant distribution $[\pi_j^{LR}(\cdot \mid \Y)]^\upsilon$.

Let $\mathcal{Z}_j = \{z_j^{(t)}\}_{t \in [T']}$ be the produced samples, $w_j(z)= [\pi_j^{LR}(z \mid \Y)]^\upsilon / \sum_{z' \in \mathcal{Z}_j} [\pi_j^{LR}(z' \mid \Y)]^\upsilon$ the self-normalised weights for $z \in \mathcal{Z}$,
and $W_j= \sum_{z \in \mathcal{Z}_{j}} w_j(z) \delta_{z}$ the corresponding self-normalized distribution.
}
\For{$t=1,\ldots,T$}{
Sample $j \in [p]$ uniformly at random, and let $\Omega_{-j,-j}=\Omega_{-j,-j}^{(t-1)}$, 
$\Sigma_{-j,-j|j}= [\Omega_{-j,-j}]^{-1}$.

Sample $u_2^* \sim \pi(u_2 \mid \Y) = \text{Ga}(u_2; \frac{n}{2}+1, \frac{S_{jj} + \lambda}{2})$.

Propose $z^* \sim W_j$  with probability $p_{glob}$, and $z^* \sim h_j(Z_{.j}^{(t-1)}, \Omega_{-j,-j}^{(t-1)})$ with probability $1 - p_{glob}$.

Set $Z_{-j,j}^{(t)}=z^*$ with probability $b_{acc}=$ 
\begin{align*}
    &  1 \wedge \frac{\pi( z^* \mid \Y, \Omega_{-j,-j})}{\pi(z \mid \Y, \Omega_{-j,-j})}
    \frac{p_{glob} w_j(Z_{-j,j}^{(t-1)}) + (1 - p_{glob}) h_j(Z_{-j,j}^{(t-1)} \mid z^*, \Omega_{-j,-j}^{(t-1)})}{p_{glob} w_j(z^*) + (1 - p_{glob}) h_j(z^* \mid Z_{-j,j}^{(t-1)}, \Omega_{-j,-j}^{(t-1)}) }
\end{align*}
and otherwise set $Z_{-j, j}^{(t)}= Z_{-j, j}^{(t-1)}$

Sample $\Omega_{.j}^{(t)} \sim \pi(\Omega_{.j} \mid Z_{-j,j}^{(t)}, \Omega_{-j,-j}^{(t-1)}, \Y)$ and set $\Omega_{-j,-j}^{(t)}= \Omega_{-j,-j}^{(t-1)}$ as in Algorithm \ref{alg:metropolis_gibbs}.
}
\end{algorithm}

The second option that we consider is Algorithm \ref{alg:global_proposal_dependent}. It is based on defining an augmented model where variable inclusion indicators for the linear regression posteriors are sampled jointly with $(\Omega,Z)$, and using a proposal that allows swapping $Z$ with the variable inclusion indicators.
Specifically, we denote by $\gamma_j \in \{0,1\}^{p-1}$ the variable inclusion indicators for the regression posterior $[\pi_j^{LR}(\gamma_j \mid \Y)]^\upsilon$ associated to column $j$, and we let $\gamma= (\gamma_1,\ldots,\gamma_p)$. We define an augmented model
\begin{align}
 \pi(\Omega, Z, \gamma \mid \Y) \propto \pi(\Omega, Z \mid \Y) Q^\upsilon(\gamma \mid \Y),
\nonumber
\end{align}
where $Q^\upsilon(\gamma \mid \Y)= \prod_{j=1}^p [\pi_j^{LR}(\gamma_j \mid \Y)]^\upsilon$ is the product of tempered linear regression proposals.

We define an MCMC algorithm to sample from $\pi(\Omega, Z, \gamma \mid \Y)$, which in particular produces samples from $\pi(\Omega, Z \mid \Y)$, as desired. 
We describe a random scan version of the algorithm that updates a randomly selected column at each iteration, but one may also use serial Gibbs updating columns sequentially, or random sequence Gibbs updating them in random order. At iteration $t$, the algorithm first updates $\gamma_j^{(t)}$ using a kernel that targets $[\pi_j(\gamma_j \mid \Y)]^\upsilon$ such as Gibbs, birth-death-swap or LIT, akin to the pre-computation step in Algorithms \ref{alg:metropolis_gibbs} and \ref{alg:local_global}. Subsequently, a swap move between $Z_{.j}$ and the newly sampled $\gamma_j^{(t)}$ is proposed, and accepted-rejected via the usual MH probability. Note that the latter does not require evaluating the proposal density, since it is equal to 1 (the proposal is a point mass).

The fact that $\gamma_j^{(t)}$ is always updated before the swap proposal allows the marginal chain on $\gamma_j$ to move at each iteration, hence the proposed $Z_{-j,j}^*= \gamma_j^{(t)}$ may be far from $Z_{-j,j}^{(t-1)}$, allowing for global moves. 
We remark that Algorithm \ref{alg:global_proposal_dependent} is only interesting when sampling from $\pi_j^{LR}$ is simpler, either in terms of computational cost or mixing, than sampling from $\pi(Z_{-j, j} \mid \Omega_{-j,-j},\Y)$. For example, if the linear regression proposal $h_j(\gamma_j \mid \gamma_j^{(t-1)}, \Y)$ employs a local kernel such as Gibbs, birth-death-swap or LIT, then the number of required operations is essentially the same as that for applying that same local kernel to directly propose $Z_{-j,j}^*$ from $Z_{-j,j}^{(t-1)}$ directly.
In contrast, suppose that the linear regression proposal $h_j(\gamma_j \mid \gamma_j^{(t-1)}, \Y)$ is based on local-global moves of the type described in Algorithm \ref{alg:local_global}. That is, one runs an MCMC algorithm on $\gamma_j$ as a pre-computation step in parallel across $j$, and $h_j(\gamma_j \mid \gamma_j^{(t-1)}, \Y)$ either samples from their renormalised distribution with probability $p_{glob}$ or from a local kernel otherwise. Then, $h_j(\gamma_j \mid \gamma_j^{(t-1)}, \Y)$ allows for global moves that may mix differently relative to performing local updates that directly propose $Z_{-j,j}^*$ from $Z_{-j,j}^{(t-1)}$.

\begin{algorithm}
\caption{Dependent global proposal Metropolis-Hastings algorithm}\label{alg:global_proposal_dependent}

\For{$t=1,\ldots,T$}{

Choose a column $j \in \{1,\ldots,p\}$ uniformly at random. 

Sample $\gamma_j^{(t)} \sim h_j(\gamma_j \mid \gamma_j^{(t-1)}, \Y)$ where $h_j$ is an MCMC kernel targeting $[\pi_j^{LR}(\gamma_j \mid \Y)]^\upsilon$.

Propose a swap move $(Z_{-j, j}^*, \gamma_j^*)= (\gamma_j^{(t)}, Z_{-j, j}^{(t-1)})$.
Set $(Z_{-j, j}^{(t)}, \gamma_j^{(t)})= (Z_{-j, j}^*, \gamma_j^*)$ with probability 
\begin{align}
1 \wedge
\frac{\pi(Z_{-j, j}^* \mid \Y, \Omega_{-j,-j}^{(t-1)}) [\pi_j^{LR}(\gamma_j^*  \mid \Y)]^\upsilon}{\pi(Z_{-j, j}^{(t-1)} \mid \Y, \Omega_{-j,-j}^{(t-1)}) [\pi_j^{LR}(\gamma_j^{(t)} \mid \Y)]^\upsilon}
,
\nonumber
\end{align}
and otherwise set $Z_{-j, j}^{(t)}= Z_{-j, j}^{(t-1)}$.

Sample $\Omega_{\cdot j}^{(t)} \sim \pi(\Omega_{\cdot j} \mid \Y, Z_{\cdot j}^{(t)}, \Omega_{-j,-j}^{(t-1)})$ as in Algorithm \ref{alg:metropolis_gibbs}.

}

\end{algorithm}

\section{Empirical Bayes estimation of the prior inclusion probability}
\label{sec:ebayes}

Consider the Binomial prior $Z_{ij} \mid \theta \sim \mbox{Bern}(\theta)$.
Let $\eta= \log(\theta/[1-\theta])$ and consider a minimally informative prior (described below) $\pi(\eta)$.
We outline an algorithm to obtain a marginal posterior mode estimate
$\hat{\eta}= \arg\max_\eta \pi(\eta \mid \Y)=
\arg\max_\eta \log \pi(\Y \mid \eta) + \log \pi(\eta)$,
where
\begin{align}
\pi(\Y \mid \eta)= \sum_{Z} \pi(\Y \mid Z, \eta) \pi(Z \mid \eta)
\label{eq:marglhood_eta}
\end{align}
is the marginal likelihood of $\eta$,
$\pi(\Y \mid Z, \eta)$ the marginal likelihood for a given graphical structure $Z$, and $\pi(Z \mid \eta)= \prod_{i,j} (1/[1 + e^{-\eta}])^{z_{ij}} (1/[1+e^{\eta}])^{1-z_{ij}}$ the prior probability of $Z$.

Obtaining $\hat{\eta}$ may feel like a daunting task, given that \eqref{eq:marglhood_eta} involves a sum over $2^{p(p-1)/2}$ graphical structures.
Fortunately, \cite{rognon:2026} (Theorem 3) leveraged the fact that $\pi(Z \mid \eta)$ factors across $(i,j)$ to express $\nabla_\eta \log \pi(\eta \mid \Y)$ as a sum across $p(p-1)/2$ terms.
Their result holds in more general situations than ours, allowing $\eta$ to depend on meta-covariates. 
In our setting, if one sets a prior $\eta \sim N(0, g_\eta)$, Proposition 1 in \cite{rognon:2026} gives a particularly simple Expectation-Maximisation algorithm.
Specifically, let $\hat{\eta}^{(b)}$ be the current estimate at iteration $b$. 
\begin{enumerate}
\item Obtain an estimate $\hat{\pi}(Z_{ij}=1 \mid \Y, \hat{\eta}^{(b)})$ of the posterior inclusion probabilities given $\eta= \hat{\eta}^{(b)}$, for example using a small number $L$ of MCMC iterations.
Denote their average by $\tilde{\theta}= \frac{2}{p(p-1)} \sum_{i>j} \hat{\pi}(Z_{ij}=1 \mid \Y, \hat{\eta}^{(b)})$.

\item Set $\hat{\eta}^{(b+1)}= h(\tilde{\theta}, g_\eta p(p+1)/2)$, where $h(\tilde{\theta}, g_\eta p(p+1)/2)$ is the value of $\eta$ solving
$$
\frac{1}{1 + e^{-\eta}} + \frac{\eta}{g_\eta p(p+1)/2}= \tilde{\theta}.
$$
\end{enumerate}

Consider the case $g_\eta= \infty$, where one seeks to maximise the marginal likelihood $\pi(\Y \mid \eta)$.
Then, Step 2 sets $\hat{\eta}^{(b+1)}= \log (\tilde \theta / [1- \tilde \theta])$, the logit of the estimated posterior average inclusion probability given $\hat{\eta}^{(b)}$.
An important pitfall with $g_\eta= \eta$ is that one may obtain $\hat{\eta} \in \{-\infty, \infty\}$, resulting in degenerate prior probabilities $\hat{\theta}= 1/[1+e^{-\eta}] \in \{0, 1\}$. 
Specifically, \cite{rognon:2026} (Theorem 3) shows that $\nabla_\eta \log \pi(\Y \mid \eta)=0$ at $\eta = -\infty$ and at $\eta= \infty$, that is these degenerate solutions are modes of the log marginal likelihood, including the possibility that one of them is the global mode.
A natural strategy to avoid such degenerate modes is to set $g_\eta > 0$ to a minimally-informative (that is, sufficiently large) value.
\cite{rognon:2026} suggested setting $g_\eta$ such that $\pi(\theta \in [0.001, 0.999])= 0.95$, to ensure that $\theta$ takes a non-extreme value with high prior probability.
Simple algebra shows that this results in 
$$g_\eta= \left( \frac{\log(0.001/0.999)}{\Phi^{-1}(0.025)} \right)^2.$$

We remark that this is a stochastic gradient algorithm, using $L$ samples from $\pi(Z \mid \Y, \hat{\eta}^{(b)})$ to obtain $\tilde \theta$.
Although in our software we set $L=1000$ to be on the safe side,
the examples in \cite{rognon:2026} showed that fairly small $L=100$ gave indistinguishable results from $L=1000$, at a fraction of the computational cost (aligning with the good empirical performance of mini-batch stochastic gradient methods in many machine learning tasks).

\color{black}

\section{Background on the efficiency of MCMC}\label{sec:background-mcmc}

We recall some standard notions to analyse the computational efficiency of Markov chain algorithms, see \cite{guruswami2016rapidlymixingmarkovchains} for a review.
The mixing time of an MCMC algorithm
measures the iterations needed to approximately reach the invariant distribution of the Markov chain, denoted by $\pi$ in this section.
For a Markov kernel $P(\cdot, \cdot)$ defined on a state space $ \mathcal{X}$ and $\epsilon > 0$, the mixing time at level $\epsilon$ of the Markov chain is 
\begin{align*}
    \tau_x(P, \epsilon) := \inf \{ t > 0; \| P(x, \cdot)^t - \pi \|_{TV}  \leq \epsilon \} 
\end{align*}
where $x \in \mathcal{X}$ is the state at iteration $t=0$ and $|| \cdot ||_{TV}$ is the total variation distance. We recall that for two probability measures $\mu$ and $\nu$ defined on $\mathcal{X}$, the total variation distance is defined as
$$\|\mu - \nu \|_{TV} = \sup_{A \subset \mathcal{X}} |\mu(A) - \nu(A)|.$$ 



A popular technique for bounding the mixing time is via spectral gaps. 
The spectral gap can be defined using the variational formulation
\begin{align}\label{eq:def-gap}
    \text{Gap}(P) = \inf \left\{ \frac{E(f,f)}{Var_\pi(f)} : f \in L_2(\pi), Var_\pi(f) > 0 \right\},
\end{align}
where $L_2(\pi)$ is the space of $\pi$-square-integrable functions on $\mathcal{X}$ and for any $f \in L_2(\pi)$,
\begin{align*}
    E(f,f) = \frac{1}{2} \int_{\mathcal{X}} (f(y) - f(x))^2 \pi(dx)P(x,dy), \quad Var_\pi(f)  = E_\pi(f^2) - E_\pi^2(f), 
\end{align*}
where we use the notation $E_\pi(f) = \int_{\mathcal{X}} f(x) \pi(dx)$. 
If the Markov kernel provides independent samples from $\pi$, i.e., $P(x,dy)=\pi(dy)$, then it is easy to see that $\text{Gap}(P) = 1$.


To analyse mixing times, it is useful to consider the so-called ``lazy'' kernel $P^L = \frac{1}{2} I + \frac{1}{2} P$, which guarantees positive-definiteness and whose spectral gap differs at most by a factor 2 from that of $P$.
The  
mixing time of $P^L$
can be bounded using its spectral gap by (see Proposition 2.3 in \cite{guruswami2016rapidlymixingmarkovchains}):
\begin{align}\label{eq:bound-mx-gap}
    \tau_x(P^L, \epsilon) \leq - \frac{\log  \pi(x) + \log \epsilon}{\text{Gap}(P^L)}.
\end{align}
Hence lower-bounds on the spectral gap of $P^L$ and $\pi(x)$ provides an upper bound on the mixing time of the corresponding Markov chain.


Moreover, the spectral gap can be bounded by the conductance, 
which quantifies the ability of the corresponding Markov chain to escape any subset $B \subset \mathcal{X}$, with small mass under $\pi$. More precisely, the conductance  of  $P$ is 
\begin{align}\label{eq:conductance}
    \phi( P) = \inf_{B \subset \mathcal{X} : \pi(B) \in (0,1/2)} \left \{ \frac{\int_B P(x, B^c) \pi(dx)}{\pi(B)}  \right \}.
\end{align} 
A slightly different definition of the conductance is sometimes convenient and given by
\begin{align}\label{eq:other-conductance}
    \phi^*( P) = \inf_{B \subset \mathcal{X}  : \pi(B) \in (0,1)} \left \{ \frac{\int_S P(x, B^c) \pi(dx)}{\pi(B) \pi(B^c)} \right \}.
\end{align}
It can be easily shown that $ \phi( P)  \leq  \phi^*( P) \leq 2 \phi( P)$ and, for any lazy Markov kernel $P^L$, $\phi^*(P^L) \leq 1$.

For reversible chains, one can bound the spectral gap by the conductance and conversely using Cheeger bounds. In particular (see Theorem 3.1 in \cite{guruswami2016rapidlymixingmarkovchains}),
\begin{align}\label{eq:spectral-gap-sandwich}
    \frac{\phi(P)^2}{2} \leq \frac{\phi^*(P)^2}{2} \leq \text{Gap}(P) \leq \phi^*(P) \leq 2 \phi(P).
\end{align}
Note that for any lazy Markov kernel, $\text{Gap}(P^L) \leq 1$. 

Yet another strategy for bounding mixing times which is particularly relevant in high-dimensional state spaces relies on the concept of  $s$-conductance \citep{ascolani2024scalability}.
For $s \geq 0$, the $s$-conductance of $P$ is 
\begin{align*}
    \phi_s( P) = \inf_{S \subset \mathcal{X}  :  s < \pi(S) < 1/2} \left \{ \frac{\int_S P(x, S^c) \pi(dx)}{\pi(S)}  \right \}.
\end{align*}
The $s$-conductance also bounds the total variation distance, when given a ``warm'' start from an initial distribution $\mu$.
More precisely, if $\mu(x) \leq K \pi(x)$ for some $K>1$ and for any $x \in \mathcal{X}$, 
then (see Lemma 1 in \cite{ascolani2024scalability}) 
\begin{align}\label{eq:tv-conductance}
    \| \mu P^L(\cdot , \cdot)^t - \pi \|_{TV} \leq K s + K \left( 1 - \frac{\phi_s(P^L)^2}{2}\right)^t, \qquad t \geq 1,
\end{align}
where $P^L$ is a lazy kernel.
Since the $s$-conductance also verifies $\phi_s(P^L) \leq \sqrt{2}$, the above bound implies that the total variation is small whenever  $s$ is sufficiently small, $\phi_s(P^L) > 0$, and $t$ is sufficiently large. Moreover, it is straightforward that  $\phi_s(P^L) \geq \phi(P^L)$ for any $s \geq 0$, thus  it is sufficient that $\phi(P^L) > 0$ or that $\phi_s(P^L)$ does not go to 0 too quickly with $s$ to obtain a non-vacuous bound.
Using \eqref{eq:tv-conductance}, the mixing time of the Markov chain with kernel $P^L$ can be bounded, e.g., by choosing $s = \frac{\epsilon}{2K}$, and one obtains that $  \| \mu P^L(\cdot , \cdot)^t - \pi \|_{TV}  \leq \epsilon$ for any $t \geq \tau_\mu(P^L,\epsilon)$ with
\begin{align*}
    \tau_\mu(P^L, \epsilon) = \frac{\log (2K) - \log \epsilon}{-\log(1 - \phi_s(P^L)^2/2)}.
\end{align*}

The latter fact and \eqref{eq:tv-conductance} is the motivation for establishing lower bounds on the $s$-conductances of our Metropolis-within-Gibbs algorithms (BDMH and GIMH).


\section{$s$-conductance of Metropolis-within-Gibbs vs. ideal Gibbs sampler}
\label{sec:MwG-theory}

In this section we consider the lazy versions of the Markov kernels of the ideal Gibbs sampler $P_{Gibbs}$
and of our MH algorithms (BDMH and GIMH), respectively $P_{MH}^M$ and $P_{GI}^M$.
We recall that we consider lazy kernels 
since they are positive definite 
and their spectral gaps differ from the original kernels by a factor $\leq 2$.
We also denote by \deb{$P_{MH}^{j, \Omega_{-j,-j}, M}(z,z')$} the \deb{lazy} MH kernels 
that updates $z$ given $\Omega_{-j,-j}$ \deb{with $M$ updates} (global or local). 

Proposition \ref{prop:conductance-bound} relates the $s$-conductance of \deb{$P^{M}_{MH}$} to that of $P_{Gibbs}$ and to the spectral gap $\text{Gap}(P_{MH}^{j, \Omega_{-j,-j}, M})$.
This is useful in Sections \ref{sec:spectral-gap-bdmh}-\ref{sec:gimh} of the main paper, where we lower-bound the spectral gap for our BDMH and GIMH kernels and use Proposition \ref{prop:conductance-bound} to control their $s$-conductance.

\begin{prop} \label{prop:conductance-bound}
    For any subset $B \subset \PSD$ and $s > 0$, and all our MH algorithms (i.e., BDMH, GIMH or LIT-MH),
    \begin{align}\label{eq:conductance-bound}
    \phi_s(P_{MH}^{M}) &\geq \frac{1}{2} \phi_s(P_{Gibbs}) \min_{j \in [p]} \inf_{\Omega \in B}  \text{Gap}(P_{MH}^{j, \Omega_{-j,-j}, M}) - \frac{\pi(B^c |\Y)}{s}.
\end{align}
\end{prop}

Proposition \ref{prop:conductance-bound} implies that, if $\pi(B^c | \Y)/s$ is  small, the loss of $s$-conductance of our algorithms relative to 
the ideal Gibbs sampler is bounded by their spectral gap. 
In particular, if the posterior concentrates sufficiently on a neighborhood $B$ of $\Omega^0$ then $\pi(B^c \mid \Y)/s$ is small. 




\section{Regularity conditions for linear regression posterior}
\label{sec:regularity_linear_regression}

We state regularity conditions required by Theorem \ref{thm:spectral-gap-gimh}, which guarantee that the linear regression posterior $\pi_j^{LR}(Z_{-j, j} \mid \Y)$ associated to regressing $\Y_{\cdot j}$ onto $\Y_{\cdot z}$ concentrates on the data-generating truth.
This is the posterior on variable inclusion indicators associated to the regression model $Y_{\cdot j} \sim N(Y_{\cdot z} \beta_z, \sigma^2 I)$, setting Zellner's prior
$
\pi(\beta_z, \sigma^2 \mid z)= N(\beta_z; 0, \tau^{-1} \sigma^2 (\Y_{\cdot z}^T \Y_{\cdot z}/n)^{-1}) \times \mbox{IG}(\sigma^2; a/2, \lambda/2),
$
where 
$\beta_z = - \Omega_{zj} / \Omega_{jj}$, $\sigma^2 = (\Omega_{jj})^{-1}$, $ a > 0$, and $\tau > 0$ is a precision parameter.
Recall that by default $\tau=1$ and $a=2$, so that $\sigma^{-2} \sim \mbox{Exp}(\frac{\lambda}{2})$ matches the prior in \eqref{eq:prior-z}.
The prior on $z$ can either be the Binomial prior in \eqref{eq:prior-z} with edge inclusion probability $\theta= 1/p^{\alpha}$, or the Beta-Binomial in \eqref{eq:prior-z-bbin} taking $a_\theta=1$ and $b_\theta= p^\alpha$. We denote by $\tilde{\alpha}:=\alpha$ for the Binomial prior and $\tilde{\alpha} := \max\{2, \alpha\}$ for the Beta-Binomial, and we make generic statements involving $\tilde{\alpha}$ that apply to both priors.

\begin{itemize}
\item [(D1)] There is a constant $\epsilon_1 \in (0,1)$ such that $S= \Y^T \Y$ verifies, for any subset $z \in \mathcal{Z}(\bar d + 1)$,
$
n \epsilon_1 \leq \lambda_{min}(S_{zz}) \leq \lambda_{max}(S_{zz}) \leq  n \epsilon_1^{-1}.
$

\item [(D2)] The prior parameter $\tau > 0$ is fixed. 

\item [(D3)] The maximum edge degree $\bar d = o\left( \min\{n, p, n/\tau \} \right)$,    $\log \bar d= o(\tilde{\alpha} \log p)$ and $|z^0|_0 = o(\bar d)$.


\item [(D4)] The dimension $p$ and $\bar d \geq 4$ satisfy $\bar d \log p = o\left(\frac{n}{(\log n)^2}\right)$.
\end{itemize}

These assumptions are mild and common, see, e.g., \cite{rossell2022concentration} for a detailed discussion.
Briefly, D1 is an eigenvalue condition on sample covariances. 
Assumption D2 can be relaxed to allow for $\tau$ such that $\tau = o(n)$ and $1/\tau= o(1/n)$, but it simplifies the exposition and is satisfied by our default $\tau=1$ (unit information prior).
Assumption D3 is a mild sparsity condition on node degrees and the true model size.
Assumption D4 
is not strictly necessary but makes the proof simpler.

\section{Proofs}

\subsection{Proof of Proposition \ref{prop:posterior-consistency}}\label{app:proof-consistency}


Recall that $Z^0$ denotes the true graphical model corresponding to the true precision matrix $\Omega^0$ and our goal is to show that $\pi(Z^0 \mid \Y) \stackrel{\mathbb P}{\longrightarrow} 1$ in probability as $n \to \infty$.

   Consider first the case where $\pi(Z)$ is the Binomial prior. Then, \color{black}
from Bayes formula
\begin{align*}
    \pi(Z^0 \mid \Y) = \frac{L(\Y \mid Z^0) \pi(Z^0)}{\sum_{Z} L(\Y \mid Z) \pi(Z)},
\end{align*}
where $L(\Y \mid Z)$ is the integrated likelihood
\begin{align*}
    &L(\Y \mid Z)=\int L(\Y \mid \Omega) d\Pi(\Omega \mid Z),
\end{align*}
which we can re-write by defining $h(\Omega) = \frac{1}{2} \left[ \frac{\lambda}{n} \sum_{i} \Omega_{ii} + \frac{1}{n} \sum_{i<j} \frac{\Omega_{ij}^2}{g_1^2} - \log |\Omega| + \frac{1}{n} Tr(S \Omega) \right]$ as
\begin{align*}
   & L(\Y \mid Z)= C_1(Z,g_1,\lambda)    \tilde{L}(\Y \mid Z), \color{black}
\end{align*}
where
\begin{align*} 
     &\tilde{L}(\Y \mid Z)= \color{black} \frac{(\lambda/2)^p}{(2 \pi)^{\frac{pn}{2}} (2 \pi g_1^2)^{|Z|_0/2}} \int e^{- n h(\bar{\Omega}_Z)} d\bar{\Omega}_Z.
\end{align*}
where $\bar{\Omega}_Z$ is an $p \times p$ matrix with $(i,j)$ entry equal to zero if and only if $Z_{ij}=0$. Recall that
\begin{align*}
     \pi(Z) \propto \frac{1}{C_1(Z,g_1,\lambda)} \theta^{|Z|_0}(1-\theta)^{p(p-1)/2 - |Z^0|_0} \mathds{1}_{d(Z) \leq \bar d}.
\end{align*}
Since by assumption $d(Z^0) \leq \bar d$, we have
\begin{align*}
\pi(Z^0 \mid \Y)
&= \frac{   \tilde{L}(\Y \mid Z^0) \color{black} \theta^{|Z^0|_0} (1 - \theta)^{p(p-1)/2 - |Z^0|_0}}{\sum_{Z}    \tilde{L}(\Y \mid Z) \color{black} \theta^{|Z|_0} (1 - \theta)^{p(p-1)/2 - |Z|_0}} \\
&= \frac{1}{1 + \sum_{Z \neq Z^0}  \frac{\tilde{L}(\Y \mid Z)}{\tilde{L}(\Y \mid Z^0)} \left(\frac{\theta}{1-\theta}\right)^{|Z|_0 - |Z^0|_0} },
\end{align*}
Since $p$ is fixed, the number of terms in the summation over $Z \neq Z^0$ is also fixed, and it suffices to show that
$$
 \frac{\tilde{L}(\Y \mid Z)}{\tilde{L}(\Y \mid Z^0)} \left(\frac{\theta}{1-\theta}\right)^{|Z|_0 - |Z^0|_0} \stackrel{\mathbb P}{\longrightarrow} 0,
$$
for each $Z \neq Z^0$. Further noting that $\theta$ is fixed and that $|Z|_0-|Z^0|_0 \leq p(p+1)/2$, we have that $(\theta/[1-\theta])^{|Z|_0-|Z^0|_0}=O(1)$, hence it suffices to show that 
\begin{align}
\frac{\tilde{L}(\Y \mid Z)}{\tilde{L}(\Y \mid Z^0)}
= (2 \pi g_1^2)^{\frac{|Z^0|_0 - |Z|_0}{2}} 
\frac{\int e^{-n h(\bar{\Omega}_Z)} d\bar{\Omega}_Z}
{\int e^{-n h(\bar{\Omega}_{Z^0})} d\bar{\Omega}_{Z^0}}
\stackrel{\mathbb P}{\longrightarrow} 0.
\label{eq:consistency_fixedp_sufficientcond}
\end{align}


This also suffices to prove the result when $\pi(Z)$ is the Beta-Binomial prior because, reasoning analogously gives that one must show that
$$
\frac{\tilde{L}(\Y \mid Z)}{\tilde{L}(\Y \mid Z^0)}
\frac{\Gamma(a_\theta + |Z|_0) \Gamma(b_\theta + p(p-1)/2 - |Z|_0)}{\Gamma(a_\theta + |Z^0|_0) \Gamma(b_\theta + p(p-1)/2 - |Z^0|_0)}
\stackrel{\mathbb P}{\longrightarrow} 0,
$$
where the second term is bounded above and below by constants (recall that we assume that $p$, and hence also $\bar d$, are fixed).
\color{black}

The remainder of the proof is dedicated to showing that \eqref{eq:consistency_fixedp_sufficientcond} holds.
To do so, we use a Laplace approximation to both integrals, which was shown by \cite{banerjee2015bayesian} to consistently estimate the GGM marginal likelihood as $n \to \infty$, provided that $(p + |Z|_0)^5 \log p = o(n)$, which holds here since $p$ is fixed.
Let $I(Z)= \int e^{-n h(\bar{\Omega}_Z)} d\bar{\Omega}_Z$
and denote by $\hat \Omega_Z$ the maximum a posteriori under model $Z$,
the Laplace approximation gives that the left-hand side in \eqref{eq:consistency_fixedp_sufficientcond} is consistently approximated by
\begin{align}
   \left(2\pi g_1^2 \right)^{\frac{(|Z^0|_0 - |Z|_0)}{2}} \color{black}
\frac{\hat{I}(Z)}{\hat{I}(Z^0)}
= 
e^{-n [h(\hat \Omega_Z) - h(\hat \Omega_{Z^0})]} 
\left(\frac{1}{   g_1^2 \color{black} n}\right)^{\frac{(|Z|_0 - |Z^0|_0)}{2}} \frac{|H(\hat \Omega_{Z^0})|^{\frac{1}{2}}}{|H(\hat \Omega_Z)|^{\frac{1}{2}}},
\nonumber
\end{align}
where 
\begin{align}
H(\Omega_Z)= 
\nabla_{\Omega_Z}^2 h(\Omega_Z)=
\frac{1}{2} \left[ \frac{1}{n g_1^2} I - \nabla_{\Omega_Z}^2 \log|\Omega_Z| \right], \label{eq:hessian}
\end{align}
and $I$ is the $|Z|_0 \times |Z|_0$ identity matrix.
The rest of this proof consists of:
\begin{enumerate}
    \item Proving that $\frac{|H(\hat \Omega_{Z^0})|^{1/2}}{|H(\hat \Omega_Z)|^{1/2}}$ converges in probability to a constant. 
    This is immediate: using that $\hat{\Omega}_Z \stackrel{\mathbb P}{\longrightarrow} \Omega_Z^*$ 
    where $\Omega^*_Z = \arg \min_{\Tilde \Omega \in \PSD, \mathcal{G}(\Tilde \Omega) = Z} \lim_{n \to \infty} \mathbb E[h(\Tilde \Omega)]$,
    and the continuous mapping theorem, we have that 
    $|H(\hat \Omega_{Z})| \stackrel{\mathbb P}{\longrightarrow} |H(\Omega_Z^*)|$.
    Similarly, $\hat \Omega_{Z^0} \stackrel{\mathbb P}{\longrightarrow} \Omega_0$ and hence
    $|H(\hat \Omega_{Z^0})| \stackrel{\mathbb P}{\longrightarrow} |H(\Omega_{Z^0})|$.
    Hence
    $$
    \frac{|H(\hat \Omega_{Z^0})|^{1/2}}{|H(\hat \Omega_Z)|^{1/2}} 
    \stackrel{\mathbb P}{\longrightarrow} 
    \frac{|H(\Omega_{Z^0})|^{1/2}}{|H(\Omega_Z^*)|^{1/2}},
    $$
    which is a constant by assumption.
        
    \item Showing that $n (h_Z(\hat \Omega_Z) - h_{Z^0}(\hat \Omega_{Z^0})) = O_P(1)$ if $Z \supset Z^0$ and  $h_Z(\hat \Omega_Z) - h_{Z^0}(\hat \Omega_{Z^0})$ converges in probability and its limit is positive if $Z \not \supset Z^0$.
\end{enumerate}

 Note that if 1. and 2. hold, then in the case where $Z \supset Z^0$, we   obtain that
    \begin{align*}
        \frac{\hat I(Z)}{\hat I(Z^0)} = O_P(1) \times e^{O_P(1) } \times (   g_1^2 \color{black} n)^{-(|Z|_0 - |Z^0|_0)} = o_P(1),
    \end{align*}
       where we used that $\lim_{n \to \infty} n g_1= \infty$, since $1/g_1^2 = o(n)$ by assumption. \color{black}
    Further, in the case where $Z \supset Z^0$,
    \begin{align*}
       \frac{\hat I(Z)}{\hat I(Z^0)} = O_P(1) \times e^{- \frac{n}{2} (c + o_P(1))} \times (   g_1^2 \color{black} n)^{-(|Z|_0 - |Z^0|_0)} = o_P(1),
    \end{align*}
    for some $c>0$, 
       where we used that $\lim_{n \to \infty} 0.5 n c + (|Z|_0 - |Z^0|_0) \log(g_1^2)= \infty$, since $|\log g_1| = o(n)$ by assumption.  \color{black}
    Since $\frac{\hat I(Z)}{\hat I(Z^0)} = o_P(1)$ for any $Z \neq Z^0$, the result of Proposition \ref{prop:posterior-consistency} holds since the error of the Laplace approximation vanishes. 

\textbf{Proving 1.}

Since $\hat \Omega_Z, \hat \Omega_{Z^0}$ are the MAP respectively in model $Z$ and $Z^0$, then we claim that
\begin{align}
     &\hat \Omega_{Z^0} \xrightarrow[n\to \infty]{\mathbb P} \Omega^0, \label{eq:cv-map}\\
     &\hat \Omega_{Z} \xrightarrow[n\to \infty]{\mathbb P} \Omega^*_Z,\label{eq:cv-mapZ}
\end{align}
where $\Omega^*_Z = \arg \min_{\Tilde \Omega \in \PSD, \mathcal{G}(\Tilde \Omega) = Z} \lim_{n \to \infty} \mathbb E[h(\Tilde \Omega)]$.
We can obtain \eqref{eq:cv-map} using similar arguments as \cite{banerjee2015bayesian} who proved the concentration rate of the posterior distribution (implying the MAP consistency) for the spike-and-slab prior with Laplace slab. The convergence of $\hat \Omega_{Z}$ in \eqref{eq:cv-mapZ} is a result of the strict convexity of $h_Z(\Omega)$ and the \jack{point-wise} convergence of $h_Z(\Omega)$ towards $$\lim_{n \to \infty}\mathbb E[h_Z(\Omega)]$$ (the log prior density vanishes at the limit while the empirical covariance converges to $\Sigma^0$). Specifically, the so-called convexity lemma implies that  $h_Z(\Omega)$ converges uniformly towards $\lim_{n \to \infty}\mathbb E[h_Z(\Omega)]$ on any compact neighborhood of $\Omega_{Z}^*$.  Moreover, given the Hessian \eqref{eq:hessian} and by the continuous mapping theorem, 
we also have
\begin{align*}
        &H(\hat \Omega_{Z^0}) \xrightarrow[n\to \infty]{\mathbb P} H^*(\Omega^0), \\
     &H(\hat \Omega_{Z}) \xrightarrow[n\to \infty]{\mathbb P} H^*(\Omega^*_Z),
\end{align*}
where $H^*(\Omega) = \lim_{n \to \infty} \mathbb E[\nabla^2 h(\Tilde \Omega)]$ is the limit of the expected Hessian. Note that the expected Hessian is negative definite at $\Omega^*_Z$ by convexity of  $\lim_{n \to \infty}\mathbb E[h_Z(\Omega)]$. We can then conclude that
\begin{align*}
    \frac{|H(\hat \Omega_{Z^0})|^{1/2}}{|H(\hat \Omega_Z)|^{1/2}} \xrightarrow[n\to \infty]{\mathbb P} \frac{|H^*( \Omega^0)|^{1/2}}{|H^*(\Omega^*_Z)|^{1/2}} = O(1).
\end{align*}

\textbf{Proving 2.}

We separately consider the cases $Z \supset Z^0$ and $Z \not \supset Z^0$.

\begin{itemize}
    \item \textbf{Case 1: $Z \supset Z^0$.} Then $\Omega_Z^* = \Omega^0$ and note that
    \begin{align*}
       \frac{n}{2} (h_Z(\hat \Omega_Z) - h_Z(\hat \Omega_{Z^0}) ) = \log L( \Y | \hat \Omega_{Z^0}) - \log L( \Y | \hat \Omega_Z ) + O(1),
    \end{align*}
    where $L(\Y | \Omega)$ is the GGM likelihood. Then from standard asymptotic theory (see, e.g., the proof of Proposition 3 in \cite{rossell2023additive}),  
    \begin{align*}
        \log L( \Y | \hat \Omega_{Z^0}) - \log L(\Y | \hat \Omega_Z ) \xrightarrow[n\to \infty]{d} \chi^2_{|Z|_0-|Z^0|_0} = O_P(1),
    \end{align*}
    thus $\frac{n}{2} (h_Z(\hat \Omega_Z) - h_Z(\hat \Omega_{Z^0}) )= O_P(1)$.

    \item \textbf{Case 2: $Z \not \supset Z^0$.} Then  the log-likelihood converges uniformly in probability to its expectation and
    \begin{align*}
        \frac{n}{2} (h_Z(\hat \Omega_Z) - h_Z(\hat \Omega_{Z^0}) ) = \mathbb E[\log L( \Y | \Omega^0) - \log L(  \Y | \Omega_Z^*) ] + O_{P}(1).
    \end{align*}
    Note that  $  \mathbb E[\log L( \Y | \Omega^0) - \log L(  \Y | \Omega_Z^*) ]$ is the negative Kullback-Leibler (KL) divergence between the GGM distributions with parameter $\Omega_Z^*$ and $\Omega^0$, and since $Z \not \supset Z^0$, the KL divergence is positive, implying that the limit of $\frac{n}{2} (h_Z(\hat \Omega_Z) - h_Z(\hat \Omega_{Z^0}) )$ is negative as we wished to prove. 
\end{itemize}

\subsection{Proof of Proposition \ref{prop:wang_exactzeroes_notation}}\label{app:proof-cond-post}

   We consider the case where $\pi(Z)$ is the Binomial prior. The case where $\pi(Z)$ is the Beta-Binomial$(a_\theta,b_\theta)$ prior follows immediately by replacing $\prod_{i<j} \theta^{Z_{ij}} (1-\theta)^{1 - Z_{ij}}$ by 
\begin{align}
    \Gamma(a_\theta + |Z|_0) \Gamma(b_\theta + p(p-1)/2 - |Z|_0).
\nonumber
\end{align}

\color{black}
Without loss of generality, we prove Proposition \ref{prop:wang_exactzeroes_notation} for $j=p$. The conditional posterior distribution on $\Omega_{\cdot p} = (\Omega_{-p,p}, \Omega_{pp}) \mid \Y, \Omega_{-p,-p}$ can be written as
\begin{align*}
\pi(\Omega_{-p,p}, \Omega_{pp} \mid \Y, \Omega_{-p,-p}) &\propto \pi(\Omega \mid \Y) \propto |\Omega|^{\frac{n}{2}} \exp \left\{ -\frac{\mbox{tr}(S \Omega)}{2} \right\} \pi(\Omega).
\end{align*}

Recall that for any $\Omega$, we define the graphical model $Z=(Z_{ij})_{i,j}$ where $Z_{ij}= \mathds{1}_{\Omega_{ij} \neq 0}$. The prior on $\Omega$ can be re-written as
\begin{align}
    \pi(\Omega) &= \pi(\Omega, Z) = \pi(Z) \pi(\Omega \mid Z) \nonumber \\
    &=  C_2(g_1, \theta, \lambda)  \prod_{i<j} \theta^{Z_{ij}} (1 - \theta)^{1 - Z_{ij}} \prod_{i<j, Z_{ij }= 1} \text{N}(\Omega_{ij}; \: 0, g_1^2)   \prod_{i<j, Z_{ij}  = 0} \delta_0(\Omega_{ij}) \prod_{i} \text{Exp}(\Omega_{ii}; \lambda) \mathds{1}_{\Omega \in \PSD} \mathds{1}_{d(Z) \leq \bar d}, \label{eq:full-prior}
\end{align}
where the normalising constant $C_2(g_1, \theta, \lambda) > 0$ only depends on the prior parameters $(g_1, \theta, \lambda)$    (or, in the case of the Beta-Binomial prior, on $(g_1, a_\theta, b_\theta, \lambda)$). \color{black} Let $z = Z_{\cdot p} \in \{0,1\}^{p-1}$ be the indicator vector of the non-zero entries in $\Omega_{-p,p}$. 
From \eqref{eq:full-prior} and using the fact that $\mathds{1}_{\Omega \in \PSD}= \mathds{1}_{\Omega_{-p,-p} \in \PSD} \mathds{1}_{\Omega_{pp} - \Omega_{-p,p}^T [\Omega_{-p,-p}]^{-1} \Omega_{-p,p} > 0}$,
we obtain the conditional prior density
\begin{align*}
    &\pi(\Omega_{-p,p}, \Omega_{pp} \mid \Omega_{-p,-p}) = \pi(\Omega_{-p,p}, \Omega_{pp},z \mid \Omega_{-p,-p})  \\
    &\propto \frac{\theta^{|z|_0} (1 - \theta)^{p-1-|z|_0}}{g_1^{|z|_0}} \exp \left \{ - \frac{1}{2g_1^2} \Omega_{zp}^T \Omega_{-zp} - \frac{\lambda}{2} \Omega_{pp} \right \} \mathds{1}_{\Omega_{pp} - \Omega_{zp}^T [[\Omega_{-p,-p}]^{-1}]_{zz} \Omega_{zp} > 0} \prod_{k : z_k=0} \delta_0(\Omega_{kp}) \mathds{1}_{|z|_0 \leq \bar d},
\end{align*}
for any $\Omega_{-p,-p} \in \PSD$ and using in the second equality that $\Omega_{-p,p}^T [\Omega_{-p,-p}]^{-1}\Omega_{-p,p} = \Omega_{z p}^T [[\Omega_{-p,-p}]^{-1}]_{zz} \Omega_{z p}$. We then combine the above expression of the conditional prior density with the likelihood. Using that $|\Omega|=|\Omega_{-p,-p}| |\Omega_{pp} - \Omega_{-p,p}^T [\Omega_{-p,-p}]^{-1} \Omega_{-p,p}|$
and that $\mbox{tr}(S \Omega)= \mbox{tr}(S_{-p,-p} \Omega_{-p,-p}) + 2 S_{-p,p}^T \Omega_{-p,p} + S_{pp} \Omega_{pp}$ we obtain that 
\begin{align}
\pi(\Omega_{-p,p}, \Omega_{pp},z \mid \Y, \Omega_{-p,-p})  &\propto 
\frac{\theta^{|z|_0} (1 - \theta)^{p-1-|z|_0}}{ g_1^{|z|_0}}
\mathds{1}_{\Omega_{pp} - \Omega_{-p,p}^T [\Omega_{-p,-p}]^{-1} \Omega_{-p,p} > 0} |\Omega_{pp} - \Omega_{-p,p}^T [\Omega_{-p,-p}]^{-1} \Omega_{-p,p}|^{\frac{n}{2}}
\\
&\times e^{ - \frac{2 S_{-p,p}^T \Omega_{-p,p} + g_1^{-2} \Omega_{-p,p}^T \Omega_{-p,p} + (S_{pp}+ \lambda) \Omega_{pp}}{2}} \prod_{k: z_{k} = 0} \delta_0(\Omega_{kp})
\nonumber \\
&\propto  \frac{\theta^{|z|_0} (1 - \theta)^{p-1-|z|_0}}{ g_1^{|z|_0}}
\mathds{1}_{\Omega_{pp} - \Omega_{zp}^T [[\Omega_{-p,-p}]^{-1}]_{zz} \Omega_{zp} > 0} |\Omega_{pp} - \Omega_{zp}^T [[\Omega_{-p,-p}]^{-1}]_{zz} \Omega_{zp}|^{\frac{n}{2}} \\
&\times e^{ - \frac{2 S_{zp}^T \Omega_{zp} + g_1^{-2} \Omega_{zp}^T \Omega_{zp} + (S_{pp}+ \lambda) \Omega_{pp}}{2} }   \prod_{k: z_{k} = 0} \delta_0(\Omega_{kp})
\nonumber,
\end{align}
using in the last equation that $S_{-p,p}^T \Omega_{-p,p} = S_{zp}^T \Omega_{zp}$.
Now consider the change of variables $u_1= - \Omega_{z p}$, $u_2= \Omega_{pp} - \Omega_{z p}^T [[\Omega_{-p,-p}]^{-1}]_{zz} \Omega_{zp} = \Omega_{pp} - \Omega_{z p}^T \Sigma_{zz|p} \Omega_{zp}$. 
We then obtain that
\begin{align}
\pi(u_1, u_2,z \mid \Y ,\Omega_{-p,-p}) \propto u_2^{\frac{n}{2}}
\exp \left\{ - \frac{u_1^T U_z u_1 -2 S_{zp}^T u_1 + (S_{pp}+ \lambda) u_2}{2}   \right\}
\frac{\theta^{|z|_0} (1 - \theta)^{p-1-|z|_0}}{ g_1^{|z|_0}} \mathds{1}_{u_2>0},
\label{eq:cond-post-u}
\end{align}
where $U_z=(S_{pp}+ \lambda)\Sigma_{zz|p} + g_1^{-2} I_z$ and we used the fact that the inverse transformation $\Omega_{z p}= - u_1$, $\Omega_{pp}= u_2 + u_1^T \Sigma_{zz|p} u_1$ has unit Jacobian. 
Defining $m_z :=U_z^{-1} S_{zp}$ and rearranging the terms in \eqref{eq:cond-post-u}, we obtain
\begin{align}
&\pi(u_1, u_2,z \mid \Y ,\Omega_{-p,-p}) \\
&\propto u_2^{\frac{n}{2}} e^{-\frac{(S_{pp}+\lambda)u_2}{2}}
\exp \left\{ - \frac{(u_1 - m_z)^T U_z (u_1-m_z)}{2}   \right\}
e^{\frac{m_z^T U_z m_z}{2}}
\frac{\theta^{|z|_0} (1 - \theta)^{p-1-|z|_0}}{g_1^{|z|_0}}  \mathds{1}_{u_2>0}
\nonumber \\
&\propto \text{N}(u_1; m_z, U_z^{-1}) \text{Ga}\left(u_2; \frac{n}{2} + 1, \frac{S_{pp} + \lambda}{2}\right) \frac{e^{\frac{m_z^T U_z m_z}{2}}}{g_1^{|z|_0}|U_z|^{\frac{1}{2}}}
\theta^{|z|_0} (1 - \theta)^{p-1-|z|_0}. 
\nonumber
\end{align}
This implies that one can re-write $\pi(u_1, u_2,z \mid \Y ,\Omega_{-p,-p}) = \pi(u_1 \mid z, \Y ,\Omega_{-p,-p}) \pi(u_2 \mid \Y) \pi(z \mid \Y ,\Omega_{-p,-p}) $ where
\begin{align*}
    &\pi(z \mid \Y ,\Omega_{-p,-p}) \propto \frac{e^{\frac{m_z^T U_z m_z}{2}}}{g_1^{|z|_0}|U_z|^{\frac{1}{2}}}
\theta^{|z|_0} (1 - \theta)^{p-1-|z|_0}  \\
&\pi(u_1 \mid z, \Y ,\Omega_{-p,-p}) = \text{N}(u_1; m_z, U_z^{-1}) \\
&\pi(u_2 \mid \Y) = \text{Ga}\left(u_2; \frac{n}{2} + 1, \frac{S_{pp} + \lambda}{2}\right),
\end{align*}
and this ends the proof of Proposition \ref{prop:wang_exactzeroes_notation}.

\subsection{Proof of Proposition \ref{prop:conductance-bound}}

Recall that we denote the lazy Markov kernel of the ideal Gibbs sampler by $ P_{Gibbs} = \frac{1}{2} I + \frac{1}{2p} \sum_{j = 1}^p P_{Gibbs}^j$, where $P_{Gibbs}^j$ is the conditional Markov kernel updating $\Omega_{.j}$ given $\Omega_{-j,-j}$, i.e., 
\begin{align*}
    P_{Gibbs}^j(\Omega, \Omega') =  P^{j, \Omega_{-j,-j}}(\Omega_{\cdot j}, \Omega_{\cdot j}')\delta_{\Omega_{-j,-j}} (\Omega_{-j,-j}'), \qquad \Omega, \Omega' \in \PSD.
\end{align*}
In the equation above, $P^{j, \Omega_{-j,-j}}(\Omega_{\cdot j}, \cdot)$ is the conditional posterior distribution
\begin{align*}
    P^{j, \Omega_{-j,-j}}(\Omega_{\cdot j}, \Omega_{\cdot j}') = \pi(\Omega_{\cdot j}' \mid \Y, \Omega_{-j,-j}) = \pi(\Omega_{\cdot j}' \mid z', \Y, \Omega_{-j,-j}) \pi( z' \mid \Y, \Omega_{-j,-j}).
\end{align*}
Recall that we denote  the lazy kernel of our conditional MH algorithms (BDMH, GIMH or LIT) with $M$ updates by
\ds{\begin{align*}
  P^{M}_{MH}(\Omega, \Omega') = \frac{1}{2} \delta_{\Omega} (\Omega') + \frac{1}{2p} \sum_{j=1}^p P^{j,\Omega_{-j,-j}, M}_{MH} (z, z')  \pi(\Omega_{\cdot j}' \mid z', \Y, \Omega_{-j,-j}) \delta_{\Omega_{-j,-j}} (\Omega_{-j,-j}'),
\end{align*}}
with 
$P^{j,\Omega_{-j,-j}, M}_{MH} (z, z') = P^{j,\Omega_{-j,-j} \otimes M}_{MH} (z, z')$ and
\begin{align}
    P^{j,\Omega_{-j,-j}}_{MH} (z, z') = \frac{1}{2}  \delta_{z}(z') + \frac{1}{2} \left \{  b(z,z') Q^j(z' | z) + \delta_{z}(z') (1 - \bar b(z)) \right\}.\label{eq:mwG-cond-kernel}
\end{align}
In the equation above, $Q^j(z' \mid z):= Q^j(z' \mid z, \Omega_{-j,-j}, \Y)$ 
is the proposal distribution of the MH, e.g., given by \eqref{eq:bdmh-proposal} (BDMH) or \eqref{eq:post-kappa-lr} (GIMH), $b(z,z') =b(z,z',\Omega_{-j,-j})$ is the acceptance probability of the proposal $z'$, given the current state $z$ and $\Omega_{-j,-j}$, defined as
\begin{align*}
    b(z,z') = 1 \wedge \frac{\pi( z' \mid \Y, \Omega_{-j,-j}) Q^j(z \mid z')}{\pi(z \mid \Y, \Omega_{-j,-j}) Q^j(z' \mid z)},
\end{align*}
and $\bar b (z) = \sum_{z''} Q^j(z'' \mid z) b(z,z'')$ is the average acceptance probability given $z$ and $\Omega_{-j,-j}$.  Recall that the form of \eqref{eq:mwG-cond-kernel} comes from the fact that after the MH step for the edge inclusion vector $z'$, the precision matrix elements $\Omega_{\cdot j}'$ are sampled from their exact posterior full conditional given $z'$. We define the conditional Markov kernel that updates $\Omega_{\cdot j}$ by
\begin{align}\label{eq:mh-cond-kernel}
  \Tilde  P^{j,\Omega_{-j,-j}, M}_{MH} (\Omega, \Omega') =   P^{j,\Omega_{-j,-j}, M}_{MH} (z, z')  \pi(\Omega_{\cdot j}' \mid z', \Y, \Omega_{-j,-j}) \delta_{\Omega_{-j,-j}} (\Omega_{-j,-j}')
\end{align}

We use the concept of conditional conductance introduced by \cite{ascolani2024scalability}, which quantifies the closeness between the Metropolis-within-Gibbs kernel \deb{$P_{MH}^{M}$} and  the ideal Gibbs kernel $P_{Gibbs}$. For a subset $B \subset \PSD$, the conditional conductance of \deb{$P_{MH}^{M}$} is 
\begin{align*}
    \phi^*( P_{MH}^{M} , B) \geq  \min_{j \in [p]} \inf_{\Omega \in B} \phi^*_j( \Tilde P_{MH}^{j, \Omega_{-j,-j}, M}, B)
\end{align*}
where 
\begin{align*}
    &\phi^*_j( \Tilde P_{MH}^{j, \Omega_{-j,-j}, M}, B) = \inf \left \{ \frac{\int_D  \Tilde P_{MH}^{j, \Omega_{-j,-j}, M}(\Omega_{\cdot j}, D^c) d \pi(\Omega_{\cdot j}|\Omega_{-j,-j} , \Y) }{\pi(D|\Omega_{-j,-j}, \Y) \pi(D^c|\Omega_{-j,-j}, \Y)} \right \},
\end{align*}
and the infimum is taken over $D$ such that
$\pi(D|\Omega_{-j,-j}, \Y) > 0$ and $D \subset C_j(B, \Omega_{-j,-j}) := \{\Omega_{\cdot j} \in \mathbb{R}^{p} : \Omega  \in B \}$.
Note that 
$\phi^*_j(\Tilde P_{MH}^{j, \Omega_{-j,-j}, M}, B)$ is the conductance (as defined in \eqref{eq:other-conductance}) of the conditional kernel  $\Tilde P_{MH}^{j, \Omega_{-j,-j}, M}$ on $C_j(B, \Omega_{-j,-j})$ and since $\Tilde  P_{MH}^{j, \Omega_{-j,-j}, M}$ is a reversible and positive definite Markov kernel, 
$\phi^*_j(\Tilde  P_{MH}^{j, \Omega_{-j,-j}, M}, B) \leq 1$. 
Using \eqref{eq:spectral-gap-sandwich}, it is lower-bounded by the spectral gap of $\Tilde  P_{MH}^{j, \Omega_{-j,-j}, M}$, i.e.,
\begin{align}
\phi^*_j(\Tilde  P^{j, \Omega_{-j,-j}}, B) \geq
    \text{Gap}(\Tilde  P_{MH}^{j, \Omega_{-j,-j}, M}).
    \nonumber
\end{align}
Moreover, by the definition \eqref{eq:mh-cond-kernel} of $\Tilde P_{MH}^{j, \Omega_{-j,-j}, M}(\Omega_{.j}, \Omega_{.j}')$, with probability $\frac{1}{2}$, the value of $\Omega_{z'j}'$ is sampled from its exact conditional posterior $\pi(\Omega_{z'j} \mid z',\Y,\Omega_{-j,-j})$. That is, the Metropolis-Hastings step applies only to $z'$. 
Hence, the spectral gap of $\Tilde  P_{MH}^{j, \Omega_{-j,-j}, M}$ is lower bounded by half the spectral gap of $ P_{MH}^{j,\Omega_{-j,-j}, M}$,
see Lemma \ref{lem:equal_spectral_gap} for a formal proof. That is, 
\begin{align*}
    \text{Gap}(\Tilde  P_{MH}^{j, \Omega_{-j,-j}, M}) \geq \frac{1}{2} \text{Gap}( P_{MH}^{j, \Omega_{-j,-j}, M}).
\end{align*}
Hence,
\begin{align}\label{eq:ineq-sg}
    \phi^*( P_{MH}^{M} , B) \geq \frac{1}{2} \min_{j\in [p]} \inf_{\Omega \in B} \text{Gap}( P_{MH}^{j, \Omega_{-j,-j}, M}).
\end{align}


Now, using Corollary 1 in \cite{ascolani2024scalability}, we can bound the $s$-conductance of $P_{MH}^{M}$ using the  $s$-conductance of $P_{Gibbs}$ and the conditional conductance as
\begin{align*}
    \phi_s( P^{M}_{MH}) &\geq \phi_s(P_{Gibbs}) \phi^*( P_{MH}^{M} , B) - \frac{\pi(B^c |\Y)}{s} \left(\frac{1}{p} \sum_{j=1}^p  \inf_{\Omega \in B} \phi_j^*( \Tilde  P_{MH}^{j, \Omega_{-j,-j}, M}, B) \right) \nonumber \\
    &\geq \phi_s(P_{Gibbs})  \phi^*( P_{MH}^{M} , B) - \frac{\pi(B^c |\Y)}{s}  \nonumber \\
    &\geq \frac{1}{2} \phi_s(P_{Gibbs})  \min_{j \in [p]} \inf_{\Omega \in B}  \text{Gap}( P_{MH}^{j, \Omega_{-j,-j}, M}) - \frac{\pi(B^c |\Y)}{s},
\end{align*}
where in the second inequality, we have used the fact that $\phi^*_j( \Tilde  P_{MH}^{j, \Omega_{-j,-j}, M}, B) \leq 1$ and \eqref{eq:ineq-sg} for the last inequality. 

\subsection{Proofs of Section \ref{sec:spectral-gap-bdmh}}
\label{ssec:proofs_bdmh}

Lemma \ref{lem:exp-max} gives an auxiliary result regarding certain random variables $d_{k,z}$ that feature in the ratio of posterior probabilities $\pi(z \mid \Y, \Omega_{-j,-j}) / \pi(z' \mid \Y, \Omega_{-j,-j})$ where $(z,z')$ differ by at most one edge.
This result and its proof help give intuition for the proofs of
Lemma \ref{lem:conditions-ass} given in Section \ref{sssec:proof_conditions-ass}
and of Theorem \ref{thm:bdsgap_chang_omegageneral}, given in Section \ref{app:proof-theorem-1}.
Finally, Section \ref{app:technical-lemmas} gives further technical lemmas used in the proofs of this section.

\subsubsection{Lemma \ref{lem:exp-max} and its proof}\label{app:proof-lem-exp-max}

We recall our notation of conditional variance
\begin{align*}
    \Sigma_{AB|C} = \Sigma_{AB} - \Sigma_{AC}\Sigma_{CC}^{-1}\Sigma_{CB},
\end{align*}
for any triplet of subsets $A,B,C \subset [p]$.

\begin{remark}\label{rem:cond-var}
Note that if $A \subseteq C$ or $B \subseteq C$ then $\Sigma_{AB|C} = 0$.
\end{remark}


\begin{lemma}\label{lem:exp-max}
Assume that $\Omega_{-j,-j}=\Omega_{-j,-j}^0$ and let $s := S_{-j,j}$. For any $k \in [p]$ and $z' \in \{0,1\}^{p-1}$, we define $d_{k,z'} = s_k -  \Sigma_{kz'|j} \Sigma_{z'z'|j}^{-1} s_{z'}$. 
Then for each $z \supset z^0$ and $z' = z \backslash \{k\}$ with $k \in z \backslash z^0$, $\frac{d_{k, z'}}{\sqrt{\Sigma^0_{kk \mid z' \cup j} S_{jj}}} \mid S_{jj}\sim N(0,1)$.
Moreover,
\begin{align*}
    \mathbb E \left[ \max_{z \supset z^0,k \in z \backslash z^0} \frac{|d_{k, z'}|}{\sqrt{\Sigma^0_{kk \mid z' \cup j} S_{jj}}} \mid S_{jj} \right]  \leq \sqrt{2\bar d \log p}.
\end{align*}
\end{lemma}

\begin{proof}
Without loss of generality, we prove Lemma \ref{lem:exp-max} for $j=p$, however, note that the proof applies similarly to any other $j=1, \dots, p-1$.  Let $z \supset z^0, k \in z \backslash z^0, z' = z\backslash k$. 
Recall that $s = S_{-p,p}$ and $d_{k,z'} = s_k -  \Sigma_{kz'|p} \Sigma_{z'z'|p}^{-1} s_{z'} =  s_k -  \Sigma_{kz'|p}^0 \Sigma_{z'z'|p}^{0,-1} s_{z'}$ since $\Omega_{-p,-p} = \Omega_{-p,-p}^0$. From \cite{eaton2007wishart}, the distribution of $s$ conditionally on $S_{pp}$ is
\begin{align*}
    s | S_{pp} \sim N(S_{pp} \frac{\Sigma_{-pp}^0}{\Sigma_{pp}^0}, S_{pp} \Sigma_{-p,-p|p}^{0}), 
\end{align*}
where we recall that $\Sigma_{-p,-p|p}^{0} = [\Omega_{-p,-p}^{0}]^{-1}$.
Using simple algebra (see the proof of Lemma \ref{lem:dist-quadratic} for the precise steps), we  deduce that
\begin{align}
    &d_{k, z'}| S_{pp}  \sim N( S_{pp} \mu_{k,z'}^0, S_{pp} \Sigma^{0}_{kk
    |z' \cup p}), \label{eq:dist-d}
\end{align}
with $\mu_{k,z'}^0 := \frac{\Sigma_{kp}^0 - \Sigma_{k z' \mid p}^0 \Sigma_{z' z' \mid p}^{0,-1} \Sigma_{z'p}^0}{\Sigma_{pp}^0}$. 
We now show that $\mu_{k,z'}^0 = 0$. For this we use that
\begin{align*}
    & \Sigma_{  k z' |p}^0 = \Sigma_{  k z'}^0 - \frac{\Sigma_{kp}^{0} \Sigma_{ z'p}^{0T}}{\Sigma_{pp}^{0}} \\
    &\Sigma_{z'z' |p}^{0, -1} = \Sigma_{ z' z'}^{0, -1} + \frac{\Sigma_{ z' z'}^{0, -1}   \Sigma_{ z'p}^{0} \Sigma_{ z' p}^{0T} \Sigma_{  z' z'}^{0, -1}}{\Sigma_{pp}^{0} -  \Sigma_{ z'p}^{0T} \Sigma_{z'  z'}^{0, -1}  \Sigma_{z'p}^{0}},
\end{align*}
which gives
\begin{align}
     \Sigma_{  k z' |p}^0 \Sigma_{z' z'|p}^{0, -1} \Sigma_{ z' p}^0  
     & = \left( \Sigma_{  k z'}^0 - \frac{\Sigma_{kp}^{0} \Sigma_{ z' p}^{0T}}{\Sigma_{pp}^{0}} \right) \left(  \Sigma_{ z' z'}^{0, -1} \Sigma_{ z' p}^0 + (\Sigma_{ z' p}^{0T} \Sigma_{  z' z'}^{0, -1}\Sigma_{ z' p}^0 )\frac{\Sigma_{  z'z'}^{0, -1}   \Sigma_{ z' p}^{0} }{\Sigma_{pp}^{0} -  \Sigma_{ z' p}^{0T} \Sigma_{  z'}^{0, -1}  \Sigma_{ z' p}^{0}} \right) \nonumber \\
      &=  \Sigma_{  k z'}^0 \Sigma_{  z'z'}^{0, -1} \Sigma_{ z' p}^0 - \frac{ \Sigma_{ z' p}^{0T}  \Sigma_{ z' z'}^{0, -1} \Sigma_{ z' p}^0}{\Sigma_{pp}^{0}} \Sigma_{k p}^{0} + (\Sigma_{ z' p}^{0T} \Sigma_{  z'z'}^{0, -1}\Sigma_{ z' p}^0 )\frac{ \Sigma_{  k z'}^0 \Sigma_{ z' z'}^{0, -1}   \Sigma_{ z' p}^{0} }{\Sigma_{pp}^{0} -  \Sigma_{ z' p}^{0T} \Sigma_{  z'}^{0, -1}  \Sigma_{z' p}^{0}} \nonumber \\
      &-  (\Sigma_{ z' p}^{0T} \Sigma_{ z' z'}^{0, -1}\Sigma_{ z' p}^0 )^2 \frac{\Sigma_{kp}^{0} }{\Sigma_{pp}^{0}(\Sigma_{pp}^{0} -  \Sigma_{ z' p}^{0T} \Sigma_{  z' z'}^{0, -1}  \Sigma_{ z' p}^{0})}.
\label{eq:ncp_equals0_step1}
\end{align}
Consider the expression $\Sigma_{  k z'}^0 \Sigma_{ z' z'}^{0, -1} \Sigma_{ z' p}^0$ featuring in the first and third terms of \eqref{eq:ncp_equals0_step1}.
Using that $\Sigma_{z' p}^0 = - \frac{1}{\Omega_{pp}^0} \Sigma^0_{z' z^0} \Omega^0_{z^0 p}$, we have 
\begin{align*}
    \Sigma_{kp}^0 - \Sigma_{k z'}^{0} \Sigma_{z'z'}^{0, -1} \Sigma_{ z' p}^0 &= - \frac{1}{\Omega_{pp}^0} (\Sigma_{  k z^0}^0 - \Sigma_{k z'}^{0} \Sigma_{z' z'}^{0, -1} \Sigma_{z' z^0}^{0} ) \Omega_{ z^0 p}^0 = - \frac{1}{\Omega_{pp}^0} \Sigma_{  k z^0|z' }^0  \Omega_{ z^0 p} ^0 = 0,
\end{align*}
using that  $\Sigma^0_{ k z^0 \mid z'}= 0$ since $z' = z \backslash \{k \} \supseteq z^0$ (see Remark \ref{rem:cond-var}). 
Therefore, $\Sigma_{  k z'}^0 \Sigma_{ z' z'}^{0, -1} \Sigma_{ z' p}^0= \Sigma_{kp}^0$ and \eqref{eq:ncp_equals0_step1} is
\begin{align*}
& \Sigma_{kp}^0 - \frac{ \Sigma_{ z' p}^{0T}  \Sigma_{  z' z'}^{0, -1} \Sigma_{ z'p}^0}{\Sigma_{pp}^{0}} \Sigma_{kp}^0 
    + \frac{\Sigma_{ z' p}^{0T} \Sigma_{ z' z'}^{0, -1}\Sigma_{ z' p}^0}{\Sigma_{pp}^{0} -  \Sigma_{ z' p }^{0T} \Sigma_{z'  z'}^{0, -1}  \Sigma_{z' p}^{0}} \Sigma_{kp}^0 
    - \frac{(\Sigma_{ z' p}^{0T} \Sigma_{ z' z'}^{0, -1}\Sigma_{ z'p}^0 )^2}{\Sigma_{pp}(\Sigma_{pp}^{0} -  \Sigma_{ z' p}^{0T} \Sigma_{ z' z'}^{0, -1}  \sigma_{ z}^{0})} \Sigma_{kp}^{0}
    \nonumber \\
&= \Sigma_{kp}^0 \left( 1  + \frac{ - \Sigma_{z' p}^{0T} \Sigma_{z'  z'}^{0, -1}  \Sigma_{z' p}^{0} (\Sigma_{pp}^{0} -  \Sigma_{z' p}^{0T} \Sigma_{z' z'}^{0, -1}  \Sigma_{z' p}^{0} - \Sigma_{pp}^{0} ) - (\Sigma_{ z' p}^{0T} \Sigma_{ z' p}^{0, -1}  \Sigma_{ z' p}^{0})^2}{\Sigma_{pp}^0 (\Sigma_{pp}^{0} -  \Sigma_{ z'p }^{0T} \Sigma_{ z' p}^{0, -1}  \Sigma_{ z' p}^{0})}\right) \\
   &=\Sigma_{k p}^0,
\end{align*}
where the right-hand side follows from simple algebra. Hence,
\begin{align}
   \Sigma_{k p}^0= \Sigma_{k z' \mid p}^0 \Sigma_{z' z' \mid p}^{0,-1} \Sigma_{z'p }^0, \label{eq:proof-sigma}
\end{align}
 and $\mu_{k,z'}^0=0$, implying that $d_{k, z'}| S_{pp}  \sim N( 0, S_{pp} \Sigma^{0}_{kk|z' \cup p})$.


Therefore, from \eqref{eq:dist-d}, $\frac{d_{k, z'}}{\sqrt{\Sigma^0_{kk \mid z' \cup p} S_{pp}}} \sim N( 0, 1)$ which proves the first statement of Lemma \ref{lem:exp-max}. 

We now prove the second statement. Since there are less than $p^{\bar d}$ different models in $\mathcal{Z}(\bar d)$, thus
\begin{align*}
    \mathbb E \left[ \max_{z \supset z^0,k \in z \backslash z^0} \frac{|d_{k, z'}|}{\sqrt{\Sigma^0_{kk \mid z' \cup p} S_{pp}}} \mid S_{pp} \right]  &\leq \sqrt{2 \bar d \log p},
\end{align*}
using that the expected maximum of $m$ standard Gaussian variables is bounded by $\sqrt{2\log m}$.

\end{proof}

\subsubsection{Proof of Lemma \ref{lem:conditions-ass}}
\label{sssec:proof_conditions-ass}

Recall that Lemma \ref{lem:conditions-ass} states that our Assumptions  \ref{ass-maxexpec-omegageneral-overfitted} and \ref{ass-betamin_omegageneral} hold for $z^* = z^0 = Z_{-j,j}^0$, for any column index $j$, under conditions C0-C4. For simplicity, we prove Assumptions  \ref{ass-maxexpec-omegageneral-overfitted} and \ref{ass-betamin_omegageneral} hold for $j=p$, though the same proof applies to any $j \in [p]$. 

We first recall Assumption \ref{ass-maxexpec-omegageneral-overfitted} and prove that it holds: 
\begin{align}
&\lim_{n \to \infty} \sqrt{\frac{\Sigma_{kk \mid z' \cup p}}{\Tilde \Sigma_{kk \mid z' \cup p}}}
\sqrt{2[(\Tilde \alpha - 1) \log p + \log (g_1 \sqrt{n}) - \log (\bar d +1)]} 
\nonumber \\
& 
- (1 + \epsilon) \frac{\sqrt{n \Sigma_{pp}^0} |\mu_{k, z'}|}{\sqrt{\Tilde \Sigma_{kk \mid z' \cup p}}}
- \mathbb E \left[ \max_{z \supset z^*} \frac{|d_{k, z'}|  - \mathbb E[|d_{k, z'}|]}{\sqrt{\Tilde \Sigma_{kk \mid z' \cup p} S_{pp}}} \right]
 - \sqrt{2 \log p} = \infty,
\label{eq:lim-lem1}
\end{align}
where $k \in z \backslash z^*$, $z^* = z \backslash k$,  $d_{k, z'} = s_k - s_{z'}^T \Sigma_{z'z' \mid p}^{-1} \Sigma_{k z' \mid p}$,
$\mu_{k, z'}= (\Sigma_{kp}^0 - \Sigma_{k z' \mid p} \Sigma_{z' z' \mid p}^{-1} \Sigma_{z' k}^0) / \Sigma_{pp}^0$ and
\begin{align}
\Tilde \Sigma_{kk \mid z' \cup p} :=  \Sigma_{kk \mid p}^{0} + \Sigma_{k z' \mid p} \Sigma_{z'z' \mid p}^{-1}  \Sigma_{z'z' \mid p}^{0} \Sigma_{z'z' \mid p}^{-1}   \Sigma_{ z' k \mid p} - \Sigma_{k z' \mid p}^{0} \Sigma_{z'z' \mid p}^{-1} \Sigma_{z' k \mid p} - \Sigma_{k z' \mid p} \Sigma_{z'z' \mid p}^{-1} \Sigma_{z' k \mid p}^0.
\nonumber
\end{align}
Recall also that
\begin{align*}
     \Sigma_{kk \mid z' \cup p} =  \Sigma_{kk} -  \Sigma_{k z' \cup p} \Sigma_{z' \cup p, z' \cup p}^{-1} \Sigma_{z' \cup p k} =  \Sigma_{kk \mid p}  - \Sigma_{k z' \mid p} \Sigma_{z'z' \mid p}^{-1} \Sigma_{z' k \mid p},
\end{align*}
and that if $\Omega_{-p,-p} = \Omega_{-p,-p}^0$, then $\Tilde \Sigma_{kk \mid z' \cup p} = \Sigma_{kk \mid z' \cup p} = \Sigma_{kk \mid z' \cup p}^0$. 


The first step in this proof consists of finding an upper-bound of the term
\begin{align*}
     I_1: = \mathbb E \left[ \max_{z \supset z^0} \frac{|d_{k, z'}|  - \mathbb E[|d_{k, z'}|]}{\sqrt{\Tilde \Sigma_{kk \mid z' \cup p} S_{pp}}} \right],
\end{align*}
appearing in \eqref{eq:lim-lem1}. Then in the second and third steps, we upper-bound $I_2 := \frac{\sqrt{n \Sigma_{pp}^0} |\mu_{k, z'}|}{\sqrt{\Tilde \Sigma_{kk \mid z' \cup p}}}$ and lower-bound $ \sqrt{\frac{\Sigma_{kk \mid z' \cup p}}{\Tilde \Sigma_{kk \mid z' \cup p}}}$. Finally, in the fourth step, we prove that \eqref{eq:lim-lem1} holds if $\alpha$ is sufficiently large, and this is the case under C4.

\textbf{Bounding $I_1$.} We use the same argument as in the proof of Lemma \ref{lem:exp-max}. Using Lemma \ref{lem:dist-quadratic},
we have
\begin{align*}
    d_{k, z'} | S_{pp}  \sim \text{N}\left(S_{pp}\mu_{k, z'}, S_{pp} \Tilde \Sigma_{kk |z'\cup p}\right),
\end{align*}
which implies that 
\begin{align*}
    \frac{d_{k, z'}  - \mathbb E [d_{k, z'}]}{\sqrt{\Tilde \Sigma_{kk \mid z' \cup p} S_{pp}}} \sim N(0,1)
\end{align*}
for any $z \in \mathcal{Z}(\bar d)$ and $k$. Therefore, since there are less than $p^{\bar d}$ models in $\mathcal{Z}(\bar d)$ 
and using the upper bound on the expectation of the maximum of Gaussian variables, we obtain
\begin{align}\label{eq:bound-i1}
  \mathbb E \left[ \max_{z \in \mathcal{Z}(\bar d) : z \supset z^0} \frac{|d_{k, z'}  - \mathbb E [d_{k, z'}]|}{\sqrt{\Tilde \Sigma_{kk \mid z' \cup p} S_{pp}}} \right] \leq \sqrt{2 \bar d \log p},
\end{align} 
and since $|d_{k, z'}  - \mathbb E [d_{k, z'}]| \geq |d_{k, z'}|  - |\mathbb E [d_{k, z'}]|  \geq |d_{k, z'}|  - \mathbb E[|d_{k, z'}|]$ we conclude that $I_1 \leq \sqrt{2 \bar d \log p}$.

\color{black}

\textbf{Bounding $I_2$.} We upper-bound 
    $I_2 = \frac{\sqrt{n \Sigma^0_{pp}}|\mu_{k,z'}|}{\sqrt{\Tilde \Sigma_{kk \mid z' \cup p} }} $ for any 
    $z \supset z^0$, $k \in z \backslash z^0$ and $z' = z \backslash \{k\}$. 
    Recall our notation $A := [\Omega_{-p,-p}]^{-1}$ and
    \begin{align*}
        \mu_{k,z'} = \frac{\Sigma_{kp}^0 - A_{k z'}A_{z'z'}^{-1}\Sigma^0_{z'p}}{\Sigma_{pp}^0} = \frac{\Sigma_{kp}^0 - \Sigma_{k z'| p}\Sigma_{z'z'|p}^{-1}\Sigma^0_{z'p}}{\Sigma_{pp}^0}.
    \end{align*}
    We first upper bound $|\mu_{k,z'}|$ and we do this by first bounding the square of the numerator of $\mu_{k,z'}$. We have
\begin{align*}
     (\Sigma_{kp}^0 - A_{kz'}A_{z'z'}^{-1}\Sigma^0_{z'p})^2 &= (\Sigma_{kp}^0 - A_{kz'}^0 A_{z'z'}^{0,-1}\Sigma^0_{z'p} + A_{kz'}^0 A_{z'z'}^{0,-1}\Sigma^0_{z'p} - A_{kz'}A_{z'z'}^{-1}\Sigma^0_{z'p})^2 \\
&\leq 2(\Sigma_{kp}^0 - A_{kz'}^0 A_{z'z'}^{0,-1}\Sigma^0_{z'p})^2  + 2( A_{kz'}^0 A_{z'z'}^{0,-1}\Sigma^0_{z'p} - A_{kz'}A_{z'z'}^{-1}\Sigma^0_{z'p})^2 \\
&\leq 2(\Sigma_{kp}^0 - A_{kz'}^0 A_{z'z'}^{0,-1}\Sigma^0_{z'p})^2  + 2 \|A_{kz'}^0 A_{z'z'}^{0,-1}- A_{kz'}A_{z'z'}^{-1}\|^2 \|\Sigma^0_{z'p}\|^2.
\end{align*}
We now bound each of the term on the RHS of the previous equation. Using \eqref{eq:proof-sigma} from the proof of Lemma \ref{lem:exp-max}, 
we have that $\Sigma_{kp}^0 - A_{kz'}^0 A_{z'z'}^{0,-1}\Sigma^0_{z'p} = \Sigma_{kp}^0 - \Sigma_{kz'|p}^0 \Sigma_{z'z'|p}^{0,-1}\Sigma^0_{z'p} = 0$. Moreover, using \eqref{eq:delta-akz} from the proof of Lemma \ref{lem:tech-3}, $\| A_{k z'} A_{z' z'}^{-1}  -  A_{k z'}^0 A_{z' z'}^{0,-1} \| \leq 6 \delta \epsilon_0^{-5}$.
Thus,
\begin{align*}
    |\mu_{k,z'}| \leq 6\sqrt{2} \delta\epsilon_0^{-5} \frac{\|\Sigma_{z'p}^0\|}{\Sigma_{pp}^0}.
\end{align*}
It thus remains to bound $\|\Sigma^0_{z'p}\|/\Sigma_{pp}^0$. Note that
\begin{align}
    &|\Sigma^0| = |\Sigma^0_{-p,p}| (\Sigma^0_{pp} - (\Sigma^0_{-p,p} )^T\Sigma^{0,-1}_{-p,-p} \Sigma^0_{-p,p}) > 0 \nonumber\\
    &\implies \Sigma^0_{pp} > (\Sigma^0_{-p,p} )^T\Sigma^{0,-1}_{-p,-p} \Sigma^0_{-p,p} \geq \lambda_{min}(\Sigma^{0,-1}_{-p,-p})^2 \|\Sigma^0_{-p,p}\|^2 \geq \epsilon_0 \|\Sigma^0_{-p,p}\|^2, \label{eq:ineq-norm-diag}
\end{align}
using that $\lambda_{min}(\Sigma^{0,-1}_{-p,-p}) = 1/\lambda_{max}(\Sigma^{0}_{-p,-p}) \geq 1/\lambda_{max}(\Sigma^{0}) = \lambda_{min}(\Omega^0) \geq \epsilon_0$.
Since \eqref{eq:ineq-norm-diag} and (C0) imply that
\begin{align*}
     \frac{\|\Sigma_{z'p}^0\|}{\Sigma_{pp}^0} \leq  \frac{\|\Sigma_{-p,p}^0\|}{\Sigma_{pp}^0} \leq \epsilon_0^{-1},
\end{align*}
we obtain
$$|\mu_{k, z'}| \leq 6 \sqrt{2} \delta \epsilon_0^{-6}.$$

The final step for bounding $I_2$ consists of  bounding $\frac{\Sigma_{pp}^0}{\Tilde \Sigma_{kk \mid z' \cup p}}$. For this, we use Lemma \ref{lem:tech-3} below. Since
\begin{align*}
    \frac{\Tilde \Sigma_{kk \mid z' \cup p}}{\Sigma_{pp}^0} =  1 + \frac{\Tilde \Sigma_{kk \mid z' \cup p} - \Sigma_{pp}^0 }{\Sigma_{pp}^0} \geq 1 - \frac{|\Sigma_{pp}^0 - \Tilde \Sigma_{kk \mid z' \cup p}|}{\Sigma_{pp}^0} \geq 1 - \frac{6\delta\epsilon_0^{-6}}{\Sigma_{pp}^0} \geq 1 - 6\delta\epsilon_0^{-7},
\end{align*}
under C1 and C3. 
Moreover, since under C3, $\delta = o(1)$, thus for any $n \geq n_0$ with $n_0$ that only depends on $\epsilon_0$, we obtain
\begin{align*}
     \frac{\Sigma_{pp}^0}{\Tilde \Sigma_{kk \mid z' \cup p}} \leq 1 + 6\delta\epsilon_0^{-7} \leq 2,
\end{align*}
which leads to
\begin{align}\label{eq:bound-i2}
    I_2 = 
    \frac{\sqrt{n \Sigma_{pp}^0}|\mu_{k, z'}|}{\sqrt{\Tilde \Sigma_{kk \mid z' \cup p} }} \leq 12 \delta \epsilon_0^{-6}  \sqrt{n}.
\end{align}

\textbf{Bounding $\sqrt{\frac{\Sigma_{kk \mid z' \cup p}}{\Tilde \Sigma_{kk \mid z' \cup p}}}$.} We lower bound $\sqrt{\frac{\Sigma_{kk \mid z' \cup p}}{\Tilde \Sigma_{kk \mid z' \cup p}}}$ using again Lemma \ref{lem:tech-3}. We have
\begin{align*}
    \frac{\Tilde \Sigma_{kk \mid z' \cup p}}{\Sigma_{kk \mid z' \cup p}} &= 1  + \frac{\Tilde \Sigma_{kk \mid z' \cup p} - \Sigma_{kk \mid z' \cup p}}{\Sigma_{kk \mid z' \cup p}} \geq 1  - \frac{|\Sigma^0_{kk \mid z' \cup p} - \Tilde \Sigma_{kk \mid z' \cup p}|}{\Sigma_{kk \mid z' \cup p}} -  \frac{|\Sigma_{kk \mid z' \cup p} -  \Sigma_{kk \mid z' \cup p}^0|}{ \Sigma_{kk \mid z' \cup p}} \\
    &\geq 1 - \frac{18 \epsilon_0^{-6}\delta }{ \Sigma_{kk \mid z' \cup p}} \geq 1 - 18 \epsilon_0^{-7}\delta,
\end{align*}
under C3. Moreover since $\delta = o(1)$, we obtain that for any $\epsilon > 0$ and for $n \geq n_0$,
\begin{align}\label{eq:bound-ratio-sigma}
    \sqrt{\frac{\Sigma_{kk \mid z' \cup p}}{\Tilde \Sigma_{kk \mid z' \cup p}}} \geq \sqrt{1 - 18 \epsilon_0^{-7}\delta} \geq 1  - 9 \epsilon_0^{-6}\delta \geq 1 - \epsilon.
\end{align}

\textbf{Sufficient condition.} Given the previous steps, we observe that for Assumption \ref{ass-betamin_omegageneral} to be satisfied it is sufficient that
\begin{align*}
 &\lim_{n \to \infty} (1-\epsilon)
\sqrt{2[(\Tilde \alpha - 1) \log p + \log (g_1 \sqrt{n})} - \log (\bar d +1)]    - \sqrt{2\bar d \log p} \\
&- 12 (1+\epsilon) \epsilon_0^{-6} \delta \sqrt{n} - \sqrt{2 \log p} = +\infty.
\end{align*}

Let $\delta_0>0$ such that $\delta = \delta_0\sqrt{\frac{\log n}{2n} + \frac{\log p}{n}}$ and note that under C3, we have $\delta_0 < \frac{1}{48} \epsilon_0^{6}$. 
Using that $\sqrt{a + b} \leq \sqrt{a} + \sqrt{b}$ for any $a,b \geq 0$, we have
\begin{align*}
    \delta \sqrt{n} \leq \delta_0 \sqrt{\log p} + \delta_0 \sqrt{\frac{1}{2}\log n}.
\end{align*}
Moreover, using that $\sqrt{a} + \sqrt{b} \leq \sqrt{2a + 2b}$ for any $a,b\geq 0$, we also have
\begin{align*}
   \sqrt{2[( \ds{\Tilde \alpha} - 1) \log p + \log \ds{(g_1 \sqrt{n})} - \log (\bar d +1)]}  \geq \sqrt{(\ds{\Tilde \alpha} - 1) \log p} + \sqrt{\frac{1}{2} \log n + \ds{\log \frac{g_1}{\bar d +1 }}}.
\end{align*}
Thus, it is sufficient to prove that
\begin{align}
  &\lim_{n \to \infty}\sqrt{\log p}\left[(1 - \epsilon) \sqrt{(\ds{\Tilde \alpha} - 1)} - \sqrt{2\bar d} - 12(1+\epsilon)\epsilon_0^{-6} \delta_0 - \sqrt{2} \right] \nonumber \\
  &+ \sqrt{\frac{\log n}{2}} \left[ (1-\epsilon) - 12(1+\epsilon)\epsilon_0^{-6} \delta_0 \ds{+ \ds{\log (\frac{g_1}{\bar d +1 })\frac{1}{\log n} }} \right] = +\infty, \label{eq:equiv-ass}
\end{align}
for $\epsilon$ small enough. We prove that the constants in factor of $\sqrt{\log p }$ and  $\sqrt{\log n}$ in \eqref{eq:equiv-ass} are positive.

Since $\delta_0 < \frac{1}{48} \epsilon_0^{6}$, for any $\epsilon \in (0,1)$,
\begin{align*}
    12(1+\epsilon) \epsilon_0^{-6} \delta_0 < \ds{\frac{1}{4} (1 + \epsilon)},
\end{align*}
which implies that \ds{$(1-\epsilon) - 12(1+\epsilon) \epsilon_0^{-6} \delta_0 > \frac{1}{2}$ for any $\epsilon < \frac{1}{5}$. Besides, since by assumption,  $\frac{g_1}{\bar d +1} > \frac{1}{\sqrt{n}} \implies \log \frac{g_1}{\bar d+1} > -\frac{1}{2} \log n$, 
the term after the $\sqrt{\log n/2}$ term in \eqref{eq:equiv-ass} 
is positive.}
Moreover, \ds{since under C4, $\Tilde \alpha > 1 + (\sqrt{2\bar d} + \sqrt{2} + \frac{1}{4})^2$, for $\epsilon$ sufficiently small,}
\begin{align*}
    (1 - \epsilon) \sqrt{\ds{\Tilde \alpha} - 1} > \sqrt{2\bar d} + \sqrt{2} + \ds{\frac{1}{4}( 1 + \epsilon)}. 
\end{align*}
We thus conclude that the constant in front of the $\sqrt{\log p}$ term in \eqref{eq:equiv-ass} is positive and thus that \eqref{eq:equiv-ass} holds which implies Assumption \ref{ass-maxexpec-omegageneral-overfitted}.

We now recall  Assumption \ref{ass-betamin_omegageneral} and prove that it holds: 
\begin{align}
    \lim_{n \to \infty} (\frac{1}{2} - \epsilon) \min_{z \not\supset z^0} \frac{ n \Sigma_{pp}^0 \mu_{\ell,z}^2}{\Tilde \Sigma_{\ell \ell \mid z \cup p}}  - (\bar d + 2) \log p = +\infty.
    \label{eq:ass-4-1}
\end{align}
for some constant $\epsilon > 0$ and
\begin{align}
\left[\log(\bar d +1) + (\ds{\Tilde \alpha} + 1) \log p + \log \ds{(g_1 \sqrt{n })} \right]
= o\left(\min_{z \not\supset z^0} \frac{n \Sigma_{pp }^0 \mu_{\ell,z}^2}{\Tilde \Sigma_{\ell \ell \mid z \cup p}} \right).
\label{eq:ass-4-2}
\end{align}
Recall that above, $\ell \in z^0 \backslash z$ and $\mu_{\ell,z} = \frac{\Sigma_{\ell p}^0 - \Sigma_{\ell z | p} \Sigma_{z z | p}^{-1}\Sigma^0_{zp}}{\Sigma_{pp}^0}$. The first step of this proof is to lower bound $\mu_{\ell,z}$  under our betamin condition C2 for any $z \not \supset z^0$ with a constant independent of $z$ and $\ell$. In the second step we upper-bound $\Sigma_{pp }^0 \Tilde \Sigma_{\ell \ell \mid z \cup p}$. We then conclude that  \eqref{eq:ass-4-1}  and \eqref{eq:ass-4-2} hold using that under C0, $\bar d \log p = o(n)$.

\textbf{Bounding $\mu_{\ell,z}$.} We first lower bound the numerator of $\mu_{\ell,z} $. Using the triangular inequality and the sub-multiplicativity property of the spectral norm we have
\begin{align*}
    \Sigma_{\ell p}^0 - \Sigma_{\ell z | p} \Sigma_{z z | p}^{-1}\Sigma^0_{zp} &\geq \Sigma_{\ell p}^0 - \Sigma_{\ell z | p}^0 \Sigma_{z z | p}^{0,-1}\Sigma^0_{zp} - | \Sigma_{\ell z | p} \Sigma_{z z | p}^{-1}\Sigma^0_{zp} - \Sigma_{\ell z | p}^0 \Sigma_{z z | p}^{0,-1}\Sigma^0_{zp}| \\
    &\geq  \Sigma_{\ell p}^0 - \Sigma_{\ell z | p}^0 \Sigma_{z z | p}^{0,-1}\Sigma^0_{zp} - \| \Sigma_{\ell z | p} \Sigma_{z z | p}^{-1} - \Sigma_{\ell z | p}^0 \Sigma_{z z | p}^{0,-1} \| \|\Sigma^0_{zp}\|.
\end{align*}
Using our notation $A^0: = \Sigma^0_{-p,-p|p}$ and $A: = \Sigma_{-p,-p|p}$ and \eqref{eq:delta-akz} (which remains valid for any $z' \in \mathcal{Z}(\bar d)$ and $k\not \in z'$) 
\begin{align*}
    \| \Sigma_{\ell z | p} \Sigma_{z z | p}^{-1} - \Sigma_{\ell z | p}^0 \Sigma_{z z | p}^{0,-1} \| = \| A_{\ell z} A_{z z}^{-1} - A_{\ell z }^0 A_{z z}^{0,-1} \| \leq  6 \delta \epsilon_0^{-5}.
\end{align*}

Using Lemma \ref{lem:tech-4} below, we have
\begin{align*}
    &\Sigma_{\ell p}^0 - \Sigma_{\ell z | p}^0 \Sigma_{z z | p}^{0,-1}\Sigma^0_{zp} \geq \sqrt{\omega_{min}} \epsilon_0 \\
    &\implies \mu_{\ell,z} \geq \frac{\sqrt{\omega_{min}} \epsilon_0}{\Sigma_{pp}^0}.
\end{align*}
Therefore, 
\begin{align}
   \frac{n \Sigma_{pp}^0}{\Tilde \Sigma_{\ell \ell \mid z \cup p}} \mu_{\ell,z}^2 \geq \frac{\omega_{min} \epsilon_0^2}{\Tilde \Sigma_{\ell \ell \mid z \cup p}\Sigma_{pp}^0}, \label{eq:inter1}
\end{align}

\textbf{Bounding $\Tilde \Sigma_{\ell \ell \mid z \cup p}\Sigma_{pp}^0$.} We have, using Lemma \ref{lem:tech-3} and C3, that
\begin{align*}
    \Tilde \Sigma_{\ell \ell \mid z \cup p} \leq \Sigma_{pp}^0 + |\Sigma_{pp}^0 - \Tilde \Sigma_{\ell \ell \mid z \cup p} | \leq \Sigma_{pp}^0 +  6 \delta \epsilon_0^{-6} = \Sigma_{pp}^0  + o(1),
\end{align*}
therefore, for $n$ large enough,
\begin{align}
    \Tilde \Sigma_{\ell \ell \mid z \cup p}\Sigma_{pp}^0 \leq \Sigma_{pp}^0 + o(1) \leq 2\epsilon_0^{-1}.\label{eq:inter2}
\end{align}

\textbf{Conclusion.} From \eqref{eq:inter1} and \eqref{eq:inter2}, we conclude that
\begin{align}
    \min_{z \not\supset z^0} \max_{\ell \in z^0 \backslash z}\frac{ n \Sigma_{pp}^0 \mu_{\ell,z}^2}{\Tilde \Sigma_{\ell \ell \mid z \cup p}}  \geq \frac{\omega_{min} \epsilon_0^3}{2} n. \label{eq:min-mul}
\end{align}

We then directly conclude that \eqref{eq:ass-4-1} and \eqref{eq:ass-4-2} hold for any fixed $\epsilon > 0$ under C0 \ds{and C4  since $\bar d \log p \leq \tilde{\alpha} \log p  = o(n)$}. 

We now state and prove the intermediary results used in the proof of Lemma \ref{lem:conditions-ass}.




\begin{lemma}\label{lem:tech-3}
    Under the assumptions of Lemma \ref{lem:conditions-ass}, for any $z \in \mathcal{Z}(\bar d)$, $k  \in z \backslash z^0$, $z' = z\backslash \{k\}$, it holds that
    \begin{align*}
       |\Tilde \Sigma_{kk|z' \cup p} - \Sigma_{kk|z'\cup p}^{0}|  &\leq  6 \delta \epsilon_0^{-6} \\ 
        |\Sigma_{kk|z' \cup p} - \Sigma_{kk|z' \cup p}^0| &\leq 12 \delta \epsilon_0^{-6}.
    \end{align*}
    where 
    \begin{align*}
        &\Sigma_{kk | z' \cup p}^0 = \Sigma_{kk|p}^0 - \Sigma_{kz' | p}^0 \Sigma_{z'z' | p}^{0,-1} \Sigma_{z'k | p}^0\\
        &\Sigma_{kk | z' \cup p} = \Sigma_{kk|p} - \Sigma_{kz' | p} \Sigma_{z'z' | p}^{-1} \Sigma_{z'k | p} \\
        &\Tilde \Sigma_{kk \mid z' \cup p} =  \Sigma_{kk \mid p}^{0} + \Sigma_{k z' \mid p} \Sigma_{z'z' \mid p}^{-1}  \Sigma_{z'z' \mid p}^{0} \Sigma_{z'z' \mid p}^{-1}   \Sigma_{ z' k \mid p} - \Sigma_{k z' \mid p}^{0} \Sigma_{z'z' \mid p}^{-1} \Sigma_{z' k \mid p} - \Sigma_{k z' \mid p} \Sigma_{z'z' \mid p}^{-1} \Sigma_{z' k \mid p}^0.
    \end{align*}
\end{lemma}

\begin{proof}
Recall our notation $A = [\Omega_{-p,-p}]^{-1} =\Sigma_{-p,-p| p}$ and $A^0 =  \Omega_{-p,-p}^{0,-1} = \Sigma_{-p,-p| p}^{0}$. We have 
\begin{align}
   &|\Tilde \Sigma_{kk|z' \cup p} - \Sigma_{kk|z'\cup p}^{0}| \nonumber \\
   &= |\Sigma_{kk \mid p}^{0} + \Sigma_{k z' \mid p} \Sigma_{z'z' \mid p}^{-1}  \Sigma_{z'z' \mid p}^{0} \Sigma_{z'z' \mid p}^{-1}   \Sigma_{ z' k \mid p} - \Sigma_{k z' \mid p}^{0} \Sigma_{z'z' \mid p}^{-1} \Sigma_{z' k \mid p} - \Sigma_{k z' \mid p} \Sigma_{z'z' \mid p}^{-1} \Sigma_{z' k \mid p}^0 \nonumber \\
   &\qquad - (\Sigma_{kk|p}^{0} - \Sigma_{kz'|p}^{0} \Sigma_{z'z'|p}^{0,-1}  \Sigma_{z'k|p}^{0}) | \nonumber \\
   &= | A_{k z'} A_{z' z'}^{-1} A_{z' z'}^{0} A_{z' z'}^{-1}  A_{ z' k} - A_{k z'}^0 A_{z' z'}^{-1}  A_{z' k} - A_{k z'} A_{z' z'}^{-1} A_{ z' k}^0 + A_{kz'}^0  A_{z' z'}^{0,-1} A_{z'k}^0 | \nonumber  \\
   &\leq |(A_{k z'} A_{z' z'}^{-1} A_{z' z'}^{0}  - A_{k z'}^0)  A_{z' z'}^{-1}  A_{z' k}| + |( A_{kz'}^0  A_{z' z'}^{0,-1} - A_{k z'} A_{z' z'}^{-1} ) A_{z'k}^0| \nonumber \\
   &\leq \|A_{k z'} A_{z' z'}^{-1} A_{z' z'}^{0}  - A_{k z'}^0 \| \|  A_{z' z'}^{-1}  A_{z' k}\| + \|A_{kz'}^0  A_{z' z'}^{0,-1} - A_{k z'} A_{z' z'}^{-1}\| \| A_{z'k}^0\|. \label{eq:delta-sigma-tilde}
\end{align}
We now upper bound each term in the RHS of the previous inequality. Note that under C3, $\|A\| \leq \|\Sigma \| \leq  \epsilon_0^{-1}$ and $\|A^{-1}\| \leq \|\Omega\| \leq  \epsilon_0^{-1}$. Thus, since $A_{z' z'}^{-1}, A_{z' k}  $ and $A_{z'k} $ are sub-matrices of $A$ and $A^{-1}$,
\begin{align}
    &\|A_{z' z'}^{-1}A_{z' k} \| \leq \|A_{z' z'}^{-1}\| \|A_{z' k} \| \leq  \|A^{-1}\|  \|A\| \leq \epsilon_0^{-2} \label{eq:bound-akz} \\
    &\|A_{z'k}\| \leq \|A^{-1}\| \leq \epsilon_0^{-1}. \nonumber
\end{align}
Moreover, 
\begin{align*}
    \|A_{k z'} A_{z' z'}^{-1} A_{z' z'}^{0}  - A_{k z'}^0 \| &= \|A_{k z'} -  A_{k z'}^0 + A_{k z'}A_{z' z'}^{-1} (A_{z' z'}^{0} - A_{z' z'})  \| \\
    &\leq \|A_{k z'} -  A_{k z'}^0\| + \|A_{k z'} A_{z' z'}^{-1}\| \|A_{z' z'}^{0} - A_{z' z'}\| \\
    &\leq \|A_{zz} - A_{zz}^0\| ( 1 + \epsilon_0^{-2}),
\end{align*}
using in the second inequality that $\|A_{k z'} -  A_{k z'}^0\| \leq \|A_{zz} - A_{zz}^0\|$ and $ \|A_{z'z'} - A_{z'z'}^0\| \leq  \|A_{zz} - A_{zz}^0\|$ 
and \eqref{eq:bound-akz}. 
Moreover,
\begin{align*}
    \|A_{zz} - A_{zz}^0\| &= \| \Sigma_{zz|p} - \Sigma_{zz|p}^0 \| \\
    &\leq \| \Sigma_{zz} - \Sigma_{zz}^0\| + \|\sigma_{zp}\sigma_{zp}^T/\sigma_{pp} - \sigma_{zp}^0(\sigma_{zp}^0)^T/\sigma_{pp}^0  \| \\
    &= \| \Sigma_{zz} - \Sigma_{zz}^0\| + \|(\sigma_{zp}\sigma_{zp}^T - \sigma_{zp}^0 (\sigma_{zp}^0)^T)/\sigma_{pp} + \sigma_{zp}^0 (\sigma_{zp}^0)^T(1/\sigma_{pp} -1/\sigma_{pp}^0) \| \\
    &\leq \| \Sigma_{zz} - \Sigma_{zz}^0\| + \sigma_{pp}^{-1} \|\sigma_{zp} - \sigma_{zp}^0\| +  \| \sigma_{zp}^0 (\sigma_{zp}^0)^T \| |1/\sigma_{pp} -1/\sigma_{pp}^0| \\
    &= \| \Sigma_{zz} - \Sigma_{zz}^0\| + \sigma_{pp}^{-1} \|\sigma_{zp} - \sigma_{zp}^0\| + \sigma_{pp}^{-1} (\sigma_{pp}^{0})^{-1}\| \sigma_{zp}^0 (\sigma_{zp}^0)^T \| |\sigma_{pp} -\sigma_{pp}^0| \\
    &\leq \delta + \epsilon_0^{-1} \delta + \epsilon_0^{-2} \delta  \leq 3 \delta \epsilon_0^{-2},
\end{align*}
under C0 and C3. 


Next, we prove that $\| A_{k z'} A_{z' z'}^{-1}  -  A_{k z'}^0 A_{z' z'}^{0,-1} \| \leq 6 \epsilon_0^{-4} \delta$. We have
    \begin{align}
    \| A_{k z'} A_{z' z'}^{-1}  -  A_{k z'}^0 A_{z' z'}^{0,-1} \| &= \| A_{k z'}^0 (A_{z' z'}^{0,-1} - A_{z'z'}^{-1}) +  A_{z'z'}^{-1} ( A_{k z'}^0 - A_{kz'} ) \| \nonumber \\
    &\leq \|A_{k z'}^0 \| \|A_{z'z'}^{0,-1} - A_{z'z'}^{-1}\| + \|A_{z'z'}^{-1} \| \|A_{k z'} - A_{k z'}^0\|\nonumber \\
    &\leq \|A_{zz}^{0,-1} - A_{zz}^{-1}\| (\|A_{k z'}^0 \|  + \|A_{z'z'}^{-1} \|). \nonumber
\end{align}
Note that $\|A_{k z'}^0 \| \leq \|A^0\| \leq \|\Sigma^0\| \leq \epsilon_0^{-1}$ under C1 and similarly $\|A_{z'z'}^{-1} \| = 1/\lambda_{min}(A_{z'z'}) \leq 1/\lambda_{min}(A) \leq 1/\lambda_{min}(\Sigma) \leq \epsilon_0^{-1}$ under C3. Using that for any two non-singular  matrices $M$ and $N$, $\|M^{-1} - N^{-1}\| \leq \|M^{-1}\| \|N^{-1}\| \|M-N\|$, we can prove that
\begin{align*}
    \|A_{zz'}^{0,-1} - A_{zz}^{-1}\| \leq \|A_{zz}^{0,-1}\| \| A_{zz}^{-1}\|\|A_{zz}^{0} - A_{zz}\| \leq 3 \delta \epsilon_0^{-4}.
\end{align*}
Finally, using the block-decomposition of the determinant and the positive-definiteness of $A^0$, we have 
\begin{align*}
    &|A_{zz}^0| = |A_{z'z'}^0| (A^0_{kk} - A^0_{k z'}A_{z'z'}^{0,-1}A^0_{z'k}) > 0 \\
    &\implies  A^0_{kk} > A^0_{k z'}A_{z'z'}^{0,-1}A^0_{z'k} \geq \lambda_{min}(A^0_{z'z'}) \|A^0_{k z'}\|^2 \geq \|A^0_{k z'}\|^2  \epsilon_0 \\
    &\implies \|A^0_{k z'}\| \leq \sqrt{A_{kk}^0}\epsilon_0^{-1/2} \leq \epsilon_0^{-1},
\end{align*}
using C1. Therefore, from the previous computation, we obtain that
\begin{align}
    \| A_{k z'} A_{z' z'}^{-1}  -  A_{k z'}^0 A_{z' z'}^{0,-1} \|  \leq 6 \delta \epsilon_0^{-5}. \label{eq:delta-akz}
\end{align}

Reporting into \eqref{eq:delta-sigma-tilde},  we obtain the desired result
\begin{align*}
    |\Tilde \Sigma_{kk|z' \cup p} - \Sigma_{kk|z'\cup p}^{0}| \leq  6 \delta \epsilon_0^{-6} 
\end{align*}

To prove the second statement we use similar computations. From the proof of the first statement,
\begin{align*}
     |\Sigma_{kk|z' \cup p} - \Sigma_{kk|z' \cup p}^0| &=  |\Sigma_{kk|p} - \Sigma_{kk|p}^0 +  \Sigma_{kz|p} \Sigma_{zz|p}^{-1}\Sigma_{zk|p} - \Sigma_{kz|p}^0 \Sigma_{zz|p}^{0,-1}\Sigma_{zk|p}^0| \\
     &\leq |A_{kk} - A_{kk}^0| + | A_{kz} A_{zz}^{-1}A_{zk} - A_{kz}^0 A_{zz}^{0,-1}A_{zk}^0| \\
     &\leq |A_{kk} - A_{kk}^0| + \| A_{kz}\| \| A_{zz}^{-1}A_{zk} - A_{zz}^{0,-1} A_{zk}^0 \| + \|A_{kz}^0 -A_{kz} \| \|A_{kz}^0 \| \| A_{zz}^{0,-1}\| \\
     &\leq 3\delta \epsilon_0^{-2} + 6\epsilon_0^{-5} \delta +  3\epsilon_0^{-6} \delta \leq 12\epsilon_0^{-6} \delta.
\end{align*}


\end{proof}

\begin{lemma}\label{lem:tech-4}
    Under the assumptions of Lemma \ref{lem:conditions-ass}, for any $z \not \supset z^0$, there exists  $\ell \in z^0 \backslash z$ such that
    \begin{align*}
    \Sigma_{\ell p}^0 - \Sigma_{\ell z | p}^0 \Sigma_{z z | p}^{0,-1}\Sigma^0_{zp} \geq \sqrt{\omega_{min}} \epsilon_0.
\end{align*}
\end{lemma}

\begin{proof}
Defining $\kappa = z^0 \backslash z$ and using simple algebra, we have 
\begin{align*}
    \Sigma_{\kappa p}^0 - \Sigma_{\kappa z | p}^0 \Sigma_{z z | p}^{0,-1}\Sigma^0_{zp}  = \frac{\Sigma_{pp}^0}{\Sigma_{pp}^0 - \Sigma_{pz}^{0} \Sigma_{ z z}^{0, -1}\Sigma_{zp}^0} (\Sigma_{ \kappa p}^0 - \Sigma_{\kappa z}^0 \Sigma_{zz}^{0,-1} \Sigma_{zp}^0).
\end{align*}
The expression above implies that
\begin{align*}
    \| \Sigma_{\kappa p}^0 - \Sigma_{\kappa z | p}^0 \Sigma_{z z | p}^{0,-1}\Sigma^0_{zp}\| \geq \|\Sigma_{ \kappa p}^0 - \Sigma_{\kappa z}^0 \Sigma_{zz}^{0,-1} \Sigma_{zp}^0\|,
\end{align*}
using that 
$\Sigma_{pp}^0 - \Sigma_{pz}^{0} \Sigma_{ z z}^{0, -1}\Sigma_{zp}^0\leq \Sigma_{pp}^0$. Besides, for any $\ell \in \kappa$, using the same argument we can obtain
\begin{align}
    \Sigma_{\ell p}^0 - \Sigma_{\ell z | p}^0 \Sigma_{z z | p}^{0,-1}\Sigma^0_{zp} \geq \Sigma_{ \ell p}^0 - \Sigma_{\ell z}^0 \Sigma_{zz}^{0,-1} \Sigma_{zp}^0. \label{eq:ineq-sigma-kappap}
\end{align}

We now lower-bound $\|\Sigma_{ \kappa p}^0 - \Sigma_{\kappa z}^0 \Sigma_{zz}^{0,-1} \Sigma_{zp}^0\|$. Using Lemma \ref{lem:decomposition} and the fact that  $z^0 = Z_{\cdot p}^0$ is an indicator vector of column $\Omega_{-p,p}^0$, we have
\begin{align*}
    &\Sigma_{ \kappa p}^0 = \frac{1}{\Omega_{pp}^0} \Sigma_{ \kappa,-p}^0 \Omega_{-p,p}^0 =\frac{1}{\Omega_{pp}^0} \Sigma_{ \kappa z^0}^0 \Omega_{z^0p}^0 \\
    &\Sigma_{ z p}^0 = \frac{1}{\Omega_{pp}^0} \Sigma_{ z,-p}^0 \Omega_{-p,p}^0 =\frac{1}{\Omega_{pp}^0} \Sigma_{ z z^0}^0 \Omega_{z^0p}^0,
\end{align*}
where in the RHS of the latter expressions we have used the fact that $\Omega^0_{-p,p} = [0,\Omega^0_{z^0 p}]$.
Therefore, 
\begin{align*}
    \Sigma_{ \kappa p}^0 - \Sigma_{\kappa z}^0 \Sigma_{zz}^{0,-1} \Sigma_{zp}^0 = \frac{1}{\Omega_{pp}^0}   (\Sigma_{  \kappa z^0}^0 - \Sigma_{  \kappa z}^0 \Sigma_{ z z}^{0, -1} \Sigma_{ z z^0}^0) \Omega_{ z^0 p}^0  = \frac{1}{\Omega_{pp}^0}(\Sigma_{  \kappa z^0 | z}^0  \Omega_{ z^0 p}^0) = \frac{1}{\Omega_{pp}^0}(\Sigma_{  \kappa \kappa | z}^0  \Omega_{\kappa p}^0),
\end{align*}
where in the last equality we have used the fact that $\Sigma_{  \kappa z^0 | z}^0 = [\Sigma_{  \kappa \kappa  | z}^0, \Sigma_{  \kappa z^0\cap z | z}^0] =  [\Sigma_{  \kappa  | z}^0, 0]$ and thus,
\begin{align}
     \Sigma_{ \ell p}^0 - \Sigma_{\ell z}^0 \Sigma_{zz}^{0,-1} \Sigma_{zp}^0 = \frac{1}{\Omega_{pp}^0}(\Sigma_{  \ell \ell | z}^0  \Omega_{\ell p}^0). \label{eq:ineq-sigma-kappap2}
\end{align}
Under  C1 and C2, we have
\begin{align}\label{eq:low}
    \frac{\|\Sigma_{  \kappa \kappa | z}^0  \Omega_{ z^0 p}^0\|^2}{(\Omega_{pp}^0)^2} \geq \frac{\| \Omega_{ z^0 p}^0\|^2}{(\Omega_{pp}^0)^2} \lambda_{min}(\Sigma_{  \kappa \kappa | z}^0)^2 \geq | \kappa|_0 \omega_{min} \epsilon_0^2,
\end{align}
using that $\Sigma_{ \kappa \kappa |z}^{0,-1}$ is a submatrix of $\Sigma_{  \kappa \cup z, \kappa \cup z}^{0,-1}$ which is also a sub-matrix of $\Omega^{0}$, therefore,
\begin{align*}
    \lambda_{min} ( \Sigma_{ \kappa \kappa |z}^{0} ) = \lambda_{max} ( \Sigma_{  \kappa \kappa |z}^{0,-1} )^{-1}  \geq \lambda_{max} ( \Sigma_{  \kappa \cup z, \kappa \cup z}^{0,-1} )^{-1} \geq \lambda_{max} ( \Omega^{0} )^{-1} \geq \epsilon_0.
\end{align*}
Observing that
\begin{align*}
    \frac{\|\Sigma_{  \kappa \kappa | z}^0  \Omega_{  \kappa p}^0\|^2}{(\Omega_{pp}^0)^2} = \sum_{\ell \in  \kappa} \frac{(\Sigma_{  \ell \ell | z}^0  \Omega_{\ell p}^0)^2}{(\Omega_{pp}^0)^2},
\end{align*}
therefore from \eqref{eq:low} there must exist $\ell \in \kappa = z^0 \backslash z$ such that
\begin{align*}
    \frac{(\Sigma_{  \ell \ell  | z}^0  \Omega_{\ell p}^0)^2}{(\Omega_{pp}^0)^2} \geq \omega_{min} \epsilon_0^2.
\end{align*}
From \eqref{eq:ineq-sigma-kappap} and \eqref{eq:ineq-sigma-kappap2}, this implies that for this $\ell$,
\begin{align*}
    \Sigma_{\ell p}^0 - \Sigma_{\ell z | p}^0 \Sigma_{z z | p}^{0,-1}\Sigma^0_{zp} \geq \sqrt{\omega_{min}} \epsilon_0,
\end{align*}
and this terminates this proof.
    
\end{proof}

\subsubsection{Proof of Theorem \ref{thm:bdsgap_chang_omegageneral}}\label{app:proof-theorem-1}

We first prove the first statement of Theorem \ref{thm:bdsgap_chang_omegageneral}: for any $\epsilon > 0$,
\begin{align*}
    \lim_{n \to \infty} \mathbb P \left[ \bigcap_{j=1}^p \left \{ \mbox{Gap}( P_{BD}^{j,\Omega_{-j-j},M=1})^{-1} \leq (4 + \epsilon) p (\bar d + 1) \right\} \right] = 1
\end{align*}
under Assumptions \ref{ass-eigen0}, \ref{ass-maxexpec-omegageneral-overfitted} and \ref{ass-betamin_omegageneral}.

Recall that we denote by $z^*$ the model in $\mathcal{Z}(\bar d)$ with largest conditional posterior probability given $\Omega_{-j,-j}$, i.e.,
\begin{align}\label{eq:z-star}
    z^* = \arg \max_{z \in \mathcal{Z}(\bar d)} \mathbb E \left[\frac{1}{2}\frac{s_{z}^T \Sigma_{z z \mid j}^{-1} s_{z}}{S_{jj}} - |z|_0 \left(\frac{1}{2}\log n     + \log g_1 + \tilde{\alpha} \color{black} \log p \right) \right],
\end{align}
where the expectation is over $\Y \sim N(0, [\Omega^0]^{-1})$,
    and $\tilde{\alpha}=\alpha$ when $\pi(Z)$ is the Binomial prior with $\theta=1/p^\alpha$ and $\tilde{\alpha}= \max\{\alpha, 2\}$ when $\pi(Z)$ is the Beta-Binomial prior with $a_\theta=1$, $b_\theta=p^\alpha$. \color{black}
    Under Assumption \ref{ass-eigen0}, which recall includes that $\bar d / (n g_1^2)= o(1)$, \color{black}
the expression inside the expectation equals $\log \pi(z \mid \Y, \Omega_{-j,-j})$ up to an $O_{P}(1)$ term as $n \to \infty$, $p \to \infty$ (see Lemma \ref{lem:tech1}).

The proof strategy is to use Theorem 1 in \cite{chang2024dimensionfreerelaxationtimesinformed}, which we now state.
Suppose that, for any model $z \neq z^*$,
\begin{align}
R= \min_{z \in \mathcal{Z}(\bar d)} \max_{z' \in \mathcal{N}(z)} \frac{\pi(z' \mid \Y, \Omega_{-j,-j})}{\pi(z  \mid \Y, \Omega_{-j,-j})} > T := \max_{z \in \mathcal{Z}(\bar d)} |\mathcal{N}(z)|
\label{eq:min_ratiopp}
\end{align}
where $\mathcal{N}(z)$ is the set of neighbors of $z$, meaning states $z'$ that are reachable by a single step of a kernel $P$. 
Then, if the kernel $P$ gives uniform proposal probability to each neighbor, Theorem 1 in \cite{chang2024dimensionfreerelaxationtimesinformed} gives that
\begin{align}
 \mbox{Gap}(P)^{-1} \leq \frac{4 T}{(1 - (T/R)^{1/2})^3}.
\nonumber
\end{align}

We will show that $R > C T$ with high probability for any constant $C$, which then implies $\mbox{Gap}(P)^{-1} \leq \frac{4 p (\bar d + 1)}{(1 - C^{-1/2})^3} \leq (4 + \epsilon) p (\bar d + 1)$ for any $\epsilon > 0$ since for the birth-death-swap sampler, $T\leq p(\bar d + 1)$ (recall that for any model $z$, there are $|z|_0$ possible death moves, $p-|z|_0$ possible birth moves and $|z|_0(p-|z|_0)$ possible swap moves, thus the number of neighbors is $p + |z|_0(p-|z|_0) \leq p(\bar d + 1)$). 
Equivalently, we will show that
\begin{align}
\lim_{n \to \infty} \mathbb P \left( \bigcup_{j=1}^p \left \{ \max_{z \in \mathcal{Z}(\bar d)} \min_{z' \in \mathcal{N}(z)} \frac{\pi(z  \mid \Y, \Omega_{-j,-j})}{\pi(z'  \mid \Y, \Omega_{-j,-j})} > \frac{1}{CT} \right \} \right)= 0.
\label{eq:max_ratiojj}
\end{align}
For this we will show that for any $j$,
\begin{align}
\mathbb P \left( \max_{z \in \mathcal{Z}(\bar d)} \min_{z' \in \mathcal{N}(z)} \frac{\pi(z  \mid \Y, \Omega_{-j,-j})}{\pi(z'  \mid \Y, \Omega_{-j,-j})} > \frac{1}{CT} \right)= o(p^{-1}),
\label{eq:max_ratiopp}
\end{align}
which implies by a union bound argument that \eqref{eq:max_ratiojj} holds.

We show that \eqref{eq:max_ratiopp} holds for $j=p$ without loss of generality. Consider two models $(z_1,z_2)$ such that $z_2= z_1 \cup \{k\}$. Lemma \ref{lem:tech1} in Section \ref{app:technical-lemmas} gives that, under Assumption \ref{ass-eigen0},
\begin{align}
    \log \frac{\pi(z_1 \mid \Y, \Omega_{-p,-p})}{\pi(z_2 \mid \Y, \Omega_{-p,-p})}  = \gamma - \frac{1}{2 }\frac{d_{j, z_1}^2  \Sigma_{kk \mid z_1 \cup p}^{-1}  }{S_{pp} + \lambda} + O_P(1) = \gamma [1  + o_{P}(1)] - \frac{1}{2 }\frac{d_{j, z_1}^2  \Sigma_{kk \mid z_1 \cup p}^{-1}  }{S_{pp} + \lambda} \label{eq:ratio-zz},
\end{align}
where $\gamma = \frac{1}{2}\log n     + \log g_1 + \tilde{\alpha} \color{black} \log p$,
$d_{k, z_1} = s_k - s_{z_1}^T \Sigma_{z_1 \mid p}^{-1} \Sigma_{j z_1 \mid p}= \Y_{.p}^T [\Y_{.k} - \Y_{z_1} \Sigma_{z_1 \mid p}^{-1} \Sigma_{k z_1 \mid p}]$,
$\Sigma_{kk \mid z_1 \cup p} = \Sigma_{kk \mid p} - \Sigma_{kz_1 \mid p} \Sigma_{z_1 \mid p}^{-1}\Sigma_{z_1k \mid p}$, and the $o_{P}(1)$ term is independent of $(z_1,z_2)$.  

Further, by Lemma \ref{lem:dist-quadratic} in Section \ref{app:technical-lemmas},
we have that
\begin{align}\label{eq:dist-skappa}
 d_{k, z_1} \mid S_{pp}  \sim \text{N}\left( S_{pp} \mu_{k, z_1}, S_{pp} \Tilde \Sigma_{kk \mid z_1 \cup p} \right),
\end{align}
with 
\begin{align}
&\mu_{k, z_1}= \frac{(\Sigma_{kk}^0 - \Sigma_{k z_1 \mid p} \Sigma_{z_1 z_1 \mid p}^{-1} \Sigma_{z_1 p}^0)}{\Sigma_{pp}^0} 
\nonumber \\
& \Tilde \Sigma_{kk \mid z_1 \cup p} :=  \Sigma_{kk \mid p}^{0} + \Sigma_{k z_1 \mid p} \Sigma_{z_1z_1 \mid p}^{-1}  \Sigma_{z_1z_1 \mid p}^{0} \Sigma_{z_1z_1 \mid p}^{-1}   \Sigma_{ z_1 k \mid p} - \Sigma_{k z_1 \mid p}^{0} \Sigma_{z_1z_1 \mid p}^{-1} \Sigma_{z_1 k \mid p} - \Sigma_{k z_1 \mid p} \Sigma_{z_1z_1 \mid p}^{-1} \Sigma_{z_1 k \mid p}^0.
\nonumber
\end{align}
To gain intuition, for the particular case where $\Omega_{-p,-p}= \Omega_{-p,-p}^0$, then $\Tilde \Sigma_{k \mid z_1 \cup p}= \Sigma_{k \mid z_1 \cup p}$. 
Further, from Lemma \ref{lem:exp-max}, if $\Omega_{-p,-p}= \Omega_{-p,-p}^0$ and $z^0 \subseteq z_1$ then $\mu_{k, z_1}=0$.
(i.e. for overfitted models $d_{k,z_1}$ has mean zero).

To show \eqref{eq:max_ratiopp}, we consider separately the set of overfitted and non-overfitted models, relative to the model $z^*$. 
Throughout the proof, we implicitly condition on $S_{pp}$, i.e., we treat it as known. Later, one can use that $S_{pp}/n$ concentrates around $\Sigma_{pp}^0$ to complete the proof without conditioning on $S_{pp}$.

\textbf{Overfitted models.}

Consider the set of models $z$ such that $z \supset z^*$ (overfitted models).
For each such $z$, consider a model $z'= z \setminus \{k\}$ that drops an inactive $k \in z \setminus z^*$ from $z$.
From \eqref{eq:ratio-zz}, the goal is to show that there exist $z'$ such that
\begin{align}
\mathbb P \left( \max_{z \supset z^*} \left| \frac{d_{k, z'}}{\sqrt{S_{pp} \Sigma_{kk \mid z' \cup p}}} \right| > \sqrt{2[\gamma - \log(CT)]} \right) = o(p^{-1}).
\label{eq:max_ratiopp_asymp_overfit_omegageneral}
\end{align}
Above we used that, under Assumption \ref{ass-maxexpec-omegageneral-overfitted}, we have $\gamma - \log(CT) = 0.5 \log n +     \log g_1 + \tilde{\alpha} \color{black} \log p - \log(CT) > 0$ for large enough $n$ (recall that $T \leq p(\bar d + 1)$, and $C$ is a constant) and hence the square root is well-defined. 
From \eqref{eq:dist-skappa}, $d_{k, z'} \mid S_{pp}  \sim \text{N}( S_{pp} \mu_{k, z'}, S_{pp} \Tilde \Sigma_{kk \mid z' \cup p})$ and thus
\begin{align}
\frac{d_{k, z'}}{\sqrt{\Sigma_{kk \mid z' \cup p}S_{pp}}}=
\frac{S_{pp}^{\frac{1}{2}} \mu_{k,z'}}{\Sigma_{kk \mid z' \cup p}^{\frac{1}{2}}} +
U_z \frac{\Tilde \Sigma_{kk \mid z' \cup p}^{\frac{1}{2}}}{\Sigma_{kk \mid z' \cup p}^{\frac{1}{2}}},
\nonumber
\end{align}
where $U_z \sim N(0,1)$.
To ease notation let $W_z= |d_{k, z'}|/ \sqrt{\Sigma_{kk \mid z' \cup p}S_{pp}}$.
Using that $f(u)= |a + bu|$ is a $b$-Lipschitz function and Theorem 5.5 in \cite{boucheron:2013} we have that 
$W_z$ is sub-Gaussian with pseudo-variance parameter $\leq \Tilde \Sigma_{kk \mid z' \cup p} / \Sigma_{kk \mid z' \cup p}$,
and hence $W_z'= [W_z - \mathbb E W_z] \Sigma_{kk \mid z' \cup p}^{1/2} / \Tilde \Sigma_{kk \mid z' \cup p}^{1/2}$ is a zero-mean sub-Gaussian with pseudo-variance $\leq 1$.
Further, using the standard formula for the mean of the absolute value of a Gaussian gives
\begin{align}
&\mathbb  E( |d_{k,z'}| )= \sqrt{\frac{2}{\pi}} S_{pp}^{\frac{1}{2}} \Tilde \Sigma_{kk \mid z' \cup p}^{\frac{1}{2}} 
e^{- \frac{S_{pp} \mu_{k,z'}^2}{2\Tilde \Sigma_{kk \mid z' \cup p}} }
 + S_{pp} \mu_{k, z'} \left[ 1 - 2 \Phi \left(- \frac{S_{pp}^{\frac{1}{2}} \mu_{k,z'}}{\Tilde \Sigma_{kk \mid z' \cup p}^{\frac{1}{2}} } \right) \right]
\nonumber \\
\Rightarrow
&\mathbb E( W_z )= \sqrt{\frac{2 \Tilde \Sigma_{kk \mid z' \cup p}}{\pi \Sigma_{kk \mid z' \cup p}}}
e^{- \frac{S_{pp} \mu_{k,z'}^2}{2\Tilde \Sigma_{kk \mid z' \cup p}} }
 + \frac{S_{pp}^{1/2} \mu_{k, z'}}{\Sigma_{kk \mid z' \cup p}^{1/2}} \left[ 1 - 2 \Phi \left(- \frac{S_{pp}^{\frac{1}{2}} \mu_{k,z'}}{\Tilde \Sigma_{kk \mid z' \cup p}^{\frac{1}{2}} } \right) \right]
\nonumber \\
&\leq \sqrt{\frac{2 \Tilde \Sigma_{kk \mid z' \cup p}}{\pi \Sigma_{kk \mid z' \cup p}}}
 +  \frac{S_{pp}^{1/2} |\mu_{k, z'}|}{\Sigma_{kk \mid z' \cup p}^{1/2}} = \sqrt{\frac{2 \Tilde \Sigma_{kk \mid z' \cup p}}{\pi \Sigma_{kk \mid z' \cup p}}}
 + \frac{S_{pp}^{1/2}}{\sqrt{n \Sigma_{pp}^0}} \frac{\sqrt{n \Sigma_{pp}^0} |\mu_{k, z'}|}{\Sigma_{kk \mid z' \cup p}^{1/2}}.
\label{eq:mean_folded_gaussian_overfitted}
\end{align}
Note that in the previous expression, $\frac{S_{pp}^{1/2}}{\sqrt{n \Sigma_{pp}^0}} \xrightarrow[n \to \infty]{a.s.} 1$.
Now, straightforward algebra show that the left-hand side of \eqref{eq:max_ratiopp_asymp_overfit_omegageneral} is
\begin{align}
&\leq \mathbb P \left( \max_{z \supset z^*} \frac{W_z - \mathbb E W_z + \mathbb E W_z}{\sqrt{\Tilde \Sigma_{kk \mid z' \cup p} / \Sigma_{kk \mid z' \cup p}}}   > \max_{z \supset z^*} \frac{\Sigma_{kk \mid z' \cup p}^{1/2}}{\Tilde \Sigma_{kk \mid z' \cup p}^{1/2}} \sqrt{2[\gamma - \log(CT)]}  \right)
\nonumber \\
&\leq \mathbb   P \left( \max_{z \supset z^*} \frac{W_z - \mathbb E W_z}{\sqrt{\Tilde \Sigma_{kk \mid z' \cup p} / \Sigma_{kk \mid z' \cup p}}}   > \max_{z \supset z^*} \frac{\Sigma_{kk \mid z' \cup p}^{1/2}}{\Tilde \Sigma_{kk \mid z' \cup p}^{1/2}} \sqrt{2[\gamma - \log(CT)]}
- \max_{z \supset z^*} \mathbb E(W_z) \frac{\Sigma_{kk \mid z' \cup p}^{1/2}}{\Tilde \Sigma_{kk \mid z' \cup p}^{1/2}}
  \right)
\leq e^{- \frac{t^2}{2}}
\label{eq:max_ratiopp_asymp_overfit_omegageneral2}
\end{align}
the right-hand side following from the fact that the maximum of sub-Gaussians is again sub-Gaussian, and where we defined
\begin{align}
t= \max_{z \supset z^*} \frac{\Sigma_{kk \mid z' \cup p}^{1/2}}{\Tilde \Sigma_{kk \mid z' \cup p}^{1/2}} \sqrt{2[\gamma - \log(CT)]}
- \max_{z \supset z^*} \mathbb E(W_z) \frac{\Sigma_{kk \mid z' \cup p}^{1/2}}{\Tilde \Sigma_{kk \mid z' \cup p}^{1/2}}
- \mathbb E \left[ \max_{z \supset z^*} \frac{W_z - \mathbb E W_z}{\sqrt{\Tilde \Sigma_{kk \mid z' \cup p} / \Sigma_{kk \mid z' \cup p}}} \right]
\nonumber
\end{align}
For \eqref{eq:max_ratiopp_asymp_overfit_omegageneral2} to be $o(p^{-1})$, it is enough    that $t - \sqrt{2 \log p} \to \infty$. 
Plugging in the expression for $\gamma$ and using that for the BDMH
sampler $T \leq p (\bar d +1)$, it suffices that
\begin{align}
\lim_{n \to \infty} \frac{\Sigma_{kk \mid z' \cup p}^{1/2}}{\Tilde \Sigma_{kk \mid z' \cup p}^{1/2}}
\sqrt{2[(    \tilde{\alpha} \color{black} - 1) \log p + \frac{1}{2} \log n     + \log g_1 \color{black} - \log(C) - \log (\bar d +1)]} 
\nonumber \\
- \max_{z \supset z^*} \mathbb  E(W_z) \frac{\Sigma_{kk \mid z' \cup p}^{1/2}}{\Tilde \Sigma_{kk \mid z' \cup p}^{1/2}}
- \mathbb  E \left[ \max_{z \supset z^*} \frac{W_z - \mathbb  E W_z}{\sqrt{\Tilde \Sigma_{kk \mid z' \cup p} / \Sigma_{kk \mid z' \cup p}}} \right] - \sqrt{2 \log p}
= \infty.
\nonumber
\end{align}

Using \eqref{eq:mean_folded_gaussian_overfitted}, we have that
\begin{align}
 \max_{z \supset z^*} \mathbb  E(W_z) \frac{\Sigma_{kk \mid z' \cup p}^{1/2}}{\Tilde \Sigma_{kk \mid z' \cup p}^{1/2}}
\leq 
\sqrt{\frac{2}{\pi}}
 + \frac{S_{pp}^{1/2}}{\sqrt{n \Sigma_{pp}^0}} \frac{\sqrt{n \Sigma_{pp}^0} |\mu_{k, z'}|}{\Tilde \Sigma_{kk \mid z' \cup p}^{1/2}}.
\nonumber
\end{align}
Combining this bound and noting that $C$ is a constant and that $S_{pp} / (n \Sigma_{pp}^0)$ concentrates on 1 as $n \to \infty$, we have that a sufficient condition for $t - \sqrt{2 \log p} \to \infty$ is
\begin{align}
&\lim_{n \to \infty} \frac{\Sigma_{kk \mid z' \cup p}^{1/2}}{\Tilde \Sigma_{kk \mid z' \cup p}^{1/2}}
\sqrt{2[(    \tilde{\alpha} \color{black} - 1) \log p + \frac{1}{2} \log n     + \log g_1 \color{black} - \log (\bar d +1)]} 
\nonumber \\
&- \sqrt{\frac{2}{\pi}}
- (1 + \epsilon) \frac{\sqrt{n \Sigma_{pp}^0} |\mu_{k, z'}|}{\Tilde \Sigma_{kk \mid z' \cup p}^{1/2}}
- \mathbb  E \left[ \max_{z \supset z^*} \frac{W_z - \mathbb  E W_z}{\sqrt{\Tilde \Sigma_{kk \mid z' \cup p} / \Sigma_{kk \mid z' \cup p}}} \right] -  \sqrt{2 \log p}
= \infty.
\nonumber
\end{align}
for some constant $\epsilon > 0$.
Plugging in that $W_z= |d_{k, z'}|/ \sqrt{\Sigma_{kk \mid z' \cup p}S_{pp}}$, this latter condition is identical to Assumption \ref{ass-maxexpec-omegageneral-overfitted}, completing the proof.

\textbf{Non-overfitted models.}

Let $z \not\supset z^*$ be a non-overfitted model. For each such $z$, consider a neighbor $z'= z \cup \{k\}$ adding a parameter $k \in z^* \setminus z$. 
The goal is to show that
\begin{align}
\lim_{n \to \infty}  \mathbb   P \left( \max_{z  \not\supset z^*} \min_{k \in z^* \setminus z} \frac{\pi(z \mid \Y, \Omega_{-p,-p})}{\pi(z' \mid \Y, \Omega_{-p,-p})} > \frac{1}{CT} \right) = o(p^{-1}).
\nonumber
\end{align}
From \eqref{eq:ratio-zz}, we can equivalently show that the following probability are $o(p^{-1})$:
\begin{align}
\mathbb P \left( \max_{z \not\supset z^*}  
    -\frac{1}{2} \frac{d_{k , z}^2}{S_{pp} \Sigma_{kk \mid z \cup p}} + \gamma > - \log(CT) \right)=
\mathbb P \left(\min_{z \not\supset z^*} \frac{d_{k , z}^2}{S_{pp} \Sigma_{kk \mid z \cup p}} < 2 [\log(CT) + \gamma] \right).
\nonumber
\end{align}

Using the union bound, the right-hand side is $\leq$
\begin{align}
\sum_{z \not\supset z^*}
\mathbb P \left( \frac{d_{k , z}^2}{S_{pp} \Sigma_{kk \mid z \cup p}} < 2 [\log(CT) + \gamma] \right)
=
\sum_{z \not\supset z^*}
\mathbb P \left( \frac{d_{k , z}^2}{S_{pp} \Tilde \Sigma_{kk \mid z \cup p}} < \frac{2 \Sigma_{kk \mid z \cup p}}{\Tilde \Sigma_{kk \mid z \cup p}} [\log(CT) + \gamma] \right)
\label{eq:chang_zhang_ubound_nonoverfitted}
\end{align}

Recall that, from \eqref{eq:dist-skappa}, $d_{k,z}^2/(S_{pp} \Tilde \Sigma_{kk \mid z \cup p}) \mid S_{pp} \sim \chi_1^2(\lambda_{k,z})$ where the non-centrality parameter is $\lambda_{k,z}= S_{pp} \mu_{k,z}^2 / \Tilde \Sigma_{kk \mid z \cup p}$.
Hence, to bound \eqref{eq:chang_zhang_ubound_nonoverfitted} we may use the non-central chi-square left tail bound in Lemma S4 of \cite{rossell2022concentration}.
Provided that 
\begin{align}
\frac{2 \Sigma_{kk \mid z \cup p}}{\Tilde \Sigma_{kk \mid z \cup p}} [\log(CT) + \gamma] = o(\lambda_{k,z})
\label{eq:condition_rossell_lemmas4}
\end{align}
for all $z \not\supset z^*$, Lemma S4 in \cite{rossell2022concentration} gives that \eqref{eq:chang_zhang_ubound_nonoverfitted} is
\begin{align}
\leq \sum_{z \not\supset z^*} e^{- (\frac{1}{2} - \epsilon) \frac{S_{pp} \mu_{k,z}^2}{\Tilde \Sigma_{kk \mid z \cup p}}}
\leq p^{\bar d}
e^{- (\frac{1}{2} - \epsilon) S_{pp} \min_{z \not\supset z^*} \frac{\mu_{k,z}^2}{\Tilde \Sigma_{kk \mid z \cup p}}}
\nonumber
\end{align}
for any constant $\epsilon > 0$ and large enough $n$, where in the right-hand side we used that the number of non-overfitted models is $\leq {p \choose \bar d} \leq p^{\bar d}$.
Using that $S_{pp}/n$ concentrates on $\Sigma_{pp}^0$, we have that the right-hand side is $o(p^{-1})$ (as we wished to prove) if
\begin{align}
    \lim_{n \to \infty} (\frac{1}{2} - \epsilon) \min_{z \not\supset z^*} \frac{ n \Sigma_{pp}^0 \mu_{k,z}^2}{\Tilde \Sigma_{kk \mid z \cup p}}  - (\bar d + 2) \log p = \infty,
    \nonumber
\end{align}
which holds by Assumption \ref{ass-betamin_omegageneral}.

To complete the proof of the first statement we verify that \eqref{eq:condition_rossell_lemmas4} holds. Noting that for the BDMH sampler, 
$T \leq (\bar d +1) p$, plugging in the expressions of $\gamma$ and $\lambda_{k,z}$, and using again that $S_{pp}/n$ concentrates on $\Sigma_{pp}^0$, gives that a sufficient condition for \eqref{eq:condition_rossell_lemmas4} to hold for all $z \not\supset z^*$ is that
\begin{align}
2 \left[\log(C) + \log(\bar d +1) + (    \tilde{\alpha} \color{black} + 1) \log p + \frac{1}{2} \log n      + \log g_1 \color{black} \right]
= o\left(\min_{z \not\supset z^*} \frac{n \Sigma_{pp}^0 \mu_{k,z}^2}{\Tilde \Sigma_{kk \mid z \cup p}} \right),
\nonumber
\end{align}
which holds under Assumption \ref{ass-betamin_omegageneral}.

The second statement relative to general BDS transition kernels with $M \geq 1$ steps  follows from the definition of the spectral gap and simple algebra: for any Markov transition kernel $P$, $\text{Gap}(P) = 1 - \lambda_2(P)$ where $\lambda_2(P)$ is the second-largest eigenvalue of $P$. Since $\lambda_2(P^{\otimes M}) =\lambda_2(P)^M$, then $\text{Gap}(P^{\otimes M}) = 1 - \lambda_2(P)^M$. Then using that $\log (1-x) \leq -x$ for $x \in [0,1)$,
\begin{align*}
    \log (\lambda_2(P)^M) = M \log (1 - \text{Gap}(P)) \leq - M \text{Gap}(P).
\end{align*}
Hence 
we obtain that, for any $j \in [p]$.
$$\text{Gap}( P_{BD}^{j, \Omega_{-j,-j}, M}) = \text{Gap}( P_{BD}^{j,\Omega_{-j,-j},M=1\otimes M}) \geq 1 - e^{- M \text{Gap}( P_{BD}^{j,\Omega_{-j,-j}, M=1})}.$$
Hence on the event where 
\begin{align*}
     \mbox{Gap}( P_{BD}^{j,\Omega_{-j,-j},M=1})^{-1} \leq (4 + \epsilon) p (\bar d + 1), \qquad \forall j \in [p],
\end{align*}
we also have
\begin{align*}
     \mbox{Gap}( P_{BD}^{j,\Omega_{-j,-j},M}) \geq 1 - e^{-\frac{M}{(4 + \epsilon) p (\bar d + 1)}}, \qquad \forall j \in [p],
\end{align*}
which finishes the proof of Theorem \ref{thm:bdsgap_chang_omegageneral}.


\subsubsection{Technical lemmas for the proofs of Section \ref{sec:spectral-gap-bdmh}} \label{app:technical-lemmas}

We state and prove two useful lemmas.
Lemma \ref{lem:tech1} gives an approximation to the log-ratio of posterior model probabilities $\pi(z \mid \Y, \Omega_{-j,-j})/ \pi(z' \mid \Y, \Omega_{-j,-j})$ for two models $z \neq z'$, where model refers to edge configurations for variable $j$. 
The expression simplifies particularly when $(z,z')$ differ by only one edge, where it only depends on the data through a scalar random variable $d_{k, z}$ defined in the lemma, and on the marginal sample variance $S_{jj}$.
Lemma \ref{lem:dist-quadratic} gives the distribution of $d_{k, z}$ which, combined with Lemma \ref{lem:tech1}, allows us to characterize the behavior of $\pi(z \mid \Y, \Omega_{-j,-j})/ \pi(z' \mid \Y, \Omega_{-j,-j})$.

\begin{lemma}\label{lem:tech1}
Let $j \in [p]$ be a column index. Suppose that Assumption \ref{ass-eigen0} holds    , which recall includes $\bar d/(g_1^2 n)= o(1)$. \color{black} Then for any $z \in \mathcal{Z}(\bar d)$ and $z' = z\cup \{k\}$ with $k \not \in z$,
    \begin{align*}
     \log \frac{\pi(z | \Y, \Omega_{-j,-j})}{\pi(z' | \Y, \Omega_{-j,-j})}  = \gamma +  \frac{1}{2} \frac{s_z^T A_{zz}^{-1} s_z}{S_{jj} + \lambda} -  \frac{1}{2} \frac{s_{z'}^T A_{z'z'}^{-1} s_{z'}}{S_{jj} + \lambda}  + O_{P}(1),
\end{align*}
where the $O_P(1)$ term is independent of $z$, $s := S_{-j,j}$, $A := [\Omega_{-j,-j}]^{-1} = \Sigma_{-j,-j|j}$ and     $\gamma = \frac{1}{2}\log n + \log g_1 + R(z,z')$, where
\begin{align*}
R(z,z')= \begin{cases}
\alpha \log p - \log(1 - 1/p^\alpha)= \alpha \log p + c_p \mbox{, for the Binomial prior with } \theta=1/p^\alpha
\\
\log\left( \frac{p^\alpha + p(p-1)/2 - |z|_0 -1}{1 + |z|_0} \right)=
\max\{\alpha, 2\} \log(p) + c'_p
\mbox{, for the Beta-Binomial}(a_\theta=1,b_\theta=p^\alpha) \mbox{ prior}
\end{cases}
\end{align*}
where $\lim_{p \to \infty} c_p= \lim_{p \to \infty} c'_p=0$.
\color{black}
 
Moreover, 
 \begin{align*}
    \log \frac{\pi(z| \Y, \Omega_{-j,-j})}{\pi(z'| \Y, \Omega_{-j,-j})}  
    = \gamma  - \frac{1}{2 }\frac{d_{k, z}^2  \Sigma_{kk|z \cup j}^{-1}  }{S_{jj} + \lambda}  + O_P(1) 
\end{align*}
with $d_{k,z} = s_k - A_{kz}A_{z z}^{-1}s_{z}$ and $\Sigma_{kk|z\cup j} = A_{kk} - A_{kz}A_{zz}^{-1}A_{zk} = \Sigma_{kk|j} - \Sigma_{kz|j} \Sigma_{zz|j}^{-1}\Sigma_{zk|j}$.
\end{lemma}

\begin{proof}
Without loss of generality, we prove this lemma for $j=p$.
From Proposition \ref{prop:wang_exactzeroes_notation} and \eqref{eq:post_submodel}, we have
\begin{align*}
     \log \frac{\pi(z |\Y, \Omega_{-p,-p})}{\pi(z' |\Y, \Omega_{-p,-p})} 
     = \log g_1 
         + R(z,z') \color{black}
     + \frac{1}{2}\log \frac{|U_{z'}|}{|U_{z}|} + \frac{s_{z}^T U_{z}^{-1} s_{z}}{2} - \frac{s_{z'}^T U_{z'}^{-1} s_{z'}}{2}.
\end{align*}
where $U_z = (S_{pp}+ \lambda ) \Sigma_{zz|p} + \frac{1}{g_1^2} I_z$,     and $R(z,z')$ is obtained by plugging in the expressions for the Binomial prior with $\theta=1/p^\alpha$ and the Beta-Binomial$(a_\theta,b_\theta)$. \color{black}
The proof strategy is to show that
\begin{align}
    \log \frac{|U_{z'}|}{|U_{z}|} &=  \log n + O_P(1) \nonumber \\
    s_z U_z^{-1} s_z &= \frac{s_z A_{zz}^{-1} s_z}{S_{pp} + \lambda}+ O_P(1), \label{eq:decomp-sus}
\end{align}
from which the first statement in the lemma follows immediately.
Recall that $A_{zz} = [[\Omega_{-p,-p}]^{-1}]_{zz} = \Sigma_{zz|p}$.
We have 
\begin{align*}
    \log |U_z| &= \log \left \{ (S_{pp} + \lambda)^{|z|_0} |\Sigma_{zz|p}| \left|I_{z} + \frac{1}{g_1^{2}(S_{pp} + \lambda)} \Omega_{zz}\right| \right \} \\
    &= |z|_0 \log (S_{pp} + \lambda) + \log |\Sigma_{zz|p}| + \log \left|I_{z} + \frac{1}{g_1^{2}(S_{pp} + \lambda)} \Omega_{zz}\right|.
\end{align*}

Since $z' = z \cup \{k\}$, we have
\begin{align*}
    \log \frac{|\Sigma_{zz|p}|}{|\Sigma_{z'z'|p}|} &= - \log \frac{|A_{z'z'}|}{|A_{zz}|} = - \log \frac{|A_{zz}|| A_{kk} - A_{kz} A_{zz} A_{zk}|}{|A_{zz}|} \\
    &= - \log | A_{kk} - A_{kz} A_{zz} A_{zk}| = \log |\Sigma_{kk|z \cup p}| \leq - \log \epsilon_0,
\end{align*}
using in the last inequality that $\Sigma_{kk|z \cup p} \leq \Sigma_{kk} \leq \|\Sigma\| \leq \epsilon_0^{-1}$  under Assumption \ref{ass-eigen0}.
Therefore, since $\epsilon_0$ is a constant independent of $n$ and $p$, $ \log \frac{|\Sigma_{zz|p}|}{|\Sigma_{z'z'|p}|}  = O(1)$.

Moreover,  using that for two $d \times d$ matrices $M,M'$,
\begin{align*}
    \| |M| - |M'|| \leq d \max(\|M\|, \|M'\|)^d \|M - M'\|,
\end{align*}
we have
\begin{align*}
   1 < \left|I_{z} + \frac{1}{g_1^{2}(S_{pp} + \lambda)} \Omega_{zz}\right| &\leq 1 + |z|_0 \left(1 \vee \| \frac{1}{g_1^{2}(S_{pp} + \lambda)} \Omega_{zz}\|\right)^{|z|_0} \left \|\frac{1}{g_1^{2}(S_{pp} + \lambda)} \Omega_{zz}\right \|  \\
   &\leq 1 + \bar d \left(1 \vee \frac{1}{g_1^{2}(S_{pp} + \lambda)} \|\Omega_{zz}\|\right)^{\bar d} \frac{1}{g_1^{2}(S_{pp} + \lambda)} \|\Omega_{zz} \| \\
   &\leq 1 + \left(1 \vee \frac{\epsilon_0^{-1}}{g_1^{2}(S_{pp} + \lambda)}\right)^{\bar d}  \frac{\bar d}{g_1^{2}(S_{pp} + \lambda)} \epsilon_0^{-1}= 1 + o_P(1),
\end{align*}
where in the last equality we have used \ds{that $\frac{\bar d\epsilon_0^{-1}}{g_1^{2}(S_{pp} + \lambda)} =  o_P(1)$, which holds since $S_{pp}/\Sigma_{pp}^0 \sim \chi_n^2$ (which implies that $S_{pp} = n (\Sigma_{pp}^0 + O_P(n^{-1/2}))$) and thus
\begin{align*}
    \frac{\bar d\epsilon_0^{-1}}{g_1^{2}(S_{pp} + \lambda)}  \sim \frac{\bar d \epsilon_0^{-1}}{g_1^2 (\Sigma_{pp}^0 \chi_n^2 + \lambda)} = o_P(1),
\end{align*}
since $\bar d/ g_1^2 = o(n)$ by assumption and $\epsilon_0^{-1}$ is a constant.
}
Thus,
\begin{align*}
    \log \frac{|U_{z'}|}{|U_{z}|} = (|z'|_0 - |z|_0) \log n +(|z'|_0 - |z|_0)  \log (\Sigma_{pp}^0 + O_P(n^{-1/2}))  + O_P(1) = \log n + O_P(1),
\end{align*}
since $|z'|_0 - |z|_0=1$. This implies that
\begin{align*}
     \log \frac{\pi(z |\Y, \Omega_{-p,-p})}{\pi(z' |\Y, \Omega_{-p,-p})} 
    &= \gamma +  \frac{s_{z}^T U_{z}^{-1} s_{z}}{2} - \frac{s_{z'}^T U_{z'}^{-1} s_{z'}}{2} + O_P(1).
\end{align*}
We now show \eqref{eq:decomp-sus}. 
Using that, by the Sherman-Morrison formula,
\begin{align*}
    U_z^{-1} &= \left( (S_{pp} + \lambda ) A_{zz} + \frac{1}{g_1^2} I_z  \right)^{-1} \\
    &= \frac{1}{S_{pp} + \lambda} A_{zz}^{-1} - \frac{1}{g_1^2(S_{pp} + \lambda)^2}  A_{zz}^{-1} \left(I + \frac{1}{g_1^2 (S_{pp} + \lambda)}  A_{zz}^{-1} \right)^{-1}  A_{zz}^{-1}.
\end{align*}
 we have
\begin{align*}
    s_{z} U_{z}^{-1} s_{z} &= \frac{s_{z}^T A_{zz}^{-1} s_{z}}{S_{pp} + \lambda}   -   \frac{1}{g_1^2(S_{pp} + \lambda)^2} s_{z}^T A_{zz}^{-1} \left(I + \frac{1}{g_1^2 (S_{pp} + \lambda)}  A_{zz}^{-1} \right)^{-1}  A_{zz}^{-1}s_{z}.
\end{align*}
We show that the second term on the RHS of the previous equation is $O_P(1)$. By the Cauchy-Schwarz inequality,
\begin{align*}
    \left| \frac{1}{g_1^2(S_{pp} + \lambda)^2} s_{z}^T A_{zz}^{-1} \left(I + \frac{1}{g_1^2 (S_{pp} + \lambda)}  A_{zz}^{-1} \right)^{-1}  A_{zz}^{-1}s_{z} \right| &\leq \frac{\|s_{z}\|^2 }{g_1^2(S_{pp} + \lambda)^2} \left\| A_{zz}^{-1} \left(I + \frac{1}{g_1^2 (S_{pp} + \lambda)}  A_{zz}^{-1} \right)^{-1}  A_{zz}^{-1} \right\| \\
    &\leq\frac{\|s_{z}\|^2 }{g_1^2(S_{pp} + \lambda)^2} \| A_{zz}^{-1} \|^2 \leq \frac{\|s\|^2 }{g_1^2(S_{pp} + \lambda)^2} \epsilon_0^{-2} = O_P(1),
\end{align*}
where we have used that $\|A_{zz}^{-1}\| = \lambda_{min}(A_{zz})^{-1} \leq \lambda_{min}(A)^{-1} \leq \lambda_{min}(\Omega^{-1})^{-1} \leq \epsilon_0^{-1}$ under Assumption \ref{ass-eigen0}, since $A$ is a sub-matrix of $\Omega^{-1}$,     and in the last equality we used that $g_1^2(S_{pp} + \lambda) \stackrel{P}{\longrightarrow} \infty$, since $S_{pp}/\Sigma_{pp}^0 \sim \chi_n^2$ and $g_1^2 n \to \infty$ by assumption (since $\bar d / g_1^2= o(n)$ implies that $g_1^2 n / \bar d \to \infty$). \color{black}
Thus, we conclude that
\begin{align*}
     \log \frac{\pi(z |\Y, \Omega_{-p,-p})}{\pi(z' |\Y, \Omega_{-p,-p})} 
= \gamma + \frac{1}{2} \frac{ s_{z}^T A_{zz}^{-1} s_{z}}{S_{pp} + \lambda}  - \frac{1}{2} \frac{ s_{z'}^T A_{zz}^{-1} s_{z'}}{S_{pp} + \lambda} + O_P(1),
\end{align*}
which proves the first statement of Lemma \ref{lem:tech1}.

The lemma's second statement follows from the following computations for 2 nested models $z,z' \in \mathcal{Z}(\bar d)$, $z \subset z'$. With $\kappa  =z' \backslash z$ (i.e., $z' = z \cup \kappa$) the indicator vector of edges in $z'$ but not in $z$, and using standard inversion formula of block matrices we have 
\begin{align*}
    A_{z'}^{-1} = A_{z \cup \kappa}^{-1} = \begin{pmatrix}
        A_{zz} & A_{z \kappa } \\
         A_{\kappa z} & A_{\kappa \kappa}
    \end{pmatrix}^{-1} = \begin{pmatrix}
        A_{zz}^{-1} + A_{zz}^{-1} A_{z \kappa} \Sigma_{\kappa \kappa | z\cup p}^{-1} A_{\kappa z} A_{zz}^{-1} & - A_{zz}^{-1} A_{z \kappa} \Sigma_{\kappa \kappa | z\cup p}^{-1} \\
        - \Sigma_{\kappa \kappa | z\cup p}^{-1} A_{\kappa z} A_{zz}^{-1}  & \Sigma_{\kappa \kappa| z\cup p}^{-1}
    \end{pmatrix}
\end{align*}
where $\Sigma_{\kappa \kappa| z\cup p}^{-1} := (A_{\kappa \kappa} - A_{\kappa z} A_{zz}^{-1}  A_{ z \kappa})^{-1} $. 
Therefore,
\begin{align*}
    s_{z'}^T A_{z'z'}^{-1}s_{z'} &= s_{z \cup \kappa}^T A_{z \cup \kappa}^{-1}s_{ z \cup \kappa} \\
    &= s_{z}^T A_{zz}^{-1} s_{z} + (s_{\kappa} - A_{\kappa z} A_{zz}^{-1} s_{z})^T \Sigma_{\kappa \kappa | z\cup p}^{-1}  (s_{\kappa} - A_{\kappa z} A_{zz}^{-1} s_{z}).
\end{align*}
Hence, denoting by $d_{\kappa, z} := s_{\kappa} - A_{\kappa z} A_{zz}^{ -1} s_{z}$, 
we obtain
\begin{align} \label{eq:difference_tstat}
    s_{z'}^T A_{z'z'}^{-1}s_{z'} - s_{z}^T A_{zz}^{-1} s_{z}=
    \frac{d_{\kappa,  z} ^T \Sigma_{\kappa \kappa | z\cup p}^{-1} d_{\kappa, z} }{S_{pp} + \lambda} \geq 0.
\end{align}

For the case where $z' = z \cup \{k \}$ with $k \not \in z$, we have $d_{k, z} := s_{k} - A_{k z} A_{zz}^{ -1} s_{z}$ and 
\begin{align*}
     s_{z'}^T A_{z'z'}^{-1}s_{z'} - s_{z}^T A_{zz}^{-1} s_{z}=
    \frac{d_{k,z}^2  \Sigma_{kk| z\cup p}^{-1}}{S_{pp} + \lambda},
\end{align*}
which proves the second statement of Lemma \ref{lem:tech1}.

\end{proof}

\begin{lemma}\label{lem:dist-quadratic}
    Let $j \in [p]$, $ z \in \mathcal{Z}(\bar d)$ and $\kappa \in \{0,1\}^{p-1}$ such that $\kappa \cap z = \emptyset$. Define $d_{\kappa, z} := s_{\kappa} - A_{\kappa z} A_{zz}^{ -1} s_{z}$ where $A := [\Omega_{-j,-j}]^{-1} = \Sigma_{-j,-j|j}$ and $s := S_{-j,j}$. Define also $A^0 := \Omega_{-j,-j}^{0,-1} = [\Omega_{-j,-j}^0]^{-1} = \Sigma^0_{-j,-j|j}$.
    It holds that
    \begin{align*} 
  d_{\kappa, z} \mid S_{jj}  \sim \text{N}\left(S_{jj}\mu_{\kappa, z}, S_{jj} \Tilde \Sigma_{\kappa \kappa |z\cup j}\right),
\end{align*}
with $\mu_{\kappa, z} = \frac{\Sigma_{\kappa j}^0 - A_{\kappa z} A_{zz}^{-1} \Sigma_{zj}^0}{\Sigma_{jj}^0} = \frac{\Sigma_{\kappa j}^0 - \Sigma_{\kappa z |j } \Sigma_{zz|j}^{-1} \Sigma_{zj}^0}{\Sigma_{jj}^0}$ and
\begin{align*}
      \Tilde \Sigma_{\kappa \kappa |z\cup j} :&=  A_{\kappa \kappa}^{0} + A_{\kappa z}A_{zz}^{-1}  A_{zz}^{0} A_{zz}^{-1}   A_{ z \kappa} - A_{\kappa z}^{0} A_{zz}^{-1} A_{ z \kappa} - A_{\kappa z}A_{zz}^{-1} A_{ z \kappa}^0 \\
      &=\Sigma_{\kappa \kappa \mid j}^{0} + \Sigma_{\kappa z \mid j} \Sigma_{zz \mid j}^{-1}  \Sigma_{zz \mid j}^{0} \Sigma_{zz \mid j}^{-1}   \Sigma_{ z \kappa \mid j} - \Sigma_{\kappa z \mid j}^{0} \Sigma_{zz \mid p}^{-1} \Sigma_{z \kappa \mid j} - \Sigma_{\kappa z \mid j} \Sigma_{zz \mid j}^{-1} \Sigma_{z \kappa \mid j}^0.
\end{align*}

Moreover, if $\Omega_{-j,-j} = \Omega_{-j,-j}^0$ then
 $\Tilde \Sigma_{\kappa \kappa |z\cup j}  =  A_{\kappa}^{0}  - A_{\kappa z}^0 A_{zz}^{0,-1} A_{ z \kappa}^0 =  \Sigma_{\kappa \kappa |z\cup j}^0$.

\end{lemma}


\begin{proof}
      Using \cite{eaton2007wishart}, the conditional distribution of the sample covariance vector is
\begin{align*}
    S_{-j,j} | S_{jj} \sim \text{N}(\frac{S_{jj}}{\Sigma_{jj}^0}\Sigma_{-j,j}^0, S_{jj} [\Omega_{-j,-j}^{0}]^{-1})
\end{align*}
from which we directly deduce that
\begin{align*} 
    &s_{ z} | S_{jj} \sim \text{N}(\frac{S_{jj}}{\Sigma_{jj}^0}\Sigma_{z j}^0, S_{jj} A^{0}_{zz}) \\
    &s_{ \kappa} | S_{jj} \sim \text{N}(\frac{S_{jj}}{\Sigma_{jj}^0}\Sigma_{ \kappa j}^0, S_{jj} A^{0}_{\kappa\kappa}).
\end{align*}
Since $d_{\kappa, z}$ is a linear combination of $s_{ z}$ and $s_{ \kappa}$, it also follows a Gaussian distribution with mean vector directly following the definition of $d_{\kappa,z}$ and covariance matrix 
\begin{align*}
    \Tilde \Sigma_{\kappa \kappa |z\cup j} &= Cov(s_{ \kappa}) + Cov(A_{\kappa z}A_{z}^{-1} s_{ z}) - Cov(s_{ \kappa}, A_{\kappa z}A_{z}^{-1} s_{ z}) - Cov(A_{\kappa z}A_{z}^{-1} s_{ z}, s_{ \kappa}) \\
    &=A_{\kappa \kappa}^{0} + A_{\kappa z}A_{zz}^{-1}  A_{zz}^{0} A_{zz}^{-1}   A_{ z \kappa} - A_{\kappa z}^{0} A_{zz}^{-1} A_{ z \kappa} - A_{\kappa z}A_{zz}^{-1} A_{ z \kappa}^0.
\end{align*}
If $\Omega_{-j,-j} = \Omega_{-j,-j}^0$, i.e., $A = A^0$, then the previous expression simplifies to $\Sigma_{\kappa \kappa |z\cup j}^0 :=  A_{\kappa \kappa}^{0}  - A_{\kappa z}^0 A_{zz}^{0,-1} A_{ z \kappa}^0$. 
\end{proof}

\subsection{Proofs of Section \ref{sec:gimh}}
\label{ssec:proofs_gimh}

Section \ref{sssec:proof_spectral-gap-gimh} gives the proof of Theorem \ref{thm:spectral-gap-gimh}.
Said proof relies on three results.
First, it relies on Proposition \ref{prop:hp-event1} (proven in Section \ref{app:proof-bf-ratio1}),
which bounds ratios $\pi(Z_{-j, j} \mid \Y, \Omega_{-j,-j}) / \pi(Z_{-j, j}^0 \mid \Y, \Omega_{-j,-j})$, where $z^0$ is the data-generating model.
Second, it relies on Proposition \ref{prop:hp-event-lr} (proven in Section \ref{app:proof-prop-bf-ratio2}), which bounds ratios $\pi^{LR}(Z_{-j, j} \mid \Y) / \pi^{LR}(Z_{-j, j}^0 \mid \Y)$.
Finally, it relies on Lemma \ref{lem:cons-lr} (proven in Section \ref{app:proof-lem-consistency}), which shows that the linear regression proposal concentrates on $Z_{-j, j}^0$.

Section \ref{sssec:proof_tail-bounds} then proves Lemma \ref{prop:tail-bounds}, 
which is a result required by Proposition \ref{prop:hp-event-lr}.
Finally, Section \ref{app:technical-lemmas-2} gives further technical lemmas required by the proofs in this section.

\subsubsection{Proof of Theorem \ref{thm:spectral-gap-gimh}}
\label{sssec:proof_spectral-gap-gimh}


We first re-state Theorem \ref{thm:spectral-gap-gimh}.
    Suppose that Assumptions C0-C4 and D1-D4 hold. Moreover, assume that the penalty parameter     $\tilde{\alpha}^{LR}$ \color{black} and the tempering parameter $\upsilon$ verify
    \begin{align*}
        &    \tilde{\alpha}^{LR} \color{black} > \ds{4} \max(\frac{\epsilon_1}{\epsilon_0^{2}}, 1)
        ,\\
        &\frac{1}{    \tilde{\alpha}^{LR} \color{black}} < \upsilon < \min \left( \frac{c_0}{    \tilde{\alpha}^{LR}} \color{black}, \frac{1}{\ds{4}} \frac{\epsilon_0^{2}}{\epsilon_1}, \frac{1}{2} \right),
    \end{align*}
    with $c_0 =  \frac{1 +     \tilde{\alpha} \color{black} - (\ds{\sqrt{2 \bar d } + \sqrt{2} + \frac{1}{4}})^2}{2}$. 
Then, for any $\epsilon \in (0,1)$, 
with $M \geq 1$ global steps,
\begin{align*}
     \lim_{n \to \infty} \mathbb{P} \left(  \bigcap_{j=1}^p \left \{ \text{Gap}(P^{j,\Omega_{-j,-j},M}_{GI}) \geq 1 - e^{-M ( 1 - \epsilon)} \right \} \right) = 1. 
\end{align*}

We follow the proof strategy as outlined in the main paper.
The first step is to obtain bounds on the ratios $\frac{\pi(Z_{-j, j}| \Y, \Omega_{-j,-j})}{\pi(Z_{-j, j}^0| \Y, \Omega_{-j,-j})}$ and 
$\frac{Q^j_\upsilon(Z_{-j, j})}{Q^j_\upsilon(Z^0_{-j, j})}$
that hold with high probability, for any column $j \in [p]$. Recall that $Q^j_\upsilon(Z_{-j, j}) \propto (\pi^{LR}(Z_{-j,j}| \Y))^\upsilon$. We first obtain bounds on the ratios $\frac{\pi^{LR}(Z_{-j, j}| \Y)}{\pi^{LR}(Z_{-j, j}^0 | \Y)}$ and from the latter deduce similar bounds on $\frac{Q^j_\upsilon(Z_{-j, j})}{Q^j_\upsilon(Z^0_{-j, j})}$.
The second step is to show that proposal distributions $Q^j_\upsilon(Z_{-j, j})$ concentrate on the data-generating $Z_{-j, j}^0$ for all $j=1,\ldots,p$.
The final step is to show that if the bounds obtained in the first two steps hold, then the maximum importance weight $\max_{z \in \mathcal{Z}(\bar d)} w^j(z)$ is bounded for any $j$.

We proceed to the first step of the proof.
We first state our results for one column $j \in [p]$. 
Define $z := Z_{-j,j}^0$, $z^0 := Z_{-j, j}^0$ and the following subsets of $\mathcal{Z}(\bar d)$:
\begin{align*}
    &\mathcal{M}_1 = \{ z \in \mathcal{Z}(\bar d): z \supset z^0 \} \\
    &\mathcal{M}_2 = \{ z \in \mathcal{Z}(\bar d): z \not \supset z^0, \: |z|_0 < \bar d \} \\
    &\mathcal{M}_3 = \{ z \in \mathcal{Z}(\bar d): z \not \supset z^0, \: |z|_0 = \bar d \}.
\end{align*}
In words, $\mathcal{M}_1, \mathcal{M}_2, \mathcal{M}_3$ contain respectively the overfitted, non-overfitted and non-saturated, non-overfitted and saturated models for column $j$. The next proposition bounds ratios of the conditional posterior model probabilities, letting $z= Z_{-j,j}$ and $z^0= Z_{-j, j}^0$.




\begin{prop}\label{prop:hp-event1}
    Suppose that Assumptions C0-C4 hold, \ds{that the slab standard deviation $g_1$ satisfies $\log g_1 > -\frac{1}{8} \log n$
    and $\log g_1= o(n)$,} and that $\bar d \geq 4$.  For any 
    $c_1< \frac{1}{4}$ and with
    \begin{align*}
    &c_0 = \frac{1 +     \tilde{\alpha} \color{black} - ( \sqrt{2} + \sqrt{2\bar d } + \frac{1}{4})^2}{2}, \qquad
        &c_2 = \frac{\omega_{min} \epsilon_0^3}{\ds{8}},
    \end{align*}
    we define the events 
    \begin{align}
        &\mathcal{E}_2^j = \bigcap_{z \in \mathcal{M}_1} \left \{  \frac{\pi(z| \Y, \Omega_{-j,-j})}{\pi(z^0| \Y, \Omega_{-j,-j})} \leq (p^{-c_0} n^{-c_1})^{|z \backslash z^0|_0}  \right \} \nonumber \\
        &\mathcal{E}_3^j = \bigcap_{z \in \mathcal{M}_2\cup \mathcal{M}_3} \left \{   \frac{\pi(z| \Y, \Omega_{-j,-j})}{\pi(z^0 | \Y, \Omega_{-j,-j})} \leq  e^{-c_2 |z^0 \backslash z|_0 n} \right \}  \nonumber \\
        &\mathcal{E}_{CP}^j = \mathcal{E}_2^j \cap \mathcal{E}_3^j. \label{eq:event-p}
    \end{align}
    Then $\mathbb P(\mathcal{E}_{CP}^j) = 1 - o(p^{-1})$,
    where the $o(p^{-1})$ term is independent of the chosen column $j$.
\end{prop}
The proof of Proposition \ref{prop:hp-event1} is in Section \ref{app:proof-bf-ratio1}. We note that the constants $c_0,c_1,c_2$ in Proposition \ref{prop:hp-event1} characterise the convergence rates of posterior ratios. Nonetheless, we conjecture that the constraint $c_1 < \frac{1}{4}$ may be an artefact of our proof strategy, and one could potentially obtain a faster rate with $c_1 < \frac{1}{2}$. We then define the event 
$
    \mathcal{E}_{CP} = \cap_{j=1}^{p}  \mathcal{E}_{CP}^j.
$
Using Proposition \ref{prop:hp-event1} and the union bound, we obtain
\begin{align}\label{eq:proba-cp}
    \mathbb P \left( \mathcal{E}_{CP} \right) \geq 1 -  \sum_{j=1}^p \mathbb P \left( (\mathcal{E}_{CP}^j)^c \right)  = 1 - o(1).
\end{align}

The following proposition gives upper and lower bounds for ratios of linear regression posterior model probabilities $\pi^{LR}(Z_{-j, j}| \Y)/\pi^{LR}(Z_{-j, j}^0 | \Y)$. The lower bounds are used to bound the importance weight function (see below) while the upper bounds allow to prove the consistency result in Lemma \ref{lem:cons-lr} below.

\begin{prop}\label{prop:hp-event-lr}
    Suppose Assumptions C1, C2, D1-D4 hold and that 
    \begin{align*}
            \tilde{\alpha}^{LR} > 2. \color{black}
    \end{align*}
    For any arbitrarily small $\epsilon \in (0, \frac{1}{2})$ and $c \in (0,1)$ arbitrarily close to 1, define the constants
    \begin{align*}
        &r_0 = \bar r_0 =     \tilde{\alpha}^{LR}, \color{black} \\
        &r_1 = \frac{1}{2} - \epsilon, \qquad \bar  r_1 = \frac{1}{2} + \epsilon \\
        &r_2 =  \frac{1-\epsilon}{4} \epsilon_1 \epsilon_0 \omega_{min}, \qquad \bar r_2 =  \frac{1-\epsilon}{2c} \epsilon_1 \epsilon_0  \omega_{min}, 
    \end{align*}
    and the event 
    \begin{align}
        &\mathcal{E}_4^j = \bigcap_{z \in \mathcal{M}_1} \left \{ p^{-\bar r_0 |z\backslash z^0|_0} n^{-\bar r_1 |z\backslash z^0|_0}  \leq  \frac{\pi^{LR}(z| \Y)}{\pi^{LR}(z^0| \Y)} \leq p^{-r_0 |z\backslash z^0|_0} n^{-r_1 |z\backslash z^0|_0} \right \} \nonumber \\
        &\mathcal{E}_5^j = \bigcap_{z \in \mathcal{M}_2 \cup \mathcal{M}_3} \left \{ e^{-\bar r_2 |z^0 \backslash z|_0 n}   \leq  \frac{\pi^{LR}(z| \Y)}{\pi^{LR}(z^0| \Y)} \leq e^{- r_2 |z^0 \backslash z|_0 \frac{n}{\log n}}  \right \} \nonumber \\
        &\mathcal{E}_{LR}^j = \mathcal{E}_4^j \cap \mathcal{E}_5^j, \label{eq:event-lr-p}
    \end{align}
    where $z = Z_{-j,j}$ and $z^0 = Z_{-j, j}^0$.
    Then,
$
\mathbb P(\mathcal{E}_{LR}^j) = 1 - o(p^{-1}),
$
    where the $o(p^{-1})$ term is independent of the chosen column $j$. 
\end{prop}

The proof of Proposition \ref{prop:hp-event-lr} is in Section \ref{app:proof-prop-bf-ratio2}. We note that the constants $\bar r_0, \bar r_1, \bar r_2$ and $r_0, r_1, r_2$ characterise the convergence rates of the \emph{linear regression} posterior ratios. Define the event 
$\mathcal{E}_{LR} = \cap_{j=1}^{p}  \mathcal{E}_{LR}^j$.
Using Proposition \ref{prop:hp-event-lr} and the union bound, we obtain
\begin{align*}
    \mathbb P \left( \mathcal{E}_{LR} \right) \geq 1 -  \sum_{j=1}^p \mathbb P \left( (\mathcal{E}_{LR}^j)^c \right)  = 1 - o(1).
\end{align*}
Combining the above bound with \eqref{eq:proba-cp}, we also obtain
\begin{align*}
     \mathbb P \left( \mathcal{E}_{CP} \cap \mathcal{E}_{LR} \right) = 1 - o(1).
\end{align*}

We now proceed to the second step of the proof.
That is, we prove that under our working assumptions and if the event $\mathcal{E}_{LR}$ holds, as $n \to \infty$ our globally-informed proposal distributions 
$Q^j_\upsilon(Z_{-j, j})$ concentrate on the data-generating $Z_{-j, j}^0$ for all $j=1,\ldots,p$.
This is given by Lemma \ref{lem:cons-lr}, proven in Section \ref{app:proof-lem-consistency}.

\begin{lemma}\label{lem:cons-lr}

Under Assumptions D1 and D4, 
if $\mathcal{E}_{LR}$ holds and if $\ds{\frac{1}{4} \frac{\epsilon_0^2}{\epsilon_1}} > \upsilon > (r_0)^{-1}$, then for any $j \in [p]$,
\begin{align*}
    Q^j_\upsilon(Z_{-j, j}^0) \geq  1 - 3 p^{1 - \upsilon r_0} n^{- \upsilon  r_1},
\end{align*}
for $n \geq n_0$ large enough where $n_0$ only depends on constants $(\upsilon, r_0, r_1, r_2, \epsilon_0,\epsilon_1, \omega_{min})$ with $r_0,r_1,r_2$ defined in Proposition \ref{prop:hp-event-lr}. 
    
\end{lemma}

As the final step of the proof,
we show that on the high probability event $\mathcal{E}_{CP} \cap \mathcal{E}_{LR}$, the maximum importance weight
\begin{align*}
   \max_{z \in \mathcal{Z}(\bar d)} w^j(z) = \frac{\pi(z |\Omega_{-j,-j}, \Y)}{Q^j_\upsilon(z)} 
\end{align*}
is bounded for any $j$. For simplicity, we show the result for $j=p$ and  denote by $Q(z) = Q^p_\upsilon(z)$ our tempered GI proposal distribution and by $w(z)$ the importance weight function. 
We denote by $n_0$ an integer that can  only depend on $(\upsilon, r_0, r_1, r_2, \epsilon_0, \epsilon_1)$ and additional constants. For any $z$, we have
\begin{align*}
    w(z) = \frac{\pi(z |\Omega_{-p,-p}, \Y)}{\pi(z^0 |\Omega_{-p,-p}, \Y)}  \frac{\pi(z^0 |\Omega_{-p,-p}, \Y)}{Q(z^0)} \frac{Q(z^0)}{Q(z)}, 
\end{align*}
therefore,
\begin{align}\label{eq:importance-weight}
    \log w(z) = \log \frac{\pi(z |\Omega_{-p,-p}, \Y)}{\pi(z^0 |\Omega_{-p,-p}, \Y)} + \log \frac{\pi(z^0 |\Omega_{-p,-p}, \Y)}{Q(z^0)}  - \log \frac{Q(z)}{Q(z^0)}.
\end{align}
Moreover, note that
\begin{align*}
     \log \frac{Q(z)}{Q(z^0)} = \upsilon \log \frac{\pi^{LR}(z| \Y)}{\pi^{LR}(z^0| \Y)}.
\end{align*}

On $\mathcal{E}_{LR}$, by Lemma \ref{lem:cons-lr}, for any $\epsilon' > 0$ and $n \geq n_0$, we have
\begin{align*}
    Q(z^0) \geq 1- \epsilon',
\end{align*}
which implies that $\log \frac{\pi(z^0 |\Omega_{-p,-p}, \Y)}{Q(z^0)} \leq -\log (1 - \epsilon') \leq \epsilon'$.
Moreover, on $\mathcal{E}_{CP} \cap \mathcal{E}_{LR}$, it holds that:

\begin{itemize}
    \item For any $z \in \mathcal{M}_1$ (overfitted):
    \begin{align*}
     \log \frac{\pi(z |\Omega_{-p,-p}, \Y)}{\pi(z^0 |\Omega_{-p,-p}, \Y)}\leq - c_0 |z \backslash z^0|_0 \log p - c_1 |z \backslash z^0|_0  \log n,
\end{align*}
and 
\begin{align*}
    \log \frac{\pi^{LR}(z| \Y)}{\pi^{LR}(z^{0}| \Y)} \geq - \bar r_0 |z \backslash z^0|_0 \log p - \bar r_1 |z \backslash z^0|_0  \log n,
\end{align*}
which implies that
\begin{align}
    \log \frac{\pi(z |\Omega_{-p,-p}, \Y)}{\pi(z^0 |\Omega_{-p,-p}, \Y)} - \log \frac{Q(z)}{Q(z^0)} \leq - (c_0 - \upsilon \bar r_0) |z \backslash z^0|_0 \log p - (c_1 - \upsilon \bar r_1) |z \backslash z^0|_0  \log n \leq 0, \label{eq:diff-overfitted}
\end{align}
since  $\upsilon < \frac{c_0}{r_0} = \frac{c_0}{    \tilde{\alpha}^{LR} \color{black}}$ by assumption and $\upsilon < \frac{c_1}{\bar r_1} = \frac{\frac{1}{4} - \epsilon}{\frac{1}{2} + \epsilon}$ since $\upsilon < \frac{1}{2}$ and $\epsilon > 0$ can be chosen arbitrarily small.
    \item For any $z \in \mathcal{M}_2 \cup \mathcal{M}_3$ (non-overfitted)
    \begin{align*}
     \log \frac{\pi(z |\Omega_{-p,-p}, \Y)}{\pi(z^0 |\Omega_{-p,-p}, \Y)} \leq - c_2  |z^0 \backslash z|_0 n,
\end{align*}
and
\begin{align*}
    \log \frac{\pi^{LR}(z| \Y)}{\pi^{LR}(z| \Y)} \geq - \bar r_2 |z^0 \backslash z|_0 n,
\end{align*}
which implies that
\begin{align}
    \log \frac{\pi(z |\Omega_{-p,-p}, \Y)}{\pi(z^0 |\Omega_{-p,-p}, \Y)} - \log \frac{Q(z)}{Q(z^0)} \leq - (c_2 - \upsilon \bar r_2)|z^0 \backslash z|_0 n \leq 0, \label{eq:diff:nonoverfitted}
\end{align}
since  $\upsilon \leq \frac{c_2}{\bar r_2} = \frac{\frac{1}{\ds{8}} \omega_{min} \epsilon_0^3}{\frac{1 - \epsilon}{2c} \omega_{min} \epsilon_1 \epsilon_0} = \frac{1}{\ds{4}}c (1 - \epsilon)^{-1}  \frac{\epsilon_0^2}{\epsilon_1} $ since $c$ can be arbitrarily close to 1, $\epsilon$ can be arbitrarily close to 0, and $\upsilon < \frac{1}{\ds{4}} \frac{\epsilon_0^2}{\epsilon_1}$ by assumption.  
\end{itemize}

Note that \eqref{eq:diff-overfitted} and \eqref{eq:diff:nonoverfitted} allow to better understand the role of the tempering $\upsilon$: it counters the over-concentration of the linear regression posterior (characterised by the constants $\bar r_0, \bar r_1, \bar r_2$) relatively to the one of the target posterior (characterised by the constants $c_0, c_1, c_2$).

Therefore, for any $z \in \mathcal{Z}(\bar d)$ we can conclude that
\begin{align*}
    \log w(z) \leq -\log (1- \epsilon),
\end{align*}
and thus that $\max_{z \in \mathcal{Z}(\bar d)} w(z) \leq \frac{1}{1 - \epsilon'}$ for any $\epsilon' > 0$ and $n \geq n_0$ on $\mathcal{E}_{CP} \cap \mathcal{E}_{LR}$. Since the same result can be proven for each column $j$, this proves that with probability $1 - o(1)$,
\begin{align*}
   \max_{j \in [p]} \max_{z \in \mathcal{Z}(\bar d)} w^j(z) \leq \frac{1}{1 - \epsilon'}.
\end{align*}
By \cite{wang2022exact} (equation (3.9) therein), the inequality above implies that
\begin{align*}
     \min_{j \in [p]} P^{j,\Omega_{-j,-j},M=1}_{GI} (x , dy) \geq (1-\epsilon') \pi(dy),
\end{align*}
for all $(x,y)$. 
Using the definition of the spectral gap in \eqref{eq:def-gap} as the infimum over $f \in L_2(\pi)$ of $E(f,f)/Var_\pi(f)$, where
\begin{align*}
&E(f,f)= \frac{1}{2} \int_{\chi} (f(y) - f(x))^2 \pi(dx) P^{j,\Omega_{-j,-j},M=1}_{GI}(x,dy) 
\geq
\frac{(1-\epsilon')}{2} \int_{\chi} (f(y) - f(x))^2 \pi(dx) \pi(dy)
\\
&= (1 - \epsilon') [ E_\pi(f(y)^2) - E_\pi^2(f(y)) ]= (1 - \epsilon') Var_\pi(f(y)),
\end{align*}
immediately gives that
\begin{align*}
    \min_{j \in [p]} Gap(P^{j,\Omega_{-j,-j},M=1}_{GI}) \geq 1 - \epsilon'.
\end{align*}
We have thus proved that the inequality above holds for any $\epsilon' > 0$ and $n \geq n_0$ with probability $1 - o(1)$.
The result for a general transition kernel with $M \geq 1$ steps  follows from the same argument as in the proof of Theorem \ref{thm:bdsgap_chang_omegageneral}.
Using that the gap for $M$ updates is $1 - (1 - G)^M$, where $G$ is the gap for 1 updates gives that, 
under the high-probability event  $\mathcal{E}_{CP} \cap \mathcal{E}_{LR}$, for any $j \in [p]$,
$$\text{Gap}(P_{GI}^{j, \Omega_{-j,-j},M}) = 
1 - (1 - \text{Gap} ( P_{GI}^{j,\Omega_{-j,-j},1}))^M
\geq 1 - e^{- M \text{Gap}( P_{GI}^{j,\Omega_{-j,-j},M=1})}  \geq 1 - e^{-M (1 - \epsilon)}, $$
where we used that $(1-x)^M \leq e^{-x M}$ for all $x>0$,
which proves the statement of Theorem \ref{thm:spectral-gap-gimh}.

\subsubsection{Proof of Proposition \ref{prop:hp-event1}}
\label{app:proof-bf-ratio1}


We first state and prove Proposition \ref{prop:hp-event}, a stronger result that implies Proposition \ref{prop:hp-event1}. The proof of Proposition \ref{prop:hp-event1} is given after that of Proposition \ref{prop:hp-event}.

\begin{prop}\label{prop:hp-event}
    Suppose that Assumptions C0-C4 hold,     that the slab standard deviation $g_1$ satisfies 
    \ds{$\log g_1 > -\frac{1}{8} \log n$}
    and $\log g_1= o(n)$ \color{black} and that $\bar d \geq 4$. 
    Define the events
    \begin{align*}
        &\mathcal{\Tilde E}_{2}^j = \left \{  \max_{z \in \mathcal{M}_1} \min_{k \in z \backslash z^0} \frac{\pi(z| \Y, \Omega_{-j,-j})}{\pi(z \backslash \{k\}| \Y, \Omega_{-j,-j})} \leq p^{-c_0} n^{-c_1}  \right \} \\
        &\mathcal{\Tilde E}_3^j = \left \{ \max_{z \in \mathcal{M}_2} \min_{k \in z^0 \backslash z} \frac{\pi(z| \Y, \Omega_{-j,-j})}{\pi(z \cup \{k\}| \Y, \Omega_{-j,-j})} \vee \max_{z \in \mathcal{M}_3} \min_{k \in z^0 \backslash z, \ell \in z \backslash z^0} \frac{\pi(z| \Y, \Omega_{-j,-j})}{\pi(z \cup \{k\} \backslash \{\ell \}| \Y, \Omega_{-j,-j})} \leq  e^{-c_2 n} \right \} \\
        &\mathcal{\Tilde E}^j = \mathcal{\Tilde E}_2^j \cap \mathcal{\Tilde E}_3^j,
    \end{align*}
    where
    \begin{align*}
        &1 < c_0 = \frac{1 +     \tilde{\alpha} \color{black} - ( \sqrt{2} + \sqrt{2\bar d } + \ds{\frac{1}{4}})^2}{2} \\
        &c_1 < \frac{1}{4} \\
        &c_2 \leq \frac{\omega_{min} \epsilon_0^3}{\ds{8}}.
    \end{align*}
    Then it holds that
    \begin{align*}
        \mathbb P( \mathcal{ \Tilde E}^j) = 1- o(p^{-1}),
    \end{align*}
    where $o(p^{-1})$ is independent of the chosen column $j$.
\end{prop}

\begin{proof}
Without loss of generality, we consider $j=p$.  Let us define
\begin{align*}
    B(z,z') = \frac{\pi(z| \Y, \Omega_{-p,-p})}{\pi(z'| \Y, \Omega_{-p,-p})}, \qquad z,z' \in \mathcal{Z}(\bar d).
\end{align*}
We first prove that $\mathbb P[ (\mathcal{ \Tilde E}^j_2)^c] = o(p^{-1})$. Second we prove that $\mathbb P[ (\mathcal{ \Tilde E}^j_3)^c] = o(p^{-1})$.  We then conclude with a simple union bound argument.


\textbf{Bound on $\mathbb P[ (\mathcal{ \Tilde E}^p_2)^c]$.} 

Recall from Lemma \ref{lem:conditions-ass} that under C0-C4, Assumptions \ref{ass-maxexpec-omegageneral-overfitted} and \ref{ass-betamin_omegageneral} hold with $z^* = z^0$. From the proof of Theorem \ref{thm:bdsgap_chang_omegageneral} (Expression \eqref{eq:max_ratiopp_asymp_overfit_omegageneral}) we have that, for any $\bar t \geq 0$, 
\begin{align}\label{eq:proba-bound-t}
    \mathbb P \left( \max_{z\supset z^0} \min_{z' \in \mathcal{N}(z)} \frac{\pi(z  \mid \Y, \Omega_{-p,-p})}{\pi(z'  \mid \Y, \Omega_{-p,-p})} > \bar t \right) \leq \mathbb P \left( \max_{z \supset z^0} \left| \frac{d_{k, z'}}{\sqrt{S_{pp} \Sigma_{kk \mid z' \cup p}}} \right| > \sqrt{2[\gamma + \log \bar t ]}\right) \leq e^{-t^2/2},
\end{align}
with 
\begin{align}
    t= \max_{z \supset z^0} \sqrt{\frac{\Sigma_{kk \mid z' \cup p}}{\Tilde \Sigma_{kk \mid z' \cup p}}} \sqrt{2[\gamma + \log \bar t]}
- \max_{z \supset z^0} \mathbb E(W_z) \sqrt{\frac{\Sigma_{kk \mid z' \cup p}}{\Tilde \Sigma_{kk \mid z' \cup p}}}
- \mathbb E \left[ \max_{z \supset z^0} \frac{W_z - \mathbb E W_z}{\sqrt{\Tilde \Sigma_{kk \mid z' \cup p} / \Sigma_{kk \mid z' \cup p}}} \right], \label{eq:def-t}
\end{align}
$W_z= |d_{k, z'}|/ \sqrt{\Sigma_{kk \mid z' \cup p}S_{pp}}$,
   
and $\gamma=\frac{1}{2}\log n + \log g_1 + \tilde{\alpha} \log p + c_p$,
where $\tilde{\alpha}=\alpha$ for the Binomial prior with $\theta=1/p^\alpha$ and $\tilde{\alpha}= \max\{2,\alpha\}$ for the Beta-Binomial$(a_\theta=1,b_\theta=p^\alpha)$ prior, and where $\lim_{p \to \infty} c_p=0$.
\color{black}
Note that in the proof of Theorem \ref{thm:bdsgap_chang_omegageneral}, Expression \eqref{eq:max_ratiopp_asymp_overfit_omegageneral}, the bound above is used with $\bar t = \frac{1}{CT}$ where $T \leq p (\bar d + 1)$ and $C>0$ is a constant, whereas here we will apply it with $\log \bar t = -c_0 \log p - c_1 \log n$.

We now use intermediate results from the proof of Lemma \ref{lem:conditions-ass} to bound each term in \eqref{eq:def-t}. We have
\begin{align*}
    \mathbb E \left[ \max_{z \supset z^0} \frac{W_z - \mathbb E W_z}{\sqrt{\Tilde \Sigma_{kk \mid z' \cup p} / \Sigma_{kk \mid z' \cup p}}} \right] = \mathbb E \left[ \max_{z \supset z^0} \frac{|d_{k, z'}| - \mathbb E |d_{k, z'}|}{\sqrt{S_{pp} \Tilde \Sigma_{kk \mid z' \cup p}}} \right] \leq \sqrt{2 \bar d \log p},
\end{align*}
the last inequality following from \eqref{eq:bound-i1}.
Moreover, 
from (46), we have that
\begin{align*}
    \max_{z \supset z^0} \mathbb E(W_z) \sqrt{\frac{\Sigma_{kk \mid z' \cup p}}{\Tilde \Sigma_{kk \mid z' \cup p}}} 
    &\leq \frac{2}{\pi} + \max_{z \supset z^0} \frac{\mathbb E [\sqrt{S_{pp}}]}{\sqrt{n \Sigma_{pp}^0}} \frac{\sqrt{n \Sigma_{pp}^0}|\mu_{k,z'}|}{\sqrt{\Tilde \Sigma_{kk \mid z' \cup p}} } 
    \leq \frac{2}{\pi} +  \max_{z \supset z^0} \frac{\sqrt{n \Sigma_{pp}^0}|\mu_{k, z'}|}{\sqrt{\Tilde \Sigma_{kk \mid z' \cup p}}} = \frac{2}{\pi} + I_2,
\end{align*}
where we used that by Jensen's inequality, $\mathbb E[\sqrt{S_{pp}}] \leq \sqrt{\mathbb E[S_{pp}]} = \sqrt{n \Sigma_{pp}^0}$, and our notation from the proof of Lemma \ref{lem:conditions-ass}. From \eqref{eq:bound-i2}, we have
\begin{align*}
    I_2 \leq 12 \epsilon_0^{-6} \delta \sqrt{n} = 12 \epsilon_0^{-6} \delta_0 \sqrt{\log p + \frac{\log n}{2}}.
\end{align*}
Additionally, from \eqref{eq:bound-ratio-sigma},
\begin{align*}
    \sqrt{\frac{\Sigma_{kk \mid z' \cup p}}{\Tilde \Sigma_{kk \mid z' \cup p}}} \geq 1 - \epsilon
\end{align*}
for any $\epsilon > 0$ and $n \geq n_0$ with $n_0$ independent of $z$. 
Combining these observations gives that $t$ in \eqref{eq:def-t} satisfies
\begin{align*}
    t &\geq (1 - \epsilon) \sqrt{2 (\gamma + \log \bar t )} - (\frac{2}{\pi} + 12 \epsilon_0^{-6} \delta_0 \sqrt{\log p + \frac{\log n}{2}})  - \sqrt{2\bar d \log p} \\
    &=(1 - \epsilon) \sqrt{2\left[ (    \tilde{\alpha} \color{black} - c_0) \log p + (\frac{1}{2} - c_1) \log n     + \log g_1 + o(1) \color{black} \right]} - \frac{2}{\pi}- 12 \epsilon_0^{-6} \delta_0 \sqrt{\log p + \frac{\log n}{2}} - \sqrt{2\bar d \log p} 
    \nonumber \\
    &\geq (1 - \epsilon) \sqrt{2\left[ (    \tilde{\alpha} \color{black} - c_0) \log p +     \ds{\frac{1}{8}} \log n \color{black} \right]} - \frac{2}{\pi} -  12 \epsilon_0^{-6} \delta_0 \sqrt{\log p + \frac{\log n}{2}} - \sqrt{2\bar d \log p} 
    \nonumber \\
    &\geq \sqrt{\log p} \left \{(1 - \epsilon) \sqrt{    \tilde{\alpha} \color{black} - c_0} - \sqrt{2\bar d} -  12 \epsilon_0^{-6} \delta_0 \right \} + \sqrt{\log n } \left \{(1 - \epsilon) \sqrt{    \ds{\frac{1}{8}} \color{black}} - 12 \epsilon_0^{-6} \delta_0 / \sqrt{2}\right\} - \frac{2}{\pi}  =: t',
\end{align*}
    for large enough $n$, where in the first equality we used that $\gamma=\frac{1}{2}\log n + \log g_1 + \tilde{\alpha} \log p + o(1)$ and that $\log \bar t = -c_0 \log p - c_1 \log n$, in the second inequality that 
\ds{$\log g_1 \geq -\frac{1}{8}\log n$}
and $c_1<1/4$ by assumption, \ds{implying that $\frac{1}{2} - c_1 - \frac{1}{8} > \frac{1}{8}$}, (and that the $o(1)$ term is arbitrarily small), \color{black}
and in the last inequality that $\frac{1}{\sqrt{2}}(\sqrt{b_1} + \sqrt{b_2}) \leq \sqrt{b_1 + b_2} \leq \sqrt{b_1} + \sqrt{b_2}$ for all $b_1,b_2 \geq 0$.

    Our goal is to show that we can find $t'$ such that $\lim_{n \to \infty} t' - \sqrt{2 \log p}= \infty$, since this implies that the right-hand side in \eqref{eq:proba-bound-t} is 
$$
\leq e^{-\frac{(t')^2}{2}} \leq 
e^{-\frac{(t' - \sqrt{2 \log p} + \sqrt{2 \log p})^2}{2}}
\leq 
e^{-\frac{(t' - \sqrt{2 \log p})^2}{2} - \frac{(\sqrt{2 \log p})^2}{2}}
=
e^{-\frac{(t' - \sqrt{2 \log p})^2}{2}} \frac{1}{p}
=o(1/p),
$$ 
as we wish to prove.
\color{black} 

To show that $\lim_{n \to \infty} t' - \sqrt{2 \log p}= \infty$,
from the definition of $t'$ above it is enough that
\begin{align}\label{eq:condition-alpha}
   (1 - \epsilon) \sqrt{    \tilde{\alpha} \color{black} - c_0} - \sqrt{2\bar d} -  12 \epsilon_0^{-6} \delta_0  > \sqrt{2}
\end{align}
and that
\begin{align}\label{eq:condition-alpha2}
(1 - \epsilon)\sqrt{    \ds{\frac{1}{8}} \color{black}} - 12 \epsilon_0^{-6} \delta_0 / \sqrt{2} \geq 0
\end{align}
for $n$ large enough. 
While \eqref{eq:condition-alpha} could be stated as an assumption on $(    \tilde{\alpha} \color{black}, \epsilon_0, \delta_0)$, below we show that our Assumptions (C3)-(C4) are sufficient for \eqref{eq:condition-alpha} to hold.
Under C3, we have $\delta_0 < \frac{\epsilon_0^{6}}{48}$, therefore,
\begin{align}\label{eq:ineq-delta0}
    12 \epsilon_0^{-6} \delta_0 / \sqrt{2} < \frac{1}{4\sqrt{2}}.
\end{align}
This implies that for $n \geq n_0$ \ds{\eqref{eq:condition-alpha2} holds} 
since $\epsilon$ can be taken arbitrarily close to 0, which holds by assumption. \color{black}  

Moreover, \eqref{eq:ineq-delta0} also implies that for \eqref{eq:condition-alpha} to be verified it is enough that 
\begin{align}
    \sqrt{    \tilde{\alpha} \color{black} - c_0} > \sqrt{2} + \sqrt{2 \bar d} + \ds{\frac{1}{4}} \iff     \tilde{\alpha} \color{black} - c_0  > (\sqrt{2 \bar d } + \sqrt{2} + \ds{\frac{1}{4}})^2,
\end{align}
which is verified under (C4) since $    \tilde{\alpha} \color{black} > (\sqrt{2 \bar d } + \sqrt{2} + \ds{\frac{1}{4}})^2 + 1$ and $c_0 <     \tilde{\alpha} \color{black} - (\sqrt{2 \bar d } + \sqrt{2} + \ds{\frac{1}{4}})^2$, completing this part of the proof.



In fact, the previous computations only require that $k \in z \backslash z^0$, therefore the previous bound holds for any posterior ratios of $z$ vs $z'$ where $z'$ has one fewer true non-edge compared to $z$, i.e.,
\begin{align}
   \mathbb P \left( \max_{z \in \mathcal{M}^*} \min_{z' \in \mathcal{N}(z)} \frac{\pi(z  \mid \Y, \Omega_{-p,-p})}{\pi(z'  \mid \Y, \Omega_{-p,-p})} > -c_0 \log p -c_1 \log n \right) = o(p^{-1}), \label{eq:proba-mstar}
\end{align}
with $\mathcal{M}^* := \{ z \in \mathcal{Z}(\bar d) : z \backslash z^0 \neq \emptyset\}$ the set of models which have at least one true non-edge.

\textbf{Bound on $\mathbb P[ (\mathcal{ \Tilde E}^p_3)^c]$.}

First note that the union bound gives that
\begin{align*}
    &\mathbb P[ (\mathcal{ \Tilde E}^j_3)^c] \\
    &\leq   \mathbb P \left[ \max_{z \in \mathcal{M}_2} \min_{k \in z^0 \backslash z} \frac{\pi(z| \Y, \Omega_{-p,-p})}{\pi(z \backslash \{k\}| \Y, \Omega_{-p,-p})} \geq e^{-c_2 n}  \right] +  \mathbb P \left[\max_{z \in \mathcal{M}_3} \min_{k \in z^0 \backslash z, \ell \in z \backslash z^0} \frac{\pi(z| \Y, \Omega_{-p,-p})}{\pi(z \cup \{k\} \backslash \{\ell \}| \Y, \Omega_{-p,-p})} \geq   e^{-c_2 n}  \right]
\end{align*}
We first bound the first term on the RHS of the previous equation using the proofs of Theorem \ref{thm:bdsgap_chang_omegageneral} and Lemma \ref{lem:conditions-ass}. From \eqref{eq:chang_zhang_ubound_nonoverfitted}, we have
\begin{align}
 \mathbb P \left[ \max_{z \in \mathcal{M}_2} \min_{k \in z^0 \backslash z} \frac{\pi(z| \Y, \Omega_{-p,-p})}{\pi(z \backslash \{k\}| \Y, \Omega_{-p,-p})} \geq \bar t  \right] \leq 
  \sum_{z \not\supset z^0} 
\mathbb P \left( \frac{d_{k , z}^2}{S_{pp} \Tilde \Sigma_{kk \mid z \cup p}} < \frac{2 \Sigma_{kk \mid z \cup p}}{\Tilde \Sigma_{kk \mid z \cup p}} [-\log\bar t + \gamma] \right) = o(p^{-1}) \label{eq:bound-m2}
\end{align}
provided that for any $k$ and $z$ (see Lemma \ref{lem:tail-bound-chi-squared}(ii)),
\begin{align*}
\frac{2 \Sigma_{kk \mid z \cup p}}{\Tilde \Sigma_{kk \mid z \cup p}} [-\log \bar t + \gamma] <  \lambda_{k,z},
\end{align*}
where  
\begin{align*}
     \lambda_{k,z}= S_{pp} \mu_{k,z}^2 / \Tilde \Sigma_{kk \mid z \cup p}.
\end{align*}
Since $\frac{\Sigma_{kk \mid z \cup p}}{\Tilde \Sigma_{kk \mid z \cup p}} \leq 1 + \epsilon$ for any $\epsilon > 0$ and $n \geq n_0$, it is enough that
\begin{align*}
    -\log \bar t + \gamma 
    < \frac{S_{pp}}{2(1+\epsilon)}\min_{z \not\subset z^0}\frac{ \mu_{k,z}^2}{\Tilde \Sigma_{kk \mid z \cup p}}
    \leq \frac{S_{pp}}{\Sigma_{pp}^0 2(1+\epsilon)}\min_{z \not\subset z^0}\frac{ \Sigma_{pp}^0 \mu_{k,z}^2}{\Tilde \Sigma_{kk \mid z \cup p}}.
\end{align*}
To see this, note that since $S_{pp}$ concentrates on $n \Sigma_{pp}^0$, we have that $S_{pp} \geq \frac{n\Sigma_{pp}^0}{1+\epsilon}$ for $n \geq n_0$. Also note that from \eqref{eq:min-mul}, we have
\begin{align*}
    \min_{z \not\subset z^0}\frac{ \Sigma_{pp}^0 \mu_{k,z}^2}{\Tilde \Sigma_{kk \mid z \cup p}} \geq \frac{\omega_{min} \epsilon_0^3}{2}.
\end{align*}
    Hence it's enough to prove that
\begin{align*}
-\log \bar t + \gamma <
\frac{n}{2(1+\epsilon)}
\frac{\omega_{min} \epsilon_0^3}{2}  \ds{= \frac{n}{1+\epsilon}
\frac{\omega_{min} \epsilon_0^3}{4}} 
.
\end{align*}

Therefore, plugging in     $\gamma=\frac{1}{2}\log n + \log g_1 + \tilde{\alpha} \log p + o(1)$ \color{black} and $\log \bar t = -c_2 n$ where $c_2 \leq \frac{\omega_{min} \epsilon_0^3}{\ds{8}}$, we have
\begin{align*}
    -\log \bar t + \gamma = c_2n +     \tilde{\alpha} \color{black} \log p + \frac{1}{2}\log n     + \log g_1 \color{black} \leq n \frac{\omega_{min}\epsilon_0^3}{\ds{8}} + o(n)  < \frac{n}{1+\epsilon} \frac{\omega_{min} \epsilon_0^3}{\ds{4}},
\end{align*}
for $n \geq n_0$,     where we used that $\log g_1= o(n)$ by assumption \color{black} and $\log p = o(n)$ under C0.

We now bound $\mathbb P \left[  \max_{z \in \mathcal{M}_3} \min_{k \in z^0 \backslash z, \ell \in z \backslash z^0} \frac{\pi(z| \Y, \Omega_{-p,-p})}{\pi(z \cup \{k\} \backslash \{\ell \}| \Y, \Omega_{-p,-p})} \geq  e^{-c_2 n}  \right]$. Observe that for any $z \in \mathcal{M}_3$, $l \in z^0 \backslash z$ and $k \in z \backslash z^0$ and with
$z' = z \cup \{l\} \backslash \{k\}$ and $z'' = z \backslash \{k\}$, we have
\begin{align*}
    \log B(z,z') = \log B(z,z'') + \log B(z'',z').
\end{align*}
Moreover, note that $z \in \mathcal{M}^* = \{ z \in \mathcal{Z}(\bar d) : z \backslash z^0 \neq \emptyset\}$. Note, that $z,z''$ are both underfitted but $z''$ has one true non-edge less than $z$ and $z'' \in \mathcal{M}_2$.  
Therefore, using \eqref{eq:proba-mstar}, \eqref{eq:bound-m2}, and  a simple union bound argument, we obtain, with $\bar t = e^{-c_2 n}$ and any $\epsilon \in (0,1)$,
\begin{align*}
    \mathbb P \left(\max_{z \in \mathcal{M}_3} \min_{l \in z^0 \backslash z, k \in z \backslash z^0} B(z, z') \geq  \bar t  \right ) &\leq \mathbb P \left(\max_{z \in \mathcal{M}^*} \min_{k \in z \backslash z^0} B(z, z'') \geq \frac{\epsilon}{1+\epsilon} \bar t  \right ) + \mathbb P \left(\max_{z'' \in \mathcal{M}_2} \min_{\ell \in z^0 \backslash z} B(z'', z') \geq \frac{\bar t}{1+\epsilon}  \right ) \\
    &= o(p^{-1}) + o(p^{-1}) = o(p^{-1}).
\end{align*}
for $n \geq n_0$, since $\frac{\epsilon}{1+\epsilon}e^{-c_2n} < p^{-c_0}n^{c_1}$ and we can easily see that \eqref{eq:bound-m2} still holds true if we replace $\bar t$ by $ \frac{\bar t}{1+\epsilon}$ for $\epsilon$ small enough. Therefore we have proven that
\begin{align*}
    \mathbb P( \mathcal{\Tilde E}_3^p) = o(p^{-1}).
\end{align*}

\textbf{Conclusion.} Aggregating all bounds above, we obtain
\begin{align*}
    \mathbb P((\mathcal{ \Tilde  E}^p)^c) &\leq \mathbb P( \mathcal{\Tilde E}_2^p) + \mathbb P( \mathcal{\Tilde E}_3^p) = o( p^{-1}),
\end{align*}
and this concludes the proof of Proposition \ref{prop:hp-event}.

\end{proof}

Proposition \ref{prop:hp-event1} is a direct consequence of Proposition \ref{prop:hp-event} choosing 
\begin{align*}
&c_0 = \frac{1 +     \tilde{\alpha} \color{black} - (\sqrt{2} + \sqrt{2\bar d } + \frac{1}{4})^2}{2} \\
        &c_1 = \frac{1}{4} - \epsilon \\
        &c_2 = \frac{\omega_{min} \epsilon_0^3}{\ds{8}}.
\end{align*}
for any $\epsilon \in (0,1/4)$ since we can easily show that $ \mathcal{\Tilde  E}^j \subset \mathcal{E}^j_{CP}$. Assume that $\mathcal{\Tilde  E}^j $ holds. Then for each overfitted model $z \in \mathcal{M}_1$, we can re-write
\begin{align*}
     \frac{\pi(z |\Y, \Omega_{-j,-j})}{\pi(z^0 |\Y, \Omega_{-j,-j})} = \prod_{k=1}^\ell  \frac{\pi(z^{(k)} |\Y, \Omega_{-j,-j})}{\pi(z^{(k-1)} |\Y, \Omega_{-j,-j})}
\end{align*}
where $z^0 = z^{(0)}, z^{(1)}, \dots z^{(\ell)} = z$ is a sequence of overfitted models such that $z^{(k-1)} = z^{(k)} \backslash \{i\}$ with $i \in z \backslash z^0$ and $\ell = |z \backslash z^0|_0 $. Then, then for each $k$. 
\begin{align*}
    \frac{\pi(z^{(k)} |\Y, \Omega_{-j,-j})}{\pi(z^{(k-1)} |\Y, \Omega_{-j,-j})} \leq p^{- c_0} n^{-c_1},
\end{align*}
and therefore,
\begin{align*}
     \frac{\pi(z |\Y, \Omega_{-j,-j})}{\pi(z^0 |\Y, \Omega_{-j,-j})} \leq p^{-c_0 \ell} n^{-c_1 \ell}.
\end{align*}
Since this holds for any overfitted model, we have therefore proven that
\begin{align*}
    \mathcal{E}_2^j = \bigcap_{z \supset z^0} \left \{ \frac{\pi(z |\Y, \Omega_{-j,-j})}{\pi(z^0 |\Y, \Omega_{-j,-j})} \leq p^{-c_0 |z \backslash z^0|_0} n^{-c_1 |z \backslash z^0|_0} \right\},
\end{align*}
holds. 

Similarly, for any non-overfitted model $z \in \mathcal{M}_2 \cup \mathcal{M}_3$, we can re-write
\begin{align*}
     \frac{\pi(z |\Y, \Omega_{-j,-j})}{\pi(z^0 |\Y, \Omega_{-j,-j})} = \prod_{k=1}^\ell  \frac{\pi(z^{(k)} |\Y, \Omega_{-j,-j})}{\pi(z^{(k-1)} |\Y, \Omega_{-j,-j})}
\end{align*}
where $z^0 = z^{(0)}, z^{(1)}, \dots z^{(\ell)} = z$ is a sequence of models such that $z^{(k-1)} = z^{(k)} \cup \{i\}$ with $i \in z^0 \backslash z$ or $z^{(k-1)} = z^{(k)} \cup \{i\} \backslash \{m\}$ with $i \in z^0 \backslash z$ and $m \in z \backslash z^0$. Note, that if $z \subset z^0$, then all $z^{(k-1)}$ are such that $z^{(k-1)} = z^{(k)} \cup \{i\}$ while if $z \not \subset z^0$ (and  $z \not \supset z^0$) then the $z^{(k-1)}$ can be such that  $z^{(k-1)} = z^{(k)} \cup \{i\}$ or $z^{(k-1)} = z^{(k)} \cup \{i\} \backslash \{m\}$. 
Then, if $\mathcal{\Tilde E}_3^j$ holds, for any $z' \in \mathcal{M}_2$ (resp., $z' \in \mathcal{M}_3$), there exists $i \in z^0 \backslash z$ (resp., $i \in z^0 \backslash z$  and $m \in z \backslash z^0$) such that
\begin{align*}
     \frac{\pi(z' |\Y, \Omega_{-j,-j})}{\pi(z'' |\Y, \Omega_{-j,-j})} \leq e^{-c_2 n},
\end{align*}
with $z'' = z' \cup \{i\}$ (resp., $z'' = z' \cup \{i\} \backslash \{m\}$). 
Thus, we can find a sequence of at most $r + \ell$ models as described above with  $r := |z^0 \backslash z|_0$ 
and $\ell: = |z \backslash z^0|$, $ 1 \leq r \leq \bar d , 0 \leq \ell \leq \bar d$,  such that 
\begin{align*}
     \frac{\pi(z |\Y, \Omega_{-j,-j})}{\pi(z^0 |\Y, \Omega_{-j,-j})} \leq e^{-r c_2 n} p^{-c_0 \ell} n^{-c_1 \ell} \leq  e^{-r c_2 n}.
\end{align*}
Since this holds for any non-overfitted model, we have therefore proven that
\begin{align*}
    \mathcal{E}_3^j = \bigcap_{z \not \supset z^0} \left \{ \frac{\pi(z |\Y, \Omega_{-j,-j})}{\pi(z^0 |\Y, \Omega_{-j,-j})} \leq e^{-|z^0 \backslash z|_0  c_2 n} \right\},
\end{align*}
holds. Therefore, $\mathcal{E}_{CP}^j = \mathcal{E}_2^j \cap \mathcal{E}_3^j$ holds with probability greater than $1+o(p^{-1})$ and this concludes the proof of Proposition \ref{prop:hp-event1}.

\subsubsection{Proof of Proposition \ref{prop:hp-event-lr}}\label{app:proof-prop-bf-ratio2}



To prove Proposition \ref{prop:hp-event-lr}, we leverage a lemma on the tail probability of ratios of \emph{linear regression} posteriors on $z = Z_{-j, j}$ vs $z^0= Z_{-j, j}^0$, which we denote
\begin{align*}
    B^{LR}(z, z^0) :=\frac{\pi^{LR}(z| \Y)}{\pi^{LR}(z^0| \Y)}, \qquad z \in \mathcal{Z}(\bar d).
\end{align*}
We also denote by $B^{LR}(z, z')$ the ratios of posterior probabilities for 2 models $z,z' \in \mathcal{Z}(\bar d)$. We recall our notation $X = \Y_{\cdot,-j}$, $y = \Y_{\cdot j}$. 

\begin{lemma}\label{prop:tail-bounds}
Assume Assumptions D1-D4 hold. For any model $z \in \mathcal{Z}(\bar d)$, define
\begin{align}
    T(z^0, z)= \frac{(X_{\cdot z^0} \beta^0_{z^0})^T (I - H_z) X_{z^0} \beta^0_{z^0} }{(\Omega_{jj}^0)^{-1}},
    \label{eq:lm_ncp}
\end{align}
with $H_z$ the projection matrix onto the columns of $X_{\cdot z}$, i.e., $H_z = X_{\cdot z}(X{\cdot z}^T X_{\cdot z})^{-1} X_{\cdot z}^T =  X_{\cdot z} S_{zz}^{-1} X_{\cdot z}^T$ and $\beta^0 = - \Omega^0_{-j,j}/\Omega^0_{jj}$.
\begin{enumerate}[label=(\roman*)]
\item Let $z \supset z^0$ be an overfitted model. Then, for any constant $c<1$,
\begin{align}
\mathbb{P} \left( B^{LR}(z,z^0) > u \right) \leq
    \frac{(ew_u)^{(|z|_0 - |z^0|_0)}}{c u^c}
    \left( \frac{\tau}{c n p^{2\alpha}} \right)^{\frac{c(|z|_0 - |z^0|_0)}{2}} 
\nonumber
\end{align}
where $w_u= \log \left( \frac{n p^{2\alpha}}{\tau} u^{\frac{2}{|z|_0 - |z^0|_0}} \right)$, for all $n \geq n_0$ and any $u$ such that
\begin{align}
(|z|_0-|z^0|_0) \left[ \log \left( \frac{n}{\tau} \right) + 2 \alpha \log \left( p \right) \right]  + 2 \log u
= o(n).
\nonumber
\end{align}
Above, $n_0$ and $c$ do not depend on $z$, i.e., they are uniform constants across $z$.
\item Let $z \not\supset z^0$ be a non-overfitted model. If
\begin{align*}
    \frac{T(z, z^0)}{\log T(z, z^0)} = o(n),
\end{align*}
then \begin{align}
\mathbb{P} \left( B^{LR}(z,z^0) > u \right) \leq 
(ew_b)^{(|z'|_0 - |z^0|_0)} 
e^{-\frac{cT(z, z^0)}{2 (\log T(z, z^0))^2}}
\nonumber
\end{align}
where
$$
w_b= \frac{T(z, z^0)}{|z'|_0 - |z^0|_0} + (1-c) \log \left( \frac{c n (1-\theta)^2}{\tau \theta^2} \right)
$$
for any constant $c<1$, and $u$ such that 
$$
u > e^{-\frac{T(z, z^0)}{2 \log T(z, z^0)} (1 - 1/\log T(z, z^0))}.
$$
\end{enumerate}
\end{lemma}

To prove Proposition \ref{prop:hp-event-lr}, we separately consider the case of overfitted and non-overfitted models. For overfitted models, we can easily find a deterministic lower bound and the upper bound is established in high probability, drawing from Lemma \ref{prop:tail-bounds}(i) and a simple union bound argument. For non-overfitted models, we use a similar argument for the upper bound and we establish the lower bound in high probability using a technical lemma on right tail bounds for F variables from \cite{rossell2022concentration} re-stated as Lemma \ref{lem:s4-rossell} in Section \ref{app:technical-lemmas-2}.

We will denote by $n_0$ and $C$ respectively a large enough but finite integer and a constant independent of $z$ which value may change from line to line. Using simple algebra and the notation introduced in the proof of Lemma \ref{prop:tail-bounds}, we can decompose $B^{LR}(z, z^0)$ for any $z \neq z^0$ as follows:
\begin{align}
B^{LR}(z, z^0)=
\frac{\pi^{LR}(y, X \mid z) \pi(z)}{\pi^{LR}(y, X \mid z^0) \pi(z^0)}=
\left( 1 + \frac{|z|_0-|z^0|_0}{n-|z|_0} \tilde{F}_{z, z^0} \right)^{\frac{1+n}{2}}
\left(1 + \frac{n}{\tau} \right)^{\frac{(|z^0|_0-|z|_0)}{2}}
\left( \frac{\theta^{LR}}{1-\theta^{LR}} \right)^{|z|_0 - |z^0|_0}
\label{eq:bf_normalprior}
\end{align}
where
\begin{align}
0 \leq \tilde{F}_{z, z^0}= \frac{(\tilde{s}_{z^0}-\tilde{s}_z)/(|z|_0-|z^0|_0)}{\tilde{s}_z/(n-|z|_0)}
\leq
\frac{(s_{z^0}-s_z)/(|z|_0-|z^0|_0)}{s_z/(n-|z|_0)}
=F_{z, z^0},
\label{eq:ftest_stat}
\end{align}
with $\tilde{s}_z= \lambda + y^T y - (n/\tau)/[n/\tau+1] y^T X_{\cdot z}(X_{\cdot z}^T X_{\cdot z})^{-1} X_{\cdot z}^T y$
and $s_z= y^T y - y^T X_{\cdot z}(X_{\cdot z}^T X_{\cdot z})^{-1} X_{\cdot z}^T y$.

\textbf{Case 1: Overfitted models.} Let $z \in \mathcal{M}_1$.
Since $\tilde{F}_{z z^0} \geq 0$ and $\theta^{LR}= p^{-\alpha^{LR}}$, we have
\begin{align*}
   \log B^{LR}(z, z^0) &\geq - \frac{(|z|_0-|z^0|_0)}{2}\log \left(1 + \frac{n}{\tau} \right) + (|z|_0 - |z^0|_0)  \log \theta^{LR} - (|z|_0 - |z^0|_0) \log (1 - \theta^{LR}) \\
   &\geq  - \frac{(|z|_0-|z^0|_0)}{2}\log \left(1 + \frac{n}{\tau} \right) + (|z|_0 - |z^0|_0)  \log \theta^{LR} \\
   &\geq - \frac{|z|_0-|z^0|_0}{2}\left( \frac{n}{\tau} \right) - \alpha^{LR} (|z|_0 - |z^0|_0)  \log p,
\end{align*}
using in the last inequality that $\log (1 + x) \leq x$ and in the second inequality  that $- (|z|_0 - |z^0|_0) \log (1 -\theta^{LR}) \geq 0$ since $\log (1 - \theta^{LR}) < 0$ and $z \supseteq z^0$. 
 Note that for an overfitted model, $|z|_0 - |z^0|_0 = |z \backslash z^0|_0$.  Thus, for $n \geq n_0$, we obtain
 \begin{align*}
      B^{LR}(z, z^0) \geq p^{- \bar r_0 |z \backslash z^0|_0} n^{-\bar r_1 |z \backslash z^0|_0}
 \end{align*}
 with $\bar r_0 = \alpha^{LR} $ and $\bar r_1 = \frac{1}{2} + \epsilon$ with $\epsilon > 0$ arbitrarily close to $0$ since under D2, $\tau = o(n)$.
 Note that this lower bound is deterministic.
We now derive a probabilistic upper bound that holds uniformly across all overfitted $z$. The proof strategy is to use Lemma \ref{prop:tail-bounds}(i) to bound $B^{LR}(z,z^0)$ for a single $z$, and then use the union bound across all $z$. Define
\begin{align}\label{eq:u-z}
    u(z) := (p^{-\alpha^{LR}} (\frac{n}{\tau})^{-1/2})^{|z|_0-|z^0|_0}.
\end{align}
Note that
\begin{align*}
 2 \log u(z) = (|z|_0-|z^0|_0) (\log \frac{n}{\tau} + 2 \alpha^{LR} \log p )  = o(n),
\end{align*}
under D4. Using Lemma \ref{prop:tail-bounds}(i), with $w_u := 2 \log (\frac{n p^{2 \alpha^{LR}}}{\tau} )$, we have, for any $c < 1$,
\begin{align*}
    \mathbb P[ B^{LR}(z, z^0)  \geq u(z)] &\leq \frac{1}{c} \left(2 e \log (\frac{n p^{2 \alpha^{LR}}}{\tau} )\right)^{(|z|_0-|z^0|_0)}   \left(\frac{\tau}{n p^{2 \alpha^{LR}}}\right)^{\frac{(|z|_0-|z^0|_0)(1-c)}{2}} = \frac{1}{c} r^{|z|_0-|z^0|_0},
\end{align*}
defining $r := 2 e \log (\frac{n p^{2 \alpha^{LR}}}{\tau} )\left(\frac{\tau}{n p^{2 \alpha^{LR}}}\right)^{\frac{1-c}{2}}$.




Observe now that there are at most $p^{\ell}$ overfitted models $z$ such that $|z\backslash z^0|_0 = \ell$ for any $\ell \leq \bar d$. Thus,
\begin{align*}
     \mathbb P \left( \bigcap_{z \in \mathcal{M}_1} \left \{ B^{LR}(z,z^0) \leq u(z) \right \} \right) &\geq 1 - \sum_{z \in \mathcal{M}_1} \mathbb P \left( B^{LR}(z,z^0) \geq u(z) \right) \\
          &\geq 1 - \sum_{\ell = 1}^{\bar d} \sum_{z \in \mathcal{M}_1 :  |z \backslash z^0|_0=\ell} \mathbb P \left( B^{LR}(z,z^0) \geq u(z) \right) \\
               &\geq 1 -   \frac{1}{c} \sum_{\ell = 1}^{\bar d} p^{\ell} r^{\ell} = 1 - pr\frac{1 - (pr)^{\bar d}}{1- pr}\geq 1 - \frac{pr}{c(1 - pr)}.
\end{align*}
where in the last step we use that $1 - (pr)^{\bar d} \leq1$ since
\begin{align*}
    pr =  2 e p \log (\frac{n p^{2 \alpha^{LR}}}{\tau} )\left(\frac{\tau}{n p^{2 \alpha^{LR}}}\right)^{\frac{1-c}{2}} = o(p^{-1})
\end{align*}
since $\alpha^{LR} > 2$, and by choosing $c < 1 - \frac{2}{\alpha^{LR}}$ and for $n \geq n_0$. Therefore, we obtain 
\begin{align*}
      \mathbb P \left( \bigcap_{z \in \mathcal{M}_1 } \left \{ B^{LR}(z,z^0) \leq u(z) \right \} \right) &\geq 1 - o(p^{-1}).
\end{align*}
Thus, with $r_1 = \frac{1}{2} - \epsilon$, $\bar r_1 = \frac{1}{2} + \epsilon$,  $r_0 = \bar r_0 = \alpha^{LR}$, we can finally conclude  that
$
     \mathbb P((\mathcal{E}_4^p)^c) = o(p^{-1}).
$


\textbf{Case 2: Non-overfitted models.} Let $z \in \mathcal{M}_2 \cup \mathcal{M}_2$.

Denote by  $T(z,z^0)$  the non-centrality parameter of $z$ versus $z^0$ 
defined as
\begin{align}\label{eq:min-non-centrality}
    T(z,z^0) =    \frac{(X_{\cdot z^0} \beta^0_{z^0})^T (I - H_z) X_{\cdot z^0} \beta^0_{z^0} }{(\Omega_{jj}^0)^{-1}} =\frac{ \|(I - H_z)X_{\cdot z^0} \beta^0_{z^0}\|^2}{(\Omega_{jj}^0)^{-1}}.
\end{align}
with $H_z = X_{\cdot z}(X_{\cdot z}^T X_{\cdot z})^{-1} X_{\cdot z}^T$ the projection matrix onto the span of the columns in $X_{\cdot z}$ and $\beta^0 = - \Omega_{-j,j}^0 /\Omega_{jj}^0$. Note that $(I - H_z)^\top(I-H_z) = (I - H_z)$ since $I - H_z$ is the projection matrix on the orthogonal space in $\R^p$ to the span of columns in $X_z$. We first obtain lower and upper bounds on $T(z,z^0)$.
Note that
\begin{align*}
    T(z,z^0) &=  \frac{(X_{\cdot z^0 \backslash z} \beta^0_{z^0 \backslash z})^T (I - H_z) X_{\cdot z^0 \backslash z} \beta^0_{z^0 \backslash z} }{(\Omega_{jj}^0)^{-1}} =  \frac{(\beta^0_{z^0 \backslash z})^T (X_{\cdot z^0 \backslash z}^T  (I - H_z) X_{\cdot z^0 \backslash z}) \beta^0_{z^0 \backslash z} }{(\Omega_{jj}^0)^{-1}} \\
    &\geq \frac{1}{(\Omega_{jj}^{0})^{-1}}\|\beta^0_{z^0 \backslash z}\|^2 \lambda_{min}(X_{\cdot z^0 \backslash z}^T (I - H_z)X_{\cdot z^0 \backslash z} ) \\
    &\geq \frac{\ds{n}}{(\Omega_{jj}^{0})^{-1}}\|\beta^0_{z^0 \backslash z}\|^2 \epsilon_1,
\end{align*}
where in the last inequality we have used the fact that $\lambda_{min}(X_{\cdot z^0 \backslash z}^T (I - H_z)X_{\cdot z^0 \backslash z} ) \geq \lambda_{min}(X_{\cdot z^0 \backslash z}^T X_{\cdot z^0 \backslash z})$ (see, e.g., Lemma 5 in \cite{yang2016computational}) and $\lambda_{min}(X_{\cdot z^0 \backslash z}^T X_{\cdot z^0 \backslash z}) \geq n \epsilon_1$ under our eigenvalue condition D1. 
Moreover, using our  betamin condition C2, we obtain
\begin{align}
    T(z,z^0) \geq  n \frac{\epsilon_1}{(\Omega_{pp}^{0})^{-1}}\omega_{min} |z^0 \backslash z|_0 \geq n |z^0 \backslash z|_0 \epsilon_0 \epsilon_1 \omega_{min} = \eta_0 n |z^0 \backslash z|_0 , \label{eq:lower-bound-t}
\end{align}
using that $\|\beta^0_{z^0 \backslash z}\| = (\Omega_{jj}^{0})^{-1} \|\Omega_{-j,j}^{0}\|$ and $\Omega_{pp}^{0} \leq \|\Omega^{0}\| \leq \epsilon_0^{-1}$ under C1, with $\eta_0 =  \epsilon_1 \epsilon_0 \omega_{min}$. Note that we also have that $T(z,z^0) \leq \frac{ \|X_{\cdot z^0 \backslash z} \beta^0_{\cdot z^0 \backslash z}\|^2}{(\Omega_{jj}^0)^{-1}} \leq n \epsilon_0^{-1} \|[\Omega^0_{-j,j}]_{z^0 \backslash z}\|^2$.

We now define
\begin{align*}
   u'(z) = e^{-\frac{T(z,z^0)}{2 \log T(z,z^0)}(1 - (\log T(z,z^0))^{-1}) }, 
\end{align*}
 and, for some arbitrary $c \in (0,1)$,
\begin{align*}
    w'(z) = \frac{T(z,z^0)}{ |z^0 \backslash z|_0} + (1-c) \log \frac{cn p^{2\alpha^{LR}}}{\tau} = \bar T(z,z^0) + (1-c) \log \frac{cn p^{2\alpha^{LR}}}{\tau},
\end{align*}
 with $\bar T(z,z^0) = \frac{T(z,z^0)}{ |z^0 \backslash z|_0}$. Applying Lemma \ref{prop:tail-bounds}(ii) we obtain
 \begin{align*}
      \mathbb P[ B^{LR}(z, z^0)  \geq u'(z)] &\leq (e  w'(z))^{|z|_0-|z^0|_0} e^{-c \frac{T(z,z^0)}{2(\log T(z,z^0))^2}} \\
      &\leq \left(e  w'(z)) e^{-c  \frac{\bar T(z,z^0)}{2(\log T(z,z^0))^2}} \right)^{|z|_0-|z^0|_0} = r'(z)^{|z|_0-|z^0|_0}
\end{align*}
with $r'(z) := e  w'(z) e^{-c  \frac{\bar T(z,z^0)}{2(\log T(z,z^0))^2}}$. Note that in light of \eqref{eq:lower-bound-t}, $\bar T(z,z^0) \geq \eta_0 n$, therefore there exists $\bar r = o(1)$ independent of $z$ such that $r'(z) \leq \bar r$: we have
\begin{align*}
    r'(z) &= e \left( \bar T(z,z^0) + (1-c) \log \frac{cn p^{2\alpha^{LR}}}{\tau} \right)e^{-c  \frac{\bar T(z,z^0)}{2(\log T(z,z^0))^2}} \\
    &\leq e \left( \bar T(z,z^0) + (1-c) \log \frac{cn p^{2\alpha^{LR}}}{\tau} \right)e^{-c  \frac{\bar T(z,z^0)}{2(\log \bar T(z,z^0)+ \log \bar d})^2} \\
    &\leq e^{-C \frac{n}{(\log n + \log \bar d)^2}} =: \bar r.
\end{align*}
 Moreover, there are at most $\bar d^k p^{\ell}$  non-overfitted models such that $|z^0 \backslash z| = k$ and $|z \backslash z^0| = \ell$ with $\ell \in \{0,\dots, \bar d\}$ and $k \in \{1,\dots, \bar d\}$. Therefore, using a union bound argument, we obtain
\begin{align*}
     \mathbb P \left( \bigcap_{z \in \mathcal{M}_2 \cup \mathcal{M}_3} \left \{ B^{LR}(z,z^0) \leq u'(z) \right \} \right) &\geq 1 - \sum_{z \in \mathcal{M}_2 \cup \mathcal{M}_3} \mathbb P \left( B^{LR}(z,z^0) \geq u'(z) \right) \\
          &\geq 1 - \sum_{\ell = 0}^{\bar d}  \sum_{k = 1}^{\bar d} \sum_{z :  |z \backslash z^0|_0=\ell,  |z^0 \backslash z|_0=k} \mathbb P \left( B^{LR}(z,z^0) \geq u'(z) \right) \\
     &\geq 1 -    \sum_{\ell = 0}^{\bar d}  \sum_{k = 1}^{\bar d} p^{\ell} \bar d^k \bar r^k  = 1 - \frac{1 - p^{\bar d }}{1 - p}\bar d \bar r \frac{1 - (\bar d \bar r)^{\bar d}}{1 - \bar d \bar r} \\
     &\geq 1 - 2 p^{\bar d-1} \frac{ \bar d \bar r}{1 -  \bar d \bar r},
\end{align*}
where in the last inequality we have used the fact that $\bar d \bar r = o(1)$ since
\begin{align*}
    p^{\bar d -1}\bar d \bar r = p^{\bar d-1} \bar d e^{-C \frac{n}{(\log n + \log \bar d)^2}}  = o(p^{-1}).
\end{align*}
Above, we have used the fact that 
$\bar d \log p = o\left(\frac{n}{(\log n + \log \bar d)^2}\right)$,
which holds since $\bar d \log p = o\left(\frac{n}{(\log n)^2}\right)$ by Assumption D4. 
For $n \geq n_0$, we finally obtain that 
\begin{align}\label{eq:case2-ub}
    \mathbb P \left( \bigcap_{z \in \mathcal{M}_2 \cup \mathcal{M}_3} \left \{ B^{LR}(z,z^0) \leq u'(z) \right \} \right) \geq 1 - 4 p^{\bar d-1} \bar d \bar r.
\end{align}
Besides, we have
\begin{align*}
    \log u'(z) &=  -\frac{T(z,z^0)}{2 \log T(z,z^0)}(1 - (\log T(z,z^0))^{-1}) \\
    &\geq - \frac{T(z,z^0)(1-\epsilon)}{2 \log T(z,z^0)} \\
    &\geq  - n  |z^0 \backslash z|_0 \frac{\eta_0 (1-\epsilon)}{2 (\log n + \log \eta_0 + \log |z^0 \backslash z|_0))} \\
    &\geq - n  |z^0 \backslash z|_0 \frac{\eta_0(1-\epsilon)}{4\log n} =: r_2 |z^0 \backslash z|_0 \frac{n}{\log n},
\end{align*}
for any $\epsilon > 0$, $n\geq n_0$ and $r_2 = \frac{\eta_0(1-\epsilon)}{4} = \frac{\epsilon_0 \epsilon_1 \omega_{min}(1-\epsilon)}{4} $. Note that in the last inequality we have used the fact that for $n \geq n_0$, $\log \eta_0 + \log |z^0\backslash z|_0 \leq \log n$.
This demonstrates that the upper bound in $\mathcal{E}_5^j$ holds with high probability.

We next demonstrate that the lower bound in $\mathcal{E}_5^j$ holds with high probability, using the same strategy as in the proof of Lemma \ref{prop:tail-bounds}(ii). For any $z \not \supset z^0$,  we denote by $z^u = z \cup z^0$  the union model and note that $B^{LR}(z,z^0)= B^{LR}(z,z^u)B^{LR}(z^u, z^0)$. Then, for any $v, w >0$
\begin{align}
 \mathbb P \left( \bigcap_{z \in \mathcal{M}_2 \cup \mathcal{M}_3} \left \{ B^{LR}(z,z^0) > v \right \} \right) \geq 1 - \sum_{z \in \mathcal{M}_2 \cup \mathcal{M}_3}\mathbb P \left(B^{LR}(z,z^0) < v  \right) \nonumber \\
\geq 1 - \sum_{z \in \mathcal{M}_2 \cup \mathcal{M}_3} \left[ \mathbb{P} \left( B^{LR}(z,z^u) < w \right)  
+ \mathbb{P} \left(  B^{LR}(z^u,z^0) < \frac{v}{w}  \right) \right] 
\label{eq:lm_lowbound_overfitted}. 
\end{align}

We first show that by choosing  $ \frac{v}{w}  < \left[\left( \frac{n}{\tau} \right)^{- \frac{1}{2}} p^{- \alpha^{LR}} \right]^{|z^u|_0 - |z^0|_0}$ we can eliminate the second probability term in \eqref{eq:lm_lowbound_overfitted}
since from Case 1,
\begin{align}
B^{LR}(z^u, z^0) \geq \left[ \left( \frac{c n}{\tau} \right)^{- \frac{1}{2}} p^{- \alpha^{LR}} \right]^{|z^u|_0 - |z^0|_0} = \bar u(z).
\nonumber
\end{align}
with probability 1 and for any constant $c<1$ and any  $n \geq n_0$. Hence if $\frac{v}{w} < \bar u(z)$, 
we have that
$ \sum_{z \in \mathcal{M}_2 \cup \mathcal{M}_3} \mathbb{P} \left( B^{LR}(z^u,z^0) < \frac{v}{w} \right)=0$ for  $n \geq n_0$. 

We bound the first term in \eqref{eq:lm_lowbound_overfitted} and for this, we consider the event 
\begin{align*}
    \mathcal{E}_3 := \bigcap_{z \in \mathcal{M}_2 \cup \mathcal{M}_3} \left \{ \frac{s_z}{\Tilde s_z} \geq \frac{1}{2} \right \}.
\end{align*}
Recall that here  $\tilde{s}_z= \lambda + y^T y - (n/\tau)/[n/\tau+1] y^T X_{\cdot z}(X_{\cdot z}^T X_{\cdot z})^{-1} X_{\cdot z}^T y$
and $s_z= y^T y - y^T X_{\cdot z}(X_{\cdot z}^T X_{\cdot z})^{-1} X_{\cdot z}^T y$.
Thus, from the expression of $\Tilde F_{z,z^u}$ in \eqref{eq:ftest_stat}, if $ \mathcal{E}_3$ holds, then  
\begin{align*}
    F_{z,z^u} \geq \Tilde F_{z,z^u} \geq \frac{n/\tau}{2(n/\tau +1)}F_{z,z^u} \geq \frac{1}{2}(1-\epsilon) F_{z,z^u}, 
\end{align*}
for any $\epsilon > 0$ and $n \geq n_0$ since under (D2), $\tau = o(n)$.
Therefore, this  implies that
\begin{align*}
   \sum_{z \in \mathcal{M}_2 \cup \mathcal{M}_3} \mathbb{P} \left( B^{LR}(z,z^u) < w \right) &= \sum_{z \in \mathcal{M}_2 \cup \mathcal{M}_3} \mathbb{P} \left( B^{LR}(z^u,z) > 1/w \right) \\
    &= \sum_{z \in \mathcal{M}_2 \cup \mathcal{M}_3} \mathbb{P} \left( (|z^u|_0 - |z|_0) \Tilde F_{z^u, z} > A(w,z)   \right) \\
   &\leq \sum_{z \in \mathcal{M}_2 \cup \mathcal{M}_3} \mathbb{P} \left( \{ (|z^u|_0 - |z|_0) F_{z^u, z} >  A(w,z) \} \cap \mathcal{E}_3  \right) + \mathbb{P} \left( \mathcal{E}_3^c \right),
\end{align*}
with 
\begin{align*}
    A(w,z) &= (n - |z|_0) \left[ \left( \left[ 1 + \frac{n}{\tau} \right] \frac{(1-\theta^{LR})^2 w^{-\frac{2}{|z^u|_0 - |z|_0}}}{(\theta^{LR})^2} \right)^{\frac{ (|z^u|_0 - |z|_0)}{1+n}} - 1 \right].
\end{align*}
Using that $x^a - 1 = e^{a \log x} - 1\leq a\log x$, we obtain
\begin{align}\label{eq:lower-bound-a}
   A(w,z)  &\leq   (n - |z|_0) \left[ \frac{ (|z^u|_0 - |z|_0)}{1+n} \left \{ \log ( 1 + \frac{n}{\tau}) - 2\log \theta^{LR} \right \} - \frac{2}{1+n} \log w   \right] \nonumber \\
    &\leq c \left[  (|z^u|_0 - |z|_0) \left \{ \log (\frac{n}{\tau}) - 2\log \theta^{LR} \right \} - 2 \log w   \right],
\end{align}
for any $c \in (0,1)$ and $n \geq n_0$ and
using that $|z|_0 \leq \bar d = o(n)$ under (C0) and (C4) (since $\Tilde \alpha \geq \bar d$ and $\Tilde \alpha=o(n)$). 
Moreover, $F_{z,z^u}$ follows a non-central F-distribution  with $|z^u|_0- |z|_0$  and $n-|z^u|_0 $ degrees of freedom in the numerator and denominator and non-centrality parameter
\begin{align}\label{eq:lower-bound-T}
    T(z^u,z) &= \frac{\|(H_{z^u} - H_z) X_{\cdot z^0} \beta^0_{z^0}\|}{(\Omega_{jj}^0)^{-1}} = \frac{\|(I - H_z) X_{\cdot z^0} \beta^0_{ z^0}\|^2}{(\Omega_{jj}^0)^{-1}} \geq n\epsilon_0 \epsilon_1 |z^0 \backslash z|_0 \omega_{min}.
\end{align}
Applying the right-tail bound for the F-distribution (Lemma \ref{lem:s4-rossell} here and Lemma S4(ii) in \cite{rossell2022concentration}), with the following values of $s,t$ and $A$ (denoted by $w$ in Lemma \ref{lem:s4-rossell}): 
\begin{align}
    &A(w,z) := w_0 |z^0 \backslash z|_0 \bar T(z^u, z), \quad w_0 > 1, \quad \bar T(z^u, z) = \frac{ T(z^u, z)}{|z^0 \backslash z|_0 } \label{eq:value-of-A} \\
    &t = \frac{1}{2} - \frac{1}{2} \sqrt{\frac{\lambda}{w_0s}} = \frac{1}{2}(1 - \sqrt{\frac{1}{w_0 s}}) \nonumber \\
    &s = 1 - s', \quad 0 < s' < 1 - \frac{1}{2w_0}, \nonumber
\end{align}
and noting that $|z^0 \backslash z|_0 = |z^u \backslash z|_0 = |z^u|_0 - |z|_0$ we obtain
\begin{align*}
     \mathbb{P} \left( \{(|z^u|_0 - |z|_0) F_{z^u, z} > A(w,z) \}  \cap \mathcal{E}_3 \right) &\leq e^{- A(w,z) \frac{s}{2}(1 - \sqrt{\frac{1}{w_0 s}})^2 } (w_0 s)^{\frac{(|z^u|_0 - |z|_0)}{4}} + e^{- \frac{n - |z^u|_0}{6}(s')^2 } \\
     &= e^{-w_0 (|z^u|_0 - |z|_0)  \bar T(z^u, z) \frac{s}{2}(1 - \sqrt{\frac{1}{w_0 s}})^2 } (w_0 s)^{\frac{1}{4}}  + e^{- \frac{n - |z^u|_0}{6}(s')^2 } \\
     &\leq \Tilde r^{|z^u|_0 - |z|_0} + e^{- \frac{n - \bar d}{6}(s')^2 }
\end{align*}
with $\Tilde r =  e^{- n w_0  \epsilon_0 \epsilon_1 \omega_{min} \frac{s}{2}(1 - \sqrt{\frac{1}{w_0 s}})^2 } (w_0 s)^{\frac{1}{4}}$ 
using that from \eqref{eq:lower-bound-T}, $\bar T(z^u,z) \geq n \epsilon_0 \epsilon_1 \omega_{min}$.

Hence, summing over all non-overfitted models, we arrive at
\begin{align*}
     \sum_{z \in \mathcal{M}_2 \cup \mathcal{M}_3} \mathbb{P} \left( \{ (|z|_0 - |z^u|_0) F_{z^u, z} > A(w,z) \}  \cap \mathcal{E}_3 \right) &\leq \sum_{\ell = 0}^{\bar d}  \sum_{k = 1}^{\bar d} p^{\ell} \bar d ^k (\Tilde r^k + e^{- \frac{n - \bar d}{6}(s')^2 }) \\
     &\leq \frac{p^{\bar d} - 1}{p-1} \frac{\bar d \Tilde r (1 - (\bar d \Tilde r)^{\bar d})}{1 - \bar d \Tilde r}  + 4 p^{\bar d}\bar d^{\bar d}e^{- \frac{n - \bar d}{6}(s')^2 } \\
     &\leq 2 \frac{p^{\bar d-1} \bar d \Tilde r}{1 - \bar d  \Tilde r} + 4 (p \bar d )^{\bar d}e^{- \frac{n - \bar d}{6}(s')^2 } = o(p^{-1}),
\end{align*}
for $n \geq n_0$
since $\bar d \log p + \bar d \log \bar d = o(n)$ under Assumption D4 and for any fixed $(w_0, s)$,
\begin{align*}
    p^{\bar d-1} \bar d \Tilde r = p^{\bar d-1} \bar d e^{- n w_0  \epsilon_0 \epsilon_1 \omega_{min} \frac{s}{2}(1 - \sqrt{\frac{1}{w_0 s}})^2 } (w_0 s)^{\frac{1}{4}} = o(p^{-1}).
\end{align*}
Therefore, from  \eqref{eq:lower-bound-a} and \eqref{eq:value-of-A}, we can choose  $w$ such that
\begin{align*}
     w_0 |z^0 \backslash z|_0 \bar T(z^u, z) \leq c \left[  (|z^u|_0 - |z|_0) \left \{ \log (\frac{n}{\tau}) - 2 \alpha^{LR} \log p \right \} - 2 \log w   \right],
\end{align*}
and it is enough to choose it such that
\begin{align}\label{eq:w} 
    w = e^{- \frac{w_0 T(z^u, z)}{c} - \alpha^{LR} (|z|_0 - |z^u|_0) \log p + \frac{|z|_0 - |z^u|_0)}{2} \log n }.
\end{align}
Since we also want that  $\frac{v}{w} \leq \bar u(z) = \left[\left( \frac{n}{\tau} \right)^{- \frac{1}{2}} p^{- \alpha^{LR}} \right]^{|z^u|_0 - |z^0|_0}$, we choose $v$ such that
\begin{align}\label{eq:v}
    v &\leq \bar u(z) e^{- \frac{w_0 T(z^u, z)}{c} - \alpha^{LR} (|z|_0 - |z^u|_0) \log p + \frac{|z|_0 - |z^u|_0}{2} \log n } \nonumber \\
    &= e^{- \frac{w_0 T(z^u, z)}{c} - \alpha^{LR} (|z|_0 - |z^u|_0 + |z^u|_0 - |z^0|_0) \log p  + \frac{|z|_0 + |z^0|_0 - 2|z^u|_0}{2} \log n + \frac{|z^u|_0 - |z^0|_0}{2} \log \tau } \nonumber \\
    &\leq e^{- \frac{w_0 T(z^u, z)}{\ds{c'}} - \alpha^{LR} (|z|_0 - |z^0|_0) \log p},
\end{align}
for any $c' > c$,
since $|z^u|_0 \geq \max( |z|_0, |z^0|_0)$ and $\log \tau =o(n)$ under (D2), which implies from  \eqref{eq:lower-bound-T} that
\begin{align*}
    \log v &\leq  - \frac{w_0}{c'} T(z^u,z) - \alpha^{LR} (|z|_0 - |z^0|_0) \log p \\
    &=  - \frac{w_0}{c'} n (|z^u|_0 - |z|_0)\epsilon_0 \epsilon_1 \omega_{min} - \alpha^{LR} (|z|_0 - |z^0|_0) \log p \\
    &\leq - n (|z^u|_0 - |z|_0) \frac{1}{c'}\epsilon_0 \epsilon_1 \omega_{min} =: - n |z^0 \backslash z| \bar r_2 n,
\end{align*}
with $\bar r_2 = \frac{1}{\ds{c'}}\eta_0$ and for any  $n \geq n_0$ since $\bar d \log p = o(n)$ under Assumption D4  and since $w_0$ can be chosen arbitrarily close to 1 in \eqref{eq:value-of-A}.

Finally we bound $\mathbb{P} \left( \mathcal{E}_3^c \right)$. Note that with $y = \Y_{\cdot p}$, for any $z \in \mathcal{M}_2 \cup \mathcal{M}_3$, since $\lambda > 0$, we have 
\begin{align}
&\frac{s_z}{\tilde{s}_z}
= \frac{y^T (I-H_z) y}{\lambda + y^T(I-H_z)y (n /\tau)/[n/\tau+1]} < \frac{y^T (I-H_z) y}{ y^T(I-H_z)y (n /\tau)/[n/\tau+1]} = \frac{n/\tau + 1}{n/\tau} = 1 + \frac{\tau}{n}
\nonumber \\
&\frac{s_z}{\tilde{s}_z} > \frac{y^T (I-H_z) y}{\lambda + y^T(I-H_z)y}
= \left( 1 + \frac{\lambda}{y^T(I-H_z) y} \right)^{-1}.
    \nonumber
\end{align}
We now show that the upper and lower bound above converge to 1, using that $y^T (I-H_z) y / (\Omega_{jj}^0)^{-1} \sim \chi^2_{n - |z|_0}$. Using the Chernoff bound for chi-squared variables (see, e.g.,  Lemma S1 in \cite{rossell2022concentration}) with $x =  \frac{\lambda}{(\Omega_{jj}^0)^{-1}} < \frac{n - |z|_0}{\log(n - |z|_0)}$, we obtain
\begin{align}
\mathbb{P} \left( \frac{s_z}{\tilde{s}_z} < \frac{1}{2} \right) &\leq \mathbb{P} \left( \frac{1}{\lambda} y^T (I-H_z) y < 1 \right) \nonumber \\
&= \mathbb{P} \left( \frac{1}{(\Omega_{jj}^0)^{-1}} y^T (I-H_z) y < \frac{ \lambda}{(\Omega_{jj}^0)^{-1}} \right)  = \mathbb{P} \left( \frac{1}{(\Omega_{jj}^0)^{-1}} y^T (I-H_z) y < x \right) \nonumber \\
&\leq \left( \frac{e  \lambda}{(\Omega_{jj}^0)^{-1} (n - |z|_0)} \right)^{\frac{n - |z|_0}{2}} e^{-\frac{ \lambda}{2 (\Omega_{jj}^0)^{-1}} } \leq e^{-\frac{(n-|z|_0)}{3} \log (n-|z|_0)},
\nonumber
\end{align}
for $n \geq n_0$.
Hence, $s_z/\tilde{s}_z \geq \frac{1}{2}$ with probability $\geq 1 - e^{-\frac{(n-|z|_0)}{2} \log (n-|z|_0)}$. To conclude we finally use the union bound
\begin{align*}
    \mathbb{P} \left( \mathcal{E}_3^c \right) &\leq \sum_{z \in \mathcal{M}_2 \cup \mathcal{M}_3}   \mathbb{P} \left( s_z/\tilde{s}_z \leq \frac{1}{2} \right)  \leq 2 p^{\bar d} \bar d^{\bar d} e^{-\frac{(n-|z|_0)}{2} \log (n-|z|_0)}  = o(p^{-1}),
\end{align*}
using again that $\bar d \log p = o(\frac{n}{\log n})$ under (D4). Therefore, we conclude that for any $w$ and $v$ as in \eqref{eq:w} and \eqref{eq:v}, 
\begin{align*}
   \sum_{z \in \mathcal{M}_2 \cup \mathcal{M}_3} \mathbb{P} \left( B^{LR}(z,z^u) < w \right) &= o(p^{-1}),
\end{align*}
and
\begin{align*}
    \mathbb P \left( \bigcap_{z \in \mathcal{M}_2 \cup \mathcal{M}_3} \left \{ B^{LR}(z,z^0) > v \right \} \right) = o(p^{-1}).
\end{align*}
Together with \eqref{eq:case2-ub}, we finally obtain that for $n \geq n_0$,
\begin{align*}
     \mathbb P((\mathcal{E}_5^p)^c) = o(p^{-1}),
\end{align*}
with
\begin{align*}
    \bar r_2 = \frac{\eta_0(1-\epsilon)}{2c} \geq r_2 =   \frac{\eta_0(1-\epsilon)}{4},
\end{align*}
for any $c, \epsilon \in (0,1)$. The result in Proposition \ref{prop:hp-event-lr} follows from the intermediate results above:
\begin{align*}
     \mathbb P\left[(\mathcal{E}_6^p)^c) \right]\leq   \mathbb P((\mathcal{E}_4^p)^c) +   \mathbb P((\mathcal{E}_5^p)^c) =  o(p^{-1}).
\end{align*}




\subsubsection{Proof of Lemma \ref{lem:cons-lr}} \label{app:proof-lem-consistency}

We first re-state Lemma \ref{lem:cons-lr}. Recall that $\mathcal{E}_{LR} = \cap_{j=1}^{p}  \mathcal{E}_{LR}^j$ where $\mathcal{E}_{LR}^j$ is defined in \eqref{eq:event-lr-p}.
Under Assumptions D1 and D4, if $\mathcal{E}_{LR}$ holds and if $\upsilon > (r_0)^{-1}$, then for any $j \in [p]$,
\begin{align*}
    Q^j_\upsilon(Z_{\cdot j}^0) \geq  1 - 3 p^{1 - \upsilon r_0} n^{- \upsilon  r_1},
\end{align*}
for $n \geq n_0$ large enough where $n_0$ only depends on constants $(\upsilon, r_0, r_1, r_2, \epsilon_0,\epsilon_1)$ with $r_0,r_1,r_2$ defined in Proposition \ref{prop:hp-event-lr}. 

\begin{proof}
Without loss of generality, we consider $j=p$ and denote $Q(\cdot ):=Q_{\upsilon}^p(\cdot)$. Recall that $Q^j_\upsilon(Z_{\cdot j}) \propto (\pi^{LR}(Z_{\cdot j}| \Y))^\upsilon$. By definition,
\begin{align*}
    Q(z^0) = \frac{\pi^{LR}(z^0 |\Y)^\upsilon}{\sum_{z \in \mathcal{Z}(\bar d)} \pi^{LR}(z |\Y)^\upsilon} = \frac{1}{1 + \sum_{z \in \mathcal{Z}(\bar d): z \neq z^0} \frac{\pi^{LR}(z |\Y)^\upsilon}{\pi^{LR}(z^0|\Y)^\upsilon}}.
\end{align*}
Moreover, if $\mathcal{E}_{LR}$ holds, for any overfitted $z \in \mathcal{M}_1$, i.e.,  $z \supset z^0$,
\begin{align*}
  \left(  \frac{\pi^{LR}(z |\Y)}{\pi^{LR}(z^0|\Y)} \right)^\upsilon \leq p^{-\upsilon r_0 |z \backslash z^0|_0} n^{- \upsilon  r_1 |z \backslash z^0|_0},
\end{align*}
and for any underfitted $z \in \mathcal{M}_2$, i.e.,  $z \not \supset z^0$,
\begin{align*}
  \left(  \frac{\pi^{LR}(z |\Y)}{\pi^{LR}(z^0|\Y)} \right)^\upsilon \leq e^{- \upsilon r_2 |z^0 \backslash z|_0} \frac{n}{\log n}.
\end{align*}
Therefore, summing over all possible models with $\ell = |z \backslash z^0|_0$ and $k = |z^0 \backslash z|_0$, and using simple algebra,
\begin{align*}
    \sum_{z \in \mathcal{Z}(\bar d): z \neq z^0} \frac{\pi^{LR}(z |\Y)^\upsilon}{\pi^{LR}(z^0|\Y)^\upsilon} &= \sum_{k=1}^{\bar d} p^{k} p^{-\upsilon r_0 k} n^{- \upsilon  r_1 k} + \sum_{k=1}^{\bar d} \sum_{\ell=0}^{\bar d} p^k p^{\ell} e^{- \upsilon r_2 k \frac{n}{\log n} }  \\
    &= p^{1 -\upsilon r_0} n^{-\upsilon r_1} \frac{1 - p^{1-\bar d\upsilon r_0} n^{-\bar d \upsilon r_1}}{1 - p^{1-\upsilon r_0} n^{-\upsilon r_1}} + p e^{-\upsilon r_2 \frac{n}{\log n}} \frac{1 - p^{\bar d} e^{- \bar d \upsilon r_2 \frac{n}{\log n}}}{1 - p^{\bar d} e^{- \bar d \upsilon r_2 \frac{n}{\log n}}} \frac{p^{\bar d +1} - 1}{p-1}\\
    &\leq 2 p^{1 - \upsilon} r_0 n^{- \upsilon \bar r_1} + 2 p^{\bar d+1} e^{-  \upsilon} r_2 \frac{n}{\log n} \\
    &\leq 3 p^{1 - \upsilon  r_0} n^{- \upsilon r_1},
\end{align*}
for $n \geq n_0$. Above we used that $\upsilon r_0 > 1$ (as stated in the lemma's assumptions) and that $\bar d \log p= o(\frac{n}{\log n})$ (Assumption D4). Therefore, we obtain 
\begin{align*}
     Q(z^0) \geq \frac{1}{1 + 3 p^{1 - \upsilon r_0} n^{- \upsilon  r_1}} \geq 1 - 3 p^{1 - \upsilon r_0} n^{- \upsilon  r_1}.
\end{align*}
\end{proof}


\subsubsection{Proof of Lemma \ref{prop:tail-bounds}}
\label{sssec:proof_tail-bounds}

We first re-state Lemma \ref{prop:tail-bounds}.
Assume that Assumptions D1-D4 hold. For any model $z \in \mathcal{Z}(\bar d)$, recall that
\begin{align}
    T(z,z^0)= \frac{(X_{\cdot z^0} \beta^0_{z^0})^T (I - H_z) X_{z^0} \beta^0_{z^0} }{(\Omega_{jj}^0)^{-1}}, \nonumber
\end{align}
with $H_z$ the projection matrix onto the columns of $X_{\cdot z}$, i.e., $H_z = X_{\cdot z}(X_{\cdot z}^T X_{\cdot z})^{-1} X_{\cdot z}^T =  X_{\cdot z} S_{zz}^{-1} X_{\cdot z}^T$ and $\beta^0 = - \Omega^0_{-j,j}/\Omega^0_{jj}$.
\begin{enumerate}[label=(\roman*)]
\item Let $z \supset z^0$ be an overfitted model. Then, for any constant $c<1$,
\begin{align}
\mathbb{P} \left( B^{LR}(z,z^0) > u \right) \leq
    \frac{(ew_u)^{(|z|_0 - |z^0|_0)}}{c u^c}
    \left( \frac{\tau}{c n p^{2    \tilde{\alpha} \color{black}}} \right)^{\frac{c(|z|_0 - |z^0|_0)}{2}} 
\nonumber
\end{align}
where $w_u= \log \left( \frac{n p^{2    \tilde{\alpha} \color{black}}}{\tau} u^{\frac{2}{|z|_0 - |z^0|_0}} \right)$, for all $n \geq n_0$ and any $u$ such that
\begin{align}
(|z|_0-|z^0|_0) \left[ \log \left( \frac{n}{\tau} \right) + 2     \tilde{\alpha} \color{black} \log \left( p \right) \right]  + 2 \log u
= o(n).
\nonumber
\end{align}
Above, $n_0$ and $c$ do not depend on $z$, i.e., they are uniform constants across $z$.
\item Let $z \not\supset z^0$ be a non-overfitted model. If
\begin{align}
    \frac{T(z, z^0)}{\log T(z, z^0)} = o(n), \label{eq:non-centrality-param}
\end{align} 
then \begin{align}
\mathbb{P} \left( B^{LR}(z,z^0) > u \right) \leq 
(ew_b)^{(|z'|_0 - |z^0|_0)} 
e^{-\frac{cT(z, z^0)}{2 (\log T(z, z^0))^2}}
\nonumber
\end{align}
where
$$
w_b= \frac{T(z, z^0)}{|z'|_0 - |z^0|_0} + (1-c) \log \left( \frac{c n (1-\theta^{LR})^2}{\tau (\theta^{LR})^2} \right)
$$
for any constant $c<1$, and $u$ such that 
$$
u > e^{-\frac{T(z, z^0)}{2 \log T(z, z^0)} (1 - 1/\log T(z, z^0))}.
$$
\end{enumerate}

\begin{proof}

For simplicity in this proof, we denote by     $\tilde{\alpha} :=\tilde{\alpha}^{LR}$, \color{black} $\theta := \theta^{LR}$     and $a_\theta= a_\theta^{LR}$, $b_\theta= b_\theta^{LR}$ \color{black}

    Consider first the Binomial prior with inclusion probability $\theta$. \color{black}
Simple algebra gives the following expression for the posterior odds for $z$ over $z^0$,
\begin{align}
B^{LR}(z, z^0)=
\frac{\pi^{LR}(\Y \mid z) \pi(z)}{\pi^{LR}(\Y \mid z^0) \pi(z^0)}=
\left( 1 + \frac{|z|_0-|z^0|_0}{n-|z|_0} \tilde{F}_{z, z^0} \right)^{\frac{a+n}{2}}
\left(1 + \frac{n}{\tau} \right)^{\frac{(|z^0|_0-|z|_0)}{2}}
h(z),
\label{eq:bf_normalprior2}
\end{align}
where
    For the Beta-Binomial$(a_\theta,b_\theta)$ prior, the last term in \eqref{eq:bf_normalprior2} is replaced by
\begin{align*}
h(z)= \begin{cases}   
\left( \frac{\theta}{1-\theta} \right)^{|z|_0 - |z^0|_0}
\mbox{, for the Binomial prior}
\\
\frac{\Gamma(a_\theta + |z|_0) \Gamma(b_\theta + 0.5 p(p-1) - |z|_0)}{\Gamma(a_\theta + |z^0|_0) \Gamma(b_\theta + 0.5 p(p-1) - |z^0|_0)}
\mbox{, for the Beta-Binomial}(a_\theta, b_\theta)
\end{cases}
\end{align*}
\color{black}
and
\begin{align}
\tilde{F}_{z, z^0}= \frac{(\tilde{s}_{z^0}-\tilde{s}_z)/(|z|_0-|z^0|_0)}{\tilde{s}_z/(n-|z|_0)}
\leq
\frac{(s_{z^0}-s_z)/(|z|_0-|z^0|_0)}{s_z/(n-|z|_0)}
=F_{z,z^0},
\label{eq:ftest_stat2}
\end{align}
$\tilde{s}_z= \lambda + y^T y - (n/\tau)/[n/\tau+1] y^T X_{\cdot z}(X_{\cdot z}^T X_{\cdot z})^{-1} X_{\cdot z}^T y$ 
and $s_z= y^T y - y^T X_{\cdot z}(X_{\cdot z}^T X_{\cdot z})^{-1} X_{\cdot z}^T y$ is the sum of squared residuals under a least-squares fit for model $z$. 
Above, $\tilde{F}_{z,z^0}$ is a Bayesian analogue to $F_{z, z^0}$, the standard F-statistic to test $z^0$ versus $z$.
For precision, in the case $|z|_0=|z^0|_0$, we define $(|z|_0-|z^0|_0) \tilde{F}_{z z^0}= (\tilde{s}_{z^0} - \tilde{s}_z) (n-|z|_0) / \tilde{s}_z$, and analogously for $F_{z z^0}$.

The proof strategy is to use that $F_{z z^0}$ follows a central F-distribution when $z \supset z^0$ is an overfitted model, whereas for non-overfitted models $z \not\supset z^0$ it can be expressed as differences between non-central and central F-distributions. Hence, in both cases, one may bound $B^{LR}(z^0, z)$ by using tail inequalities for F-distributions.

Before proceeding, we simplify the expression for $h(z)$. Specifically, we show that $\log h(z) \leq (|z|_0 - |z^0|_0) (\tilde{\alpha} \log p) [1 + o(1)]$, where recall that $\tilde{\alpha}=\alpha$ for the Binomial prior and $\tilde{\alpha}= \max\{\alpha, 2 \}$ for the Beta-Binomial. Since our later derivations only depend on $\log h(z)$, we may replace $h(z)$ by $h(z)= p^{-\alpha (|z|_0 - |z^0|_0)}$.
For the Binomial prior with $\theta= 1/p^\alpha$, we have that
\begin{align*}
    h(z)= \frac{1}{p^{\alpha (|z|_0 - |z^0|_0)}} 
    \left( \frac{1}{1 - 1/p^{\alpha}} \right)^{|z|_0 - |z^0|_0}
\end{align*}
where in the case where $|z|_0 > |z^0|_0$ simple derivations show that the second term lies in $[1 , e^{(|z|_0-|z^0|_0)/ p^\alpha}]$ (and analogously for the case where $|z|_0 < |z^0|_0$). Noting that $(|z|_0-|z^0|_0) / p^\alpha \leq \bar d / p^\alpha= o(1)$ by assumption D3 gives that $h(z)= p^{-\alpha (|z|_0 - |z^0|_0)}  [1 + o(1)]$.

For the Beta-Binomial with $a_\theta=1$, $b_\theta=p^\alpha$, in the case where $|z|_0 > |z^0|_0$ we have that
\begin{align*}
h(z) = \frac{\Gamma(1 + |z|_0)}{\Gamma(1 + |z^0|_0)}
\frac{\Gamma(p^\alpha + 0.5 p(p-1) - |z|_0)}{\Gamma(p^\alpha + 0.5 p (p-1) - |z^0|_0)}=
\prod_{l=1}^{|z|_0 - |z^0|_0}
\frac{(|z|_0 - l + 1)}{p^\alpha + 0.5 p(p-1) - |z^0|_0 - l}.
\end{align*}
Hence,
\begin{align*}
h(z) 
&\leq \left( \frac{|z|_0}{p^\alpha + 0.5 p(p-1) - |z^0|_0 - 1} \right)^{|z|_0 - |z^0|_0}
 \\
&\leq \frac{1}{p^{\tilde{\alpha} (|z|_0 - |z^0|_0)}}
\left( \frac{p^{\tilde{\alpha}} \bar d}{p^\alpha + 0.5 p(p-1) - 1} \right)^{|z|_0 - |z^0|_0}=
\left( \frac{\bar d [1 + o(1)]}{p^{\tilde{\alpha}}} \right)^{|z|_0 - |z^0|_0}.
\end{align*}
Hence, using that $\log \bar d = o(\tilde{\alpha} \log p)$ by Assumption D3, we obtain that $\log h(z) \leq (|z|_0 - |z^0|_0) (\tilde{\alpha} \log p) [1 + o(1)]$.
\color{black}

From expression \eqref{eq:bf_normalprior2}, 
    and replacing $h(z)$ by $p^{-\alpha (|z|_0 - |z^0|_0)}$, \color{black}
we obtain
\begin{align}
    &\mathbb{P} \left( B^{LR}(z,z^0) > u \right) =
    \mathbb{P} \left( \left[ 1 + \frac{|z|_0 - |z^0|_0}{n - |z|_0} \tilde{F}_{z, z^0} \right]^{\frac{a+n}{2}} \left[ \left(1 + \frac{n}{\tau}\right) 
        p^{2 \tilde{\alpha}}  \color{black}
    \right]^{\frac{|z^0|_0 - |z|_0}{2}} > u \right)
    \nonumber \\
    &= \mathbb{P} \left( (|z|_0 - |z^0|_0) \tilde{F}_{z, z^0} > (n - |
    z|_0) \left[  \left( \left( 1 + \frac{n}{\tau} \right) 
        p^{2 \tilde{\alpha}} \color{black} 
    u^{\frac{2}{|z|_0 - |z^0|_0}}
    \right)^{\frac{|z|_0 - |z^0|_0}{a + n}}  - 1 \right]      \right).
    \label{eq:lm_postodds}
\end{align}

{\bf Overfitted models.}
Let $z \supset z^0$. Then $|z|_0 - |z^0| \geq 1$ and $(|z|_0 - |z^0|_0) \tilde{F}_{z z^0} > 0$. Using that $\log(x) \leq x -1$ for any $x>0$, 
\eqref{eq:lm_postodds} gives 
\begin{align}
    \mathbb{P}(B^{LR}(z,z^0) > u) \leq
    \mathbb{P} \left( (|z|_0 - |z^0|_0) \tilde{F}_{z,z^0} >
    \frac{(n - |z|_0) (|z|_0 - |z^0|_0)}{a+n} \log \left( \left[ 1 + \frac{n}{\tau} \right] 
        p^{2 \tilde{\alpha}} \color{black} u^{\frac{2}{|z|_0 - |z^0|_0}}
    \right)  \right)
    \nonumber \\
    \leq \mathbb{P} \left( (|z|_0 - |z^0|_0) F_{z,z^0} >
    \frac{(n - |z|_0) (|z|_0 - |z^0|_0)}{a+n} \log \left( \left[ 1 + \frac{n}{\tau} \right] 
        p^{2 \tilde{\alpha}} \color{black} u^{\frac{2}{|z|_0 - |z^0|_0}}
    \right)  \right), 
    \label{eq:lm_postodds_logupbound}
\end{align}
the right-hand side following from \eqref{eq:ftest_stat2}.

To upper-bound \eqref{eq:lm_postodds_logupbound}, we use the classical result that $F_{z,z^0} \sim F(0, |z|_0 - |z^0|_0, n - |z|_0)$, a central F distribution with numerator and denominator degrees of freedom $|z|_0 - |z^0|_0$ and $n - |z|_0$, respectively.
For example, Lemma S4 and Corollary S1 in \cite{rossell2022concentration} provide such tail bounds.
The intuition is that $\mathbb{P}_0((|z|_0 - |z|_0^0) F_{z,z^0} > w) \leq 2 e^{-w/2}$, up to lower terms as $w \rightarrow \infty$. A more precise reasoning follows.

To simplify the expressions, we use that $|z|_0 \leq \bar d = o(n)$ under our assumption C0,
and that $\tau= o(n)$ by Assumptions D1-D2. 
Then, for large enough $n$, \eqref{eq:lm_postodds_logupbound} can be lower and upper-bounded (uniformly across $z$) by
\begin{align}
    \mathbb{P} \left( (|z|_0 - |z^0|_0) F_{z,z^0} >
    c (|z|_0 - |z^0|_0) \log \left( c \frac{n p^{2    \tilde{\alpha} \color{black}}}{\tau} u^{\frac{2}{|z|_0 - |z^0|_0}} \right)  \right), 
    \label{eq:lm_postodds_logupbound_asymp}
\end{align}
where either $c<1$ (for the lower bound) or $c>1$ (for the upper bound) is a constant that is arbitrarily close to 1.

Lemma S4 and Corollary S1 in \cite{rossell2022concentration} give different bounds for 
$\mathbb{P}((|z|_0 - |z|_0^0) F_{z,z^0} > w)$, depending on whether $w < n - |z|_0$ (Case 1) or $w \geq n - |z|_0$ (Case 2). Specifically, we take
\begin{align}
w = c (|z|_0 - |z^0|_0) \log \left( c \frac{n p^{2    \tilde{\alpha} \color{black}}}{\tau} u^{\frac{2}{|z|_0 - |z^0|_0}} \right)
\Rightarrow
u= \left( \frac{\tau}{c n p^{2    \tilde{\alpha} \color{black}}  } \right)^{\frac{|z|_0 - |z^0|_0}{2}} e^{\frac{w}{c(|z|_0 - |z^0|_0)}}.
\label{eq:fstatbound_w2u}
\end{align}

We focus on Case 1. If $w \in ([|z|_0-|z|_0]/[2-\sqrt{3}], n - |z|_0)$, by Corollary S1 in \cite{rossell2022concentration}, we have $\mathbb{P}((|z|_0 - |z|_0^0) F_{z,z^0} > w) \leq$
\begin{align}
    &\left( \frac{e w}{|z|_0 - |z^0|_0} \right)^{|z|_0 - |z^0|_0}
    e^{-\frac{w}{2} (1 - \sqrt{2w/(n - |z|_0)})} + e^{-\frac{w}{2}} \nonumber \\
    &= (ew')^{|z|_0 - |z^0|_0} \left( \frac{\tau}{c n p^{2    \tilde{\alpha} \color{black}} u^{\frac{2}{|z|_0 - |z^0|_0}}} \right)^{\frac{c(|z|_0 - |z^0|_0)}{2} (1- \sqrt{2w/(n - |z|_0)} )} + \left( \frac{\tau}{c n p^{2    \tilde{\alpha} \color{black}} u^{\frac{2}{|z|_0 - |z^0|_0}}} \right)^{\frac{c(|z|_0 - |z^0|_0)}{2}} \nonumber \\
    &\leq \frac{(ew')^{(|z|_0 - |z^0|_0)}}{u^c}
    \left( \frac{\tau}{c n p^{2    \tilde{\alpha} \color{black}}} \right)^{\frac{c(|z|_0 - |z^0|_0)}{2}} 
    + \frac{1}{u^{c}} \left( \frac{\tau}{cn p^{2    \tilde{\alpha} \color{black}}} \right)^{c\frac{|z|_0 - |z^0|_0}{2}}.
\end{align}
since $c< 1$ 
and with $w'= c \log \left( c \frac{n p^{2    \tilde{\alpha} \color{black}}}{\tau} u^{\frac{2}{|z|_0 - |z^0|_0}} \right)$, and the right-hand side holds for large enough $n$.
Since $\log n = O(w')$, the second term in the right-hand side is asymptotically negligible, relative to the first term 
as $u$, $n$ and $p$ grow. Altogether,
\begin{align}
    \mathbb{P}((|z|_0 - |z|_0^0) F_{z,z^0} > w) \leq
    \frac{(ew')^{(|z|_0 - |z^0|_0)}}{c u^c}
    \left( \frac{\tau}{c n p^{2    \tilde{\alpha} \color{black}}} \right)^{\frac{c(|z|_0 - |z^0|_0)}{2}} 
    \label{eq:lm_postodds_logupbound_asymp_case1}
\end{align}
for sufficiently large $n$, where recall that $c<1$ is a constant that is arbitrarily close to 1.

{\bf Non-overfitted models.}

Let $z'= z \cup z^0$ be the union model that features all edges that are either included by $z$ or by $z^0$, and note that $B^{LR}(z,z^0)= B^{LR}(z,z') /B^{LR}(z^0,z')$. Then,
\begin{align}
    \mathbb{P} \left( B^{LR}(z,z^0) > u \right)&=
    \mathbb{P} \left( \frac{B^{LR}(z,z')}{B^{LR}(z^0,z')} > u \right) \\
    &\leq  \mathbb{P} \left( B^{LR}(z,z') >u b \right)
+ \mathbb{P} \left( B^{LR}(z^0,z') < b \right) \\
    &\leq
    \mathbb{P} \left( B^{LR}(z',z) < \frac{1}{u b} \right)
+ \mathbb{P} \left( B^{LR}(z',z^0) > 1/b \right)  
\label{eq:postodds_lm_nonov_bound}
\end{align}
for any $b>0$.
Above, we used that for any two random variables $W_1>0, W_2>0$, and any $b > 0$, a simple union bound-base argument gives that $\mathbb P(W_1/W_2 > u) \leq \mathbb  P(W_1 > u b) + \mathbb  P(1/W_2 > 1/b)$.


The proof strategy is as follows.
Given that $z' \supset z^0$ is an overfitted model, the second term in \eqref{eq:postodds_lm_nonov_bound} can be bounded using Lemma \ref{prop:tail-bounds}(i).
Regarding the first term in \eqref{eq:postodds_lm_nonov_bound}, from Expression \eqref{eq:lm_postodds} it can be essentially expressed in terms of the left tail of a non-central F distribution. Bounds for such tails are given in Lemma S5 in \cite{rossell2022concentration}.
Then, we set $b$ to provide a sufficiently tight bound.




Consider the second term in \eqref{eq:postodds_lm_nonov_bound}. From Lemma \ref{prop:tail-bounds}(i), 
\begin{align}
\mathbb{P} \left( B^{LR}(z',z^0) > 1/b \right) \leq
    (ew_b)^{(|z'|_0 - |z^0|_0)} b^c
    \left( \frac{\tau}{c n p^{2    \tilde{\alpha} \color{black}}} \right)^{\frac{c(|z'|_0 - |z^0|_0)}{2}} 
\nonumber
\end{align}
where $w_b= \log \left( \frac{n p^{2\alpha}}{\tau} \left[ \frac{1}{b} \right]^{\frac{2}{|z'|_0 - |z^0|_0}} \right)$, for all $n \geq n_0$ and any $b$ such that
\begin{align}
(|z'|_0-|z^0|_0) \left[ \log \left( \frac{n}{\tau} \right) +     p^{2 \tilde{\alpha}} \color{black} \right]  - 2 \log b
= o(n).
\label{eq:lm_cond_bound_fstat}
\end{align}
In particular, consider the choice 
\begin{align}
   b= e^{-\frac{T(z,z^0)}{2 (\log T(z,z^0))^2}} \left( \frac{c n     p^{2 \tilde{\alpha}} \color{black}}{\tau} \right)^{\frac{|z'|_0-|z^0|_0}{2}}, \label{eq:choice-b} 
\end{align}
for which the condition in \eqref{eq:lm_cond_bound_fstat} becomes
\begin{align}
\frac{T(z,z^0)}{(\log T(z,z^0))^2} - (|z'|_0-|z^0|_0) \log c
= o(n),
\nonumber
\end{align}
as $n = O(T(z,z^0))$
which holds under our assumption \eqref{eq:non-centrality-param}. 
Then, by Proposition \ref{prop:tail-bounds}(i), we have
\begin{align}
&\mathbb{P} \left( B^{LR}(z',z^0) > 1/b \right) \leq
(ew_b)^{(|z'|_0 - |z^0|_0)} 
e^{-\frac{cT(z,z^0)}{2 (\log T(z,z^0))^2}},
\nonumber \\
& w_b= \frac{T(z,z^0)}{|z'|_0 - |z^0|_0} + (1-c) \log \left( \frac{c n     p^{2 \tilde{\alpha}} \color{black}}{\tau} \right).
\label{eq:lm_overfitted_term1}
\end{align}

Consider now the first term in \eqref{eq:postodds_lm_nonov_bound}.
Arguing as in \eqref{eq:lm_postodds} gives
\begin{align}
\mathbb{P} \left( B^{LR}(z',z) < \frac{1}{u b} \right)=
\mathbb{P} \left( (|z'|_0 - |z|_0) \tilde{F}_{z', z} < (n - |z'|_0) \left[  \left( \left( 1 + \frac{n}{\tau} \right) \frac{    p^{2 \tilde{\alpha}} \color{black}}{(ub)^{\frac{2}{|z'|_0 - |z|_0}}}  \right)^{\frac{|z'|_0 - |z|_0}{1 + n}}  - 1 \right]      \right).
\label{eq:blr-ub}
\end{align}

For simplicity, 
for suitable $ub$ this probability is essentially equal to 
\begin{align}
    \mathbb{P} \left( (|z'|_0 - |z|_0) F_{z', z} < (|z'|_0 - |z|_0) \log \left( \left( 1 + \frac{n}{\tau} \right) \frac{    p^{2 \tilde{\alpha}} \color{black}}{(ub)^{\frac{2}{|z'|_0 - |z|_0}}}  \right)     \right).
    \label{eq:lm_noncentralF_upbound}
\end{align}

More precisely, \eqref{eq:lm_noncentralF_upbound} is asymptotically equivalent  to the probability in the RHS of \eqref{eq:blr-ub}  if 
\begin{align}
    &\lim_{n\to \infty} \left[\left( 1 + \frac{n}{\tau} \right) \frac{    p^{2 \tilde{\alpha}} \color{black}}{(ub)^{\frac{2}{|z'|_0 - |z|_0}}}\right]^{\frac{|z'|_0 - |z|_0}{a + n}} = 1 \nonumber \\
    &\iff \lim_{n \to \infty} \frac{|z'|_0 - |z|_0}{a + n} \left \{ \log \left(1 + \frac{n}{\tau} \right) + \log \left(     p^{2 \tilde{\alpha}} \color{black} \right) - \frac{2}{|z'|_0 - |z|_0}  \log(ub) \right \} = 0 \label{eq:lim-term-proba}
\end{align}
using that $\lim_{x \rightarrow 1} (x - 1)/\log(x)=1$ 
\begin{align}
    (|z'|_0 - |z|_0) \log \left( [1 + \frac{n}{\tau}]     p^{2 \tilde{\alpha}} \color{black} \right) - 2 \log(ub) = o(n)
   \Leftrightarrow \frac{T(z,z^0)}{(\log T(z,z^0))^2} - 2 \log(u) = o(n),
   \label{eq:lm_noncentralF_ucond1}
\end{align}
where the right-hand side follows by plugging in our choice of $b$.
Since $T(z,z^0)/\log T(z,z^0)= o(n)$ under our assumption \eqref{eq:non-centrality-param}, 
the condition above requires that $-2\log u = o(n)$. Below we shall see that \eqref{eq:lm_noncentralF_ucond1} is satisfied when condition \eqref{eq:lm_noncentralF_ucond2} holds, which implies it is enough to upper bound \eqref{eq:lm_noncentralF_upbound}. 

Next, suppose that
\begin{align}
    (|z'|_0 - |z'|_0) \log \left( \left( 1 + \frac{n}{\tau} \right) \frac{    p^{2 \tilde{\alpha}} \color{black}}{(ub)^{\frac{2}{|z'|_0 - |z|_0}}}  \right)
    < \frac{T(z,z^0)}{\log T(z,z^0)}
\label{eq:condition_lemmas5_rossell}
\end{align}
and that the right-hand side in \eqref{eq:condition_lemmas5_rossell} grows with $n$.
Then, applying Lemma S5 in \cite{rossell2022concentration} to \eqref{eq:lm_noncentralF_upbound} gives that
\begin{align}
\mathbb{P} \left( B^{LR}(z',z) < \frac{1}{u b} \right) \leq e^{-c T(z,z^0)/2} + e^{-c_1\frac{n - |z'|_0}{2}}
\leq 
e^{-c T(z,z^0)/2} ( 1 + e^{-c_2 n/2}).
\label{eq:lm_overfitted_term2}
\end{align}
for any constants $c_1 > c_2>0$, and sufficiently large $n$, the right-hand following from our assumption \eqref{eq:non-centrality-param}. 

All that remains is seeing that \eqref{eq:condition_lemmas5_rossell} indeed holds.
Simple algebra shows that \eqref{eq:condition_lemmas5_rossell} can be re-written as
\begin{align}
ub > 
   \left( \left( 1 + \frac{n}{\tau} \right)     p^{2 \tilde{\alpha}} \color{black} \right)^{\frac{(|z'|_0 - |z|_0)}{2} } e^{-\frac{T(z,z^0)}{2 \log T(z,z^0)}}.
\nonumber
\end{align}
Plugging in our choice of $b$ \eqref{eq:choice-b},   the above inequality is equivalent to
\begin{align*}
    \log u > \frac{(|z'|_0 - |z|_0)}{2} \log  \left(  \frac{1 + \frac{n}{\tau}}{c\frac{n}{\tau}}\right) - \frac{T(z,z^0)}{2 \log T(z,z^0)} \left(1 -\frac{1}{\log T(z,z^0)} \right),
\end{align*}
and since $\frac{(|z'|_0 - |z|_0)}{2} \log  \left( (1 + \frac{n}{\tau} ) \frac{\tau}{cn}\right) = O(\bar d)$, this requires that
\begin{align}
 \log u > - \frac{T(z,z^0)}{2\log T(z,z^0)} \left( 1 - \frac{1}{\log T(z,z^0)} \right)
\Leftrightarrow
u > e^{-\frac{T(z,z^0)}{2 \log T(z,z^0)} (1 - 1/\log T(z,z^0))}.
\label{eq:lm_noncentralF_ucond2}
\end{align}
Since $T(z,z^0)/ \log T(z,z^0) = o(n)$ by \eqref{eq:non-centrality-param}, 
condition \eqref{eq:lm_noncentralF_ucond2} implies that $-2 \log u= o(n)$, as required by condition \eqref{eq:lm_noncentralF_ucond1} above.

Summing up, for any $u$ satisfying \eqref{eq:lm_noncentralF_ucond2}, combining \eqref{eq:lm_overfitted_term1} and \eqref{eq:lm_overfitted_term2} 
we have shown that
\begin{align}
    \mathbb{P}(B^{LR}(z,z^0) > u) \leq
    (ew_b)^{(|z'|_0 - |z^0|_0)} 
e^{-\frac{cT(z,z^0)}{2 (\log T(z,z^0))^2}}
+ e^{-c T(z,z^0)/2} (1 + e^{-c_2 n/2})
\leq
(ew_b)^{(|z'|_0 - |z^0|_0)} 
e^{-\frac{c_3T(z,z^0)}{2 (\log T(z,z^0))^2}}
.
\end{align}
for any constant $c_3<c<1$ and sufficiently large $n$.
Recall that, upon eigenvalue conditions the non-centrality parameter $\lambda \geq d n$ for some constant $d$, hence in this final bound the dominant term is the first one.

\end{proof}

\subsubsection{Technical lemmas for the proofs of Section  \ref{sec:gimh}}\label{app:technical-lemmas-2}

We state several auxiliary results required by our proofs.
Lemma \ref{lem:main-event-s} states that, under the GGM assumption that each row in $\Y$ is generated from a $N(0, \Sigma^0)$, the sample variances $S_{jj}/n= \sum_{i=1}^n Y_{ij}^2/n$ are close to $\Sigma_{jj}^0$.
Lemma \ref{lem:tail-bound-chi-squared} gives standard tail bounds for chi-squared variables (see \cite{rossell2022concentration} or \cite{ghosh2021exponential}).
Lemma \ref{lem:s4-rossell} gives right tail bounds for F variables proven by \cite{rossell2022concentration} (Lemma S4).
For the proofs of Lemma \ref{lem:tail-bound-chi-squared} and \ref{lem:s4-rossell}, we refer the reader to the references above.

To state Lemma \ref{lem:main-event-s}, define for any $j \in [p]$,
\begin{align*}
     \mathcal{E}_1^j &=  \left \{ \left|\frac{S_{jj}}{n \Sigma_{jj}^0} - 1 \right| \leq \frac{1}{M_0} \right \},
\end{align*}
with $M_0 > 2$. Note that on $\mathcal{E}_1^j$ the empirical marginal variance of variable $j$, 
$S_{jj}/n$, is close to its expected value $\Sigma_{jj}^0$.

\begin{lemma}\label{lem:main-event-s}
   Under the GGM model, for any $M_0 > 2$ and $j=1,\dots, p$,
    \begin{align*}
        &\mathbb P \left( \mathcal{E}_1^j  \right) \geq 1 - 2 e^{-\frac{n}{8 M_0^2}}.
    \end{align*}
    Moreover if $\log p < \frac{n}{16 M_0^2}$, then
    \begin{align*}
       \mathbb P \left( \bigcap_{j=1}^p \mathcal{E}_1^j \right) = \mathbb P \left( \bigcap_{j=1}^p \left \{ \left|\frac{S_{jj}}{n \Sigma_{jj}^0} - 1 \right| \leq \frac{1}{M_0} \right \}\right) \geq 1 - 2 e^{-\frac{n}{16 M_0^2}}.
    \end{align*}
\end{lemma}

\begin{proof}

For simplicity, we consider $j=p$ and denote $\mathcal{E}_1 = \mathcal{E}_1^p$. Since $s_{pp} = \Y_{\cdot p}^T  \Y_{\cdot p}$ with $Y_{p} \sim \text{N}(0, \Sigma_{pp}^0)$ then $\frac{S_{pp}}{\Sigma_{pp}^0} \sim \chi^2_n$. We can then apply bounds for the $\chi^2$-squared distribution stated in Lemma \ref{lem:tail-bound-chi-squared}(i) below with  $t = M_0^{-1}$ and $M_0 > 2$ to obtain
\begin{align*}
\mathbb P \left[ \mathcal{E}_1^c \right]=
 \mathbb P \left[\left|\frac{S_{pp}}{\Sigma_{pp}^0} - n\right|\geq \frac{n}{M_0} \right] \leq 2 e^{-\frac{n}{8 M_0^2}}.
\end{align*}
The second statement is simply obtained using a union bound argument for calculating the probability of the event  $\bigcap_{j=1}^p \mathcal{E}_1^j$. Since $2 p e^{-\frac{n}{8 M_0^2}} \leq 2 e^{-\frac{n}{16 M_0^2}}$ as $\log p \leq \frac{n}{16 M_0^2} $ by assumption, we obtain the desired result.

\end{proof}

\begin{lemma}[Left and right tail bounds for chi-squared variables]\label{lem:tail-bound-chi-squared}
Let $ X \sim \chi^2_k (\lambda) $ be a chi-squared variable with $k \geq 1$ degrees of freedom and $\lambda$ as the non-centrality parameter parameter.
\begin{itemize}
    \item[(i)] If $\lambda = 0$, i.e.,  $ X $ is a central chi-squared, then for any $t>0$,
\begin{align*}
 &\mathbb P[|X-k|\geq k t ] \leq 2 e^{-kt^2/8} \\
 &\mathbb P[X - k > k t] \leq  e^{- \frac{k}{4} t }.
\end{align*}
    \item[(ii)] If $\lambda \neq 0$, for any $w < \lambda $,
    \begin{align*}
        \mathbb P[ X < w ] \leq (\frac{w}{\lambda})^{k/4} e^{- \frac{1}{2} (\sqrt{\lambda} - \sqrt{k})^2}
    \end{align*}
        \item[(ii)] If $\lambda \neq 0$, for any $w > \lambda + k$,
    \begin{align*}
        \mathbb P[ X > w ] \leq (\frac{ew}{\lambda})^{k/2} e^{- \frac{\lambda}{2} - \frac{w}{2}(1 - \frac{\lambda}{k})}.
    \end{align*}
\end{itemize}
 \end{lemma}

\begin{lemma}\label{lem:s4-rossell}
    Let $W = \frac{U_1}{\nu_1} \frac{\nu_2}{U_2}$ where $U_1 \sim \chi^2_{\nu_1}(\lambda)$ and $U_2 \sim \chi^2_{\nu_2}$ are  chi-squared variables with non-zero and zero centrality parameter respectively,  
     $\lambda > 0, \nu_1 \geq 1, \nu_2 \geq 1$. Let $w > \lambda + \nu_1$. Then for any $s \in ((\lambda + \nu_1)/w, 1)$ and $t \in (0,1/2)$,
    \begin{align*}
        \mathbb P (\nu_1 W > w ) \leq \frac{e^{\frac{\lambda t}{1 - 2t} -tws}}{(1 - 2t)^{\nu_1/2}} + (es)^{\nu_2/2} e^{-s\nu_2/2}.
    \end{align*}
\end{lemma}

\subsection{Other lemmas}\label{app:lemmas}

In this section we collect a variety of useful auxiliary results.
First, Lemma \ref{lem:decomposition} recalls a well-known formula expressing a block, column, and element 
of $\Omega$ as a function of elements of its inverse $\Sigma := \Omega^{-1}$, and conversely. 
From Lemma \ref{lem:decomposition}, one can compute $(\Sigma_{-p,p}, \Sigma_{pp})$  (resp., $(\Omega_{-p,p}, \Omega_{pp})$) from $(\Omega_{-p,p},\Omega_{pp},\Sigma_{-p,-p})$ (resp., $(\Sigma_{-p,p},\Sigma_{pp},\Omega_{-p,-p})$) without matrix inversion. Moreover, one can obtain the inverse of $\Omega$ (resp., $\Sigma$) from $([\Omega_{-p,-p}]^{-1}, \Omega_{-p,p}, \Omega_{pp})$ (resp., $(\Sigma_{-p,-p}^{-1}, \Sigma_{-p,p}, \Sigma_{pp})$) using one-rank update of inverses, another well-known result that is summarised in Lemma \ref{lem:update_inv}.


Finally, Lemmas \ref{lem:posterior_linreg_notation}-\ref{lem:equal_spectral_gap} are slightly more involved results that we prove explicitly, for completeness.
Specifically, in Lemma \ref{lem:posterior_linreg_notation} we derive the posterior distribution of a linear regression model under a discrete spike-and-slab prior.
Lemma \ref{lem:equal_spectral_gap} characterises the spectral gap of Markov chains where the parameters are split into two blocks, for the first block one uses an arbitrary proposal whereas for the second the proposal is the exact target distribution given the proposed value of the first block. 
This is useful in our setting because we propose a model $Z_{\cdot j}$ using some suitable proposal (e.g., Gibbs, BDMH, LIT or GIMH) and then we sample from $\pi(\Omega_{\cdot j} \mid Z_{\cdot j}, \Y, \Omega_{-j,-j})$ exactly.
Lemma \ref{lem:equal_spectral_gap} states that the spectral gap of the combined $(Z_{\cdot j}, \Omega_{\cdot j})$ is equal to the spectral gap associated to only $Z_{\cdot j}$.

\begin{lemma}\label{lem:decomposition}
If we decompose the covariance and precision matrices into a block with the first  $p-1$ columns and rows, the last column without the last coefficient, and the coefficient in the last row and column, i.e.,
\begin{align*}
    \Omega = \begin{pmatrix}
    \Omega_{-p,-p} & \Omega_{-p,p} \\
    \Omega_{-p,p}^T & \Omega_{pp}
     \end{pmatrix} \quad   \Sigma = \begin{pmatrix}
    \Sigma_{-p,-p} & \Sigma_{-p,p} \\
    \Sigma_{-p,p}^T & \Sigma_{pp}
     \end{pmatrix}
\end{align*}
then we have that
\begin{align}
    &\Omega_{-p,p} = - \Omega_{pp}\Sigma_{-p,-p}^{-1} \Sigma_{-p,p} \nonumber\\
    &\Omega_{-p,-p} = \left(  \Sigma_{-p,-p} - \frac{ \Sigma_{-p,p}\Sigma_{-p,p}^T }{ \Sigma_{pp}}\right)^{-1} = 
     \Sigma_{-p,-p}^{-1} + \frac{ \Sigma_{-p,-p}^{-1}\Sigma_{-p,p}\Sigma_{-p,p}^T\Sigma_{-p,-p}^{-1}}{ \Omega_{pp}^{-1}} = \Sigma_{-p,-p}^{-1} + \frac{\Omega_{-p,p} \Omega_{-p,p}^T}{ \Omega_{pp} } \\
     &[\Omega_{-p,-p}]^{-1}  = 
     \Sigma_{-p,-p} - \frac{ \Sigma_{-p,-p}\Omega_{-p,p}\Omega_{-p,p}^T\Sigma_{-p,-p}}{ \Omega_{pp} + \Omega_{-p,p}^T \Sigma_{-p,-p} \Omega_{-p,p}}  \\
    &\Sigma_{-p,p}  = - \frac{1}{ \Omega_{pp}} \Sigma_{-p,-p} \Omega_{-p,p} \\
    &\Sigma_{pp}  =  \frac{1}{ \Omega_{pp} - \Omega_{-p,p}^T [\Omega_{-p,-p}]^{-1} \Omega_{-p,p} } = \frac{1}{\Omega_{pp}} + \Sigma_{-p,p}^T \Sigma_{-p,-p}^{-1} \Sigma_{-p,p}=
    \frac{1}{\Omega_{pp}} - \frac{1}{\Omega_{pp}} \Omega_{-p,p}^T \Sigma_{-p,p} 
    \\
    &     \Omega_{-p,p}^T [\Omega_{-p,-p}]^{-1} \Omega_{-p,p} =  \Omega_{-p,p}^T \Sigma_{-p,-p} \Omega_{-p,p} + (\Sigma_{-p,p}^T \Omega_{-p,p})^2 = \Omega_{-p,p}^T \Sigma_{-p,-p} \Omega_{-p,p} + \frac{(\Omega_{-p,p}^T \Sigma_{-p,-p}^{-1} \Omega_{-p,p})^2}{\Omega_{pp} +\Omega_{-p,p}^T \Sigma_{-p,-p}^{-1} \Omega_{-p,p}}.
\end{align}

\end{lemma}

\begin{lemma}[One-rank update of inverse]\label{lem:update_inv}
Let $\Omega$ and $\Sigma = \Omega^{-1}$ decomposed as in Lemma \ref{lem:decomposition}. Then we have
\begin{itemize}
    \item[(i)] $[\Omega_{-p,-p}]^{-1} =  \Sigma_{-p,-p} - \frac{\Sigma_{-p,p} \Sigma_{-p,p}^T}{ \Sigma_{pp} }$
    \item[(ii)] If one replaces $(\Omega_{-p,p},\Omega_{pp})$ by $(\Omega_{-p,p}^*,\Omega_{pp}^*)$ given $\Omega_{-p,-p}$ (thus $[\Omega_{-p,-p}]^{-1}$) and define $\Omega^* = (\Omega_{-p,-p}, \Omega_{-p,p}^*,\Omega_{pp}^*)$, then one can obtain $\Sigma^* = [\Omega^*]^{-1}$ using
    \begin{align*}
        &v =  [\Omega_{-p,-p}]^{-1} \Omega_{-p,p}^*, \quad \gamma = \Omega_{pp}^* - v^T \Omega_{-p,p}^* \\
        &\Sigma_{-p,-p}^* = [\Omega_{-p,-p}]^{-1}  + \frac{v v^T}{\gamma} \\
        &\Sigma_{-p,p}^* = - \frac{ v}{\gamma}, \quad \Sigma_{pp}^* = \frac{ 1}{\gamma}, \\
        &\Sigma^* = (\Sigma_{-p,-p}, \Sigma_{-p,p}^*,\Sigma_{pp}^*).
    \end{align*}
\end{itemize}
\end{lemma}


\begin{lemma} \label{lem:posterior_linreg_notation}
Let $y \in \R^n$ and $X \in \R^{n \times d}, d \geq 1$ such that the $y_i$'s are i.i.d. and 
\begin{align*}
    y_i \mid \beta, \sigma^2, X_{i} \sim  N(y_i;\beta^T X_i , \sigma^2), \qquad i=1, \dots, n,
\end{align*}
where $\beta \in \R^{d}$ and $\sigma^2>0$.
Let $\gamma = (\gamma_k)_{k \in [p-1]}$ with  $\gamma_k= \mathds{1}_{\beta_k \neq 0}$ be the variable inclusion vector, and consider the prior distribution
\begin{align*}
    \pi(\beta, \sigma^{-2}, \gamma ) =  \pi(\beta \mid \sigma^{-2}, \gamma ) \pi(\sigma^{-2}) \pi(\gamma),
\end{align*}
with
\begin{align}
&\pi(\sigma^{-2})  = \mbox{Exp}(\sigma^{-2}; \frac{\lambda}{2}) \\ 
&\pi(\gamma) = \theta^{|\gamma|_0} (1 - \theta)^{d-|\gamma|_0}
\nonumber \\
&\pi(\beta \mid \gamma, \sigma^2) = \text{N}(\beta_\gamma; 0, \sigma^2 \tau^{-1} I_\gamma) \prod_{k, \gamma_k=0} \delta_0(\beta_k) \nonumber
\end{align}
where $\tau>0$ is a prior dispersion parameter. Let $s_n := y^T y$ be the empirical  variance, $X_\gamma$ be the $n \times |\gamma|_0$ matrix selecting the columns in $X$ with $\gamma_k = 1$, $W_\gamma := X_\gamma^T X_\gamma + \tau I_\gamma$, and $\mu_\gamma := W_\gamma^{-1} X_\gamma^T y$. Then the posterior distribution on $(\beta_\gamma, \sigma^2, \gamma )$ can be written as
\begin{align}
&\pi(\beta_\gamma, \sigma^2, \gamma \mid y, X) \\
&\quad \propto
 \text{N}(\beta_\gamma; \mu_\gamma, W_\gamma^{-1}) \text{Ga} \left(\sigma^2; \: \frac{n}{2} +1, \frac{\lambda + s_n - \mu_\gamma^T W_\gamma \mu_\gamma}{2} \right)
\frac{\theta^{|\gamma|_0} (1-\theta)^{d - |\gamma|_0}}{\tau^{-\frac{|\gamma|_0}{2}} |W_\gamma|^{\frac{1}{2}}} \left(\frac{1}{ \lambda + s_n - \mu_\gamma^T W_\gamma \mu_\gamma }\right)^{\frac{n}{2}+1},
\nonumber
\end{align}
which implies that
\begin{align*}
    \pi(\beta_\gamma, \sigma^2, \gamma \mid y, X) = \pi( \beta_\gamma \mid y, X, \gamma, \sigma^2)  \pi(\sigma^2 \mid \gamma, y, X) \pi(\gamma \mid y,X),
\end{align*}
with 
\begin{align}
    &\pi( \beta_\gamma \mid y, X, \gamma, \sigma^2) = N(\beta_\gamma; \mu_\gamma, W_\gamma^{-1}) \\
    &\pi(\sigma^2 \mid \gamma, y, X) = \text{Ga}\left(\sigma^2; \frac{n}{2} +1, \frac{\lambda + s_n - \mu_\gamma^T W_\gamma \mu_\gamma}{2}\right) \\
    &\pi(\gamma \mid y,X) \propto \frac{\theta^{|\gamma|_0} (1-\theta)^{d - |\gamma|_0}}{\tau^{-\frac{|\gamma|_0}{2}} |W_\gamma|^{\frac{1}{2}}} \left(\frac{1}{ \lambda + s_n - \mu_\gamma^T W_\gamma \mu_\gamma }\right)^{\frac{n}{2}+1}.
\end{align}
\end{lemma}

\begin{proof}
Let $\phi = \sigma^{-2}$. First, note that the marginal prior density of $\beta$ given the $\phi$ is
\begin{align}
p(\beta \mid \phi)= p(\gamma) p(\beta \mid \gamma, \phi)=
\theta^{|\beta|_0} (1-\theta)^{d - |\beta|_0} N(\beta_\gamma; 0, \frac{\phi^{-1}}{\tau}  I_{\gamma})
\nonumber
\end{align}
and that $|\beta|_0=|\gamma|_0$.

Therefore the posterior density is proportional to 
\begin{align}
&p(\beta, \phi \mid y, X) \\
&\propto
N(y; X_\gamma \beta_\gamma, \phi^{-1} I) \pi(\beta \mid \phi) \mbox{Exp}(\phi; \lambda/2)
\nonumber \\
&\propto \phi^{\frac{n+|\beta|_0}{2}} e^{-\frac{\phi \lambda}{2}} \exp \left\{ -\frac{\phi}{2} [(y-X_\gamma \beta_\gamma)^T (y - X_\gamma \beta_\gamma) + \tau \beta_\gamma^T \beta_\gamma ]  \right\}
\frac{\theta^{|\beta|_0} (1-\theta)^{d - |\gamma|_0}}{(2\pi \tau^{-1})^{\frac{|\beta|_0}{2}}}
\mbox{I}(\phi>0)
\nonumber \\
&=\exp \left\{ -\frac{\phi}{2} [\beta_\gamma^T ( X_\gamma^T X_\gamma + \tau I) \beta_\gamma - 2 y^T X_\gamma \beta_\gamma ]  \right\}
\frac{\theta^{|\beta|_0} (1-\theta)^{d - |\beta|_0}}{(2\pi \tau^{-1})^{\frac{|\beta|_0}{2}}}
\phi^{\frac{n+|\beta|_0}{2}} e^{-\frac{\phi (\lambda + s_n)}{2}} \mbox{I}(\phi>0).
\nonumber
\end{align}
Letting $W_\gamma= X_\gamma^T X_\gamma + \tau I$ and $\mu_\gamma= W_\gamma^{-1} X_\gamma^T y$, we obtain
\begin{align*}
&p(\beta, \phi \mid y, X) \\
&\propto \exp \left\{ -\frac{\phi}{2} [\beta_\gamma^T W_\gamma \beta_\gamma - 2 \mu_\gamma^T W_\gamma \beta_\gamma + \mu_\gamma^T W_\gamma \mu_\gamma - \mu_\gamma^T W_\gamma \mu_\gamma ]  \right\}
\frac{\theta^{|\gamma|_0} (1-\theta)^{d - |\gamma|_0}}{{(2\pi \tau^{-1})^{\frac{|\gamma|_0}{2}}}}
\phi^{\frac{n+|\gamma|_0}{2}} e^{-\frac{\phi (\lambda + s_n)}{2}} 
\nonumber \\
&= \frac{\phi^{\frac{|\gamma|_0}{2}} |W_\gamma|^{\frac{1}{2}}}{(2\pi)^{\frac{|\gamma|_0}{2}}}
\exp \left\{ -\frac{\phi}{2} (\beta_\gamma - \mu_\gamma)^T W_\gamma (\beta_\gamma - \mu_\gamma)   \right\}
\frac{e^{ \frac{\phi \mu_\gamma^T W_\gamma \mu_\gamma }{2} }}{{\tau^{-\frac{|\gamma|_0}{2}}} |W_\gamma|^{\frac{1}{2}}}
\theta^{|\gamma|_0} (1-\theta)^{d - |\gamma|_0}
\phi^{\frac{n}{2}} e^{-\frac{\phi (\lambda + s_n)}{2}} \\
&= \underbrace{\left(\frac{\lambda  +s_n - \mu_\gamma^T W_\gamma \mu_\gamma}{2}\right)^{\frac{n}{2}+1}\phi^{\frac{n}{2} } e^{ \frac{\phi ( \lambda + s_n - \mu_\gamma^T W_\gamma \mu_\gamma) }{2} }}_{\propto \pi(\phi\mid \gamma, y,X)} \times 
\underbrace{ \frac{\phi^{\frac{|\gamma|_0}{2}} |W_\gamma|^{\frac{1}{2}}}{(2\pi)^{\frac{|\gamma|_0}{2}}} \exp \left\{ -\frac{\phi}{2} (\beta_\gamma - \mu_\gamma)^T W_\gamma (\beta_\gamma - \mu_\gamma)   \right\}}_{\pi(\beta\mid \phi, \gamma, y, X)} \\
&\times
\underbrace{\frac{1}{{(2 \pi \tau^{-1} )^{\frac{|\gamma|_0}{2}}}|W_\gamma|^{\frac{1}{2}} }
\theta^{|\gamma|_0} (1-\theta)^{d - |\gamma|_0} \left(\frac{1}{\lambda + s_n - \mu_\gamma^T W_\gamma \mu_\gamma} \right)^{\frac{n}{2}+1}}_{\propto \pi(\gamma \mid y, X)}
\end{align*}
Therefore, the posterior writes as
\begin{align*}
  \pi(\beta_\gamma, \phi, \gamma \mid y) \propto
& N(\beta_\gamma; \mu_\gamma, W_\gamma^{-1}) \text{Ga}\left(\sigma^2; \frac{n}{2} + 1, \frac{\lambda + s_n - \mu_\gamma^T W_\gamma \mu_\gamma}{2}\right) \\
\times 
    & \frac{\theta^{|\gamma|_0} (1-\theta)^{d - |\gamma|_0}}{\tau^{-\frac{|\gamma|_0}{2}} |W_\gamma|^{\frac{1}{2}}} \left(\frac{1}{ \lambda + s_n - \mu_\gamma^T W_\gamma \mu_\gamma }\right)^{\frac{n}{2}+1}.
\end{align*}

Note that the joint posterior $\pi(\gamma, \phi \mid y, X)$ writes as
\begin{align*}
\pi(\gamma, \phi \mid y, X)
\propto 
\frac{e^{ \frac{\phi \mu_\gamma^T W_\gamma \mu_\gamma }{2} }}{\tau^{-\frac{|\gamma|_0}{2}} |W_\gamma|^{\frac{1}{2}}}
\theta^{|\gamma|_0} (1-\theta)^{d - |\gamma|_0}
\phi^{\frac{n}{2}} e^{-\frac{\phi (\lambda + s_n)}{2}} \mbox{I}(\phi>0).
\end{align*}
and
\begin{align*}
\pi(\gamma \mid   \phi, y, X)
\propto 
\frac{e^{ \frac{\phi \mu_\gamma^T W_\gamma \mu_\gamma }{2} }}{\tau^{-\frac{|\gamma|_0}{2}} |W_\gamma|^{\frac{1}{2}}}
\theta^{|\gamma|_0} (1-\theta)^{d - |\gamma|_0}
\end{align*}
\end{proof}

In comparison, in the GGM, we have
\begin{align*}
    \pi(z \mid \Y, \Omega_{-p,-p}) 
\propto
\frac{e^{\frac{m_z^T U_z m_z}{2}}}{g_1^{|z|}|U_z|^{\frac{1}{2}}}
\theta^{|z|_0} (1 - \theta)^{p-1-|z|_0},
\end{align*}
hence we have $\pi(\gamma, \phi \mid y, X) =  \pi(z \mid \Y, \Omega_{-p,-p})$ if $\phi=1$, $\tau = g_1^{-2}$ and $U_z = W_\gamma$.


    


\begin{lemma}
    Let $P$ a reversible and positive semi-definite Markov transition kernel on $\mathcal{X} = \mathcal{X}_1 \times \mathcal{X}_2$ defined as 
    \begin{align*}
        P((x_1, x_2), (dy_1, dy_2)) = Q(x_1, d y_1)\pi(dy_2 | y_1)
    \end{align*}
    Then $Gap(P) = Gap(Q)$.
\label{lem:equal_spectral_gap}
\end{lemma}

\begin{proof}
Recall that 
\begin{align*}
    &Gap(P) = 
    \inf_{f \in L_2(\pi)} \frac{E(f,f)}{Var_\pi(f)}=
\inf_{f \in L_2(\pi)} \frac{<f,f> + <f,f> - <f,Pf>}{<f,f> - <f,1>^2}
\\
= &   1 - \sup_{f \in L_2(\pi)} \frac{<f, Pf> - <f, 1>^2}{<f,f>},
\end{align*}
where we denote
\begin{align*}  
&<f_1,P f_2> = \int f_1(x) f_2(y) P(dy \mid x) \pi(dx)
\\
&<f_1,f_2> = \int f_1(x) f_2(x) \pi(dx).
\end{align*}

Hence we will show that 
\begin{align*}
    Gap(P)= \sup_{f \in L_2(\pi)} \frac{<f, Pf> - <f,1>^2}{<f,f>} = \sup_{g \in L_2(\pi)} \frac{<g, Qg> - <g,1>^2}{<g,g>}= Gap(Q).
\end{align*}
To prove this, we first show that $Gap(P) \leq Gap(Q)$ and subsequently that $Gap(Q) \leq Gap(P)$.
Let $f \in L_2(\pi)$. Then
\begin{align*}
    <f,Pf> &= \int  f(x_1, x_2) f(y_1, y_2)  P( (x_1, x_2), (dy_1, dy_2)) \pi(dx_1, dx_2) \\
    &= \int \left[ \int f(x_1, x_2) \pi(dx_2|x_1)  \right] \left [\int f(y_1, y_2) \pi(dy_2 | y_1) \right] Q(x_1, d y_1) \pi(dx_1) \\
    &= \int g_f(x_1) g_f(y_1) Q(x_1, d y_1) \pi(dx_1) = <g_f, Qg_f>
\end{align*}
with $g_f(x_1) = \int f(x_1, x_2) \pi(dx_2|x_1)$. Moreover, by Cauchy-Schwarz,
\begin{align*}
    <g_f, g_f> &= \int g_f(x_1)^2 \pi(dx_1) = \int \left( \int f(x_1, x_2) \pi(dx_2|x_1) \right)^2 \pi(dx_1) \\
    &\leq \int  f^2(x_1, x_2) \pi(dx_2|x_1) \pi(dx_1) = <f,f>.
\end{align*}
Also note that 
\begin{align*}
<g_f,1>= \int g_f(x_1) \pi(d x_1)=
\int f(x_1,x_2) \pi(d x_2 \mid x_1) \pi(d x_1)
= \int f(x) \pi(d x)= <f,1>.
\end{align*}

This implies that
\begin{align*}
     \sup_{f \in L_2(\pi)} \frac{<f, Pf> - <f,1>^2}{<f,f>}  \leq  
     \sup_{f \in L_2(\pi)} \frac{<g_f, Qg_f> - <g_f,1>^2}{<g_f,g_f>} \leq
     \sup_{g \in L_2(\pi)} \frac{<g, Qg> - <g,1>^2}{<g,g>},
\end{align*}
and hence that $Gap(P) \leq Gap(Q)$.

Moreover for any $g \in L_2(\pi)$, let $f_g(x_1, x_2) = g(x_1)$. Then, using that $P((x_1,x_2),(dy_1,dy_2))= Q(x_1,dy_1) \pi(d y_2 \mid y_1)$, we obtain
\begin{align*}
      <f_g,P f_g> &= \int  g(x_1) g(y_1) Q(x_1, d y_1)\pi(dy_2 \mid x_1) \pi(dx_1) = <g, Qg>
      \\
      <f_g, 1> &= \int g(x_1) \pi(d x_1)= <g,1>.
\end{align*}
This proves that
\begin{align*}
      \sup_{g \in L_2(\pi)} \frac{<g, Qg> - <g,1>^2}{<g,g>} =  \sup_{g \in L_2(\pi)} \frac{<f_g, Pf_g> - <f_g,1>^2}{<f_g,f_g>} \leq \sup_{f \in L_2(\pi)} \frac{<f, Pf> - <f,1>^2}{<f,f>}, 
\end{align*}
and hence that $Gap(Q) \leq Gap(P)$.
\end{proof}

\subsection{Proof of Theorem \ref{thm:rsgibbs_geomergodic}}
\label{ssec:proof_thm_rsgibbs_geomergodic}

Before outlining the proof, we reproduce two similar theorems from \cite{meyn:1994} and \cite{rosenthal:1995}, both of which guarantee geometric ergodicity using drift and minorization conditions.
Theorem \ref{thm:drift_min_meyn} \citep{meyn:1994} is slightly more general in allowing one to define an arbitrary set $C$, and suffices for our proof. 
However, we also state Theorem \ref{thm:drift_min_rosenthal} \citep{rosenthal:1995} because it gives a simpler form for the constants featuring in the geometric rate bound.
The proof of Theorem \ref{thm:rsgibbs_geomergodic} also requires some auxiliary results that are stated and proven in Section \ref{ssec:aux_rsgibbs_geomergodic}.

\subsubsection{Drift-minorization theorems}

\begin{thm} \cite{meyn:1994}
Suppose a Markov chain $P(x,dy)$ on a state space $\chi$ satisfies the drift condition
$$
\mathbb E \left[ V(X^{(1)}) \mid X^{(0)} =x \right] \leq \lambda V(x) + b 1_{x \in C}
$$
for some $V: \chi \to \mathbb{R}^+$, and some $\lambda < 1$ and $b < \infty$; and further satisfies a minorization condition
$$
P(x, \cdot) \geq \epsilon Q(\cdot)   \mbox{, for all } x \in C,
$$
for some $\epsilon > 0$, and some probability measure $Q()$ on $\chi$.
Then 
$$
\| P^t( \cdot \mid X^{(0)} = x) - \pi  \|_{TV} \leq M(x) (1 - \rho)^t
$$
for every $x \in \chi$ and $t \geq 1$, some $M(x) \in \mathbb{R}^+$ and some $\rho < 1$.
\label{thm:drift_min_meyn}
\end{thm}

\begin{thm} (Theorem 12, \cite{rosenthal:1995})
Suppose a Markov chain $P(x,dy)$ on a state space $\chi$ satisfies the drift condition
$$
\mathbb E \left[ V(X^{(1)}) \mid X^{(0)} =x \right] \leq \lambda V(x) + b
$$
for some $V: \chi \to \mathbb{R}^+$, and some $\lambda < 1$ and $b < \infty$; and further satisfies a minorization condition
$$
P(x, \cdot) \geq \epsilon Q(\cdot)   \mbox{, for all } x \in \chi \mbox{ with } V(x) \leq d,
$$
for some $\epsilon > 0$, some probability measure $Q(\cdot)$ on $\chi$, and some $d > \frac{2b}{1 - \lambda}$.
Then 
$$
\| P^t( \cdot \mid X^{(0)}) - \pi  \|_{TV} \leq M(x) (1 - \rho)^t
$$
for every $x \in \chi$ and $t \geq 1$, some $M(x) \in \mathbb{R}^+$ and some $\rho < 1$.
Specifically, the result holds for
$$
(1-\rho)= \max\left\{ (1-\epsilon)^r, \left( \frac{1 + 2b + \lambda d}{1 + d} \right)^{1-r} A^r   \right\}
$$
and any $r \in (0,1)$ and $M(x) = 2 + \frac{b}{1-\lambda} + \mathbb E_\nu V(X^{(0)})$, where
$A= 1 + 2 (\lambda d + b)$ and  $\nu$ is the initial distribution of $X^{(0)}$.

\label{thm:drift_min_rosenthal}
\end{thm}

\subsubsection{Main proof}

We proceed to prove Theorem \ref{thm:rsgibbs_geomergodic}.
The proof relies on verifying the conditions of Theorem \ref{thm:drift_min_meyn}.
It also relies on auxiliary Lemmas \ref{lem:finite_logDj}-\ref{lem:compact_levelset_driftfunction}, stated and proven below after the proof of Theorem \ref{thm:rsgibbs_geomergodic}.

The first part of the proof is to define a suitable drift function $V(\Omega)$ and show that 
\begin{align}
 \mathbb{E}_{P}[V(\Omega^{(t+1)}) \mid \Omega^{(t)}] \leq 
 \left( 1 - \frac{1}{p} \right) V(\Omega^{(t)}) + C + \frac{\log M + 1}{p}, \nonumber
\nonumber
\end{align}
for some universal constant $C$ and $M= \prod_{j=1}^p \Omega_{jj}$, which is a fixed constant for fixed $\mbox{diag}(\Omega)$, satisfying the drift condition in Theorem \ref{thm:drift_min_meyn} with $\lambda= 1 - 1/p < 1$ and $b= C + (\log M + 1)/p$.
A simple application of the law total expectation shows that $ \mathbb{E}_{P}[V(\Omega^{(t+m)}) \mid \Omega^{(t)}]$ also satisfies the drift condition for any $m \geq 1$, and in particular for $m=p$ (that is, a full sweep of the random scan Gibbs).
The second part of the proof is to check the minorization condition of Theorem \ref{thm:drift_min_meyn}, for $m=p$ updates.

As preliminaries, Hadamard's inequality gives that $|\Omega| \leq \prod_{j=1}^p \Omega_{jj}=M$.
Basic determinant properties also give that $|\Omega|= |\Omega_{-j,-j}| D_j(\Omega)$,
where $D_j(\Omega)= \Omega_{jj} - \Omega_{-j,j}^T [\Omega_{-j,-j}]^{-1} \Omega_{-j,j}$.
Further, using standard block-wise matrix inversion gives that $\Sigma_{jj}= 1/(\Omega_{jj} - \Omega_{-j,j}^T \Omega_{-j,-j}^{-1} \Omega_{-j,j})= 1/ D_j(\Omega)$.

We define the drift function
$$
V(\Omega)= - \log |\Omega| + \log M + 1 \geq 1,
$$
the inequality holding because, since $|\Omega| \leq M$, we have that $\log M - \log |\Omega| \geq 0$.

\vspace{0.5cm}
{\bf \underline{Part 1. Drift condition.}}
Let $P_j$ be the block Gibbs kernel updating a single column $j$. The corresponding expected drift is
\begin{align}
&\mathbb{E}_{P_j}[V(\Omega^{(t+1)}) \mid \Omega^{(t)}]=
\int (-\log |\Omega| + \log M + 1) \pi(\Omega_j \mid \Omega^{(t)}, \Y) d\Omega_j
\nonumber \\
&= -\log |\Omega_{-j,-j}^{(t)}| + \log M + 1 - \int \log D_j(\Omega) \pi(\Omega_j \mid \Omega^{(t)}, \Y) d\Omega_j
\nonumber \\
&= - \log |\Omega^{(t)}| + \log D_j(\Omega^{(t)}) + \log M + 1 - \int \log D_j(\Omega) \pi(\Omega_j \mid \Omega^{(t)}, \Y) d\Omega_j,
\nonumber
\end{align}
the second and third equalities above following from $|\Omega|= |\Omega_{-j,-j}| D_j(\Omega)$.
Lemma \ref{lem:finite_logDj} shows that the integral on the right-hand side is upper-bounded by a universal constant $C$, giving that
$ \mathbb{E}_{P_j}[V(\Omega^{(t+1)}) \mid \Omega^{(t)}] \leq$
\begin{align}
- \log |\Omega^{(t)}| + \log D_j(\Omega^{(t)}) + \log M + 1 + C
= V(\Omega^{(t)}) + \log D_j(\Omega^{(t)}) + C.
\nonumber
\end{align}

Hence, for the random scan Gibbs kernel $P$ updating a single column chosen uniformly at random, we have
\begin{align}
&\mathbb{E}_P[V(\Omega^{(t+1)}) \mid \Omega^{(t)}] = \frac{1}{p} \sum_{j=1}^p \mathbb{E}{P_j}[V(\Omega^{(t+1)}) \mid \Omega^{(t)}] 
\leq \frac{1}{p} \sum_{j=1}^p \left[ V(\Omega^{(t)}) + \log D_j(\Omega^{(t)}) + C \right] 
\nonumber \\
&= V(\Omega^{(t)}) + C + \frac{1}{p} \sum_{j=1}^p \log D_j(\Omega^{(t)}) 
 = V(\Omega^{(t)}) + C - \frac{1}{p} \sum_{j=1}^p \log (\Sigma_{jj}^{(t)})
\nonumber \\
& \leq V(\Omega^{(t)}) + C - \frac{1}{p} \log |\Sigma^{(t)}|
= V(\Omega^{(t)}) + C + \frac{1}{p} \log |\Omega^{(t)}|,
\nonumber
\end{align}
where we used that $D_j(\Omega^{(t)})=(\Sigma_{jj}^{(t)})^{-1}$, that from Hadamard's inequality we have $|\Sigma^{(t)}| \leq \prod_{j=1}^p \Sigma_{jj}^{(t)}$ 
and hence that $- \sum_{j=1}^p \log \Sigma_{jj}^{(t)} \leq -\log |\Sigma^{(t)}|$, and that $|\Sigma^{(t)}|= 1/ |\Omega^{(t)}|$.
Since $\log |\Omega^{(t)}|= -V(\Omega^{(t)}) + \log M + 1$ by definition of $V(\Omega^{(t)})$, we obtain that
\begin{align}
\mathbb{E}_{P}[V(\Omega^{(t+1)}) \mid \Omega^{(t)}] &\leq V(\Omega^{(t)}) + C + \frac{1}{p} \left( -V(\Omega^{(t)}) + \log M + 1 \right) \nonumber \\
&= \left( 1 - \frac{1}{p} \right) V(\Omega^{(t)}) + C + \frac{\log M + 1}{p}, \nonumber
\end{align}
completing the first part of the proof.

\vspace{0.5cm}
{\bf \underline{Part 2. Minorization condition.}}
Let
\begin{align}
C_d = \left\{ \Omega \in \PSD \mid \Omega_{ii} \text{ are fixed}, \, V(\Omega) \leq d \right\}.
\nonumber
\end{align}
The goal is to prove that, for any finite $d \geq 1$, there is a constant $\epsilon > 0$ and a probability measure $Q$ such that for all $\Omega \in C_d$ and any measurable set $A$:
\begin{align}
P^p(\Omega, A) \geq \epsilon Q(A)
\nonumber
\end{align}
where $P^p$ is the $p$-step transition kernel of the random scan Gibbs sampler.

To prove this, note that for any $\Omega \in C_d$ and $A$, we have that
\begin{align}
P^p(\Omega, A) &= \int_A k^{(p)}(\Omega, \Omega') \mu(d\Omega')
\geq \int_{A \cap C_d} k^{(p)}(\Omega, \Omega') \mu(d\Omega'),
\label{eq:minorization_cond_ineq}
\end{align}
where $\mu$ is the reference measure on $\Pi(\Omega \mid \Y)$ (that is, a mixture of Lebesgue measures on the sub-manifolds corresponding to different graphs $Z \in \{0, 1\}^{p(p-1)/2}$),
and $k^{(p)}(\Omega, \Omega')$ the $p$-step transition density 
associated to $P^{p}(\Omega, \Omega')$.
Given that after $p$ steps there is a strictly positive probability that every column of $\Omega$ has been updated once, we have that
$k^{(p)}(\Omega, \Omega')$ is strictly positive over $C_d \times C_d$. Given that $C_d \times C_d$ is a compact set (Lemma \ref{lem:compact_levelset_driftfunction}), we have that
\begin{align}
\inf_{\Omega, \Omega' \in C_d \times C_d} k^{(t)}(\Omega, \Omega') = \epsilon_0 > 0.
\nonumber
\end{align}
for some $\epsilon_0 > 0$. Therefore, \eqref{eq:minorization_cond_ineq} gives that
\begin{align}
P^p(\Omega, A) \geq \epsilon_0 \mu(A \cap C_d).
\nonumber
\end{align}

To complete the proof, define $Q(\cdot)$ to be the normalized reference measure $\mu$ restricted to $C_d$, that is
\begin{align}
Q(A) = \frac{\mu(A \cap C_d)}{\mu(C_d)}.
\nonumber
\end{align}
where note that $0 < \mu(C_d) < \infty$, since $C_d$ is compact.
This implies that
\begin{align}
P^{p}(\Omega, A) \geq \epsilon_0 \mu(C_d) Q(A)= \epsilon Q(A),
\nonumber
\end{align}
where $\epsilon = \epsilon_0 \mu(C_d) > 0$.

\subsubsection{Auxiliary lemmas for Theorem \ref{thm:rsgibbs_geomergodic}}
\label{ssec:aux_rsgibbs_geomergodic}

\begin{lemma} [Drift function boundedness.]
Let 
$\PSDD = \{ \Omega \in \PSD : \mbox{diag}(\Omega)=D \}$ for some fixed $D$.
Let $x = \Omega_{zj} \in \mathbb{R}^{|z|_0}$ for a given $z \in \{0, 1\}^{p-1}$ satisfying $|z|_0 \leq \bar{d}$. 
Under the full conditional density 
\begin{align}
\pi(x \mid z, \Y, \Omega_{-j,-j}) \propto \exp\left( -\frac{1}{2} (x - m_z)^T U_z (x - m_z) \right) \mathds{1}_{x^T \Sigma_{zz \mid j} x \leq \Omega_{jj}}
\nonumber
\end{align}
where $U_z = (n+\lambda) \Sigma_{zz \mid j} + g_1^{-2} I$, $\Sigma_{zz \mid j}= [\Omega_{-j,-j}^{-1}]_{zz}$
and $m_z = U_z^{-1} S_{zj}$, define the expected drift
\begin{align}
C = \max_{1 \leq j \leq p} \max_{z : |z|_0 \leq \bar{d}} \sup_{\Omega \in \PSDD} \mathbb{E} \left[ -\log(\Omega_{jj} - x^T \Sigma_{zz \mid j} x) \mid z, \Y, \Omega_{-j,-j} \right].
\nonumber
\end{align}
Then, $C < \infty$.
\label{lem:finite_logDj}
\end{lemma}

\begin{proof}
Let $D_j(x) = \Omega_{jj} - x^T \Sigma_{zz \mid j} x$. Apply the change of variables $u = \Omega_{jj}^{-1/2} \Sigma_{zz \mid j}^{1/2} x$, 
so that $\{x : x^T \Sigma_{zz \mid j} x \leq \Omega_{jj}\} \Leftrightarrow \{u \in \mathbb{R}^{|z|_0} : \|u\| \leq 1\} := \mathcal{B}$
and $-\log D_j(x) = -\log \Omega_{jj} - \log(1 - \|u\|^2)$.
Plugging $x = \Omega_{jj}^{1/2} \Sigma_{zz \mid j}^{-1/2} u$ into $\pi(x \mid z, \Y, \Omega_{-j,-j})$,
the density of $u$ is
\begin{align}
f(u) \propto \frac{|\tilde{Q}_z|^{\frac{1}{2}}}{(2\pi)^{\frac{|z|_0}{2}}} \exp\left( -\frac{1}{2} u^T \tilde{Q}_z u + \tilde{c}_z^T u \right) \mathds{1}_{u \in \mathcal{B}}
\nonumber
\end{align}
where 
\begin{align}
\tilde{Q}_z &= \Omega_{jj} \Sigma_{zz \mid j}^{-1/2} U_z \Sigma_{zz \mid j}^{-1/2} = \Omega_{jj} (n+\lambda) I + \Omega_{jj} g_1^{-2} \Sigma_{zz \mid j}^{-1} 
\nonumber \\
\tilde{c}_z &= \Omega_{jj}^{1/2} \Sigma_{zz \mid j}^{-1/2} U_z m_z = \Omega_{jj}^{1/2} \Sigma_{zz \mid j}^{-1/2} S_{zj}.
\nonumber
\end{align}
where $\Sigma_{zz \mid j} = [\Omega_{-j,-j}^{-1}]_{zz}$.
To bound the operator norm $\| \tilde{Q}_z \|$, we examine $\Sigma_{zz \mid j}^{-1} = \Omega_{zz} - \Omega_{z, -z} \Omega_{-z, -z}^{-1} \Omega_{-z, z}$. Since $\Omega$ is positive-definite, $\Sigma_{zz \mid j}^{-1} \preceq \Omega_{zz}$.
Since $\| \Omega_{zz} \| \leq \mbox{tr}(\Omega)$, we have that $\|\Sigma_{zz \mid j}^{-1}\| \leq \mbox{tr}(\Omega)$ and therefore that
\begin{align}
&\|\tilde{Q}_z\| \leq \Omega_{jj}(n+\lambda) + \Omega_{jj}g_1^{-2} \mbox{tr}(\Omega)
\nonumber \\
&\|\tilde{c}_z\|^2 = \Omega_{jj} S_{zj}^T \Sigma_{zz \mid j}^{-1} S_{zj} \leq \Omega_{jj} \|S_{zj}\|^2 \mbox{tr}(\Omega).
\nonumber
\end{align}

Since $\mbox{diag}(\Omega)$, $S$ and $\bar d$ are fixed, both $\|\tilde{Q}_z\|$ and $\|\tilde{c}_z\|$ are uniformly bounded above over all $\Omega \in \PSDD$,
and $e^{-B} \leq f(u) \leq e^B$ for some $B$ and all $u \in \mathcal{B}$.
Therefore,
\begin{align}
\mathbb{E}[-\log(1 - \|u\|^2)] = \frac{\int_{\mathcal{B}} [-\log(1 - \|u\|^2)] f(u) du}{\int_{\mathcal{B}} f(u) du} 
\leq e^{2B} \frac{\int_{\mathcal{B}} -\log(1 - \|u\|^2) du}{\text{Vol}(\mathcal{B})},
\label{eq:bound_logint_unitsphere}
\end{align}
where $\mbox{Vol}(\mathcal{B})= \pi^{|z|_0/2}/\Gamma(|z|_0/2+1)$ is the volume of the $|z|_0$-dimensional unit ball $\mathcal{B}$.
Further, taking the change of variables $r= \| u \|$ and $(q_2,\ldots,q_{|z|_0})= (u_2,\ldots,u_{|z|_0}) / \|u\|$ gives that the Jacobian is $r^{|z|_0-1}$ and hence
\begin{align}
 \int_{\mathcal{B}} [-\log(1 - \|u\|^2)] f(u) du=
\int [- \log (1 - r^2) ] r^{|z|_0-1} dr dq=
S_{|z|_0} \int_0^1 [- \log(1 - r) ] r^{|z|_0-1}  dr
\nonumber
\end{align}
where $S_{|z|_0}= (2\pi)^{|z|_0/2}/\Gamma(|z|_0/2)$ is the surface of the $|z|_0$-dimensional unit sphere.
Taking another change of variables $t=r^2$ gives that
\begin{align}
  \int_{\mathcal{B}} [-\log(1 - \|u\|^2)] du=
\frac{(2\pi)^{|z|_0/2}}{\Gamma(|z|_0/2)} \int_0^1 t^{\frac{d}{2} -1} [-\log(1-t)] dt=
\frac{(2\pi)^{|z|_0/2}}{\Gamma(|z|_0/2)} \frac{2}{|z|_0} \left[ \Psi\left(\frac{|z|_0}{2} + 1\right) + \gamma \right],
\nonumber
\end{align}
where $\Psi$ is the digamma function and $\gamma \approx 0.577$ the Euler-Mascheroni constant.
Plugging this expression into \eqref{eq:bound_logint_unitsphere} and simplifying gives that
\begin{align}
\mathbb{E}[-\log(1 - \|u\|^2)] \leq 
e^{2B} 2^{|z|_0/2} \left[ \Psi\left(\frac{|z|_0}{2} + 1\right) + \gamma \right].
\nonumber
\end{align}
Denote the right-hand side above by $K_z$, a finite constant that only depends on $|z|_0$.
We have proven that
\begin{align}
\mathbb{E}[-\log D_j(x) \mid z, \Y, \Omega_{-j,-j}] \leq -\log \Omega_{jj} + K_z.
\nonumber
\end{align}
Taking the maximum over $1 \leq j \leq p$ and all $|z|_0 \leq \bar{d}$, we obtain $C \leq \max_{j, z} \left( -\log \Omega_{jj} + e^{2B} K_z \right) < \infty$.
\end{proof}

\begin{lemma}[Compactness of Sublevel Sets]
Let $V(\Omega) = -\log |\Omega| + \log M + 1$ where $M = \prod_{i=1}^p \Omega_{ii}$ is fixed. For any finite $d \geq 1$, the set 
\begin{align}
C_d = \left\{ \Omega \in \PSD \mid \Omega_{ii} \text{ are fixed}, \, V(\Omega) \leq d \right\}
\nonumber
\end{align}
is a compact subset of the interior of $\PSD$.
\label{lem:compact_levelset_driftfunction}
\end{lemma}

\begin{proof}
Because $\mbox{diag}(\Omega)$ is fixed, $\mbox{tr}(\Omega) = \sum_{j=1}^p \Omega_{jj} := T$ is constant for any $\Omega \in C_d$.. 
Since $\Omega$ is positive-definite, 
\begin{align}
\lambda_{\max}(\Omega) \leq T. \label{eq:max_eig_bound}
\end{align}
The condition $V(\Omega) \leq d$ implies $-\log |\Omega| + \log M + 1 \leq d$, or equivalently
$|\Omega| \geq M e^{1-d}$.
Combining this latter expression with \eqref{eq:max_eig_bound}, we obtain 
\begin{align}
\lambda_{\min}(\Omega) = \frac{|\Omega|}{\prod_{j \neq \min} \lambda_j} \geq \frac{M e^{1-d}}{T^{p-1}} > 0. \label{eq:min_eig_bound}
\end{align}
From \eqref{eq:max_eig_bound} and \eqref{eq:min_eig_bound}, the eigenvalues of every $\Omega \in C_d$ lie within a finite interval. Since $C_d$ is closed and its elements have bounded matrix norms, it is a compact set and strictly in the interior of $\PSD$.
\end{proof}

\color{black}
 
\section{Block Gibbs sampling for the continuous spike-and-slab prior}\label{app:wang}

We reproduce the block Gibbs sampler of \cite{Wang2015ScalingIU} for GGMs under the continuous spike-and-slab prior. 
We first recall a useful proposition (Proposition 1 in \cite{Wang2015ScalingIU}) that provides  a closed-form formula for the conditional posterior distribution on the transformed variables $(u_1 := \Omega_{-p,p}, u_2: = \Omega_{pp} - \Omega_{-p,p}^T [\Omega_{-p,-p}]^{-1} \Omega_{-p,p})$ given $(\Omega_{-p,-p}, Z, \Y)$.

\begin{prop}[\citep{Wang2015ScalingIU}] \label{prop:wang}
Let $Z = (Z_{ij})_{1\leq i<j \leq p} \in \{0,1\}^{\frac{p(p-1)}{2}}$. We first partition the matrix $S = \Y^T \Y$ and define the $p \times p$ matrix $V$ as follows:
\begin{align*}
    &S = \begin{pmatrix}
    S_{-p,-p} & S_{-p,p} \\
    S_{-p,p}^T & S_{pp}
     \end{pmatrix}, \quad  S_{-p,-p} \in \R^{(p-1) \times (p-1)}, \quad S_{-p,p} \in \R^{p-1}, \quad   S_{pp} > 0, \\
    &V = (V_{ij})_{i,j \in [p]} = \begin{pmatrix}
    V_{-p,-p} & V_{-p,p} \\
    V_{-p,p}^T & V_{pp}
     \end{pmatrix}, \quad V_{ij} = \begin{cases}
        g_0^2 & \text{ if } Z_{ij} = 0 \\
         g_1^2 & \text{ otherwise }
     \end{cases}.
\end{align*}
We also define the transformed variables $(u_1, u_2)$ as
\begin{align*}
    u_1 := \Omega_{-p,p}, \quad u_2 := \Omega_{pp} - \Omega_{-p,p}^T [\Omega_{-p,-p}]^{-1} \Omega_{-p,p}.
\end{align*}
Then under the continuous spike-and-slab prior distribution (see   \eqref{eq:prior-cont-offdiag}), the conditional posterior distribution on $(u_1, u_2)$, $\pi(u_1,u_2 \mid\Y,\Omega_{-p,-p}, Z)$ is a  Normal-Gamma distribution
\begin{align}\label{eq:wang1}
    \pi(u_1,u_2 \mid\Y,\Omega_{-p,-p}, Z)= \text{N}(u_1; \: -C S_{-p,p}, C) \text{Ga} (u_2; \: \frac{n}{2} + 1, \frac{S_{pp} + \lambda }{2}),
\end{align}
where $C^{-1} = (S_{pp} + \lambda  )[\Omega_{-p,-p}]^{-1} + \text{Diag}(V_{-p,p})^{-1}$.
\end{prop}

 We note that the transformed variable $u_2$ corresponds to the marginal precision of $Y_p$. 
 Moreover, a consequence of the prior \eqref{eq:prior-cont-offdiag} is that indicator variables $(Z_{ij})_{i<j}$ are independent conditionally on $\Omega$, hence for  each $Z_{ij}$, $\pi(Z_{ij} \mid\Omega, \Y)$,  is a Bernoulli distribution with mean
\begin{align}\label{eq:mean-bernoulli}
        m_{ij} = \frac{\theta \text{N}(\Omega_{ij}; \: 0, g_1^2)}{\theta \text{N}(\Omega_{ij}; \: 0, g_1^2) + (1-\theta) \text{N}(\Omega_{ij}; \: 0, g_0^2)}.
\end{align}
The previous equations \eqref{eq:wang1} and \eqref{eq:mean-bernoulli} imply that one can sample the transformed variables  $(u_1,u_2) \sim  \pi(u_1,u_2 \mid\Y,\Omega_{-p,-p}, Z_{\cdot i} )$ from a $p$-variate normal distribution and a Gamma distribution. Then, from this sample one gets a sample from $ \pi(\Omega_{-p,p},\Omega_{pp}  \mid\Y,\Omega_{-p,-p}, Z_{\cdot i} )$ using the transformation $\Omega_{-p,p} = u_1$, $\Omega_{pp} = u_2 + u_1^T [\Omega_{-p,-p}]^{-1} u_1$. Finally a sample  $Z \sim \pi(Z \mid\Y,\Omega )$ can be obtained by sampling $p(p-1)/2$ independent Bernoulli distributions. The block Gibbs sampler proposed by \cite{Wang2015ScalingIU} to sample from $\pi(\Omega, Z\mid\Y)$  is summarised in  Algorithm \ref{alg:block_gibbs}. We denote by $\psi:= (\theta, g_0, g_1, \lambda)$ the hyperparameters of the prior in \cite{Wang2015ScalingIU}.

\begin{algorithm}
\caption{Block Gibbs Sampler \citep{Wang2015ScalingIU}}\label{alg:block_gibbs}
\KwIn{ $\Y$, $\psi$, number of iterations $T$, initial value $\Omega^{(0)}$.}
\KwOut{$\{(\Omega^{(t)}, Z^{(t)})\}_{t \in [T]}$, samples from $\pi(\Omega\mid\Y)$.} 
\For{$t \gets 1$ to $T$}{
\For{$j \gets 1$ to $p$}{ 
Fix $\Omega_{-j,-j} := \Omega_{-j,-j}^{(t-1)}$, the sub-matrix of $\Omega^{(t-1)}$ removing the $i$-th row and $i$-th column. \\
Sample $(u_1^*,u_2^*)  \sim \pi(u_1,u_2 \mid\Y,\Omega_{-j,-j}, Z_{\cdot j}^{(t-1)} )$ using \eqref{eq:wang1}. \\
Compute $\Omega_{-j,j}^* = u_1^*$ and $\Omega_{jj}^* = u_2^* + (u_1^*)^T \Omega_{-j,-j}^{-1} u_1^* $. \\

Set $\Omega^{(t)} = \begin{pmatrix}
    \Omega_{-j,-j} & \Omega_{-j,j}^* \\
    (\Omega_{-j,j}^*)^T & \Omega_{jj}^*
\end{pmatrix}$
}
Sample $ Z_{ij}^{(t)} \sim \text{Ber}(m_{ij})$ given by \eqref{eq:mean-bernoulli} for each $1 \leq i < j \leq p$ and set $Z^{(t)} = (Z_{ij}^{(t)} )_{1 \leq i < j\leq p}$.
}
\end{algorithm}

We note that in Algorithm \ref{alg:block_gibbs} each update of a column of $\Omega$ guarantees that the positive-definiteness constraint is met thanks to the reparametrisation. In fact, using our previous notation, it holds that
\begin{align*}
    \Omega \in \PSD \iff \Omega_{-p,-p} \in \PSD \quad \text{ and } \quad \Omega_{pp} -\Omega_{-p,p}^T [\Omega_{-p,-p}]^{-1} \Omega_{-p,p} > 0,
\end{align*}
and for the RHS in the previous equivalence to be achieved it is enough that $u_2^*$ is stricly positive. While Algorithm \ref{alg:block_gibbs} can easily be coded, it is limited by a trade-off between the inclusion of moderately small coefficients (more precisely, posterior consistency) and the computational efficiency of the Gibbs sampler. 


Following the arguments above, the claim by \cite{oya2022positive} that the Gibbs sampling algorithm by \cite{Wang2015ScalingIU} does not guarantee positive-definiteness is not valid. 
\cite{oya2022positive} in fact considers a different Gibbs sampler that alternates between $u_2 \sim \pi(u_2 \mid u_1, \Y, \Omega_{-j,-j}, Z_{\cdot j})$ and $u_1 \sim \pi(u_1 \mid u_2, \Y, \Omega_{-j,-j}, Z_{\cdot j})$. This algorithm requires sampling $u_1$ from a constrained probability distribution to guarantee positive-definiteness. Critically, however, Algorithm \ref{alg:block_gibbs} samples from the joint full conditional $u_1, u_2 \mid \Y, \Omega_{-j,-j},Z_{\cdot j}$.

\section{Additional details on the linear regression proposal}\label{Sec:SSPriorLR}


In this section we explain the intuition behind the design of our proposal distribution in our GIMH algorithm from Section \ref{sec:par-gibbs}.

Recall the expression of the GGM likelihood
\begin{align}\nonumber
    &L(\Y\mid\Omega) = (2\pi)^{-\frac{np}{2}}|\Omega|^{n/2} \prod_i \exp (-\frac{1}{2} y_i^T \Omega  y_i)  \propto |\Omega|^{n/2} \exp \left \{ - \frac{1}{2}tr(S \Omega) \right \},
\end{align}
We can re-write the likelihood in a form that makes the connection with a linear regression likelihood function explicit. For illustration purposes, we consider the last variable $Y_p$ and decompose the data matrix as $\Y = (\Y_{\cdot, -p}, \Y_{\cdot p})$. Define $\beta := - \Omega_{-p,p} / \Omega_{pp}$ and $\sigma^2 := 1 / \Omega_{pp}$. Since $(\Sigma_{-p,-p}, \Omega_{-p,p}, \Omega_{pp})$ uniquely define $\Omega$ (see Lemma \ref{lem:decomposition}) 
so does  $(\Sigma_{-p,-p}, \beta, \sigma^2)$, and $L(\Y\mid\Omega)$ can be re-written as
\begin{align}
    L(\Y\mid\Omega) &= L(\Y; \Sigma_{-p,-p}, \beta,\sigma^2) = \underbrace{(2 \pi)^{-n(p-1)/2}  |\Sigma_{-p,-p}|^{-n/2}  \exp (-\frac{1}{2} tr(S_{-p,-p} \Sigma_{-p,-p}^{-1}))}_{ L(\Y_{-p} \mid \Sigma_{-p,-p}) } \nonumber \\
     &\times \underbrace{(2 \pi \sigma^2)^{-n/2}\exp \left \{ -\frac{1}{2 \sigma^2} ( S_{pp} - 2 \beta^T S_{-p,p} + \|\beta\|^2 S_{-p,-p}) \right \} }_{ L(\Y_{\cdot p} \mid \Y_{\cdot, -p} , \beta,\sigma^2)} \nonumber \\
 L(\Y_{\cdot, -p}\mid \Sigma_{-p,-p}) &= \prod_{i=1}^n \mathcal{N}(\Y_{i,-2}; \mathbf 0_{p-1}, \Sigma_{-p,-p}),  \label{eq:ll_1} \\
  L(\Y_{\cdot p}\mid  \Y_{\cdot,-p} , \beta,\sigma^2)  &= \prod_{i=1}^n \mathcal{N}(\Y_{i p}; \beta^T \Y_{i,-p}, \sigma^2), \label{eq:lin-reg-model} 
\end{align}
Note that the previous equations result from the fact that the normal vector $Y$, decomposed into $(Y_p, Y_{-p})$ is jointly Gaussian and the conditional distribution of $Y_p \mid Y_{-p}$ is also Gaussian with mean $\beta^T Y_{-p}$ and variance $\sigma^2$ while $ Y_{-p}$ is a $(p-1)$-dimensional normal vector with zero mean and covariance matrix $\Sigma_{-p,-p}$. In other words, $L(\Y_{\cdot, -p}; \Sigma_{-p,-p})$ is the likelihood of $n$ observations from a $(p-1)$-dimensional Gaussian distribution with mean 0 and covariance matrix $\Sigma_{-p,-p}$ and $L(\Y_{\cdot p}\mid  \Y_{\cdot, -p} , \beta,\sigma^2)$ is the likelihood of $n$ observations under a linear regression model with fixed design matrix  $\Y_{\cdot, -p}$ and Gaussian noise with unknown variance. 

This decomposition implies that one can re-write the conditional posterior distribution on  $(\beta, \sigma^2)$ as
\begin{align}\label{eq:cond-post-beta}
    \pi(\beta, \sigma^2 \mid \Omega_{-p,-p}, \Y) \propto L(\Y_{\cdot, -p}\mid \Sigma_{-p,-p}) \times L(\Y_{\cdot p}\mid  \Y_{\cdot,-p} , \beta,\sigma^2)\times  \pi(\beta, \sigma^2 \mid \Omega_{ -p,-p}).
\end{align}
Note that since $\Sigma_{-p,-p}$ is a function of $(\Omega_{-p,-p}, \beta, \sigma^2)$ (see Lemma \ref{lem:decomposition}), the first term on the RHS of \eqref{eq:cond-post-beta} also depends on $(\beta, \sigma^2)$. Nonetheless the second and third terms are alike the likelihood function and spike-and-slab prior distribution in a linear regression problem. In fact, consider the following prior distribution in linear regression
\begin{align}\label{eq:lin-reg-prior}
    &\pi_z(z) = \theta^{|z|} (1- \theta)^{p-1-|z|} 
    \nonumber \\
    &\pi_\beta(\beta \mid \sigma^{-2}, z) = \prod_{k, z_k=1}  \text{N}(\beta_k; \: 0, \sigma^2 \tau^{-1}) \prod_{k, z_k=0} \delta_0(\beta_k) 
    \nonumber \\
    &\pi^{LR}(\beta, \sigma^{-2}, z \mid \Y) = \pi^{LR}(\beta \mid \sigma^{-2}, z \mid \Y)  \pi^{LR}( \sigma^{-2} \mid z, \Y) \pi^{LR}(z \mid \Y) ,
\end{align}
where $z$ is the indicator vector of the non-zero entries in $\beta$.
Using Lemma \ref{lem:posterior_linreg_notation} with $y := \Y_{\cdot p}$ and $X := \Y_{\cdot, - p}$, the posterior distribution on $z$ implied by the model \eqref{eq:lin-reg-model} is
\begin{align}\label{eq:lin-reg-post}
    \pi^{LR}(z \mid \Y) \propto 
\frac{e^{ \frac{ \mu_z^T W_z \mu_z }{2 } }}{g_1^{-|z|_0} |W_z|^{\frac{1}{2}}}
\theta^{|z|_0} (1-\theta)^{p-1 - |z|_0},
\end{align}
where $W_z := X_{\cdot z}^T X_{\cdot z} + \tau I_z = S_{zz}+ \tau I_z$, and $\mu_z := W_z^{-1} X_{\cdot z}^T y = W_z^{-1} S_{zp}$. 

\begin{remark}
    The spike-and-slab prior \eqref{eq:prior-z} does not exactly match the conjugate prior \eqref{eq:lin-reg-prior} on $(\beta, \sigma^2)$. The only difference lies in the scaling of the variance of the slab in $\pi_\beta(\beta \mid\sigma^{-2})$, which is $\sigma^4 g_1^2$ under \eqref{eq:prior-z}. However, the latter parameterisation is not conjugate to the linear regression likelihood and therefore, would not give closed-form $\pi^{LR}(z \mid \Y)$. 
\end{remark}

We call \eqref{eq:lin-reg-post} the \emph{linear regression} posterior for column $p$ as this is a valid posterior on $(\beta, \sigma^2)$ assuming model \eqref{eq:lin-reg-model} and that $\Y_{\cdot,-p}$ is fixed. Although \eqref{eq:lin-reg-post} differs from \eqref{eq:cond-post-beta} (in particular, it does not depend on $\Omega_{-p,-p}$), its expression is interestingly close (see Lemma \ref{lem:posterior_linreg_notation}).

We note that the distribution $\pi^{LR}(z \mid \Y)$ is also intractable, in the sense that if $p$ is large it is also impossible to compute $\pi^{LR}(z \mid \Y)$ for all $2^{p-1}$ models. But recently-developed MCMC algorithms for discrete spaces such as the Tempered Gibbs Sampler  by \cite{zanella2019scalable} or the Locally-Informed Birth-Death sampler by \cite{zhou2022dimension} can rapidly find a subset of models of interest $\mathcal{S} \subset \{0,1\}^{p-1}$ with a good approximation of  $\pi^{LR}(z\mid \Y)$ or its tempered version $Q_\upsilon^p(z) \propto (\pi^{LR}(z\mid \Y))^{\upsilon}$ for  each $z \in \mathcal{S}$. By subset of interest we mean that $\pi^{LR}(\mathcal{S} \mid \Y)$ (or $Q_\upsilon^p(\mathcal{S})$) should be close to one. Moreover, we note that such samplers can leverage Rao-Blackwell estimators of the posterior model probabilities $\hat \pi^{LR}(z\mid \Y)$.

\section{Simulation comparing the continuous and discrete spike-and-slab priors}\label{app:simulation-figure}


\begin{figure}[hbt!]
    \centering
     \begin{subfigure}[b]{0.49\textwidth}
    \includegraphics[width=\textwidth, trim=0.cm 0.cm 0cm  0.8cm,clip]{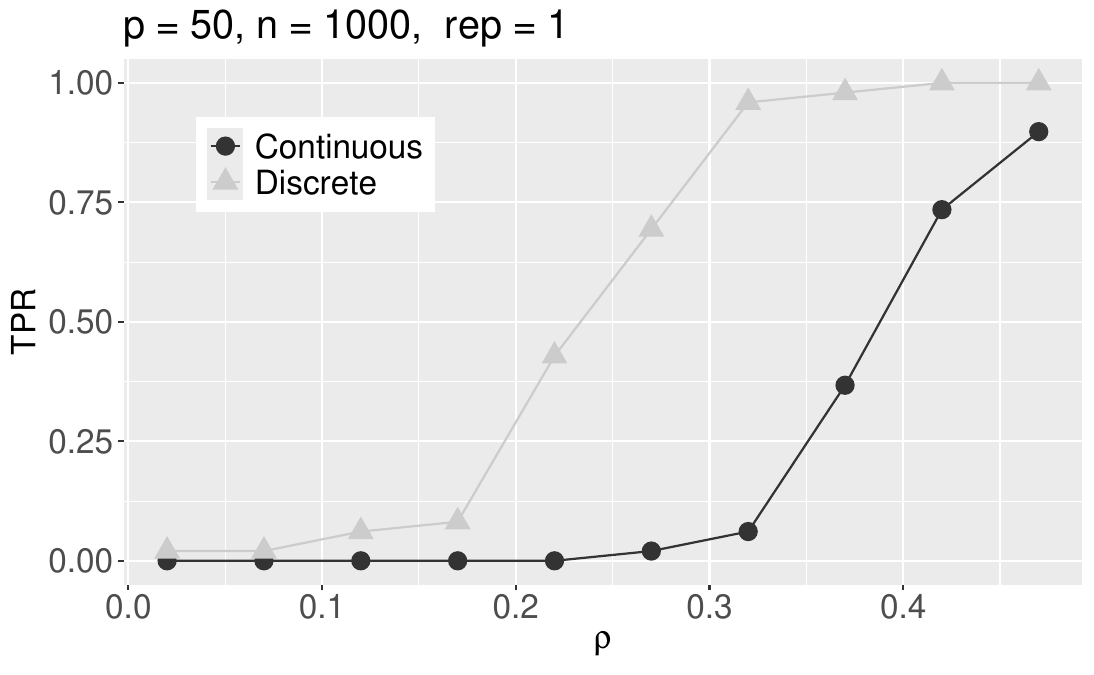}
    \end{subfigure}
    \hfill
    \begin{subfigure}[b]{0.49\textwidth}
    \includegraphics[width=\textwidth, trim=0.cm 0.cm 0cm  0.8cm,clip]{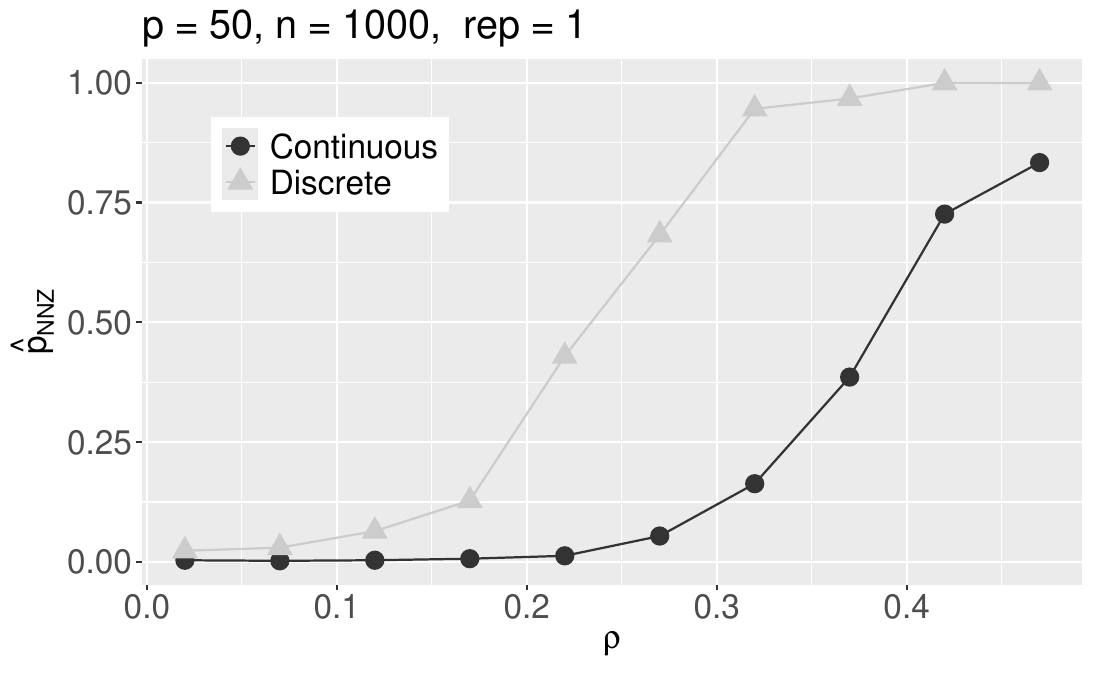}
    \end{subfigure}
\caption{Comparison of the continuous and discrete spike-and-slab priors in \textbf{Setting 1} with a tri-diagonal ground-truth precision matrix $\Omega^0$, $p=50$, $n=1000$.
We plot the true positive rate (TPR) (left panel) and $\hat p_{NNZ}$, 
the average marginal posterior inclusion probabilities of truly non-zero coefficients of $\Omega^0$, (right panel) versus $\rho$, the value of the non-zero off-diagonal coefficients, for the continuous spike-and-slab prior distribution with $g_0 = 0.05$ (black circles) and the discrete spike-and-slab prior (gray triangles). 
}
\label{fig:comparison-wang-inf-1}
\end{figure}

\begin{figure}[hbt!]
    \centering
     \begin{subfigure}[b]{0.49\textwidth}
    \includegraphics[width=\textwidth, trim=0.cm 0.cm 0cm 0.8cm,clip]{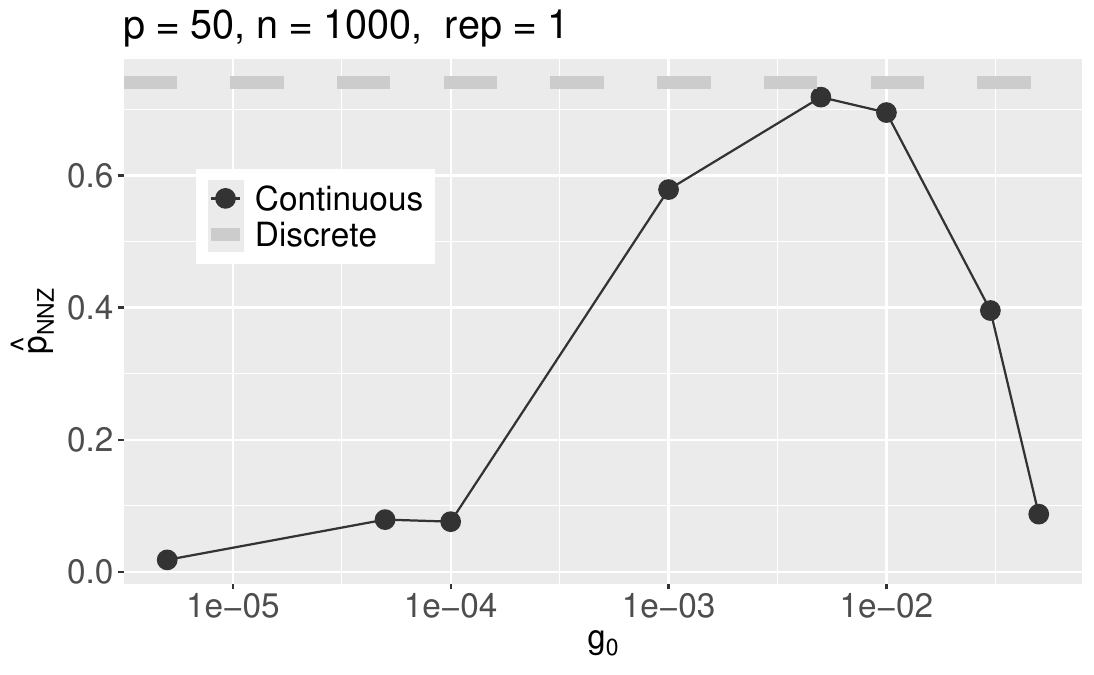}
    \end{subfigure}
    \hfill
    \begin{subfigure}[b]{0.49\textwidth}
    \includegraphics[width=\textwidth, trim=0.cm 0.cm 0cm  0.8cm,clip]{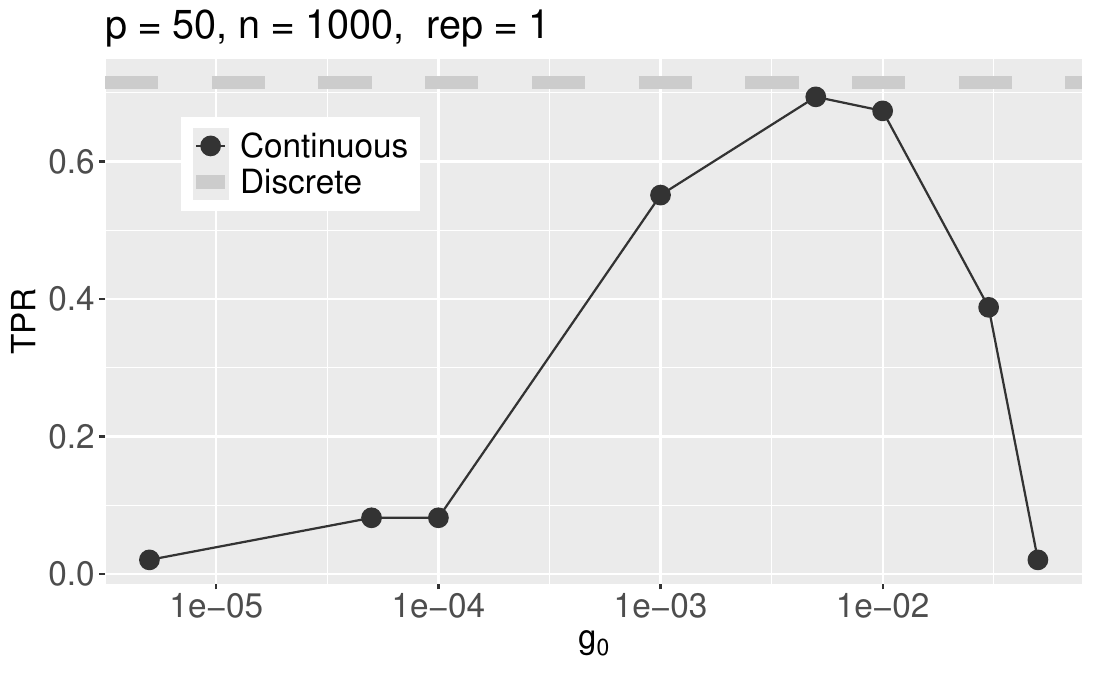}
    \end{subfigure}
    \hfill
        \begin{subfigure}[b]{0.49\textwidth}
    \includegraphics[width=\textwidth, trim=0.cm 0.cm 0cm  0.8cm,clip]{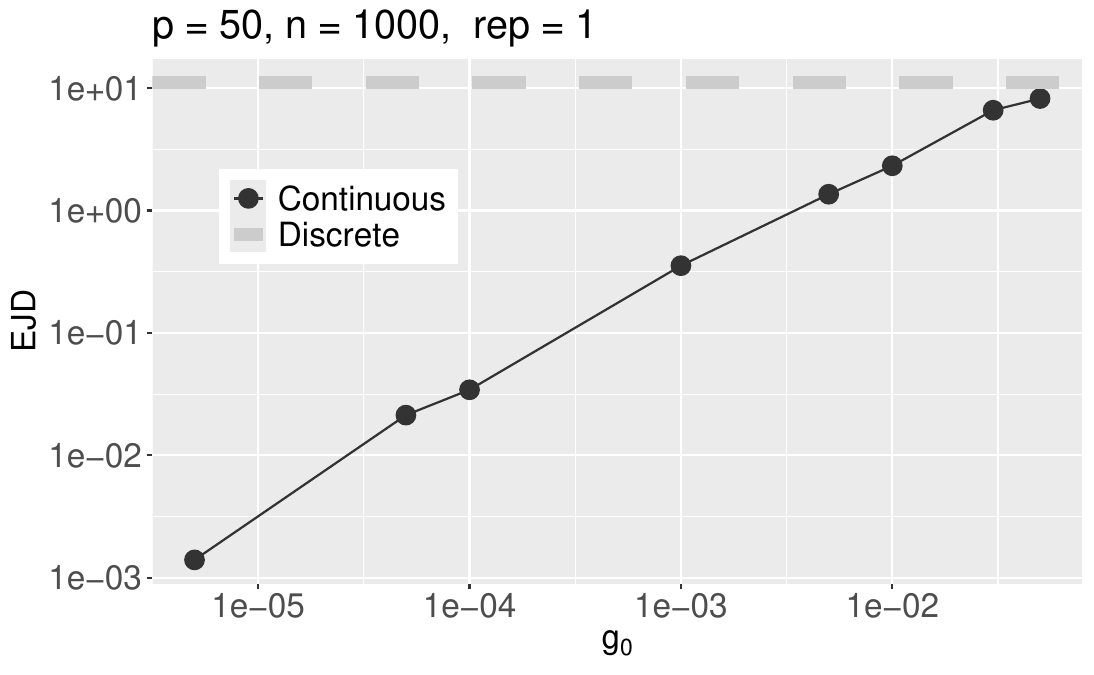}
    \end{subfigure}
\caption{Comparison of the continuous and discrete spike-and-slab priors in \textbf{Setting 2} with a tri-diagonal ground-truth precision matrix $\Omega^0$ with first off-diagonal coefficients equal to $\rho = 0.3$, $p=50$, $n=1000$. In the first row, we plot the true positive rate (TPR) (left panel) and the average marginal posterior inclusion probabilities of truly non-zero coefficients of $\Omega^0$ (right panel) versus $g_0$, the continuous spike-and-slab's spike standard deviation (black circles). In the second row, we plot the Expected Jump Distance (EJD) (higher is better) versus $g_0$. The dotted grey line corresponds to the discrete spike-and-slab prior. 
}
\label{fig:comparison-wang-inf-2}
\end{figure}

In this simulation we set $n=1000$, $p=50$ and 
\begin{align*}
    \Omega^0 = \begin{pmatrix}
        2.0 & \rho & 0.0 & \dots & \dots & 0.0 \\
        \rho & 2.0 & \rho & 0.0 & \dots & 0.0 \\
         & \ddots & \ddots & \ddots & \ddots  \\
         & \ddots & \ddots & \ddots & \ddots & \rho  \\
        0.0 & \dots & \dots & \dots &  \rho & 2.0 \\
    \end{pmatrix}
\end{align*}
where $\rho$ is a parameter we vary. For both the continuous and discrete spike-and-slab priors, we set $g_1=1$, $\lambda = 2$, $\theta = 1/p$. In \textbf{Setting 1}, we set $g_0 = 0.05$ for the continuous spike-and-slab and vary $\rho$ between 0.02 and 0.5.
In \textbf{Setting 2}, we set $\rho=0.3$ and vary $g_0$ in the continuous spike-and-slab from $10^{-6}$ to 0.05.
For each set of parameters, we generate 1 set of observations, initialise the MCMC chain at the identity matrix, and run it for 10 000 iterations. We discard the first 5000 iterations as burn-in and use next 5000 iterations for inference. For the computation of the power (or, true positive rate (TPR)), 
we threshold the marginal posterior inclusion probabilities at 0.5. 

In Figure \ref{fig:comparison-wang-inf-1}, we plot the True Positive Rate and the average marginal posterior inclusion probabilities of truly non-zero coefficients of $\Omega^0$ under the discrete and continuous spike-and-slab priors in \textbf{Setting 1}. These results illustrate that 
as the discrete prior does does not assign small $|\Omega_{ij}|$ to the spike, it is easy to find examples where it leads to higher true positive rate than the continuous spike-and-slab.
In particular, we clearly see that the discrete prior provides significantly more power than the continuous one for $0.2\leq|\Omega_{ij}|<0.4$.

In Figure \ref{fig:comparison-wang-inf-2}, we plot the same metrics and the Expected Jump Distance of the variable inclusion indicators $Z$ from one MCMC iteration to the next, a measure of their mixing abilty, versus $g_0$ in \textbf{Setting 2}. 
We see that for the continuous prior as $g_0$ decreases from 0.05 to $10^{-6}$ inference first improves as smaller truly-nonzero coefficients of $\Omega^0$ get higher marginal posterior inclusion probabilities, then worsens, due to the poor mixing of the MCMC algorithm (as shown by the decrease in Expected Jump Distance).
The low $\hat{p}_{NNZ}$ and TPR when $g_0<10^{-4}$ result from the MCMC algorithm being unable to move away from the initial value (here the identity matrix).

In summary Figure \ref{fig:comparison-wang-inf-1} motivates wanting to use a small $g_0$ for the continuous spike-and-slab in order to improve inferences, however Figure \ref{fig:comparison-wang-inf-2} demonstrates that so doing has a detrimental impact on MCMC mixing and convergence. 
In comparison our discrete prior does not suffer from this problem. 
We further see in this experiment that the  discrete spike-and-slab prior is more computational efficient that its continuous counterpart for very small values of $g_0$.

\section{Additional Simulation Results}

This section contains additional details and experimental results. All experiments were conducted on a 13th Gen Intel(R) Core(TM) i7-13700 (2.10 GHz) with 64 GB of RAM.

\subsection{Figure \ref{fig:exact_vs_pseudo}}


The right-hand plot in Figure \ref{fig:exact_vs_pseudo} shows the time required by ssgraph, bdgraph, bdgraph.mpl, Gibbs (fixed $\theta$), GIMH and regression.pl to produce 4,000 posterior samples as $p$ grows. 
Our Gibbs sampling algorithm can sample from the GGM posterior with $p = 1000$, i.e., over $\approx 500,000$ precision matrix parameters in less than 9.25 hours ($p = 500$ with $>120,000$ parameters requires $<1$ hour). While both bdgraph.mpl and regression.pl are faster than this, neither is able to produce (pseudo) posterior simulations for $\Omega$ in this time. The additional computational burden of sampling from the posterior for $\Omega$ leaves bdgraph.mpl requiring 6 hours ($p = 500$), and while the regression.pl method is able to estimate posterior means and intervals, constructing joint samples for the whole of $\Omega$ from the individual regressions is not straightforward.  

\begin{figure}
\begin{center}
\includegraphics[width =0.49\linewidth]{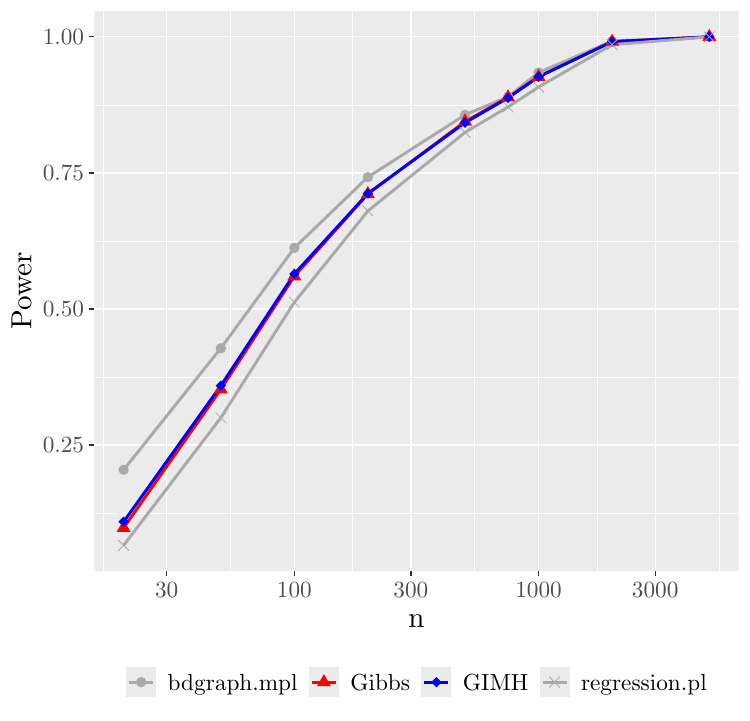} 
\includegraphics[width =0.49\linewidth]{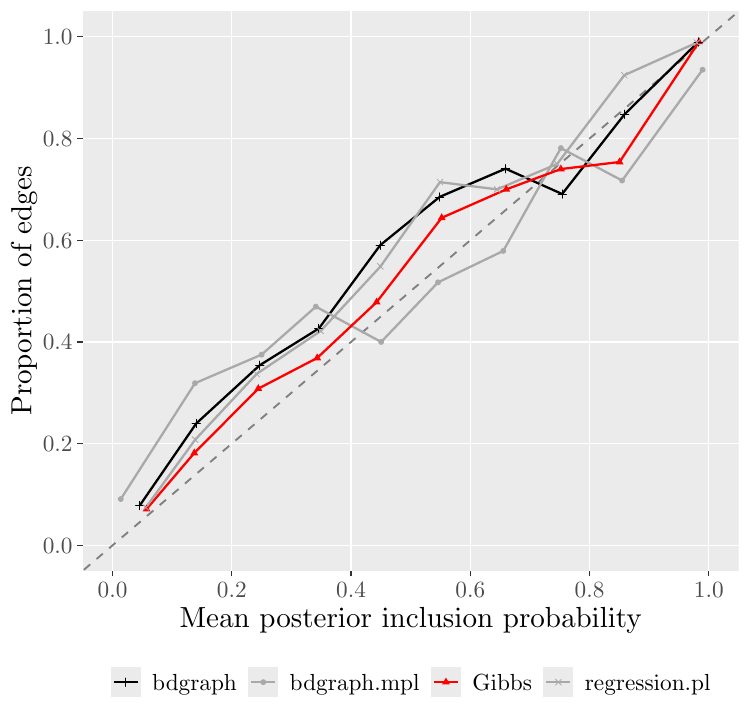}\\
\includegraphics[width =\linewidth]{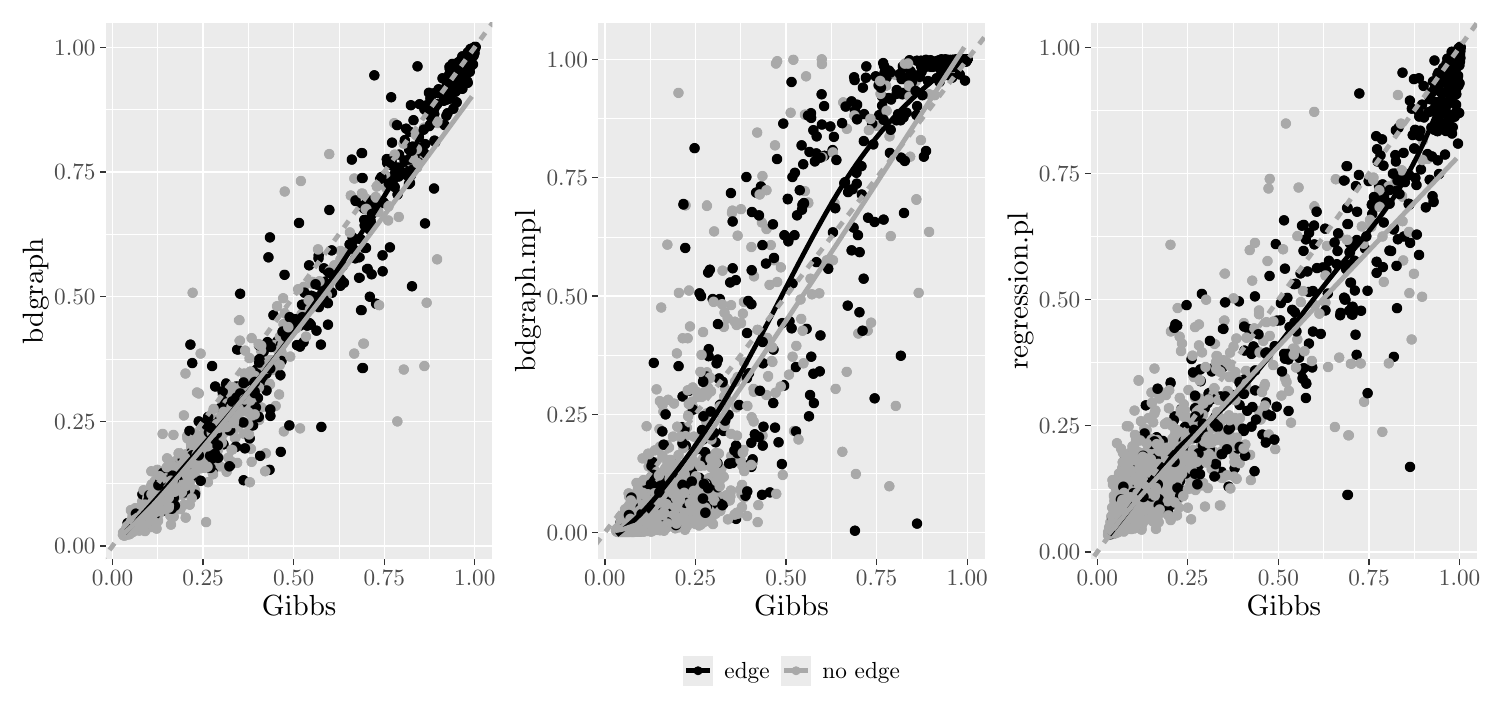}
\caption{Left: Power when thresholding inclusion probabilities to control the Bayesian False Discovery rate at 0.05 of exact and pseudo Bayesian methods for a tri-diagonal $\Omega$ with $p = 10$ and increasing $n$, repeated 100 times. 
Right: Comparison of binned average posterior inclusion probability with  observed frequency of edges  when $n = 50$. The dotted line $y = x$ corresponds to perfect calibration. 
Bottom: Pairwise comparison of posterior inclusion probabilities when $n = 50$
coloured according to whether there truly was an edge or not.
}
\label{fig:exact_vs_pseudo2}
\end{center}
\end{figure}

In addition to Figure \ref{fig:exact_vs_pseudo}, Figure \ref{fig:exact_vs_pseudo2} compares the power of bdgraph.mpl, Gibbs (fixed $\theta$), GIMH and regression.pl inference methods as the number of observation $n$ grows, 
   
along with a calibration curve comparing the binned edge frequency with the average posterior inclusion probability, 
\color{black}
and pairwise comparisons of the estimated inclusion probabilities when $n = 50$ across the 100 repeats. When evaluating power, edges were declared by thresholding  $\hat{\pi}(\Omega_{jk} \neq 0 \mid \Y)$ such that the average inclusion probability of declared edges was greater than 0.95, controlling the Bayesian FDR to be less than 0.05 \citep{mueller:2004}.

The top left of Figure \ref{fig:exact_vs_pseudo2} shows that the power of all four methods goes to 1 as $n$ increases. However, for smaller values of $n$ bdgraph.mpl has higher power and regression.pl has lower power than Gibbs and GIMH. 
   
The top right figure evaluates the calibration of the posterior inclusion probabilities estimated by each model by binning them and comparing them to the true edge frequency within each  bin. In this setting, Gibbs appears the best calibrated (closest to $y=x$), regression.pl consistently under estimates the probability of inclusion (always above $y=x$) and bdgraph.mpl under estimates small probabilities and overestimates large ones. 
\color{black}
The bottom panels provide an explanation for this. The left panel shows that the posterior inclusion probabilities estimated by Gibbs largely agree with those estimated by bdgraph. The middle panel shows that bdgraph.mpl estimates sharper probabilities (closer to either 0 or 1) than Gibbs (and by extension bdgraph). The large cluster of edges in top right hand corner above the dotted line for $y = x$ provide bdgraph.mpl with higher power than Gibbs, however there are also several non-edges at the top of the plot which cause the elevated FDR of bdgraph.mpl observed in Figure  \ref{fig:exact_vs_pseudo}. The right panel confirms that regression.pl generally estimates smaller inclusion probabilities than Gibbs which endows regression.pl with lower power. 

\subsection{Local updates}

We provide additional experimental results concerning the local proposal sampling algorithms.

\subsubsection{Gibbs, BDMH and LIT}

\begin{figure}
\begin{center}
\begin{tabular}{cc}
Mean $|\hat{E}^{(1)}(\Omega_{jk} \mid \Y) - \hat{E}^{(2)}(\Omega_{jk} \mid \Y)|$  & Mean $|\hat{\pi}^{(1)}(\Omega_{jk} \neq 0 \mid \Y) - \hat{\pi}^{(2)}(\Omega_{jk} \neq 0 \mid \Y)|$ \\
\includegraphics[trim= {0.0cm 1.2cm 0.0cm 0.0cm}, clip,width =0.475\linewidth]{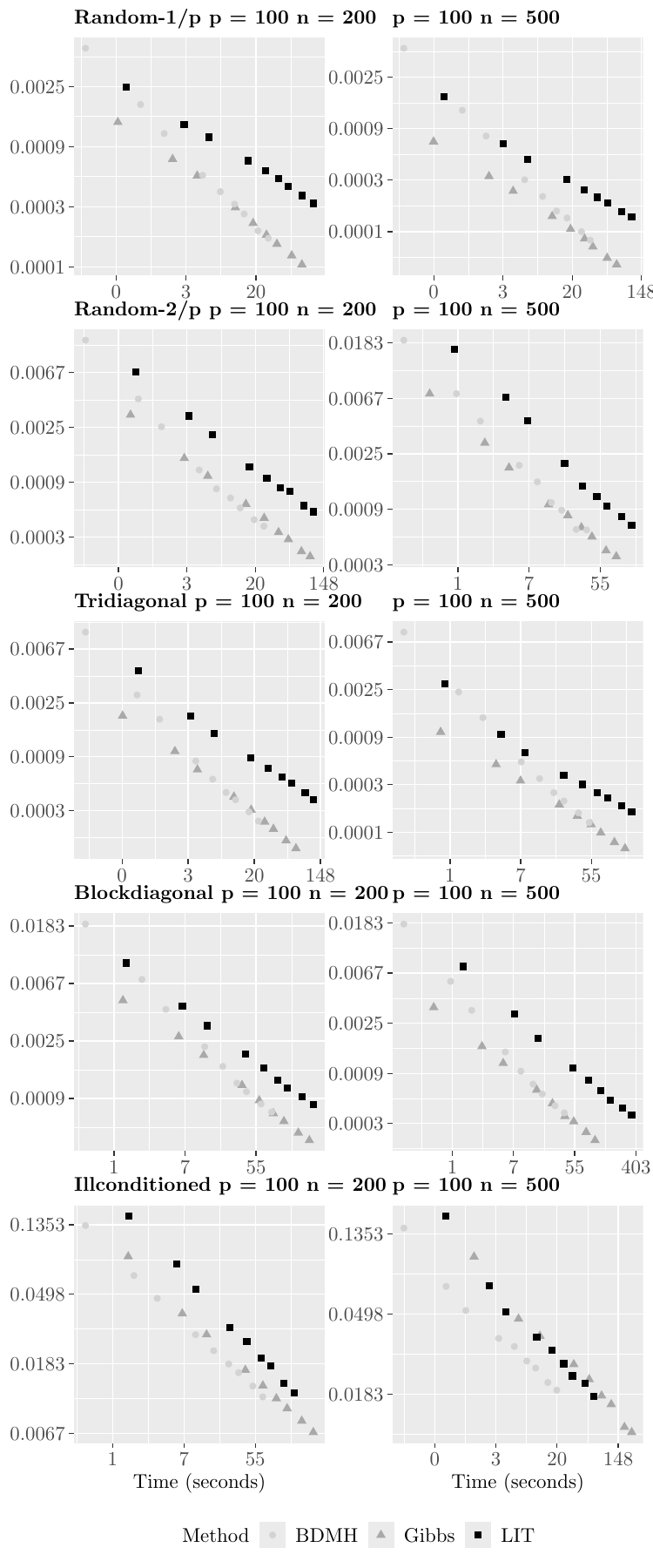} & \includegraphics[trim= {0.0cm 1.2cm 0.0cm 0.0cm}, clip,width =0.475\linewidth]{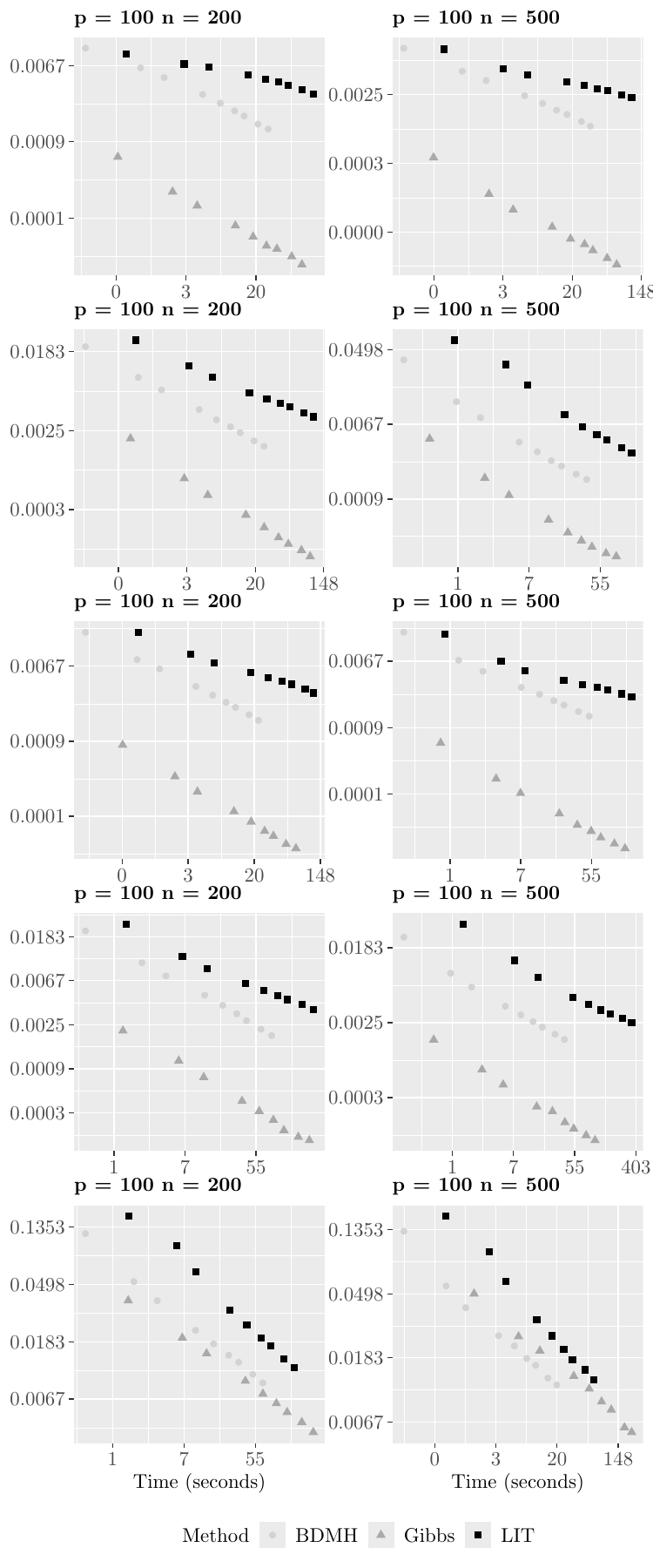}\\
\multicolumn{2}{c}{\includegraphics[trim= {0.0cm 0.4cm 0.0cm 29.2cm}, clip, width =0.95\linewidth]{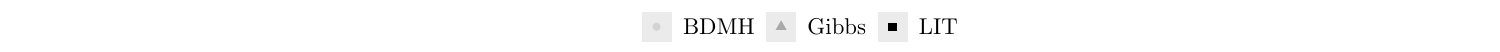}}\\
\end{tabular}
\caption{Difference between posterior mean estimates from two MCMC chains initialised at GLASSO-EBIC and $I$ vs. clock time. Comparison between Gibbs sampling, BDMH and LIT under the binomial model prior for $p = 100$ and $n \in \{200, 500\}$.
}
\label{fig:mixing_p100_Serial}
\end{center}
\end{figure}

In addition to Gibbs sampling and BDMH, we also considered the LIT algorithm of \cite{zhou2022dimension} to sample from $\pi(z \mid \Y, \Omega_{-j,-j})$. All three methods conduct a scan of all columns, Gibbs updates all $p-1$ elements of each column, BDMH makes $\lceil\sqrt{p}\rceil$ proposals per column and LIT makes 1 proposal per column, to reflect the increased computation in deciding upon that proposal.  For BDMH we set $p_{birth} = 0.75$ and $p_{death} = p_{swap} = 0.5(1-p_{birth}) $. 
Figure \ref{fig:mixing_p100_Serial} is an analogous plot to Figure \ref{fig:mixing_p100} and compares the efficiency of the posterior estimation of these three local sampling procedures under the binomial model prior for $p = 100$. The LIT sampler is much less efficient at estimating both posterior means and posterior inclusion probabilities than Gibbs and BDMH. Gibbs and BDMH appear similarly efficient for the parameters, but Gibbs sampling is more efficient for the inclusion probabilities.


\subsubsection{Rao-Blackwellisation}

\begin{figure}
\begin{center}
\begin{tabular}{cc}
Mean $|\hat{E}^{(1)}(\Omega_{jk} \mid \Y) - \hat{E}^{(2)}(\Omega_{jk} \mid \Y)|$  & Mean $|\hat{\pi}^{(1)}(\Omega_{jk} \neq 0 \mid \Y) - \hat{\pi}^{(2)}(\Omega_{jk} \neq 0 \mid \Y)|$ \\
\includegraphics[trim= {0.0cm 1.2cm 0.0cm 0.0cm}, clip,width =0.475\linewidth]{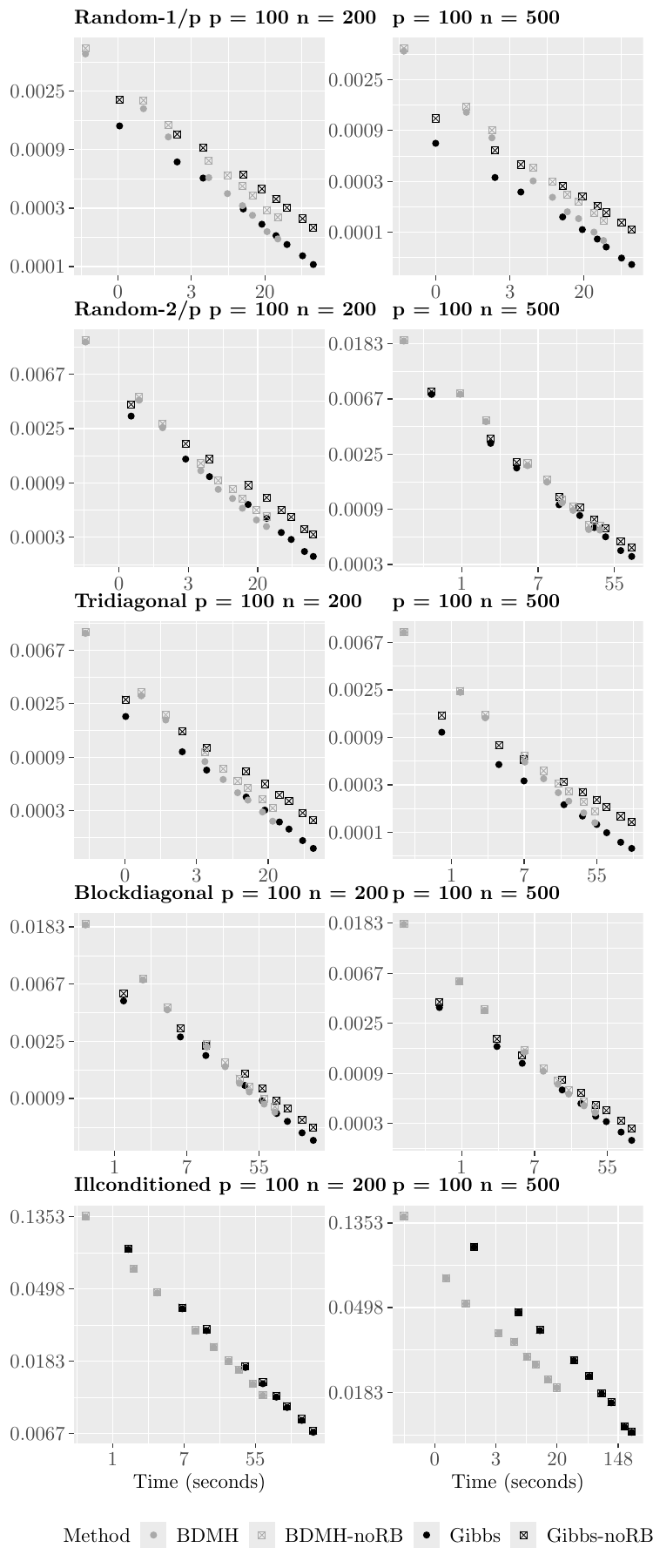} & \includegraphics[trim= {0.0cm 1.2cm 0.0cm 0.0cm}, clip,width =0.475\linewidth]{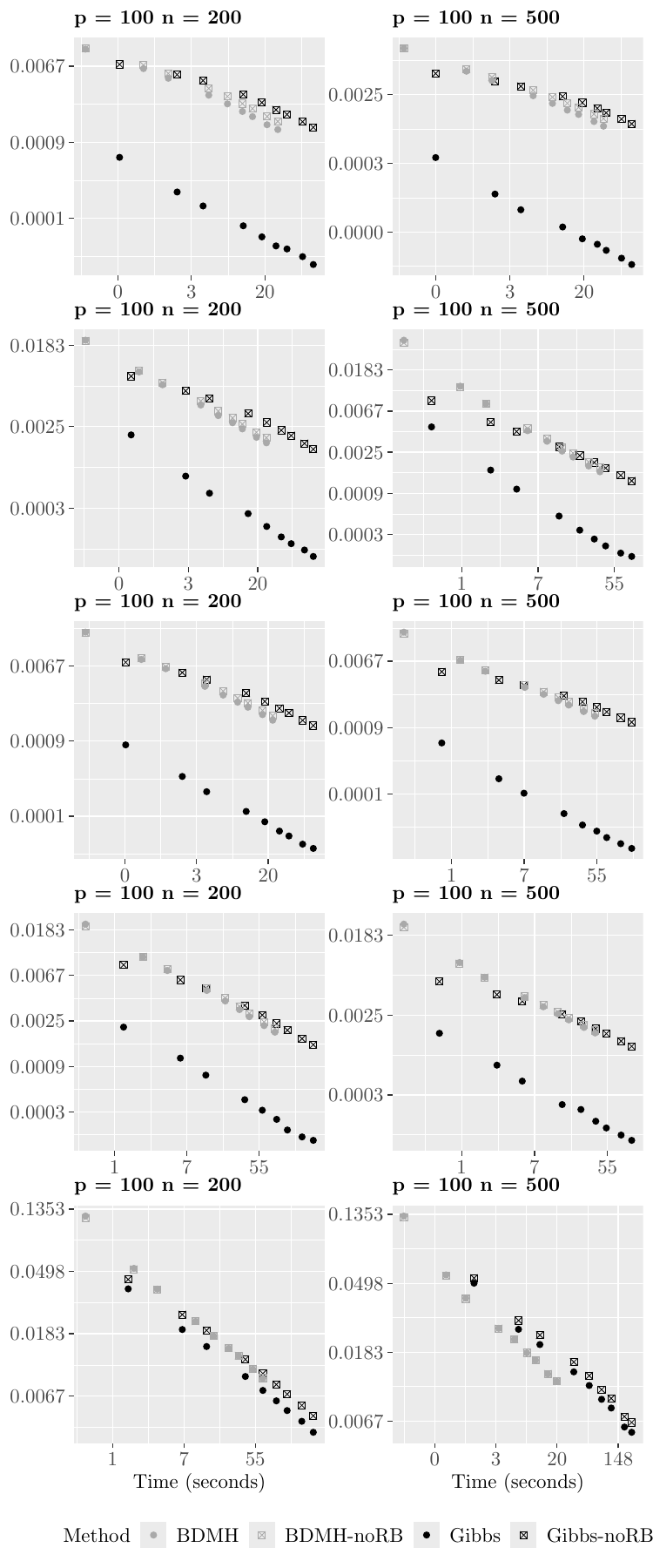}\\
\multicolumn{2}{c}{\includegraphics[trim= {0.0cm 0.4cm 0.0cm 29.2cm}, clip, width =0.95\linewidth]{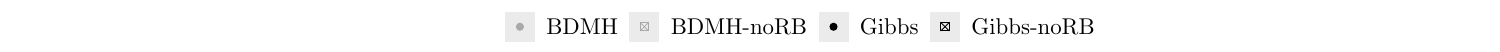}}\\
\end{tabular}
\caption{Difference between posterior mean estimates from two MCMC chains initialised at GLASSO-EBIC and diag(1) vs. clock time.  Comparison between Rao-Blackwellised and non-Rao-Blackwellised estimates from Gibbs sampling and BDMH under the binomial model prior for  $p = 100$ and $n \in \{200, 500\}$.
}
\label{fig:mixing_p100_RB}
\end{center}
\end{figure}

Figure \ref{fig:mixing_p100_RB} demonstrates that the improved efficiency of the  Gibbs algorithm for estimating the posterior probability of inclusion demonstrated in Figure \ref{fig:mixing_p100} (and Figure \ref{fig:mixing_p100_Serial}) is a result of Gibbs sampling allowing for more efficient Rao-Blackwellisation. Based only on averages of posterior samples (i.e., non Rao-Blackwellised estimation) the efficiency of Gibbs and BDMH algorithms is almost identical. However, the Rao-Blackwellisation improves the efficiency of the Gibbs sampler to a much greater extent than it improved the efficiency of BDMH, particular for the inclusion probabilities.


\subsection{Global proposal tempering parameter}

\begin{figure}
\begin{center}
\begin{tabular}{cc}
Mean $|\hat{E}^{(1)}(\Omega_{jk} \mid \Y) - \hat{E}^{(2)}(\Omega_{jk} \mid \Y)|$  & Mean $|\hat{\pi}^{(1)}(\Omega_{jk} \neq 0 \mid \Y) - \hat{\pi}^{(2)}(\Omega_{jk} \neq 0 \mid \Y)|$ \\
\includegraphics[trim= {0.0cm 1.5cm 0.0cm 0.0cm}, clip,width =0.475\linewidth]{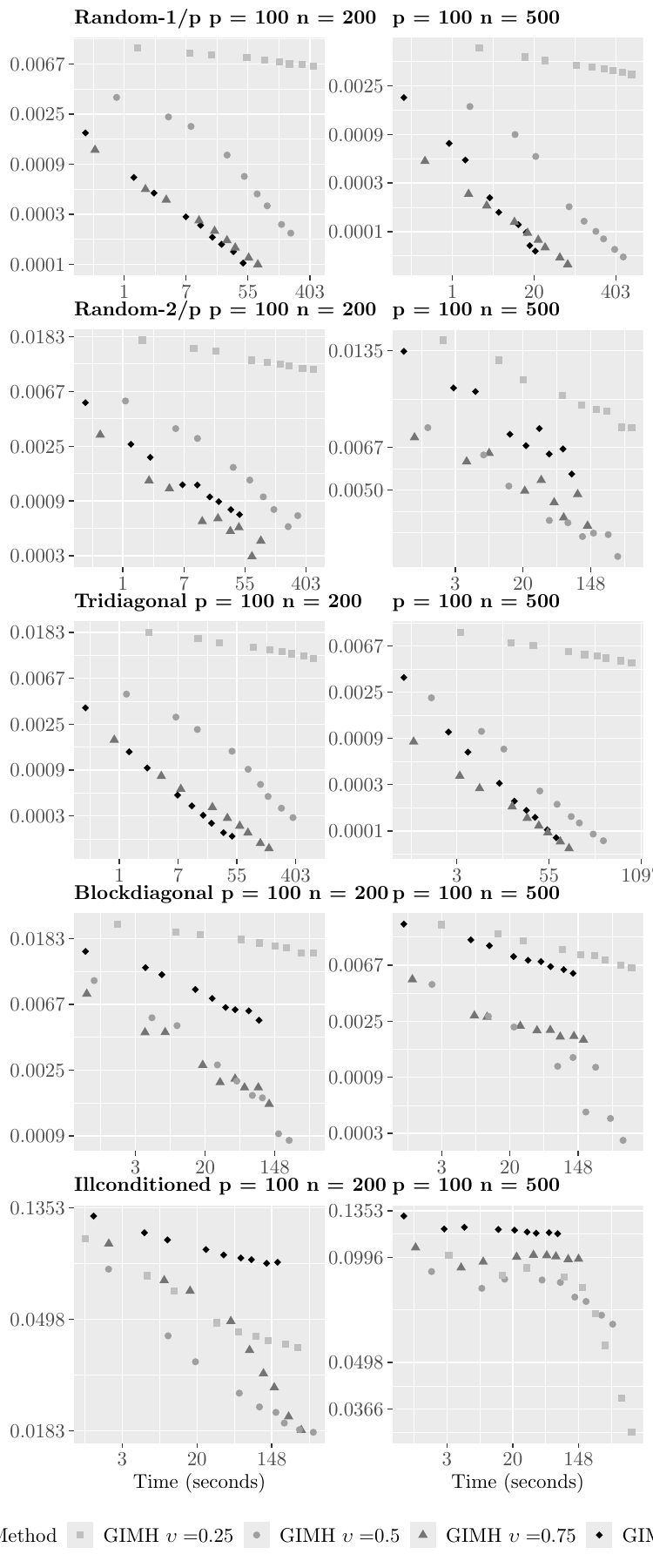} & \includegraphics[trim= {0.0cm 1.5cm 0.0cm 0.0cm}, clip,width =0.475\linewidth]{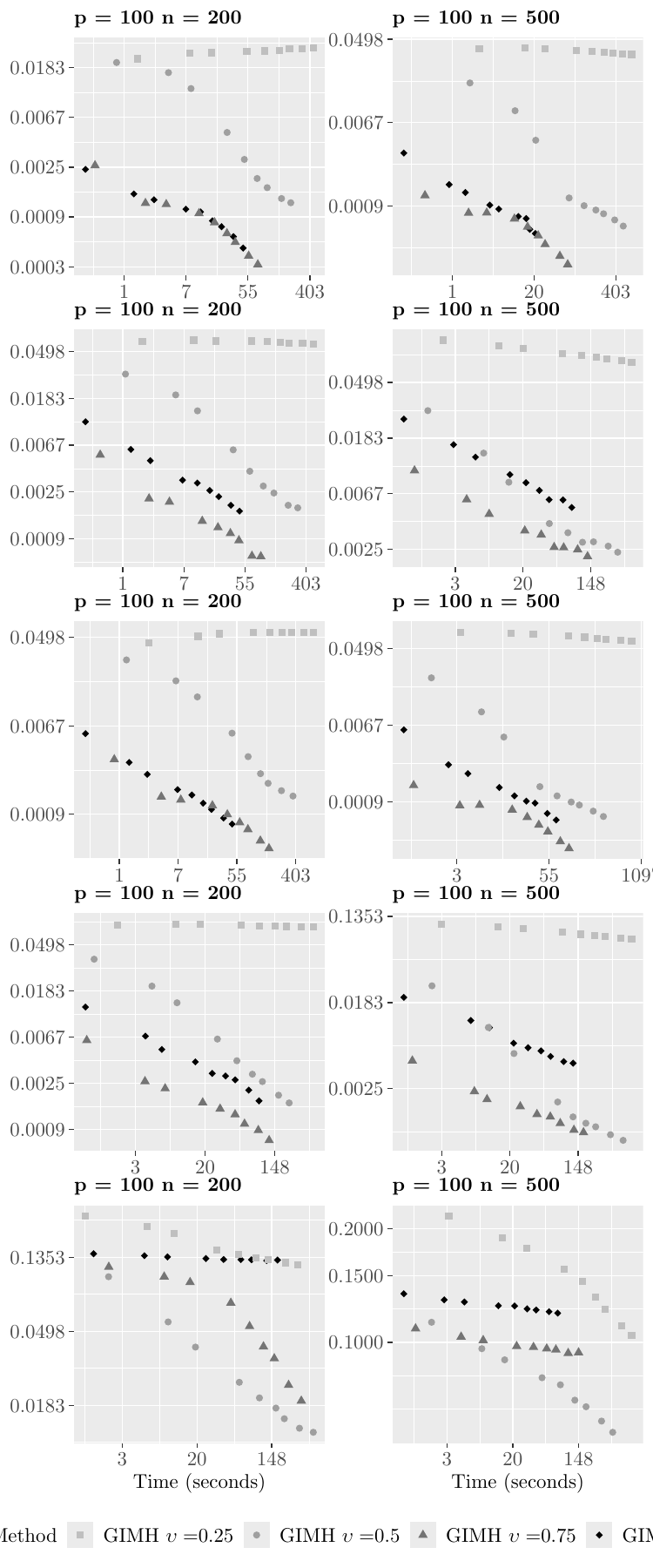}\\
\multicolumn{2}{c}{\includegraphics[trim= {0.0cm 0.4cm 0.0cm 29.2cm}, clip, width =0.95\linewidth]{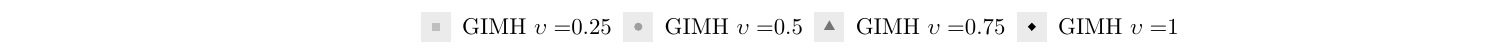}}\\
\end{tabular}
\caption{Difference between posterior mean estimates from two MCMC chains initialised at GLASSO-EBIC and diag(1) vs. clock time. Comparison between GIMH with  tempering parameter $\upsilon \in \{1, 0.75, 0.5, 0.25\}$ under the binomial model prior for $p = 100$ and $n \in \{200, 500\}$.
}
\label{fig:tempering_p100}
\end{center}
\end{figure}

In a similar manner to Figure \ref{fig:mixing_p100}, Figure \ref{fig:tempering_p100} investigates the posterior sampling efficiency of the GIMH algorithm for different values of the tempering hyperparameter $\upsilon$ under the binomial model prior. This shows that $\upsilon = 0.75$ generally provides the most efficient sampling. The exception to this is the \textit{ill-conditioned} setting where $\upsilon = 0.5$ appears to provide superior sampling efficiency.



\begin{figure}
\begin{center}
\begin{tabular}{cc}
Mean $|\hat{E}^{(1)}(\Omega_{jk} \mid \Y) - \hat{E}^{(2)}(\Omega_{jk} \mid \Y)|$  & Mean $|\hat{\pi}^{(1)}(\Omega_{jk} \neq 0 \mid \Y) - \hat{\pi}^{(2)}(\Omega_{jk} \neq 0 \mid \Y)|$ \\
\includegraphics[trim= {0.0cm 1.5cm 0.0cm 0.0cm}, clip,width =0.475\linewidth]{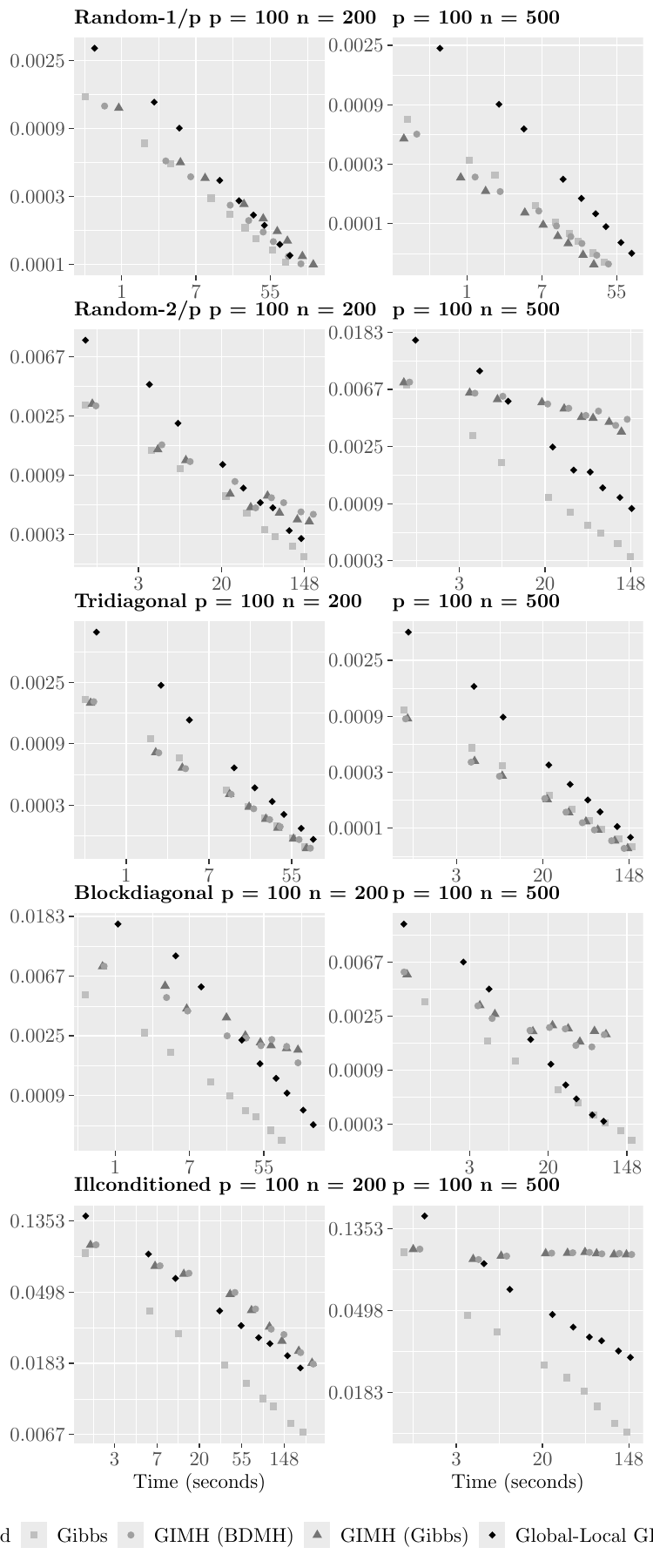} & \includegraphics[trim= {0.0cm 1.5cm 0.0cm 0.0cm}, clip,width =0.475\linewidth]{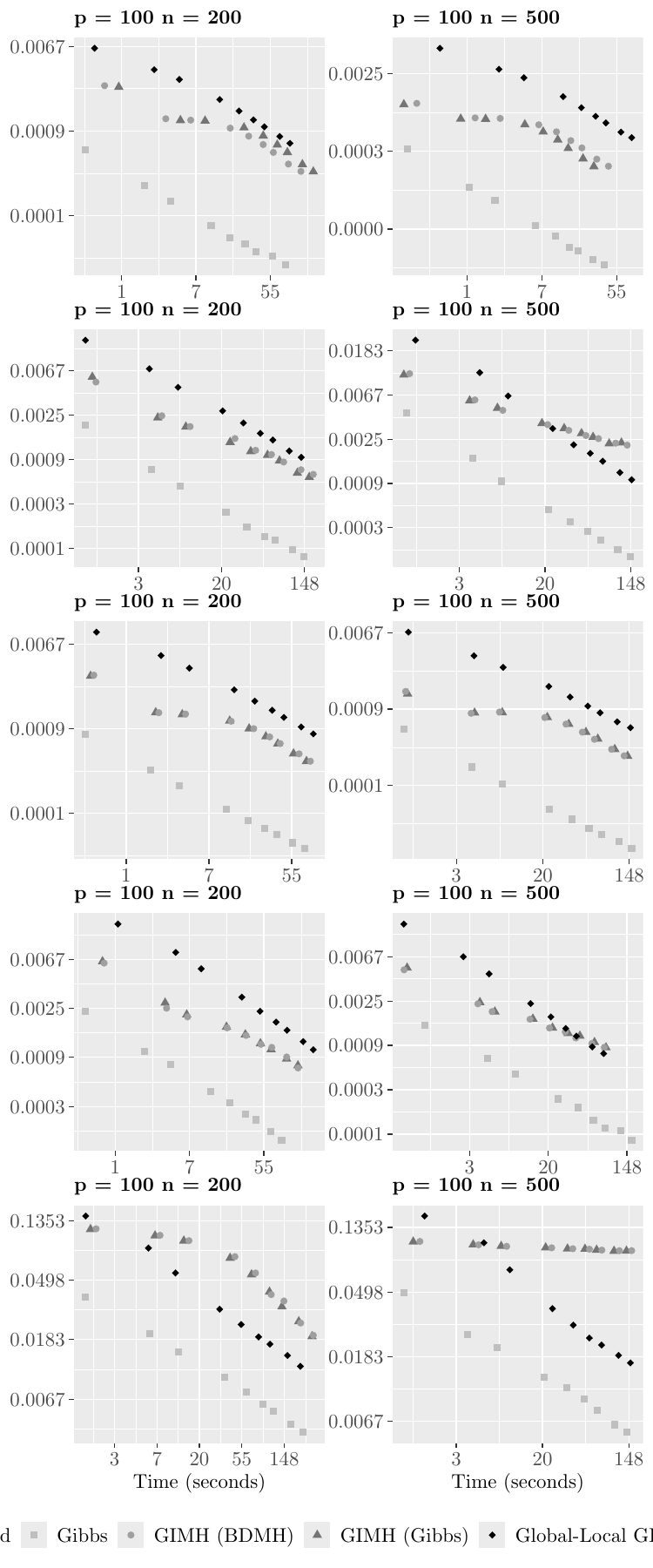}\\
\multicolumn{2}{c}{\includegraphics[trim= {0.0cm 0.4cm 0.0cm 29.2cm}, clip, width =0.95\linewidth]{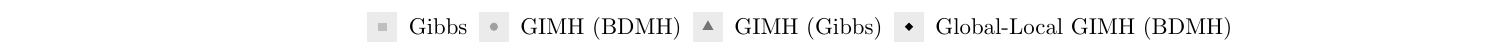}}\\
\end{tabular}
\caption{Difference between posterior mean estimates from two MCMC chains initialised at GLASSO-EBIC and diag(1) vs. clock time. Comparison between Gibbs, GIMH using Gibbs sampling in the global proposals, GIMH using BDMH in the global proposals and a mixture between local BDMH and GIMH moves under the binomial model prior for $p = 100$ and $n \in \{200, 500\}$.
}
\label{fig:GIMH_GibbsBDMH_p100}
\end{center}
\end{figure}

Figure \ref{fig:GIMH_GibbsBDMH_p100} compares Gibbs sampling with  GIMH ($\upsilon = 0.75$) using Gibbs sampling in the global proposals, GIMH ($\upsilon = 0.75$) using BDMH in the global proposals and  a mixture between 50\% local BDMH and 50\% GIMH proposals under the binomial model prior for $p = 100$. The efficiency of the GIMH algorithm is largely invariant to whether Gibbs or BDMH is used to construct the global proposals. GIMH is generally more efficient than the Global-Local mixture proposal and less efficient than Gibbs.

\color{black}

\subsection{Posterior Sampling Efficiency}

We provide plots analogous to Figure \ref{fig:mixing_p100} for $p = 50$, $p = 200$ and an additional simulation setting. 

\subsubsection{$p = 50$}{\label{Sec:mixingp50}}

\begin{figure}
\begin{center}
\begin{tabular}{cc}
Mean $|\hat{E}^{(1)}(\Omega_{jk} \mid \Y) - \hat{E}^{(2)}(\Omega_{jk} \mid \Y)|$  & Mean $|\hat{\pi}^{(1)}(\Omega_{jk} \neq 0 \mid \Y) - \hat{\pi}^{(2)}(\Omega_{jk} \neq 0 \mid \Y)|$ \\
\includegraphics[trim= {0.0cm 2.2cm 0.0cm 0.0cm}, clip,width =0.475\linewidth]{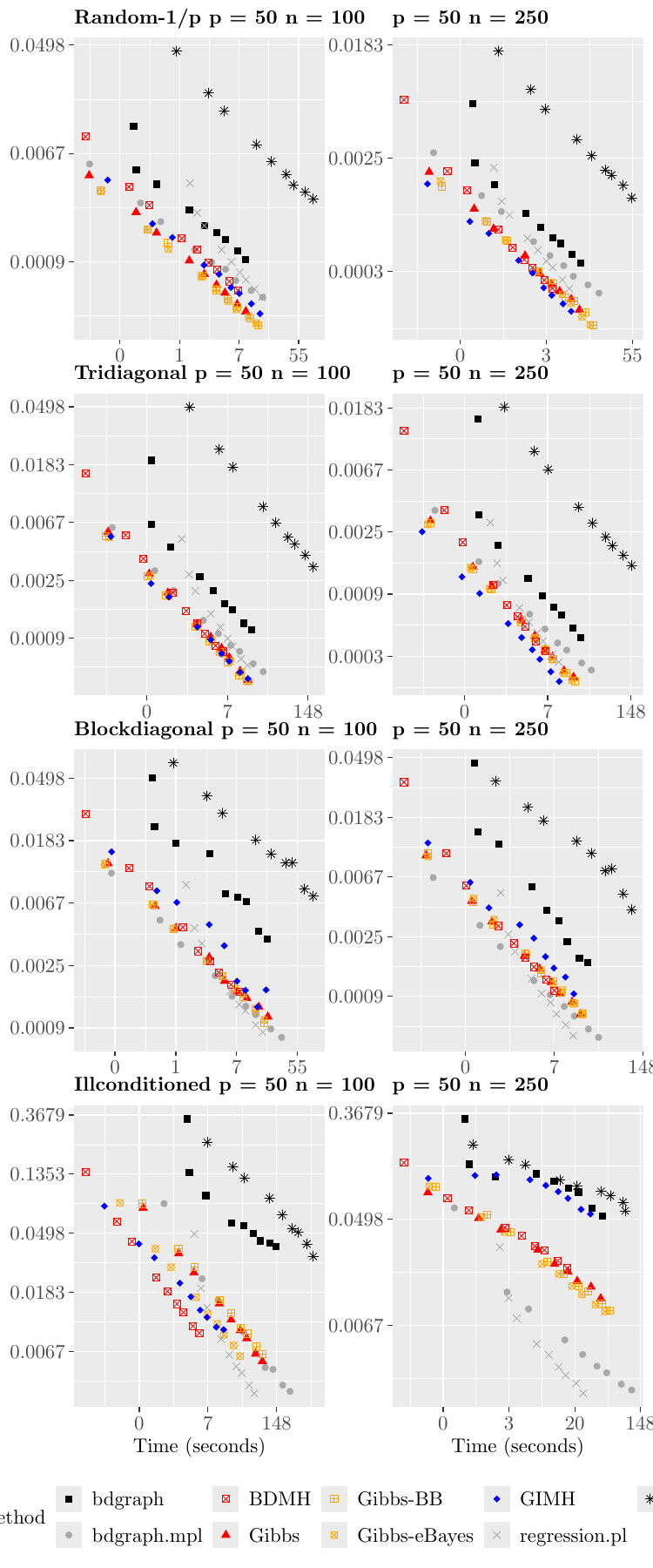} & \includegraphics[trim= {0.0cm 2.2cm 0.0cm 0.0cm}, clip,width =0.475\linewidth]{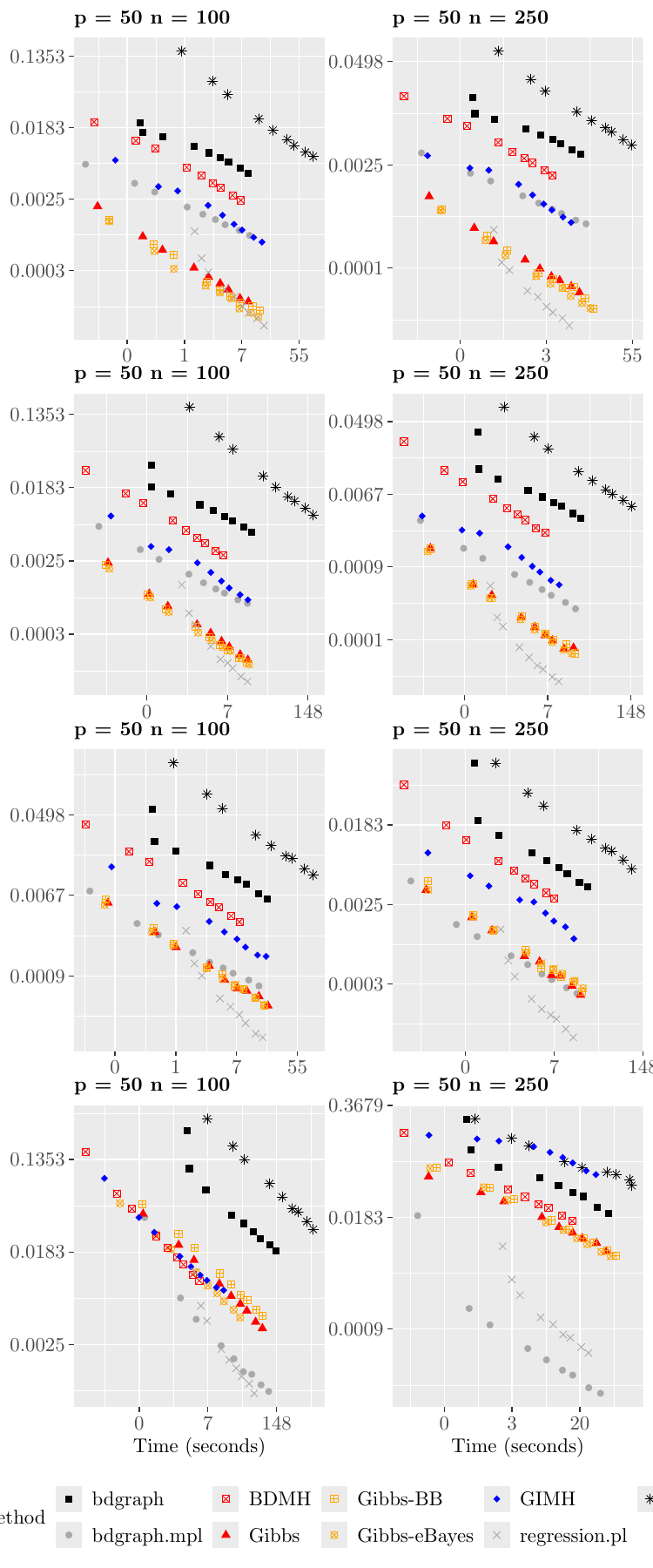}\\
\multicolumn{2}{c}{\includegraphics[trim= {0.0cm 0.4cm 0.0cm 29.2cm}, clip, width =0.95\linewidth]{figures/rev_figures/Final_Comparison_p50_diag-1.pdf}}\\
\end{tabular}
\caption{Difference in posterior mean and edge inclusion probability estimates between two chains initialised at GLASSO-EBIC and $I$ vs. clock time for $p = 50$, $n \in \{2p, 5p\}$. Black: exact Bayesian methods (ssgraph, bdgraph). Gray: pseudo-likelihood methods (bdgraph.mpl, regression.pl). Red: Gibbs with fixed $\theta$ and BDMH. Orange: Gibbs with $\theta\sim \text{Beta}(1, 1)$ and empirical Bayes. Blue: GIMH ($\upsilon = 0.75$).}
\label{fig:mixing_p50}
\end{center}
\end{figure}

The posterior sampling efficiency illustrated in Figure \ref{fig:mixing_p50} for $p = 50$ is largely consistent with the $p = 100$ case. Gibbs, BDMH and GIMH are more efficient than the exact competitors bdgraph and ssgraph, but less efficient than regression.pl in general. 
bdgraph.mpl was less efficient than Gibbs, BDMH and GIMH in the Random-1/p and Tridiagonal settings, but more efficient in the Blockdiagonal and Illconditioned settings. 


\subsubsection{$p = 200$}{\label{Sec:mixingp200}}

\begin{figure}
\begin{center}
\begin{tabular}{cc}
Mean $|\hat{E}^{(1)}(\Omega_{jk} \mid \Y) - \hat{E}^{(2)}(\Omega_{jk} \mid \Y)|$  & Mean $|\hat{\pi}^{(1)}(\Omega_{jk} \neq 0 \mid \Y) - \hat{\pi}^{(2)}(\Omega_{jk} \neq 0 \mid \Y)|$ \\
\includegraphics[trim= {0.0cm 1.75cm 0.0cm 0.0cm}, clip,width =0.475\linewidth]{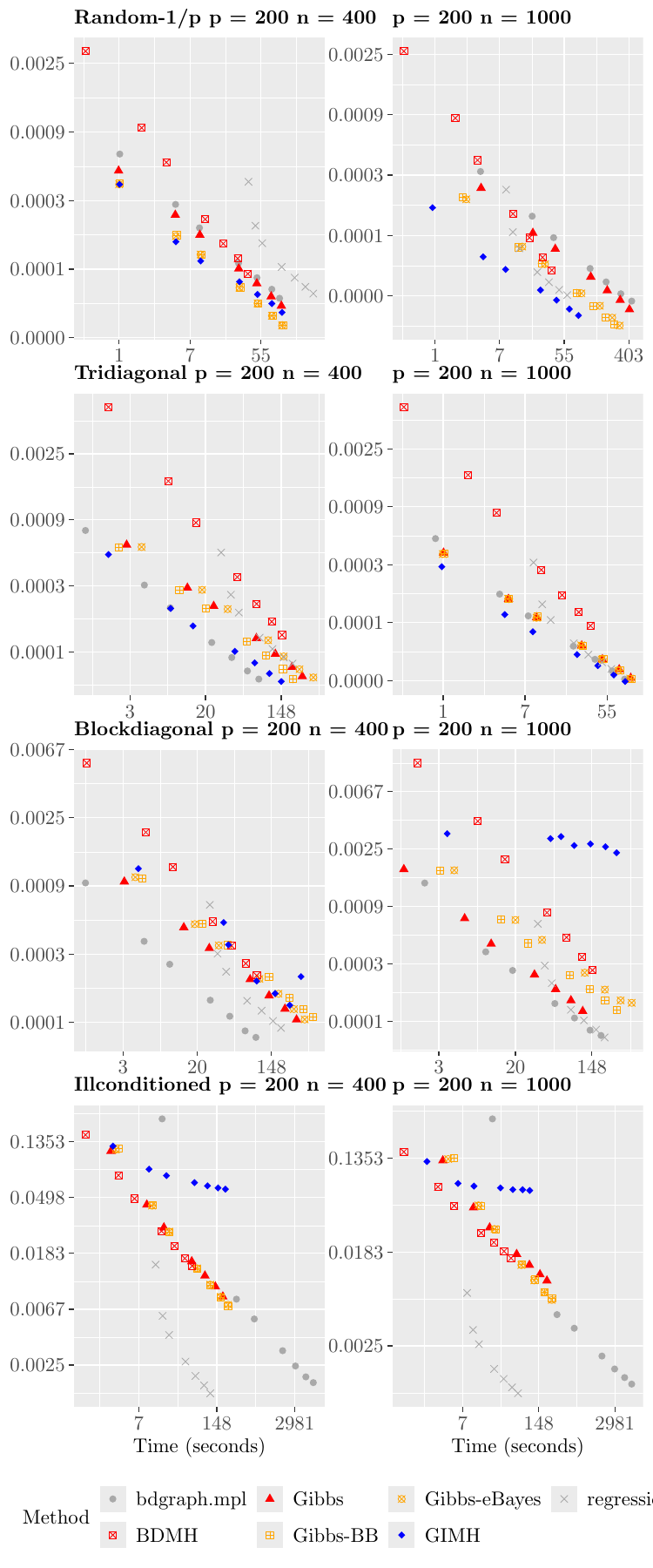} & \includegraphics[trim= {0.0cm 1.75cm 0.0cm 0.0cm}, clip,width =0.475\linewidth]{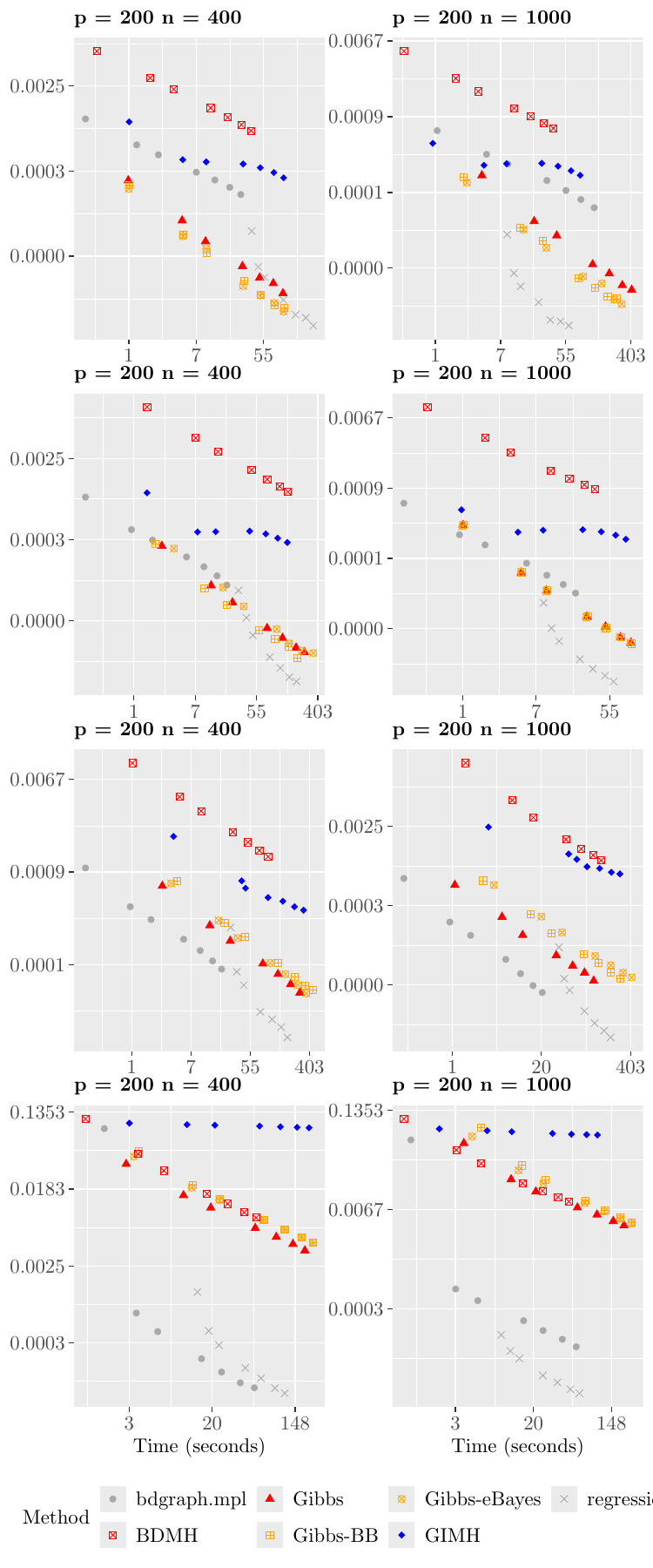}\\
\multicolumn{2}{c}{\includegraphics[trim= {0.0cm 0.4cm 0.0cm 29.2cm}, clip, width =0.95\linewidth]{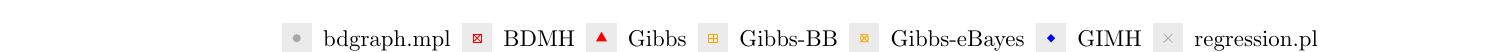}}\\
\end{tabular}
\caption{Difference in posterior mean and edge inclusion probability estimates between two chains initialised at GLASSO-EBIC and $I$ vs. clock time for $p = 200$, $n \in \{2p, 5p\}$. Gray: pseudo-likelihood methods (bdgraph.mpl, regression.pl). Red: Gibbs with fixed $\theta$ and BDMH. Orange: Gibbs with $\theta\sim \text{Beta}(1, 1)$ and empirical Bayes. Blue: GIMH ($\upsilon = 0.75$)}
\label{fig:mixing_p200}
\end{center}
\end{figure}

Figure \ref{fig:mixing_p200} compares the posterior sampling efficiency, analogously to Figure \ref{fig:mixing_p100}, of bdgraph.mpl, BDMH, Gibbs, GIMH and regression.pl for $p = 200$. bdgraph and ssgraph were not consider for this larger $p$ setting. Comparing to the $p = 50$ and $p = 100$ examples, similar behavior is observed with regression.pl continuing to be the most efficient. 

\subsubsection{Random-$2/p$}{\label{Sec:mixingpRandom2p}}

We consider an additional simulation setting
for the graph structures.
\begin{itemize}
    \item \textit{Random-$2/p$}: the presence of an edge between any two vertices is drawn uniformly at random with probability $2/p$, corresponding to $2(p-1)/p$ edges per node in expectation.
\end{itemize}
This corresponds to a less sparse version of the Random-$1/p$ setting considered in the main paper. Performance is similar to what was observed in Figures \ref{fig:mixing_p100}, \ref{fig:mixing_p50} and \ref{fig:mixing_p200}. 

\begin{figure}
\begin{center}
\begin{tabular}{cc}
Mean $|\hat{E}^{(1)}(\Omega_{jk} \mid \Y) - \hat{E}^{(2)}(\Omega_{jk} \mid \Y)|$  & Mean $|\hat{\pi}^{(1)}(\Omega_{jk} \neq 0 \mid \Y) - \hat{\pi}^{(2)}(\Omega_{jk} \neq 0 \mid \Y)|$ \\
\includegraphics[trim= {0.0cm 2.2cm 0.0cm 0.0cm}, clip,width =0.475\linewidth]{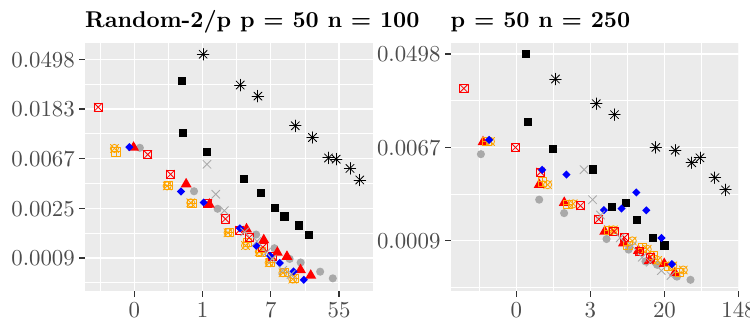} & \includegraphics[trim= {0.0cm 2.2cm 0.0cm 0.0cm}, clip,width =0.475\linewidth]{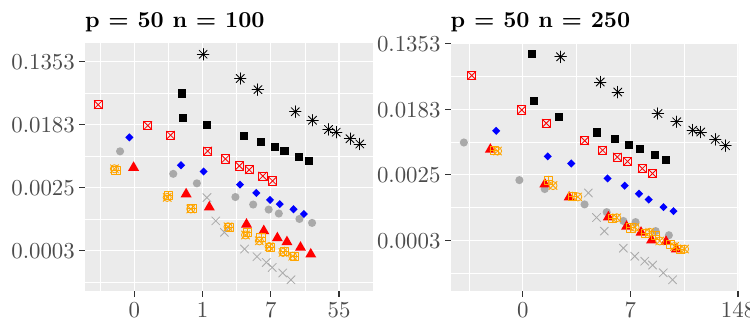}\\
\includegraphics[trim= {0.0cm 2.2cm 0.0cm 0.0cm}, clip,width =0.475\linewidth]{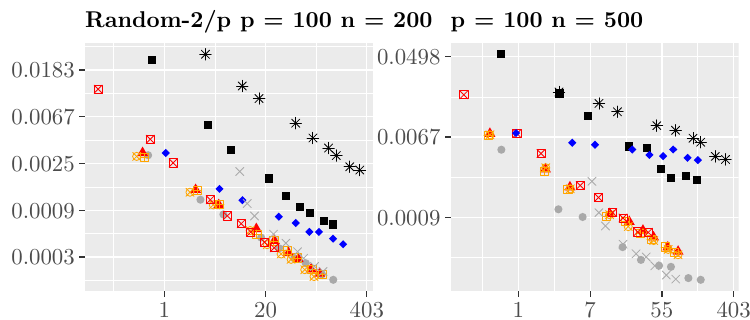} & \includegraphics[trim= {0.0cm 2.2cm 0.0cm 0.0cm}, clip,width =0.475\linewidth]{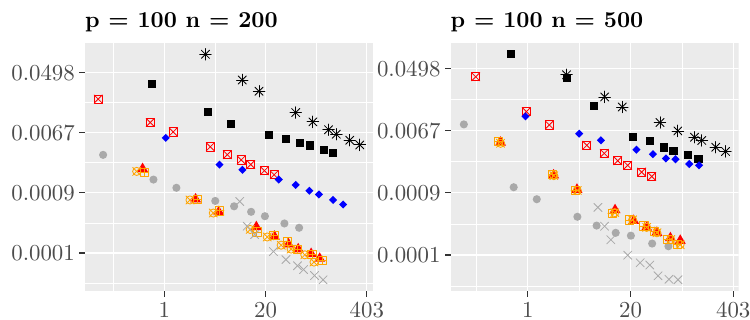}\\
\includegraphics[trim= {0.0cm 2.2cm 0.0cm 0.0cm}, clip,width =0.475\linewidth]{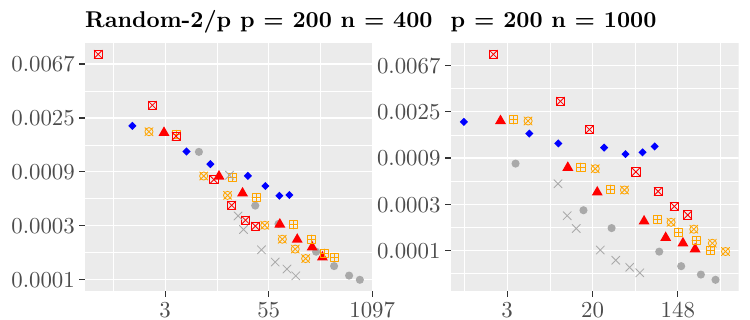} & \includegraphics[trim= {0.0cm 2.2cm 0.0cm 0.0cm}, clip,width =0.475\linewidth]{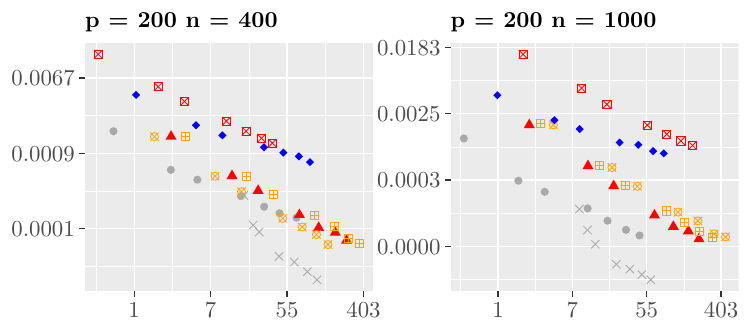}\\
\multicolumn{2}{c}{\includegraphics[trim= {0.0cm 0.4cm 0.0cm 29.2cm}, clip, width =0.95\linewidth]{figures/rev_figures/Final_Comparison_p50_diag-1.pdf}}\\
\end{tabular}
\caption{Difference in posterior mean and edge inclusion probability estimates between two chains initialised at GLASSO-EBIC and $I$ vs. clock time for $p = 50$, $n \in \{2p, 5p\}$. Black: exact Bayesian methods (ssgraph, bdgraph). Gray: pseudo-likelihood methods (bdgraph.mpl, regression.pl). Red: Gibbs with fixed $\theta$ and BDMH. Orange: Gibbs with $\theta\sim \text{Beta}(1, 1)$ and empirical Bayes. Blue: GIMH ($\upsilon = 0.75$)}
\label{fig:mixing_Random2/p}
\end{center}
\end{figure}

\subsection{Posterior Inference }

We provide plots analogous to Figure \ref{fig:Inference_p100} for $p = 50$, additional performance metrics for $p = 100$, and an additional simulation setting.





\subsubsection{$p = 50$}

\begin{figure}[hbt!]
\begin{center}
\includegraphics[width =0.95\linewidth]{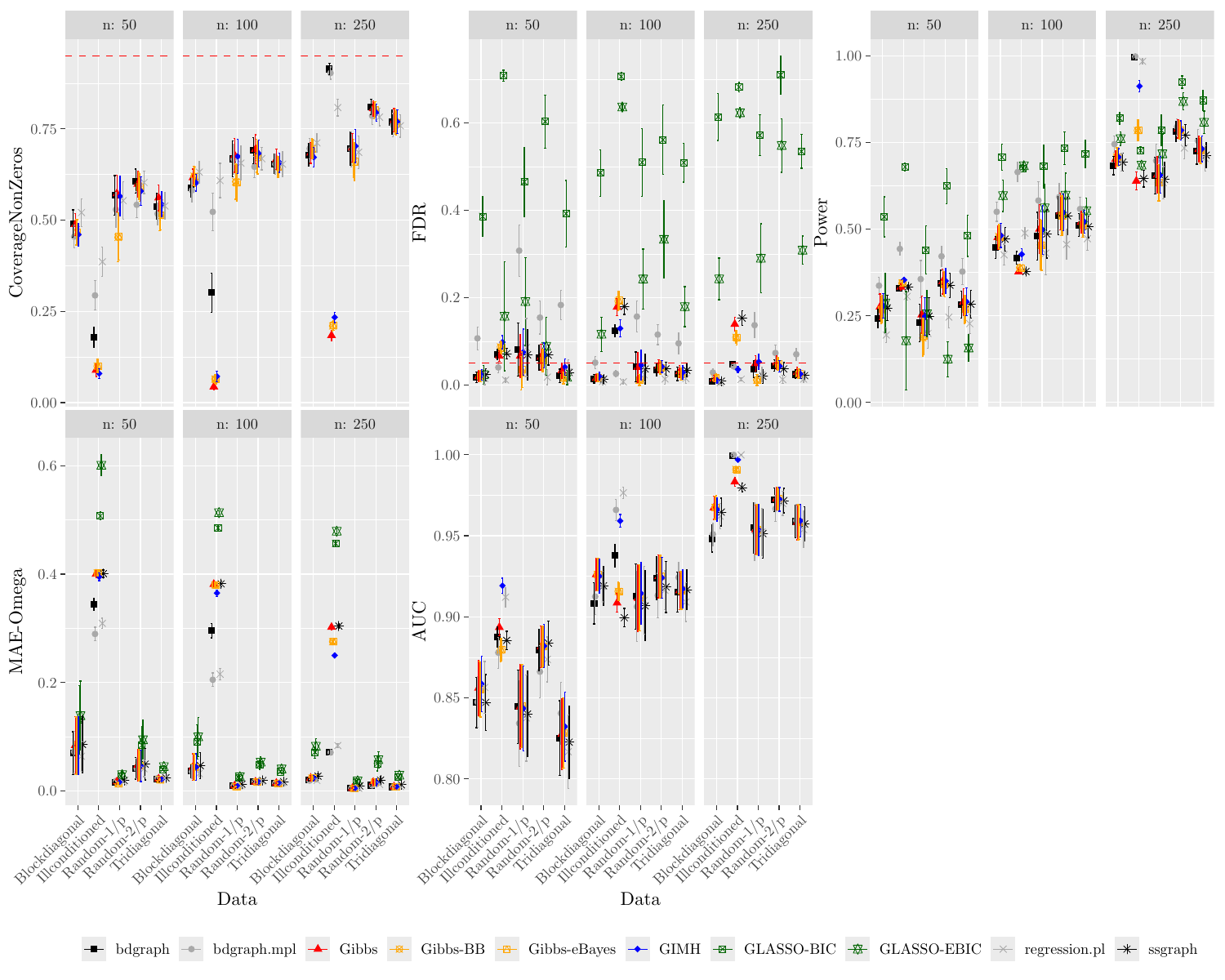}
\caption{Quality of inference comparison between GLASSO-BIC, GLASSO-EBIC, ssgraph, bdgraph, bdgraph.mpl, regression.pl, Gibbs (with fixed $\theta$, $\theta\sim \text{Beta}(1, 1)$ and empirical Bayes) and GIMH for  $p = 50$ and $n \in \{p, 2p, 5p\}$. Power, CoverageNonZeros (defined as the average coverage of all parameters with non-zero generating value) and AUC should be maximised, MAE-Omega and FDR should be minimised. 
Intervals are two standard errors. Red dotted lines indicate nominal values.
}
\label{fig:Inference_p50}
\end{center}
\end{figure}

Figure \ref{fig:Inference_p50} compares the Gibbs and GIMH algorithms with GLASSO-BIC, GLASSO-EBIC, ssgraph, bdgraph, bdgraph.mpl, and regression.pl,  for  $p = 50$ and $n \in \{p, 2p, 5p\}$ across five simulation settings. Besides the ill-conditioned example, the inference of Gibbs and GIMH is reassuringly similar with the other exact Bayesian methods, bdgraph and ssgraph. Further, there is evidence that bdgraph.mpl has an elevated FDR relative to the exact Bayesian methods and that regression.pl has reduced power. Both of these effects are most pronounced at small sample sizes.     The Beta-Binomial and empirical Bayes priors result in very similar inferences. Relative to Gibbs they result in slightly lower coverage and power, but also slightly lower FDR. \color{black}

\subsubsection{$p = 100$}

\begin{figure}[hbt!]
\begin{center}
\includegraphics[trim= {0.0cm 1.5cm 0.0cm 0.0cm}, clip, width =0.95\linewidth]{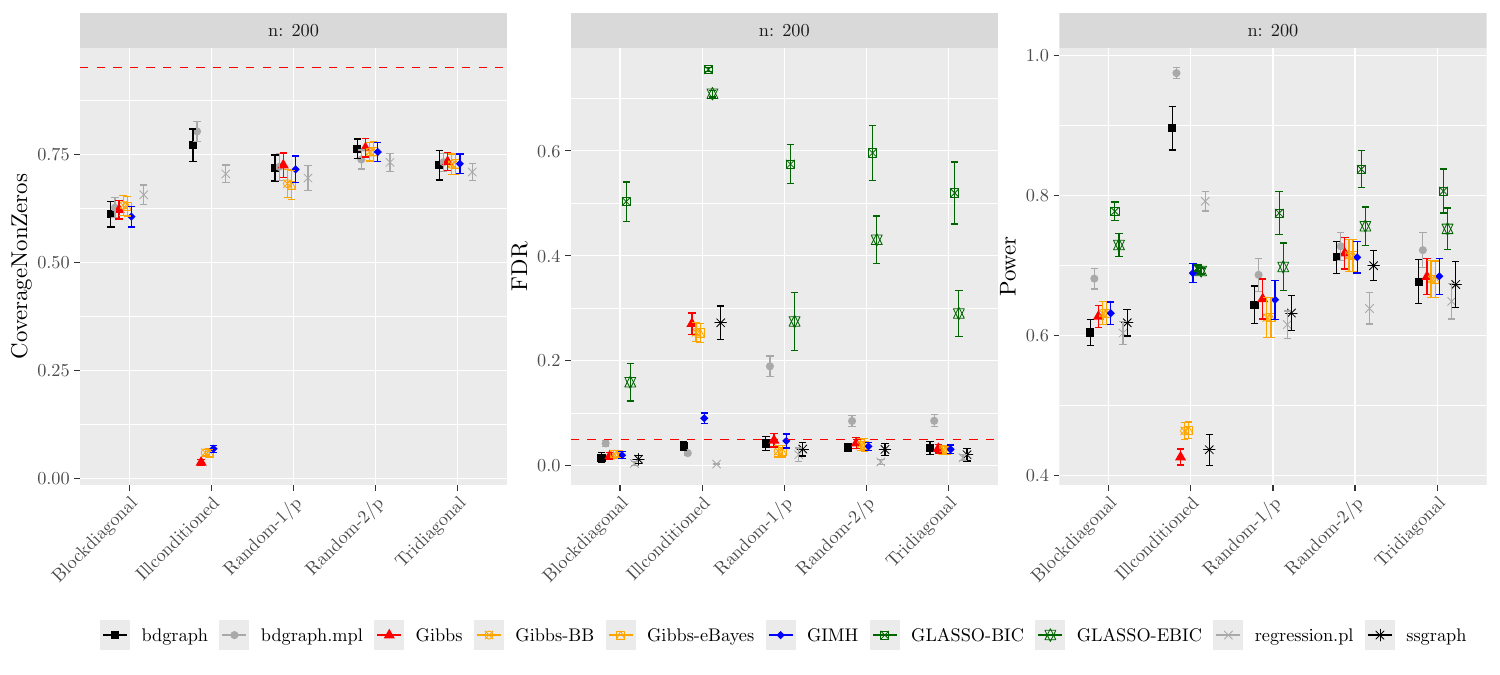}\\
\includegraphics[trim= {0.0cm 1.5cm 0.0cm 0.0cm}, clip,width =0.95\linewidth]{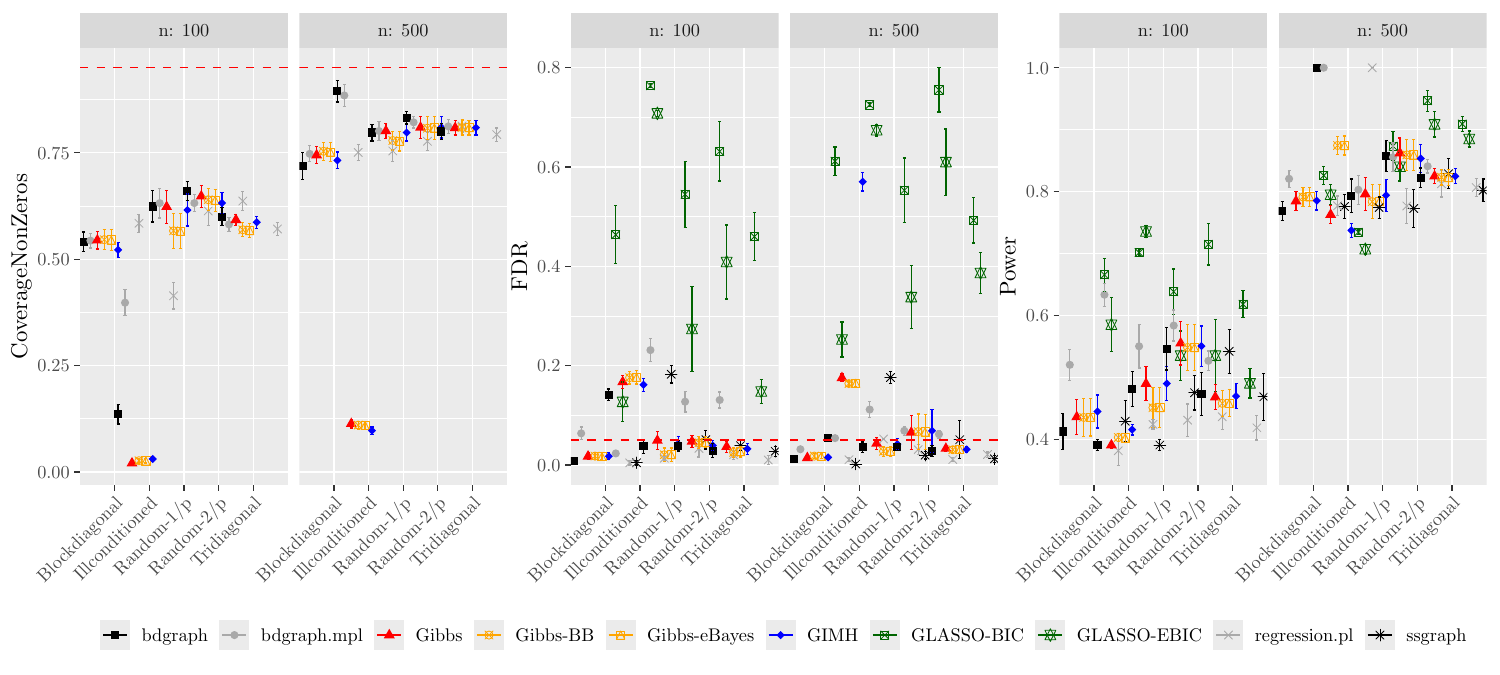}\\
\includegraphics[width =0.95\linewidth]{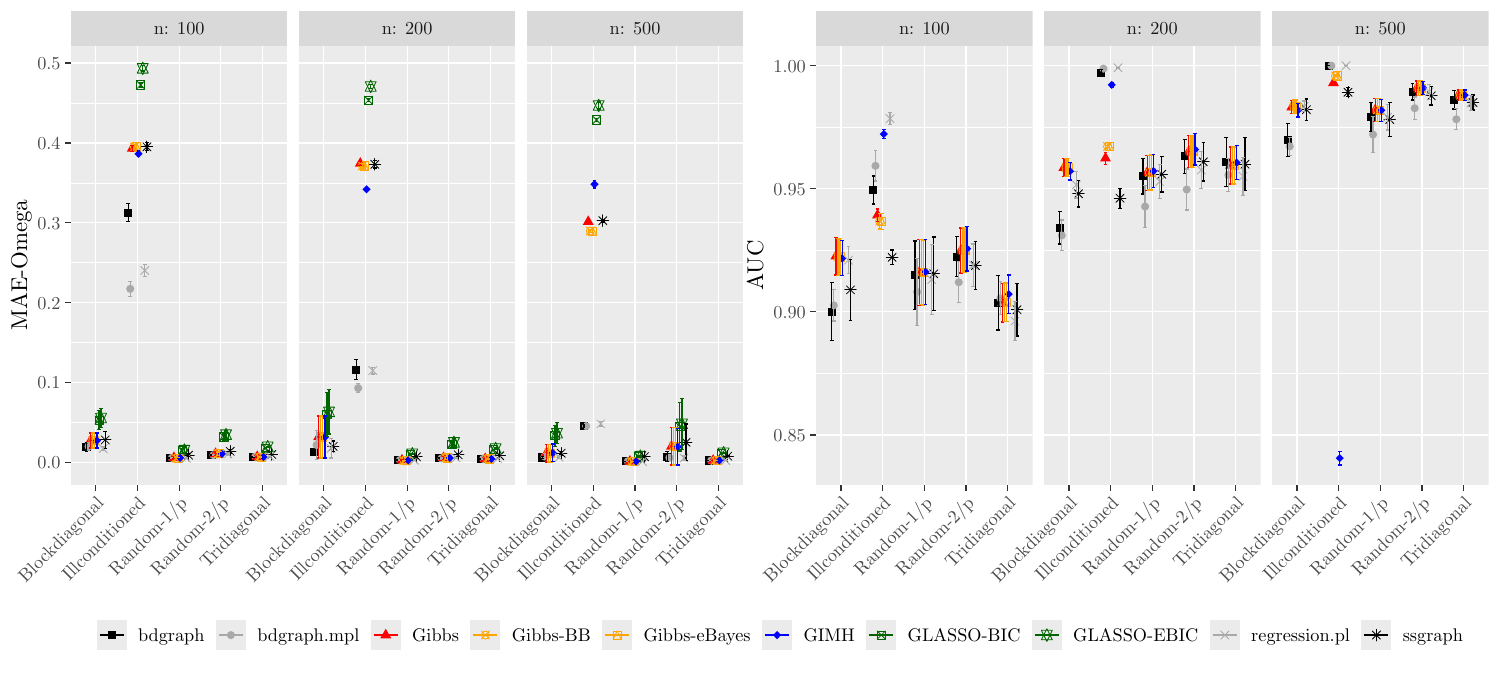}
\caption{
Quality of inference compared between GLASSO-BIC, GLASSO-EBIC, ssgraph, bdgraph, bdgraph.mpl, regression.pl, Gibbs (with fixed $\theta$, $\theta\sim \text{Beta}(1, 1)$ and empirical Bayes) and GIMH ($w = 0.75$)  for  $p = 100$ and $n \in \{p, 2p, 5p\}$. Power and CoverageNonZeros (defined as the average coverage of all parameters with non-zero generating value) and AUC should be maximised, MAE-Omega and FDR should be minimised. Intervals are 2 standard errors. Red dotted lines indicate nominal values.
}
\label{fig:Inference_p100_additional}
\end{center}
\end{figure}

Figure \ref{fig:Inference_p100_additional} supplements Figure \ref{fig:Inference_p100} from the main paper, providing simulation results for an additional sample size $n$ (top), additional simulation setting \textit{Random-$2/p$} (middle), and two additional evaluation metrics, the ROC-AUC of the estimates of $\hat{\pi}(\Omega_{jk} \neq 0 \mid \Y)$ and the average mean absolute error (MAE) of $\hat{\Omega}_{jk}$ of each method (bottom). 

\section{Sensitivity Analysis}{\label{sec:sensitivity}}

In this section we investigate the sensitivity of the posterior sampling efficiency and inference to departures from our default prior hyperparameters

\subsection{Sensitivity to $g_1$}

\begin{figure}
\begin{center}
\begin{tabular}{cc}
Mean $|\hat{E}^{(1)}(\Omega_{jk} \mid \Y) - \hat{E}^{(2)}(\Omega_{jk} \mid \Y)|$  & Mean $|\hat{\pi}^{(1)}(\Omega_{jk} \neq 0 \mid \Y) - \hat{\pi}^{(2)}(\Omega_{jk} \neq 0 \mid \Y)|$ \\
\includegraphics[trim= {0.0cm 0cm 0.0cm 0.0cm}, clip,width =0.475\linewidth]{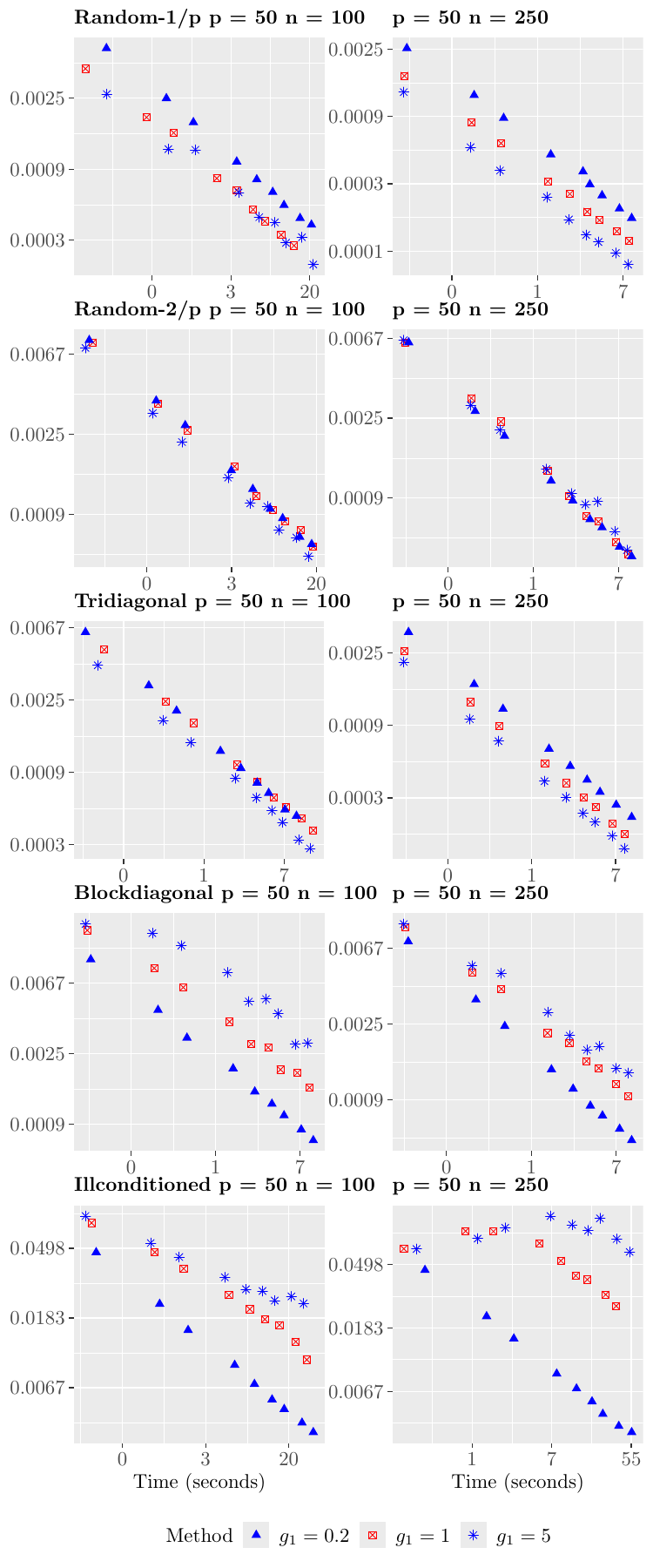} & \includegraphics[trim= {0.0cm 0cm 0.0cm 0.0cm}, clip,width =0.475\linewidth]{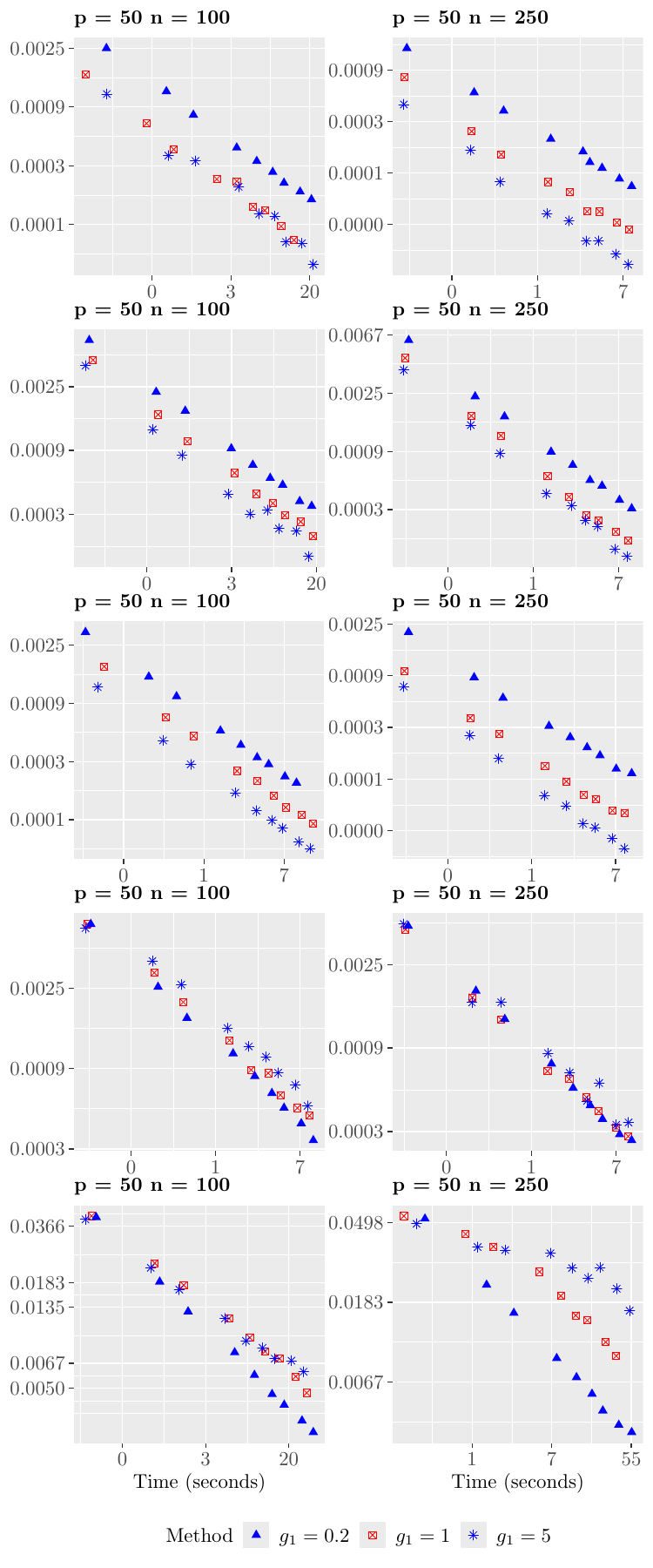}\\
\end{tabular}
\caption{Difference in posterior mean and edge inclusion probability estimates between two chains initialised at GLASSO-EBIC and diag(1) vs. clock time for $p = 50$, $n \in \{2p, 5p\}$. Gibbs sampler with $\theta = 2/(p-1)$ fixed. Red: recommended $g_1 = 1$. Blue: 5 times bigger and 5 times smaller than recommended i.e. $g_1 = 5$ and $g_1 = 0.2$}
\label{fig:mixing_g_p50}
\end{center}
\end{figure}

\begin{figure}[hbt!]
\begin{center}
\includegraphics[width =0.95\linewidth]{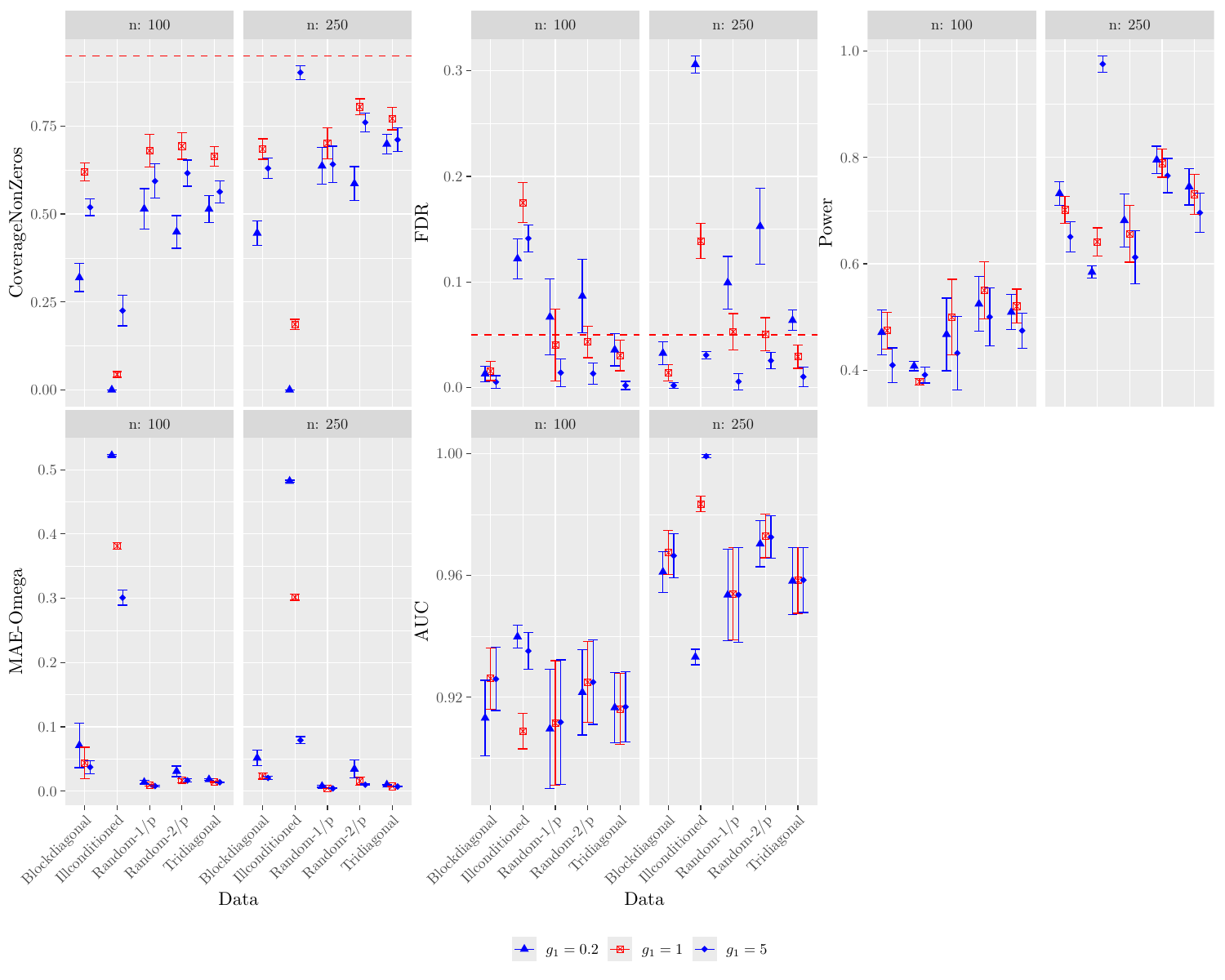}
\caption{
Simulations. Quality of inference for Gibbs sampler with $\theta = 2/(p-1)$ fixed, Red: recommended $g_1 = 1$. Blue: 5 times bigger and 5 times smaller than recommended i.e. $g_1 = 5$ and $g_1 = 0.2$, for  $p = 50$ and $n \in \{2p, 5p\}$. Higher Power, CoverageNonZeros (average coverage of all parameters with non-zero generating value), and AUC and smaller FDR and MAE-Omega are better. Intervals are 2 standard errors. Red dotted lines indicate the target coverage=0.95 and FDR=0.05
}
\label{fig:Inference_g_p50}
\end{center}
\end{figure}

We investigate the sensitivity of the sampling efficiency and quality of posterior inference of our Gibbs sampler with fixed $\theta$ to changes in $g_1$ the slab's scale parameter. Specifically we compare the default $g_1 = 1$ with larger $g_1 = 5$ and smaller $g_1 = 0.2$ values keeping $\theta = 2/(p-1)$ fixed. Figure \ref{fig:mixing_g_p50} demonstrates the posterior sampling efficiency for $p = 50$ and Figure \ref{fig:Inference_g_p50} investigates the quality of inference.

For the Random-1/p and Tridiagonal case, the MCMC efficiency seems to be slightly improved, taking $g = 5$ rather than $g = 1$, and for the Blockdiagonal and Illconditioned, the opposite occurs, where improved mixing is achieved by setting $g = 0.2$. However, in general, the performance seems stable. For inference, the Coverage Non-Zeros is best for our default $g = 1$. $g = 0.2$ sometimes achieves higher power than $g = 1$ but is also associated with a higher false discovery rate and higher MAE. $g = 5$ has lower false discovery rate than $g = 1$, but lower power.

Overall, mixing is not drastically affected by the value of $g_1$, with our default $g = 1$ doing reasonably across the board. Inference is affected by $g_1$, with our default achieving good coverage and a good balance between power and FDR.

\subsection{Sensitivity to $\lambda$}

\begin{figure}
\begin{center}
\begin{tabular}{cc}
Mean $|\hat{E}^{(1)}(\Omega_{jk} \mid \Y) - \hat{E}^{(2)}(\Omega_{jk} \mid \Y)|$  & Mean $|\hat{\pi}^{(1)}(\Omega_{jk} \neq 0 \mid \Y) - \hat{\pi}^{(2)}(\Omega_{jk} \neq 0 \mid \Y)|$ \\
\includegraphics[trim= {0.0cm 0cm 0.0cm 0.0cm}, clip,width =0.475\linewidth]{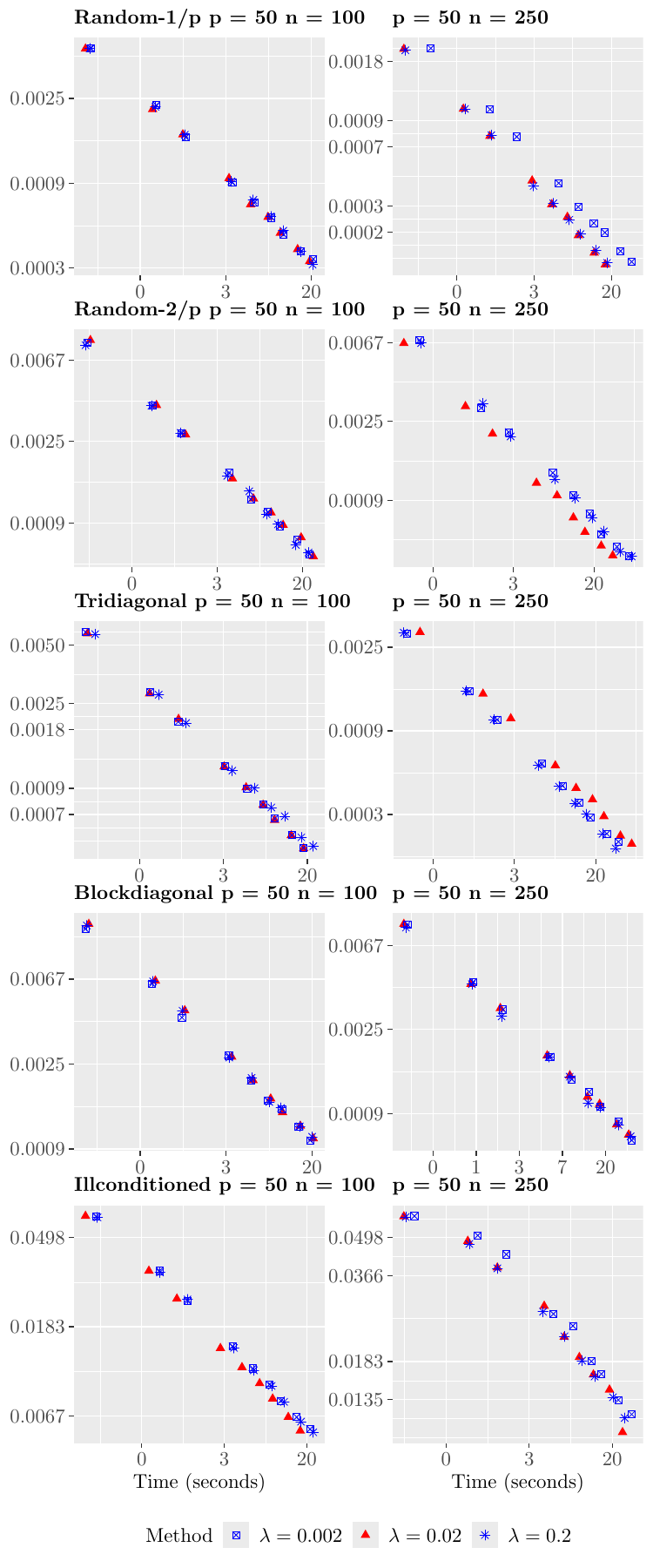} & \includegraphics[trim= {0.0cm 0cm 0.0cm 0.0cm}, clip,width =0.475\linewidth]{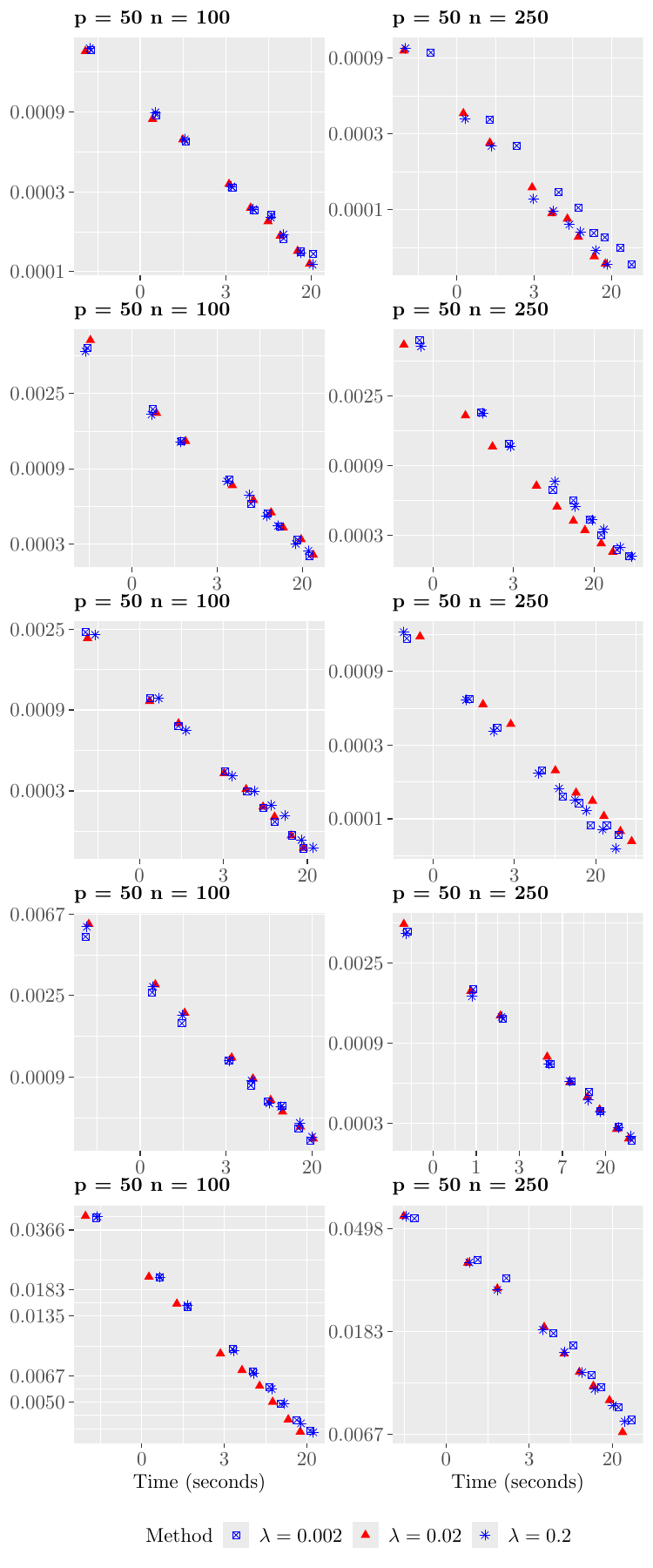}\\
\end{tabular}
\caption{Difference in posterior mean and edge inclusion probability estimates between two chains initialised at GLASSO-EBIC and diag(1) vs. clock time for $p = 50$, $n \in \{2p, 5p\}$. Gibbs sampler with $\theta = 2/(p-1)$ fixed. Red: recommended $\lambda = 0.02$. Blue: 10 times bigger and 10 times smaller than recommended i.e. $\lambda = 0.2$ and $\lambda = 0.002$}
\label{fig:mixing_lambda_p50}
\end{center}
\end{figure}

\begin{figure}[hbt!]
\begin{center}
\includegraphics[width =0.95\linewidth]{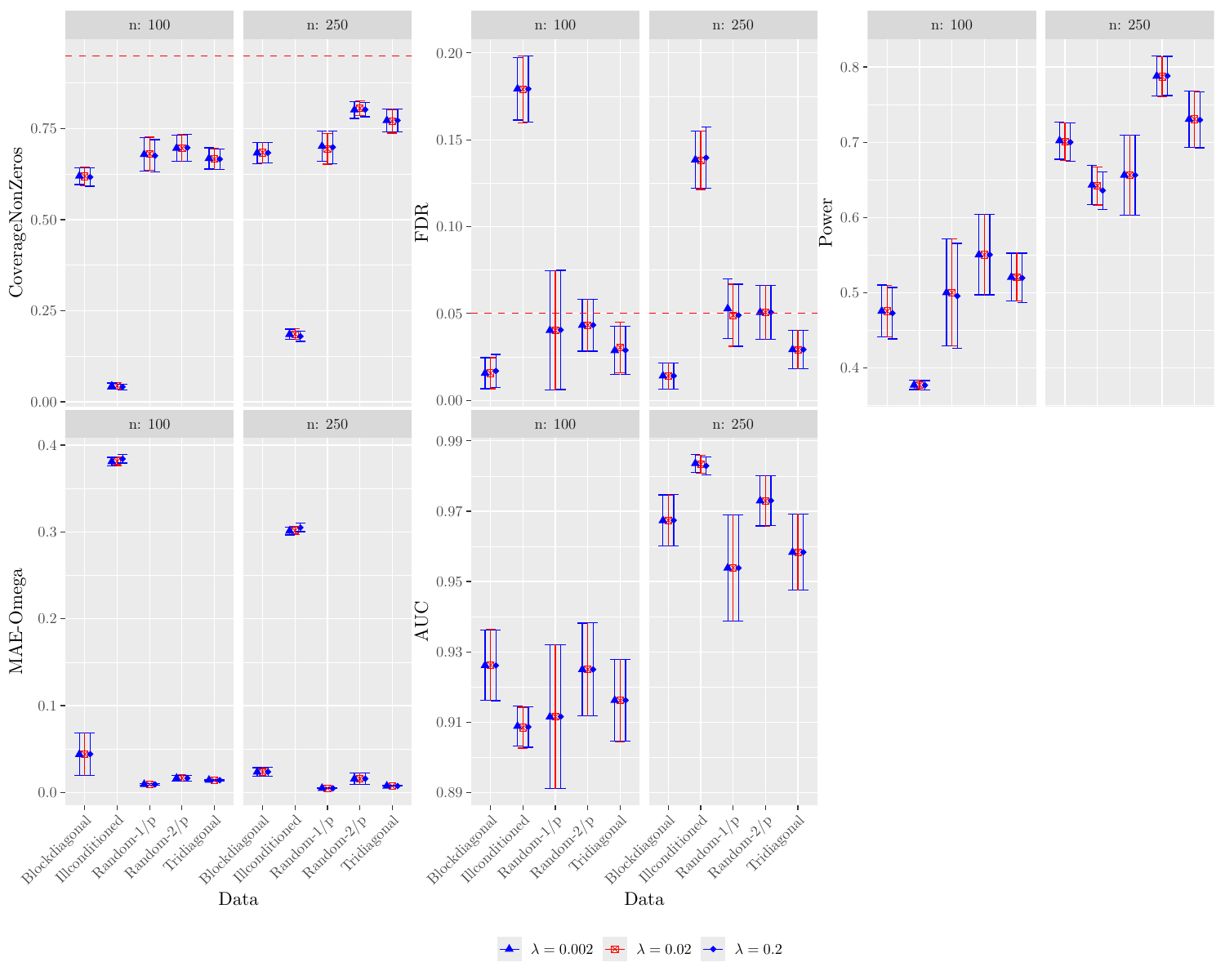}
\caption{
Simulations. Quality of inference for Gibbs sampler with $\theta = 2/(p-1)$ fixed, Red: recommended $\lambda = 0.02$. Blue: 10 times bigger and 10 times smaller than recommended i.e. $\lambda = 0.2$ and $\lambda = 0.002$, for  $p = 50$ and $n \in \{2p, 5p\}$. Higher Power, CoverageNonZeros (average coverage of all parameters with non-zero generating value), and AUC and smaller FDR and MAE-Omega are better. Intervals are 2 standard errors. Red dotted lines indicate the target coverage=0.95 and FDR=0.05
}
\label{fig:Inference_lambda_p50}
\end{center}
\end{figure}

We investigate the sensitivity of the sampling efficiency and quality of posterior inference of our Gibbs sampler with fixed $\theta$ to changes in $\lambda$, the hyperparmeter of the exponential prior for the diagonal elements of $\Omega$. Specifically we compare the default $\lambda = 0.02$ with larger $\lambda = 0.2$ and smaller $\lambda = 0.002$ values keeping $\theta = 2/(p-1)$ and $g_1 = 1$ fixed. Figure \ref{fig:mixing_lambda_p50} demonstrates the posterior sampling efficiency for $p = 50$ and Figure \ref{fig:Inference_lambda_p50} investigates the quality of inference. Both are very stable across this range of values.

\color{black}

\section{COVID-19 Data}

\subsection{Sampling Diagnostics}{\label{Sec:COVID_diagnostics}}

For the COVID-19 data analysis presented in Section \ref{sec:COVID} we investigate the convergence of the Gibbs, GIMH, bdgraph.mpl and regression.pl sampling algorithms. We follow \cite{mohammadi2015bayesian} and investigate the traceplots of the model size (number of non-zero off-diagonal precision matrix elements)  at each MCMC iteration and running estimates of the posterior probability of inclusion for each edge after each MCMC iteration. For the later we consider only 100 edges to avoid cluttering the plots. 

\begin{figure}[hbt!]
\begin{center}
\includegraphics[trim= {0.0cm 0.0cm 0.0cm 0.0cm}, clip, width =0.49\linewidth]{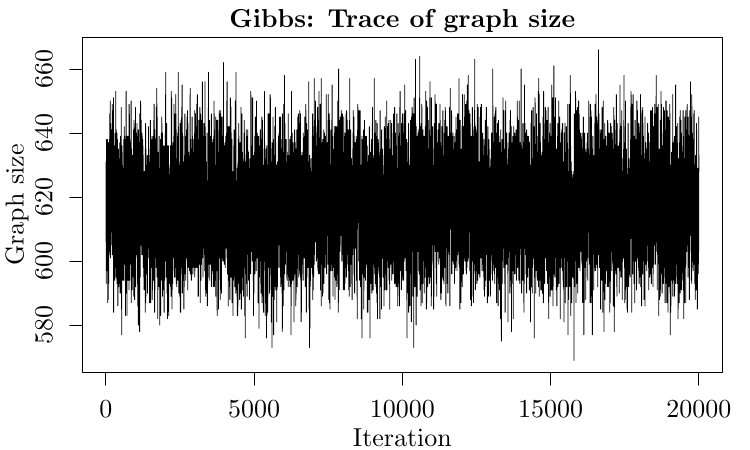}
\includegraphics[trim= {0.0cm 0.0cm 0.0cm 0.0cm}, clip, width =0.49\linewidth]{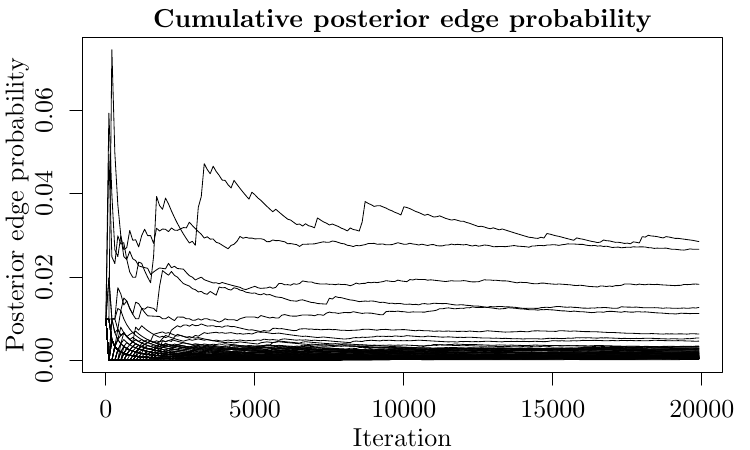}\\
\includegraphics[trim= {0.0cm 0.0cm 0.0cm 0.0cm}, clip, width =0.49\linewidth]{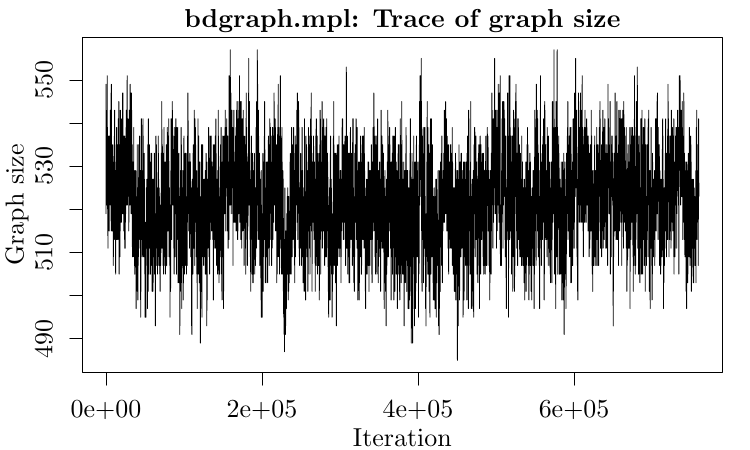}
\includegraphics[trim= {0.0cm 0.0cm 0.0cm 0.0cm}, clip, width =0.49\linewidth]{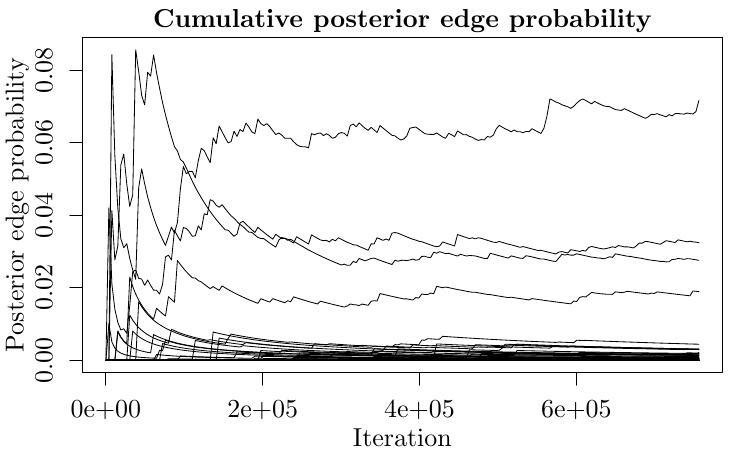}\\
\includegraphics[trim= {0.0cm 0.0cm 0.0cm 0.0cm}, clip, width =0.49\linewidth]{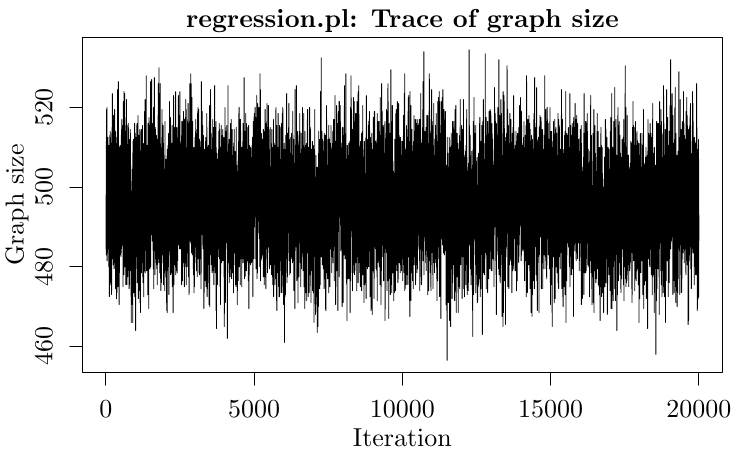}
\includegraphics[trim= {0.0cm 0.0cm 0.0cm 0.0cm}, clip, width =0.49\linewidth]{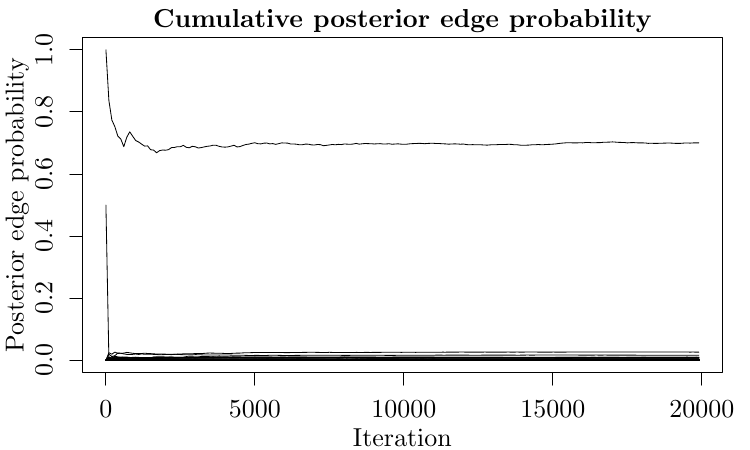}\\
\caption{Trace plot for the graph size (left) and cumulative posterior inclusion probability for Gibbs, GIMH, bdgraph.mpl, and regression.pl for the COVID-19 data.
}
\label{fig:COVID_MCMC_diagnostics}
\end{center}
\end{figure}

Gibbs was run for 5,000 warmup iterations followed by a further 20,000 iterations, where each iteration is defined as updating every column of $\Omega$ and every element of each column. Both plots in Figure \ref{fig:COVID_MCMC_diagnostics} indicate that this was sufficient for convergence. 
bdgraph.mpl was run for $2\lceil\sqrt{p}\rceil$ times more warmup and sampling iterations than Gibbs, with each iteration only changing one element of $\Omega$. This number of iterations was chosen such that the run time of bdgraph.mpl was similar to that of Gibbs (see Table \ref{Tab:Covid}). Such a comparison favours bdgraph.mpl as the time taken does not include the time required to sample parameter $\Omega$ for bdgraph.mpl, which Figure \ref{fig:exact_vs_pseudo} showed to be considerable for large $p$. The trace plot indicates some stickiness in the model dimension, and the running edge poserior inclusion probabilities indicate that more iterations are required for convergence. Lastly, regression.pl was run for the same number of iterations as Gibbs with each iteration defined as it was in Gibbs. Figure \ref{fig:COVID_MCMC_diagnostics} shows that this appears to have been a sufficient number of iterations for convergence.

\subsection{US Maps}

\begin{figure}[!ht]
\centering
\begin{subfigure}[b]{\linewidth}
    \includegraphics[trim= {0.0cm 0.5cm 0.0cm 0.5cm}, clip, width = 1\linewidth]{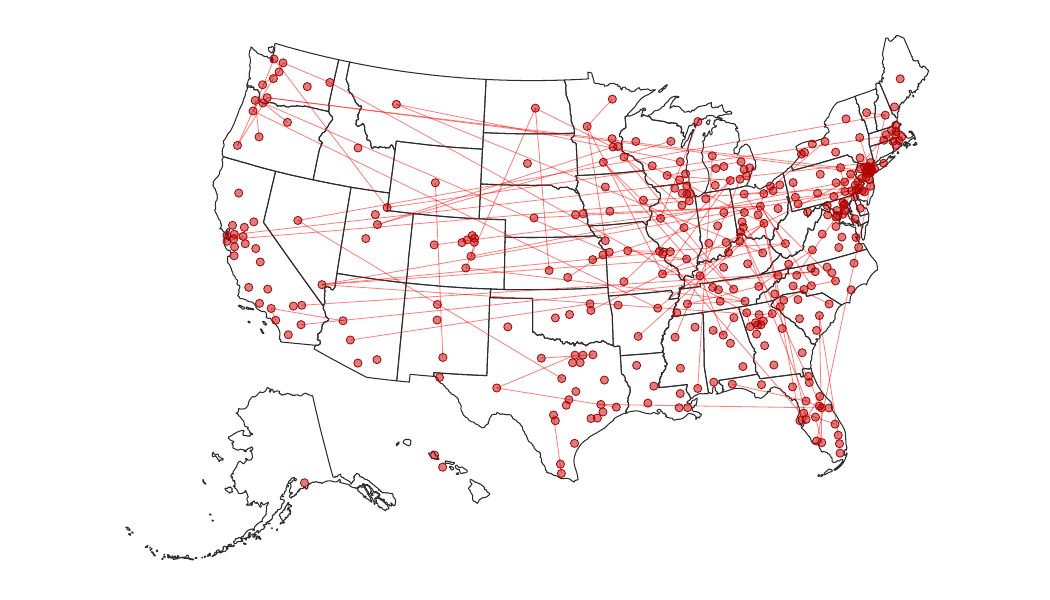}
    \caption{Edges identified by bdgraph.mpl but not by Gibbs thresholding the posterior inclusion probability at 0.95.
    }
\end{subfigure}
\begin{subfigure}[b]{\linewidth}
    \includegraphics[trim= {0.0cm 0.5cm 0.0cm 0.5cm}, clip, width = 1\linewidth]{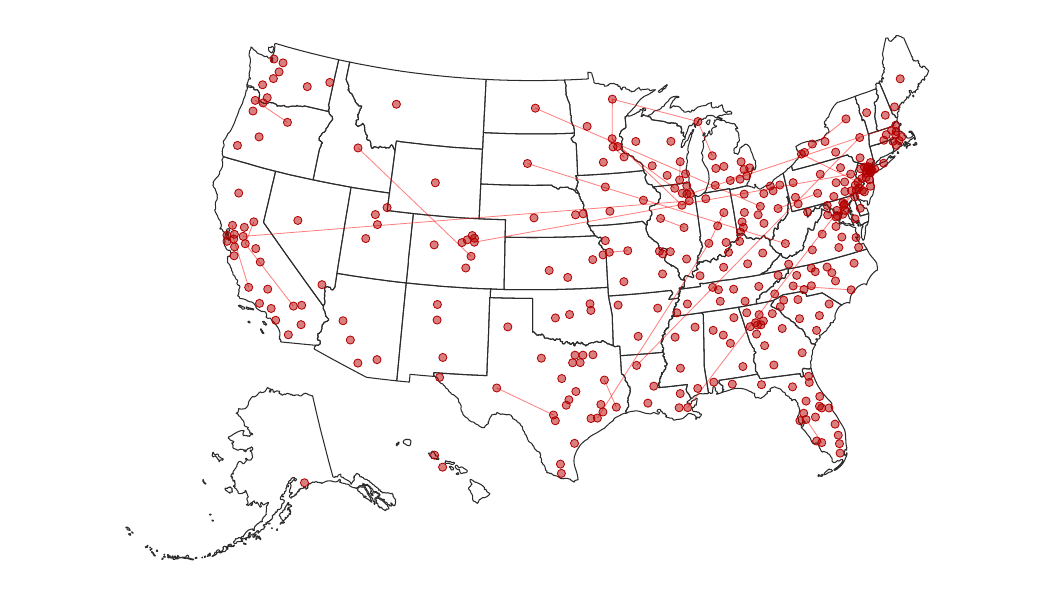}
    \caption{Edges identified by Gibbs but not by bdgraph.mpl thresholding the posterior inclusion probability at 0.95. }
\end{subfigure}
\caption{Edges identified by bdgraph.mpl. 
}
\label{fig:edges_bdgraph.mpl}
\end{figure}

\begin{figure}[!ht]
\centering
\begin{subfigure}[b]{\linewidth}
    \includegraphics[trim= {0.0cm 0.5cm 0.0cm 0.5cm}, clip, width = 1\linewidth]{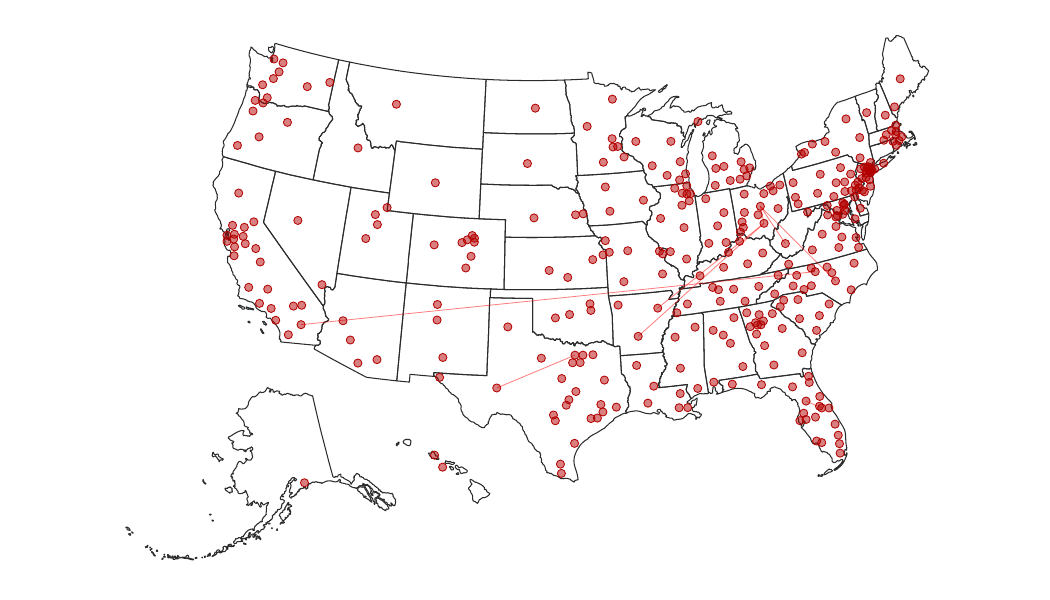}
    \caption{Edges identified by regression.pl but not by Gibbs thresholding the posterior inclusion probability at 0.95.
    }
\end{subfigure}
\begin{subfigure}[b]{\linewidth}
    \includegraphics[trim= {0.0cm 0.5cm 0.0cm 0.5cm}, clip, width = 1\linewidth]{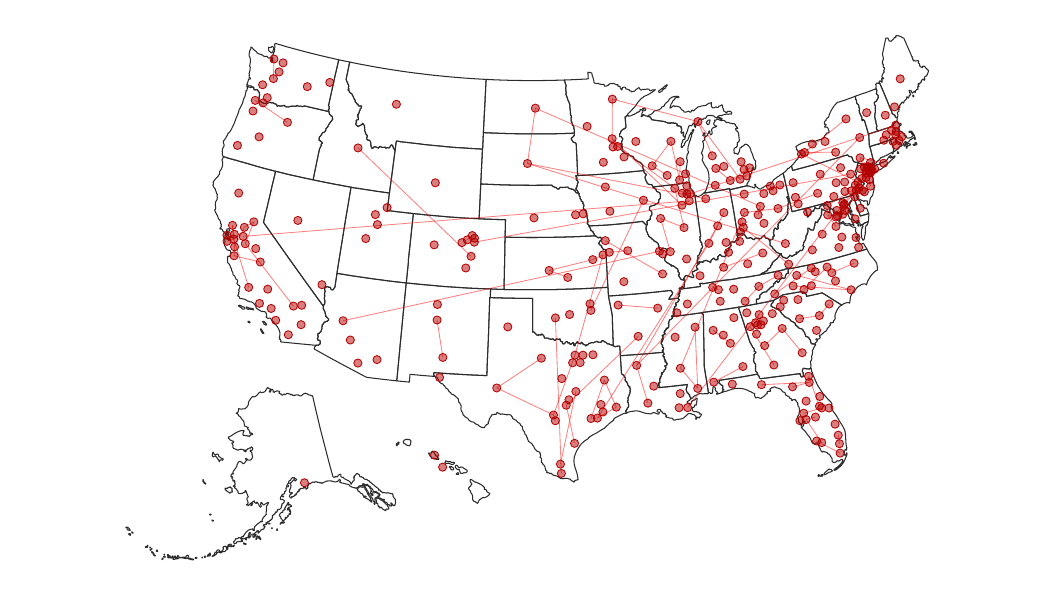}
    \caption{Edges identified by Gibbs but not by regression.pl thresholding the posterior inclusion probability at 0.95.   
    }
\end{subfigure}
\caption{Edges identified by regression.pl. 
}
\label{fig:edges_regression.pl}
\end{figure}

In addition to Figure \ref{fig:edges_SerialGibbs}, we investigate which edges were estimated differently by bdgraph.mpl and regression.pl when compared to the Gibbs algorithm. Figure \ref{fig:edges_bdgraph.mpl} shows that the edges estimated by bdgraph.mpl largely contain the edges estimated by Gibbs, with a few exceptions. Conversely, Figure \ref{fig:edges_regression.pl} shows that the edges estimated by regression.pl are largely contained within those estimated by Gibbs, but with a few exceptions.   

\FloatBarrier

\setlength{\bibsep}{0pt}

\bibliography{bib}

\end{document}